\documentclass[12pt]{cit_thesis}

\newif\ifpdf
  \ifx\pdfoutput\undefined
    \pdffalse     
  \else
    \pdfoutput=1 
    \pdftrue
  \fi

\ifpdf
  {
    \usepackage[pdftex]{graphics}
    \pdfcompresslevel=9
    \newcommand\includefigure[1]{
      \includegraphics{figures/#1.pdf}
  }
\else
  {
    \usepackage{graphics}
    \newcommand\includefigure[1]{
      \includegraphics{figures/#1.eps}
  }
\fi

\newcommand{\ignore}[1]{}       

\newcommand{\href}[2]{#2}       

\newcommand \nue{\nu_e}
\newcommand \nuebar{\overline{\nu}_e}
\newcommand \numu{\nu_\mu}
\newcommand \numubar{\overline{\nu}_\mu}
\newcommand \nutau{\nu_\tau}

\newcommand{\fourvec}[1]{\stackrel{\rightharpoonup}{#1}}
\newcommand{\poverm}[2]{^{#1}_{#2}}
\newcommand \dmm{\Delta m^2}
\newcommand \sstt{\sin^{2} 2\theta}

\newcommand{\ket}[1]{\hbox{$ \mid #1 \rangle$}}
\newcommand{\braket}[2]{\hbox{$ \langle #1 \mid #2 \rangle$}}
\newcommand{\braopket}[3]{\hbox{$\langle #1 \mid #2 \mid #3 \rangle$}}

\newcommand{\cpp}{C\raisebox{0.9ex}{\scalebox{0.5}{++~}}}
\newcommand{\gcmsq}{\mathrm{g}/\mathrm{cm^2}}

\newcommand{\differential}[1]{\mathrm{d}#1}
\newcommand{\derivative}[2]{\frac{\differential{#1}}{\differential{#2}}}

\newcommand{\cdash}{\multicolumn{1}{c|}{-}}

\newcommand{\dashmodule}[1]{{\small \bf #1}}
\newcommand{\calcon}[1]{\mathsf{#1}}    
\newcommand{\calrec}{\mathrm}    

\newcommand{\pawfig}[1]{\resizebox{6in}{!}{\includefigure{#1}}}

\newcommand{\twofig}[2]{
  {\makebox[6in]{
      \resizebox{2.95in}{!}{\includefigure{#1}}
      \resizebox{2.95in}{!}{\includefigure{#2}}}
    \makebox[6in]{\makebox[2.95in]{(a)} \makebox[2.95in]{(b)}}
  }
}

\newcommand{\picfig}[3]{
  \begin{figure}[tb!]
    \begin{center}
      #3 
    \end{center}
    \caption{\protect\label{#1}
      #2}
  \end{figure}
}

\input epsf
\def\epsfsize#1#2{\hsize } 

\begin{document}

\sloppy


\pagenumbering{roman}
\begin{center}

{\huge \bf 
Semi-contained Interactions of Atmospheric Neutrinos in the MACRO Detector
}\\[0.5in]
{\Large Thesis by} \\

{\Large Robert G. Nolty
} \\[0.2in]
{\Large In Partial Fulfillment of the Requirements} \\
{\Large for the Degree of} \\
{\Large Doctor of Philosophy}
\end{center}
\vfill
\def\epsfsize#1#2{1in }
\hspace*{\fill } \epsfbox{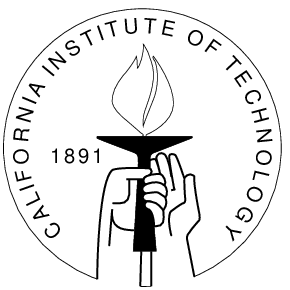} \hspace{.25in} \hspace*{\fill }
\def\epsfsize#1#2{\hsize } 
\begin{center}
{\Large California Institute of Technology} \\
{\Large Pasadena, California} \\[0.2in]

{\Large 2002} \\
{\Large (Defended 18 October 2001)}
\end{center}
\clearpage

\ 
\vfill
\begin{center}

\copyright 2002 \\
Robert G. Nolty \\
All rights reserved
\end{center}
\clearpage


\chapter*{Acknowledgments}

Personally, I would like to thank my family and my spiritual
communities, especially Urban Village, for tremendous support through
some very lean times.

Professionally, I extend thanks first of all to my advisor, Barry
Barish, for his incredible patience and generosity when the road did
not seem to be going anywhere.  I am grateful to Caltech, the U.S.
Department of Energy and the Italian Istituto Nazionale di Fisica
Nucleare for financial support.  I thank my collaborators for making
MACRO what it was, with special affection for the Italians who showed
me hospitality and for the Americans (and Greeks!) with whom I shared a
cross-cultural experience.

During my mid-grad-school crisis in 1992, when I was looking for an
opportunity to convert to experimental physics, I chose to join MACRO
largely because of the nature of the weekly meeting of the Caltech
MACRO group.  It was lively, intense, sometimes loud, usually filled
with laughter.  Ideas were proposed, shot down, and rebuilt.  It was
there that I learned to be a scientist, and it is to this group that
this thesis is dedicated:

\vskip .5in

\begin{center}
  \makebox[4.5in]{
    \begin{tabular}{lr}
      Barry Barish & Charlie Peck \\
      Gary Liu & Doug Michael  \\
      John Hong & Stephane Coutu \\
      Erotokritos Katsavounidis & Sophia Kyriazopoulou \\
      Chris Walter & Kate Scholberg \\
      Rongzhi Liu & Neal Pignatano
    \end{tabular}
  }
\end{center}

\chapter*{Abstract}

Atmospheric neutrinos arise from the decay of particles (primarily
pions, muons and kaons) produced in the collision of high energy
cosmic ray particles with the atmosphere.  The great distances
traveled by atmospheric neutrinos between their production and
detection make them useful for studying neutrino oscillations, the
predicted phenomenon of massive neutrinos changing flavor in flight.
This thesis reports a study of atmospheric neutrinos interacting in
the MACRO detector.  The results, though somewhat clouded by large
theoretical uncertainties, clearly rule out the no-oscillations
hypothesis, and are consistent with oscillations with the parameters
preferred by other MACRO neutrino analyses, as well as those of other
experiments (most notably Super-Kamiokande).  Combining this analysis
with another MACRO neutrino analysis, some of the theoretical errors
cancel, further constraining the region of allowed oscillation
parameters.


\tableofcontents
\pagebreak
\pagenumbering{arabic}

\chapter{Introduction}

\section{The Wily Neutrino\protect\label{sec:wily}}

Although much of what we know about the neutrino is based on
experimental evidence, it may still be considered the theorists'
particle; indeed, we could say the neutrino was not so much discovered
as invented.  The neutrino is so elusive that there have been no
serendipitous experimental discoveries.  It is only with extensive
guidance from theory that the experimentalists have been able to build
apparatus to catch traces of the wily neutrino, and even then it has
been hard going.

The idea that was to evolve into the neutrino was first postulated in
1930 in Pauli's famous ``Dear Radioactive Ladies and Gentlemen''
letter~\cite{pauli}.  The purely theoretical motivation was to
reconcile the spin and statistics of nuclear beta decay and to explain
the continuous electron spectrum.  Within a few years the neutrino was
on solid theoretical ground as an essential component of Fermi's
quantitative theory of beta decay~\cite{fermi}.  However, more than
two decades were to pass before the neutrino was directly observed.

By 1953 nuclear theory including neutrinos had matured despite the
non-observation of neutrinos; indeed it was only the theoretical
understanding of neutrinos that allowed Reines and Cowan to devise
their delayed coincidence experiments to make the first experimental
detection of neutrinos when they observed inverse beta decay induced
by reactor-produced electron neutrinos~\cite{reines&cowan}.

Soon after the first experimental observation of neutrinos,
experimentalists were in a position to answer some theoretical
questions about neutrinos.  The V-A Lorentz structure of the neutrino
vertex was quickly deduced by a large number of experiments within two
years of the 1956 suggestion of parity violation by Lee and
Yang~\cite{parvio}.

Another question awaiting an experimental answer concerned the
existence of two types of neutrinos.  Some neutrinos arose during the
production of electrons and some during the production or decay of
muons, but it was not known if these two types of neutrinos were
identical.  Schwarz, Lederman and Steinberger invented the neutrino
beam at Brookhaven and their results showed that neutrinos associated
with muons behaved differently than those associated with
electrons~\cite{twoneu}.  The existence of multiple types of neutrinos
became a central assumption of the electroweak theories of the 1960s.

Soon the theorists were again out in front with the prediction of
neutral currents.  A number of experiments in the early-1970s
confirmed the existence of neutral
currents~\cite{garg,benvenuti,barish} and bolstered confidence in the
emerging theories.

By this time neutrinos were understood well enough that they could be
used as an experimental tool to study other physics.  In particular,
deep inelastic scattering of neutrinos off nucleons confirmed the
picture of nucleons consisting of fractionally-charged partons which
had been emerging from electron scattering
experiments~\cite{barishX,aubertX}.  Also in the 1960s Ray Davis began
his chlorine experiment~\cite{davis} which attempted to study the solar
interior by observing the rate of solar neutrino interactions on
earth.  In 1987, observations of neutrinos from Supernova
1987A~\cite{sn1987MB,sn1987Kam,sn1987IMB} were analyzed in attempts to
understand the interior of a collapsing star.

Extensive high energy neutrino scattering experiments throughout the
1970s and 1980s confirmed that neutrinos behave very much as
predicted by the Standard Model they helped to create.  However, by
the end of the 1980s intriguing hints were emerging that neutrinos
that travel large distances may depart from the behavior predicted by
the Standard Model.  Many interpreted these hints as evidence that the
neutrinos are not massless as the Standard Model assumes.

\section{Massive Neutrinos}

\subsection{Physics Beyond the Standard Model}

The Standard Model of particle physics has been very successful in
quantitative calculations of particle phenomenology but is rather
unsatisfying as a fundamental theory.  It is widely believed that the
Standard Model is a low-energy effective theory deriving from a more
elegant high-energy theory.  In the 1990s, after the spectacular
validation of the Standard Model in precision measurements at LEP,
attention of both theorists and experimentalists turned to phenomena
beyond the Standard Model in the hopes of constraining explorations of
fundamental theories.  Many sectors of particle physics may give a
glimpse beyond the Standard Model, but three in particular have
received attention as the possible next step in experimental particle
physics: studying the origin of mass by direct observation of the
Higgs particle (awaiting completion of the LHC experiment at CERN);
studying the origin of CP violation by precision measurements of quark
mixing (currently underway at SLAC and KEK); and finally, mixing and
mass in the neutrino sector.   Of these three, neutrino experiments
have been the first to yield exciting results pointing beyond the
Standard Model.

In the Standard Model, neutrinos are assumed to be massless.  However,
{\it a priori} there is no compelling theoretical reason to expect
neutrinos to be massless, and the Standard Model may be easily
extended to include small neutrino masses without contradicting any
known observations.  In fact, if one assumes the Standard Model is a
low-energy effective theory of a non-renormalizable quantum gravity
field theory, neutrino masses arise generically~\cite{weinberg}.
However, simple extensions to the Standard Model are only satisfying
if they explain why the neutrino masses are so small (compared to the
typical mass scales in the theories).

\subsection{Direct Searches}
Upper limits on neutrino mass come from observations that rely
directly on the mass properties of neutrinos.  In the next section,
the more subtle effect of neutrino oscillations will be considered.

The most straightforward experiments attempt to detect neutrino mass
by examining the energy spectrum of observable particles in decays
involving neutrinos; if neutrinos have mass the maximum possible
invariant mass for the sum of other particles is reduced.  Such
examinations have yielded no evidence for neutrino masses, and have
set the upper limits shown in Table~\ref{tab:mnulim}.

\begin{table}[htb]
  \leavevmode
  \begin{center}
    \begin{tabular}{|l|l|} \hline
      $m_{\nue}   $ & $<$ 3 eV     \\
      $m_{\numu} $ & $<$ 0.19 MeV \\
      $m_{\nutau}$ & $<$ 18.2 MeV \\ \hline
    \end{tabular}
    \caption{\protect\label{tab:mnulim}
      Direct limits on neutrino masses from PDG\protect~\cite{rpp}.}
  \end{center}
\end{table}

Analyses of neutrino mass from the arrival time of SN1987A
neutrinos~\cite{sn1987MB,sn1987Kam,sn1987IMB} are somewhat difficult
to interpret, and under the skeptical judgment of the Particle Data
Group~\cite{rpp}, they do not improve the limits shown above.

If we adopt standard Big Bang Cosmology, in the current universe there
is a high density of relic neutrinos created during the
radiation-dominated epoch.  From the current observation that the
density of the universe is not far from the critical density, we may
rule out neutrino masses greater than a few tens of
MeV~\cite{boehm&vogel}.

\subsection{Neutrino Oscillations}

If neutrinos have mass, there may be no reason to expect the mass
eigenstates to be the same as the weak force eigenstates.  Thus, the
particles created and destroyed in weak interactions (the electron
neutrino, muon neutrino and tau neutrino) may be superpositions of
mass eigenstates; conversely the mass eigenstates would be
superpositions of the flavor eigenstates.  As a particle propagates
according to the mass eigenstates, its flavor content could evolve, a
process known as neutrino oscillation.  We can gain insight into this
process by considering a two-neutrino model, with flavor eigenstates
\ket{\nu_a} and \ket{\nu_b}, and mass eigenstates \ket{\nu_1} and
\ket{\nu_2}.

In general,

\[ \ket{\nu_a}~=~\ket{\nu_1}\cos\theta~+~\ket{\nu_2}\sin\theta \]
\[ \ket{\nu_b}~=~-\ket{\nu_1}\sin\theta~+~\ket{\nu_2}\cos\theta \]

\noindent
where $\theta$ is called the mixing angle.  

\subsubsection{Vacuum Oscillations}

If a flavor eigenstate \ket{\nu_a} is created at $x~=~0$ with
momentum $p$, the two mass components of the particle will have
slightly different velocities, and the flavor content of the
propagating particle will vary in time.  Invoking the quantum
mechanical equivalence of energy to frequency, the state at time $t$
of a particle that is pure \ket{\nu_a} at $t~=~0$ is

\[ \ket{\nu_a}_t~=~\exp(-iE_{1}t)\ket{\nu_1}\cos\theta~+~
                     \exp(-iE_{2}t)\ket{\nu_2}\sin\theta~\]

\noindent
Invoking the orthogonality of the mass eigenstates
(\braket{\nu_1}{\nu_2}~$=~0$), we find the probability the particle
will be observed as \ket{\nu_b} at time $t$ is\footnote{NOTE: No cats
  were harmed in the collapse of this wavefunction.}

\[ \mid\braket{\nu_b}{\nu_a}_t\mid^2~=~
   \sin^{2} 2\theta\sin^{2}({1 \over 2}(E_{2}-E_{1})t) \]

For ultrarelativistic neutrinos of given momentum,

\begin{eqnarray*}
E_2~-~E_1~& =       & \sqrt{p^2+m_{2}^2}~-~\sqrt{p^2+m_{1}^2} \\
             & \approx & \frac{m_{2}^2-m_{1}^2}{2 p}
\end{eqnarray*}

\noindent
so that (equating $p$ with $E$ and $t$ with $x$ for ultrarelativistic
neutrinos)

\[ \mid\braket{\nu_b}{\nu_a}_t\mid^2~=~
   \sin^{2} 2\theta\sin^{2}\left(\frac{\Delta m^2 x}{4 E}\right) \]

This is the fundamental equation of neutrino oscillations.  (This
simple treatment sweeps a lot of details under the rug, but more
rigorous treatments involving wave packets do not change the basic
phenomenology of oscillations~\cite{kayser}.)  The unknown parameters
of the theory are the mixing angle $\theta$ and the mass difference
$\Delta m^2$.  If we restore the factors of $c$ and $\hbar$ that have
been suppressed, we find the half-wavelength for oscillation is

\begin{equation}\protect\label{eq:lvac}
  L_{vac} = \frac{4 \pi E \hbar}{\Delta m^2 c^3}
          = 2.48 \left(\frac{E}{\mathrm{GeV}}\right)
            \left(\frac{\mathrm{eV^2}}{\Delta m^2} \right)~\mathrm{km}
\end{equation}

We have little theoretical guidance for the expected value of the
parameters $\theta$ or $\Delta m^2$.  We do know that the analogous
mixing angles in the quark sector are relatively small (for example,
for the Cabbibo Angle $\theta_c$, $\sin^2\theta_c \approx 0.05$).
Without unexpected fine tuning, the mass difference is expected to be
on the order of the neutrino masses.

Atmospheric neutrinos (described below), with typical energies of a
few GeV and with typical path lengths of thousands of kilometers, are
in principle sensitive to mass differences less than 0.1~eV\@.

\subsubsection{Oscillations in Matter}

The above arguments apply for neutrinos propagating in vacuum.
However, when neutrinos propagate through matter, their evolution is
altered by weak interactions with the medium, which may be considered
a perturbation to the mass hamiltonian.  All three known types of
neutrinos (electron, muon and tau neutrinos) will feel the effects of
the neutral current weak force, typified by the Feynman diagram in
Figure~\ref{fig:weakfeyn}a.  If these are the only types of neutrinos
that participate in mixing, this force will shift the phase of all
components of a mass eigenstate equally, so the nature of oscillations
is unchanged by it.  However, only the electron weak eigenstate
participates in the charged current weak force shown in
Figure~\ref{fig:weakfeyn}b.  This changes the phase evolution of the
electron component of a mass eigenstate and alters the oscillation
behavior.  It has also been postulated that a sterile neutrino exists
and mixes with the known neutrino flavors, but does not participate in
neutral current or charged current weak interactions.  If so, the
absence of reaction (a) will cause its phase to shift relative to the
other types of neutrinos.

\picfig{fig:weakfeyn}{Feynman diagrams for weak interactions in
  matter.  (a)~Neutral current interactions, affecting all known
  neutrino flavors.  (b)~Charged current, affecting only electron
  neutrinos.}  {\twofig{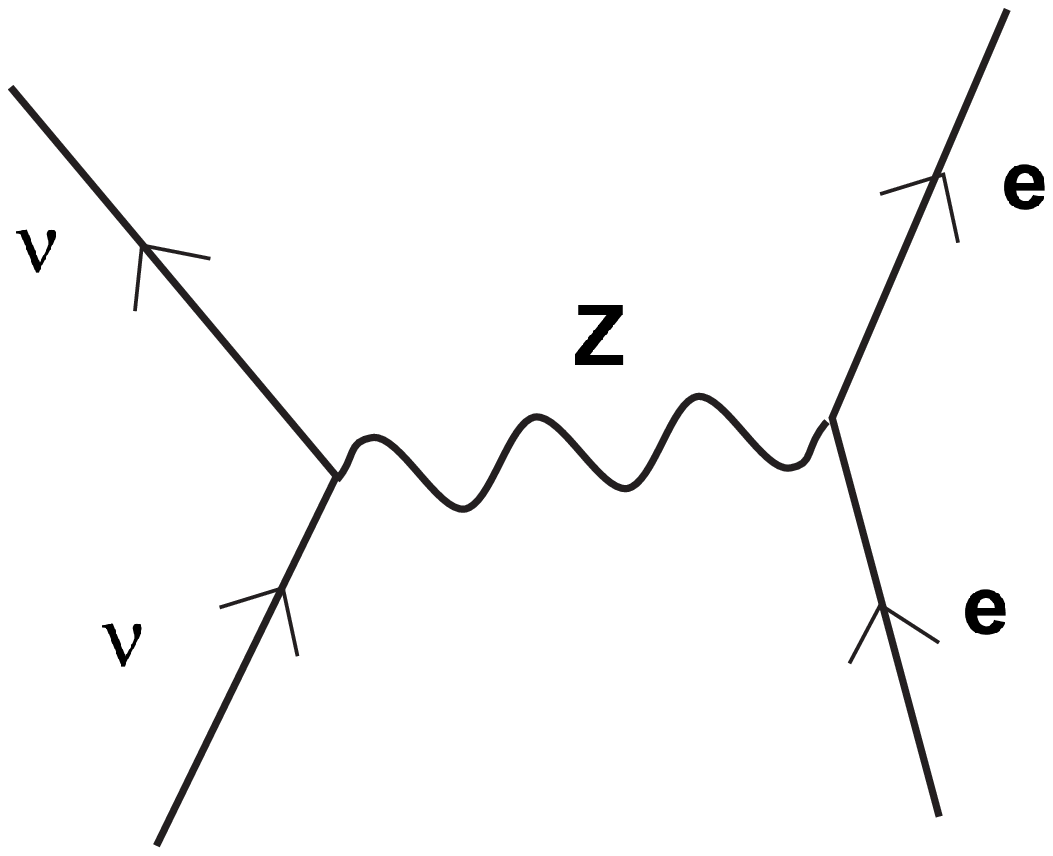}{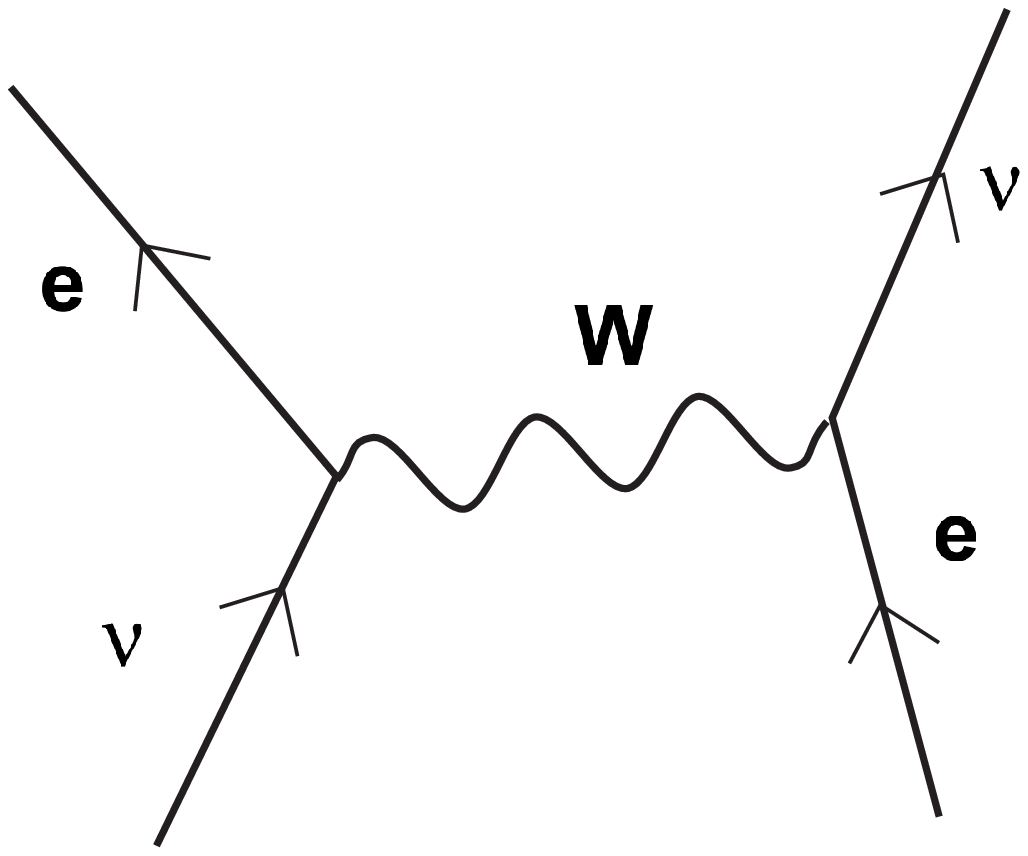}}

Wolfenstein~\cite{wolf} was the first to make quantitative
calculations of matter effects, in 1978\footnote{There were errors in
  Wolfenstein's original formula, but they were corrected in the
  subsequent literature and here.}.  The energy perturbation is
\braopket{\nue}{H_{eff}}{\nue} where $H_{eff}$ arises from the
$W$-exchange of Figure~\ref{fig:weakfeyn}b.  For low energy neutrinos
we may use the old Fermi theory for the diagram; thus the
magnitude of the effect will be proportional to the Fermi constant
$G_F$.  The effect will also be proportional to the number of
electrons with which the neutrino can interact.  The result of
Wolfenstein's exact calculation is

\[ \Delta E~=~\sqrt{2}G_F N_e \]

\noindent
where $N_e$ is the density of electrons.  Thus, electron neutrinos
acquire an additional phase $e^{-i \Delta E t}$ relative to the
muon and tau components of a mass eigenstate, and the phase depends on
the density of electrons but is independent of neutrino energy (at
least in the range where the Fermi approximation is valid).  For
ultrarelativistic neutrinos one may obtain the ``matter oscillation
length'' $L_m$, which is the length over which the phase of the
electron neutrinos changes by $2\pi$ relative to the vacuum
oscillation formula.  For propagation through the earth, with

\[ N_e~\approx~3 \frac{\mathrm{g}}{\mathrm{cm^3}}~*\
                     N_A \frac{\mathrm{nucleon}}{\mathrm{g}}~*\
                     \frac{1}{2} \frac{\mathrm{electron}}{\mathrm{nucleon}}~
         =~9.9 \times 10^{23} \frac{\mathrm{electron}}{\mathrm{cm^3}} \]

\noindent
we find $L_m \approx 10^4$~km, so matter effects are an important
correction to vacuum oscillation calculations.  (For solar neutrinos,
in the center of the sun, $L_m \approx 200$~km, and matter effects
lead to some surprising phenomenology.)

\subsection{Current Oscillation Experiments and Limits}

Regions of parameter space excluded or preferred by various
experiments to be discussed below are shown in
Figure~\ref{fig:pdgosc}.

\picfig{fig:pdgosc}{Figure and caption on neutrino oscillation limits
  from the latest Review of Particle Physics~\protect\cite{rpp}.}{
  \makebox{\rule{0pt}{1.2in}\hspace{\textwidth}}
  \framebox{\resizebox{6in}{!}{\rotatebox{-90}{\includefigure{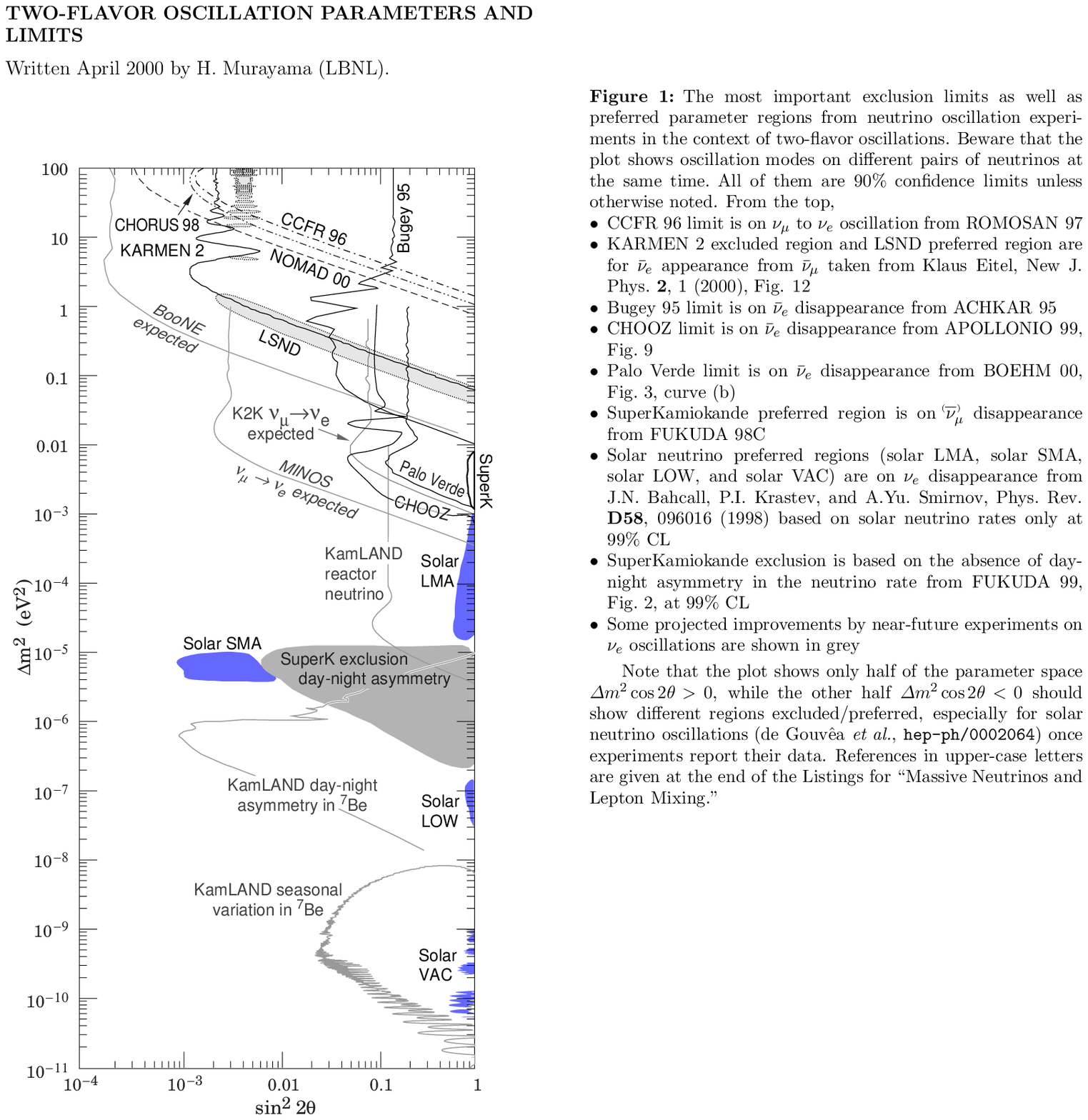}}}}}

\subsubsection{Accelerator Neutrinos}

An accelerator neutrino beam consists almost entirely of $\numu$ or
almost entirely of $\numubar$.  Some detectors that are not efficient
to detect $\nue$ or $\nutau$ measure only $\numu$ disappearance.
Others, also sensitive to $\nue$ or $\nutau$ appearance, are able to
probe much smaller mixing angles.  Accelerator experiments from the
1970s through the end of the century did not see evidence of muon
neutrino disappearance.  Given the energies (typically tens to
hundreds of GeV) and baselines from neutrino creation to detection
(hundreds of meters), typical experiments could only see oscillations
if $\Delta m^2$ were a few $\mathrm{eV}^2$ or greater.

An innovation on the traditional beam experiments is K2K, an ongoing
experiment with a baseline from accelerator to detector of 250~km, and
a detector very sensitive to both $\numu$ and $\nue$ interactions.
With low statistics (44~interactions in the detector in two years,
with 64 expected in the absence of oscillations) they see evidence for
$\numu$ disappearance and no evidence for $\nue$ appearance~\cite{k2k}.

A completely different type of accelerator-based experiment brings
accelerator-created pions to rest in a dense medium where they decay.
(Some pions that decay in flight also contribute to the experiment.)
Detailed reconstructions of neutrino interactions in a target a few
tens of meters away can show evidence of
$\numubar\,\rightarrow\,\nuebar$ or
$\numu\,\rightarrow\,\nue$.  Two experiments of this type have
been performed.  LSND~\cite{lsnd} sees evidence of electron neutrino
appearance while Karmen~\cite{karmen} does not.  Because of differences
in geometry, much of the LSND-preferred region is not excluded by
Karmen, at least at the 90\% confidence level.  The parameter space
preferred by LSND (and not ruled out by beam experiments) is
$0.04~\mathrm{eV}^2 < \dmm < 2~\mathrm{eV}^2$ with $\sstt > 0.002$.
(See Figure~\ref{fig:pdgosc}.)

\subsubsection{Reactor Neutrino Experiments}

Nuclear reactors produce copious numbers of electron antineutrinos
with energies of a few MeV\@.  Because the neutrino flux can be
calculated precisely when the power level is known, detectors tens to
hundreds of meters from a reactor can serve as electron antineutrino
disappearance experiments.  None has seen any evidence for
oscillations.

\subsubsection{Solar Neutrino Experiments}

One solar neutrino detector, the chlorine detector at Homestake, has
been running almost continuously since 1967, and has consistently seen
a flux of electron neutrinos around one-third that predicted by the
Standard Solar Model.  However, due to large theoretical uncertainties
and some experimental questions, the result did not garner much
attention at first.  By the 1990s the errors were under control and
then a series of new experiments confirmed the deficit in additional
energy ranges.  Today, there is no doubt that electron neutrinos are
disappearing due to unknown neutrino physics, but it is difficult to
prove the deficit is due to oscillations.\footnote{During the
  preparation of this thesis, the SNO collaboration~\protect\cite{sno}
  reported evidence that the missing electron neutrinos are present at
  earth in other, non-sterile flavors (presumably $\numu$ or
  $\nutau$).}  Assuming the deficit is due to oscillations, four
disjoint regions of parameter space are allowed, all with $\dmm <
0.001~\mathrm{eV}^2$.  (See Figure~\ref{fig:pdgosc}.)

\subsubsection{Atmospheric Neutrinos}

When high energy cosmic ray particles (protons or heavier nuclei with
typical energies in the GeV - TeV range) collide with the atmosphere,
a hadronic shower develops.  Charged pions and kaons in the shower can
decay in flight yielding high energy muons and muon neutrinos.  If the
muon also decays in flight, it can yield another muon neutrino and an
electron neutrino.  Flux of these so-called atmospheric neutrinos is
on the order of several hundred per square centimeter per second.
(Flux models will be discussed in detail in Chapter~\ref{chap:rate}.)
The energy and flight pathlength of these neutrinos are not as
well-determined as accelerator neutrinos.  However, the detected
neutrinos come from a large variety of energies and pathlengths so
one can potentially explore more oscillation parameter space with
atmospheric neutrinos than with accelerator neutrinos.

Prior to 1996, a number of experiments built primarily for other
purposes (including MACRO) were able to make measurements of
atmospheric neutrino fluxes.  The early results were confusing, with
some measurements favoring, some disfavoring, oscillations.  Even the
positive experiments did not have enough precision to show that the
deviations were best explained by oscillations.  In 1996 the first
experiment built primarily to study oscillations in atmospheric
neutrinos, Super-Kamiokande (\mbox{Super-K}), turned on and quickly amassed
the dominant data set in the field.  By 1998, \mbox{Super-K} was able to
claim the discovery of neutrino oscillations~\cite{sk98}.  The result
is convincing because \mbox{Super-K} derives oscillation parameters from many
different analyses involving different event sets with different
energy ranges, some utilizing ratios in which much of the systematic
uncertainty cancels, and all the analyses give results consistent with
a single oscillation hypothesis.

\mbox{Super-K} results~\cite{sk00} indicate muon neutrino disappearance,
without a corresponding appearance of electron neutrinos, with large
mixing angle ($\sstt > 0.88$) and $0.0015~\mathrm{eV}^2 < \dmm <
0.005~\mathrm{eV}^2$.  Other recent experiments with positive results
(MACRO~\cite{upmu98} and Soudan~\cite{soudan}) have results consistent
with \mbox{Super-K} but with a less-precise preferred region in parameter
space.

This thesis presents an analysis of atmospheric neutrinos interacting
inside the MACRO detector.

\subsubsection{Future Experiments}

Energized by the discovery that neutrinos are oscillating at least in
the atmospheric neutrino sector, a number of new experiments are in
the works to resolve the outstanding qualitative questions (How many
types of neutrinos participate in oscillations?  Which neutrinos mix
with which?  Is the solar neutrino problem due to oscillations?) and
to make more precise measurements of the oscillation behavior.

One cannot simultaneously accommodate the LSND, atmospheric and solar
results with only the three known types of neutrinos.  Perhaps there
is a ``sterile neutrino'' -- a fourth type of light neutrino that does
not participate in weak interactions but does mix with the well-known
neutrinos through oscillations.  Or, perhaps the LSND result is wrong,
or either the LSND or solar neutrino problems are not due to
oscillations.

Also, the results so far presented have been within the framework of
2-flavor oscillations.  However, the reality probably involves a
$3\times3$ or $4\times4$ matrix of mixing parameters.  Perhaps,
for example, muon neutrinos mix primarily with tau neutrinos but have
a small mixing angle with electron neutrinos.

A few of the most significant planned experiments include:

\begin{itemize}
\item Sudbury Neutrino Observatory - sensitive to neutral current
  interactions of solar neutrinos, SNO should be able to determine if
  the disappearing electron neutrinos are turning into active or
  sterile neutrinos.
  
\item NuMI and CNGS - long baseline (730~km) experiments in the U.S.
  and Europe respectively which are designed to explore the parameter
  space around the atmospheric neutrino preferred parameters.

\item Boone - a short baseline accelerator experiment that should
  definitively confirm or refute the LSND result.

\item Kamland - the first long baseline reactor neutrino experiment
  which will extend down to mass differences relevant to the solar
  neutrino problem.
\end{itemize}

\section{MACRO as a Detector of Atmospheric Neutrinos}

In the years leading up to such next-generation experiments, we may
use existing detectors (most built for other purposes) to further
elicit the nature of the flux discrepancies and guide the developers
of the new experiments.  The use of the MACRO detector in atmospheric
neutrino physics is an example of such a fortuitous and timely
analysis.

Located deep underground at the Italian Gran Sasso National Lab
(LNGS), the Monopole, Astrophysics and Cosmic Ray Observatory (MACRO)
was a large area detector of ionizing particles.  In the neutrino
realm, it was most efficient for detecting muons from the charged
current interaction of muon neutrinos and antineutrinos.  The
experimental technique is to look for upward-going muons, thus
eliminating the large background due to downgoing primary muons
produced by cosmic rays in the atmosphere above MACRO\@.  MACRO could
detect upgoing muons that were produced within its volume or those
that were produced in the rock below.  MACRO also had some sensitivity
to electrons from charged current interactions of electron neutrinos
within the detector, or neutral current interactions of any flavor
neutrino within the detector.  However, MACRO was not optimized for
these experimental signatures.

This thesis reports a study of the interactions within the MACRO
detector of upgoing atmospheric neutrinos.


\chapter{The MACRO Detector \protect\label{chap:macro}}

The MACRO experiment is a collaboration of about 100 physicists from
Italian and U.S. institutions.  Originally conceived to detect
slow-moving supermassive particles in the cosmic radiation (e.g.,\ GUT
monopoles)~\cite{mono,monoST,mono97,nucl,nucl00}, MACRO has also
produced results in the study of penetrating muons from high energy
cosmic
rays~\cite{anisot,comp,deco,time,intensity,hecrI,hecrII,seasonal,TRD,deco99}
(sometimes in coincidence with a surface
detector~\cite{eastop,eastop2,grace}), searches for astronomical point
sources of penetrating muons~\cite{muastro}, searches for low-energy
neutrino bursts from collapsing stars~\cite{gc,gc98}, a search for
fractionally-charged particles~\cite{lip}, as well as neutrino
physics~\cite{upmu95,upmu98,wimps,lowenu,nuastro,nusterile}\footnote{An
  up-to-date list of MACRO publications may be found at \hfill \break
  \href{http://www.df.unibo.it/macro/pub1.htm}{\tt
    http://www.df.unibo.it/macro/pub1.htm}.}.  MACRO began taking data
in 1989 while it was still under construction.  The detector was
completed in 1994, and stable running lasted until the detector was
retired in December, 2000.

MACRO was located in Hall B of the Italian Gran Sasso National
Laboratory (LNGS) in the Apennine mountain range east of Rome (see
Figure~\ref{fig:italy}).  Located in the middle of a mountain, Hall B 
is covered by at least 3150~m.w.e.\ of rock.  (1 meter of water
equivalent (m.w.e.)\ $= 100~\gcmsq$.)  At this depth,
electrons and hadrons from cosmic ray showers cannot penetrate and the
muon flux from cosmic rays is about $10^{-6}$ of the flux at the
surface.

\picfig{fig:italy}{Location of Gran Sasso.}
  {\resizebox{4 in}{!}{\includefigure{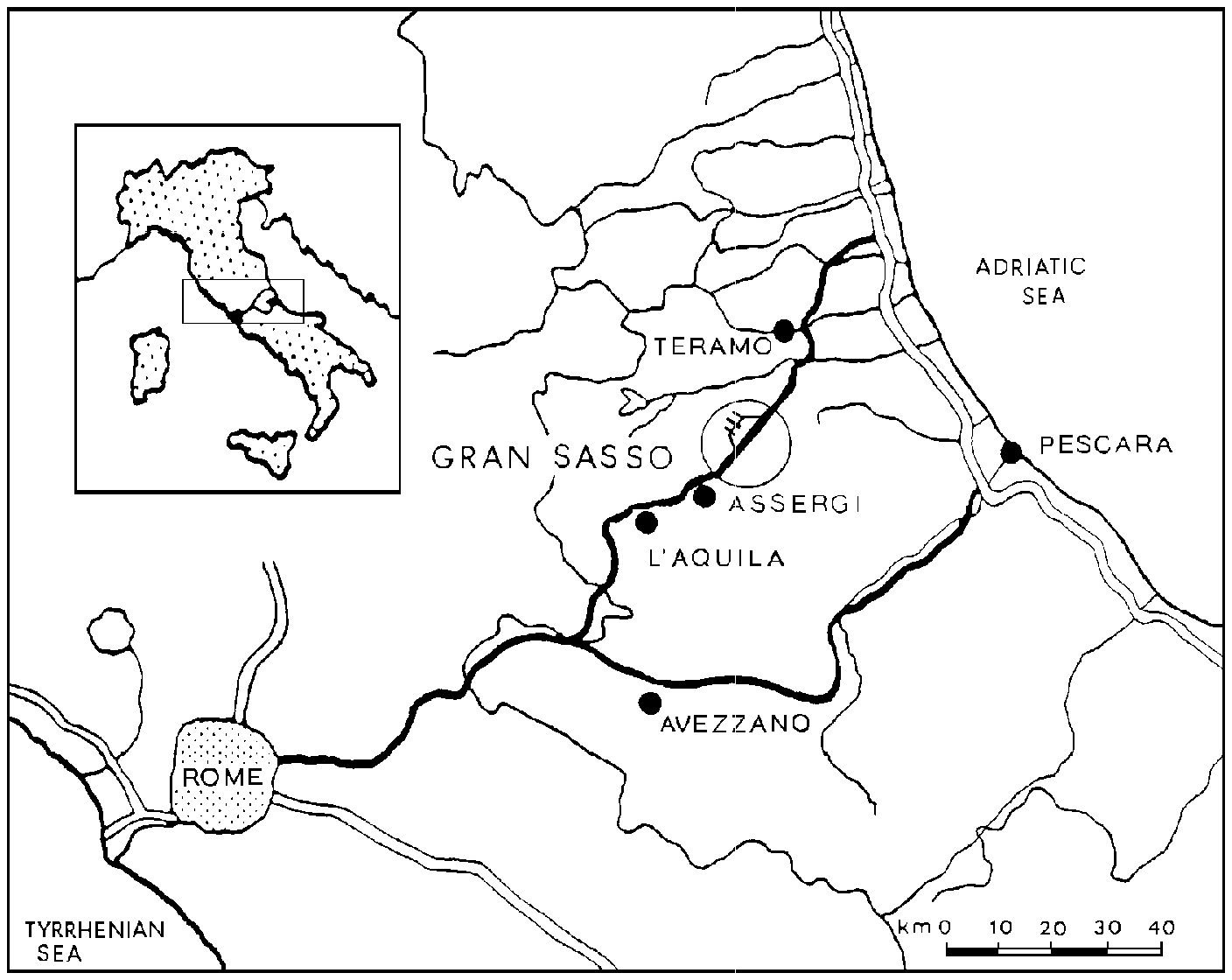}}}

MACRO (Figure~\ref{fig:macro}) was a very large detector
(approximately 72~m $\times$ 12~m $\times$ 10~m).  The long axis consisted of six
identical {\it supermodules}, often abbreviated ``SM.''  The walls and
central layer of the detector were instrumented with planes of
streamer tubes (ST) and boxes of liquid scintillator (SC).  The upper
half of MACRO (the {\it attico}) was hollow while the lower half
consisted of layers of uninstrumented rock absorber and additional
streamer tube planes.  Also in the lower half was one plane of passive
plastic track etch detector in which exotic particles were expected to
leave distinctive tracks that could be analyzed months later.  (See
Figure~\ref{fig:xsec}.)

\picfig{fig:macro}{The MACRO detector.}
  {\resizebox{4 in}{!}{\includefigure{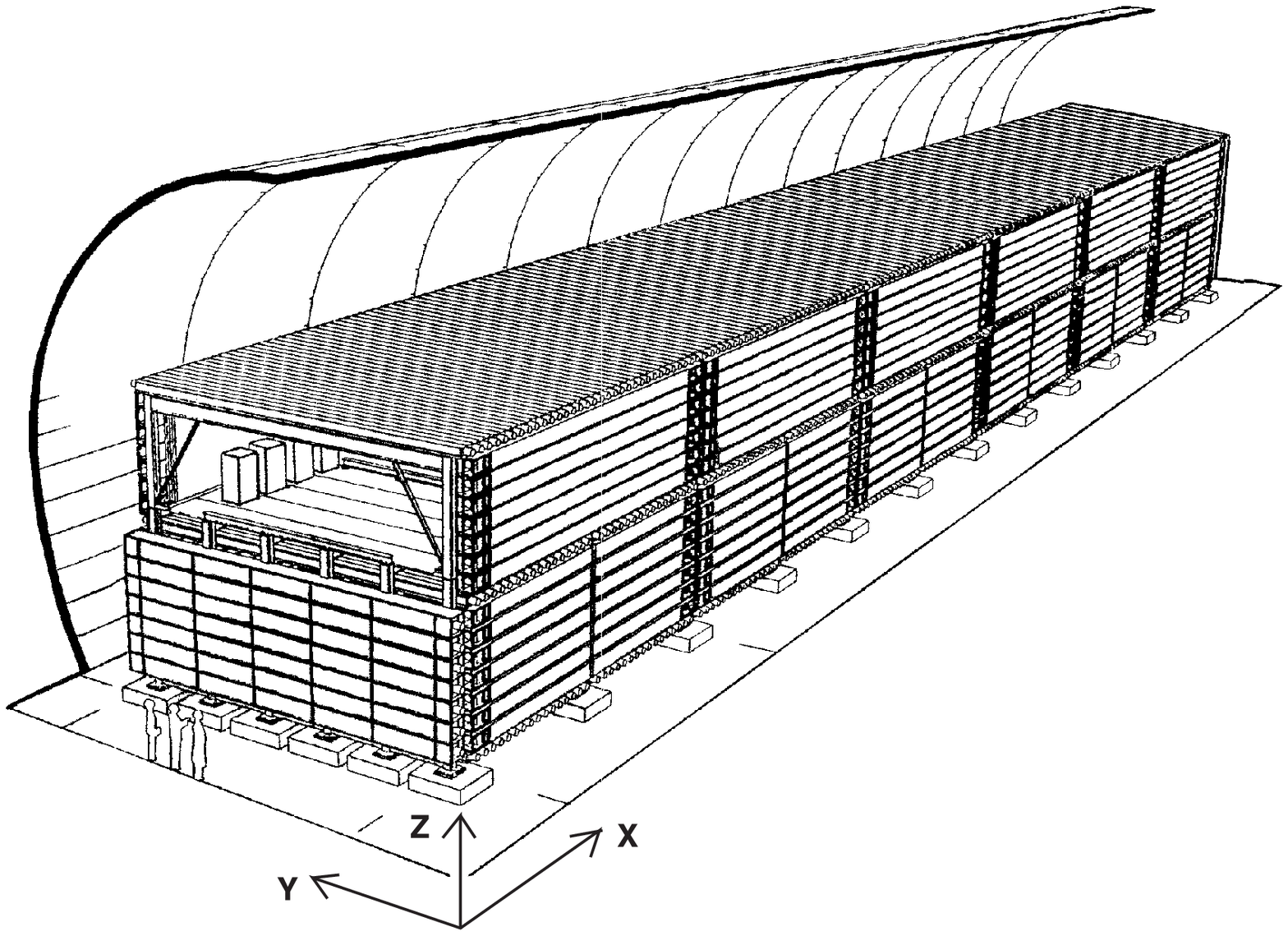}}}

\picfig{fig:xsec}{Cross section of MACRO\@.}
  {\resizebox{5 in}{!}{\includefigure{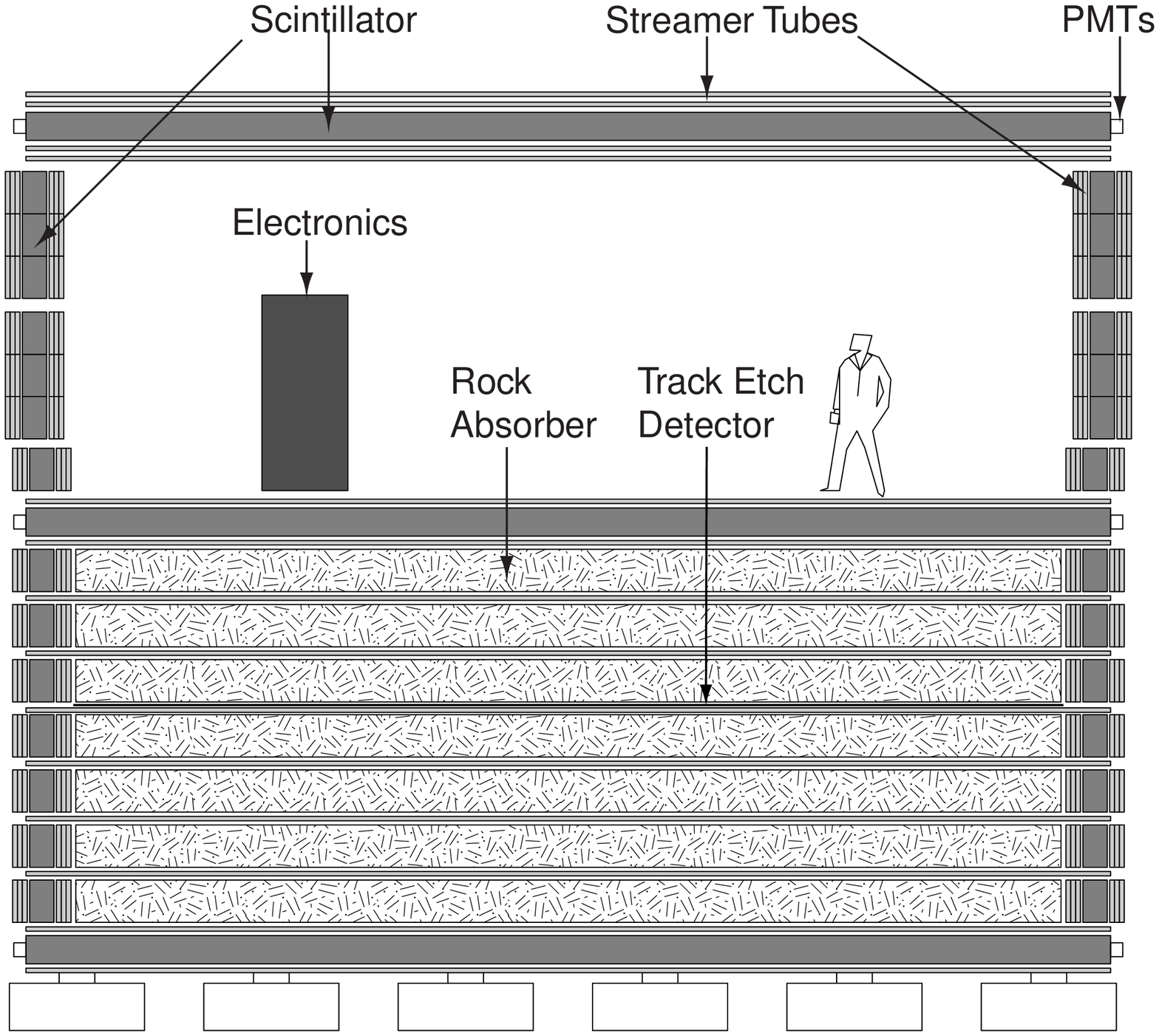}}}
  
We define the Z-axis to be vertical upward, the X-axis along the long
dimension of MACRO (north to south), and the Y-axis along the
horizontal short dimension (west to east).  An additional coordinate
in the X-Y plane is defined; D is the coordinate perpendicular to the
streamer tube pickup strips (described below).

The two main electronic detector systems of MACRO, streamer tubes and
scintillation counters, are very complementary.  The amount of
scintillation light produced was proportional to the energy loss in the
scintillator and the light was produced promptly, so the scintillation
counters had good energy and timing resolution.  However, the high
cost of the photomultiplier tubes used to detect the light limited the
designers of MACRO to about 500 scintillator boxes.

The streamer tubes had poor timing resolution because of the drift
time in the gas and poor energy resolution because of the digital
nature of streamer formation.  However, the low cost of plastic
streamer tubes allowed MACRO to employ about 50,000 channels of
streamer tube readout.  In each streamer tube plane the wire spacing
was about 3~cm.

Therefore, to reconstruct a particle traversing the MACRO detector,
the streamer tubes provide most of the position information, and
scintillators provide the timing and energy information.

\section{The Scintillator System}

\subsection{Scintillation Counters}
The scintillation counters were boxes of opaque PVC, filled with a liquid
scintillator consisting of a base of mineral oil with a 6.2\%
admixture of scintillator concentrate.  The concentrate consisted
primarily of pseudocumene (1,2,4-trimethylbenzene) with small
amounts of PPO (2,5-diphenyl-oxazole) and bis-MSB
(p-bis[o-methylstyryl]benzene).  The mineral oil had high transparency
and purity, allowing scintillation light to propagate several meters.
At the end of each box was an endchamber separated from the
scintillator by a clear PVC window.  One or more photomultiplier tubes
(PMTs) were placed in the endchamber to detect scintillation light.
Each endchamber was filled with pure mineral oil to optically couple
it to the scintillator and to reduce sparking of the
phototube high voltage system.  A mirror of shaped, aluminized PVC
around each phototube increased its light-collection efficiency.  (See
Figure~\ref{fig:tankend}.)

\picfig{fig:tankend}{A scintillator box endchamber.}
  {\resizebox{5 in}{!}{\includefigure{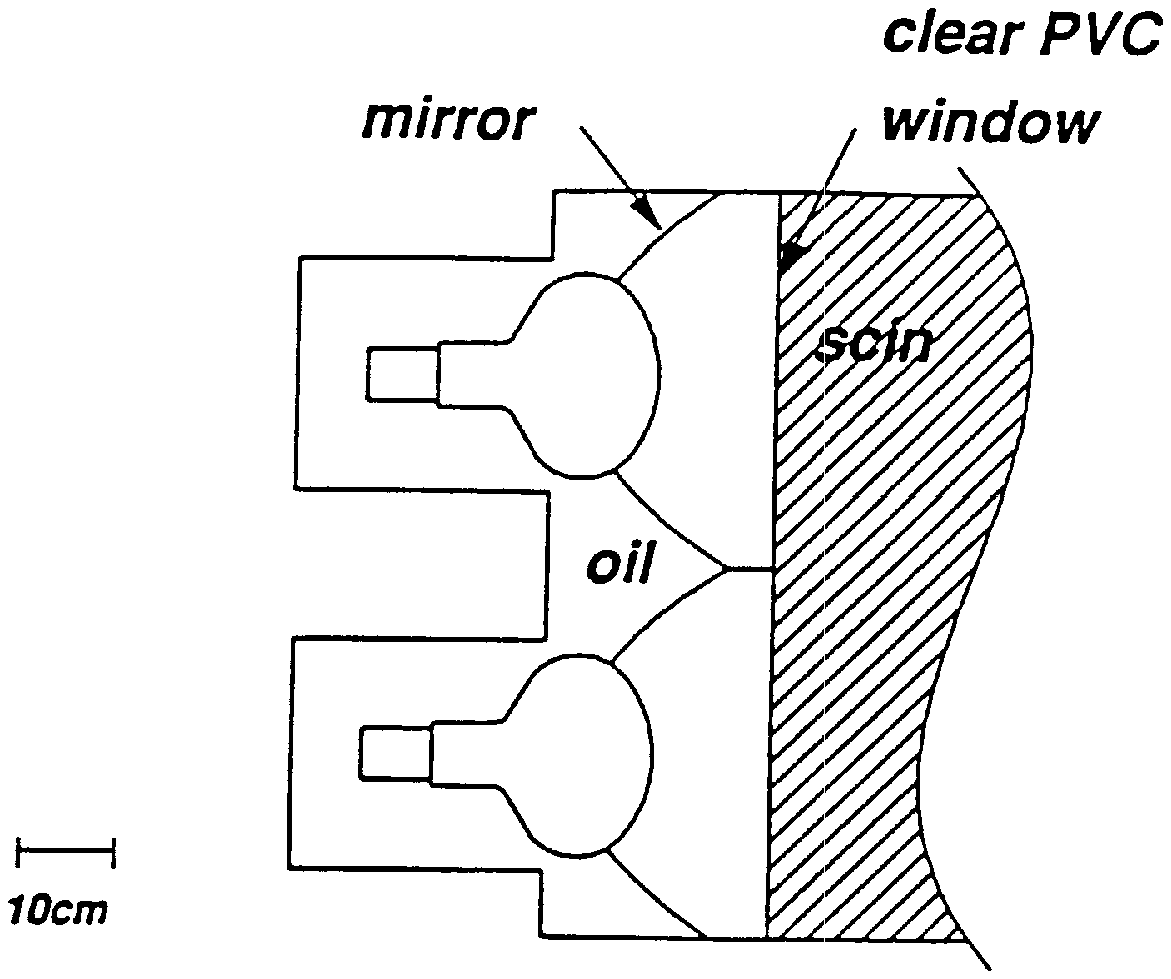}}}

The boxes of scintillator lived in layers.  {\it Horizontal} boxes
were found in the Bottom (B), Center (C) and Top (T) layers.
Horizontal boxes were about 12~m long, 75~cm wide and filled to a
depth of 19~cm with liquid scintillator.  Horizontal boxes had two
PMTs in each end, but the outputs were immediately summed and treated
as a single channel in the electronics.  {\it Vertical} boxes lived in
the vertical faces.  The West (W) and East (E) faces were along the
long axis of MACRO\@.  The North (N) and South (S) faces were
instrumented only on the lower part of the detector to allow
convenient access to the hollow part of the upper detector (refer back
to Figure~\ref{fig:macro}).  Vertical
boxes were about 12~m long, 21~cm wide and filled to a depth of 43~cm
with scintillator.  They had only one phototube in each end.  The
speed of light in the scintillator was about $0.7\,c$ so light could
travel from one end of the box to the other in about 60~ns.

Scintillator boxes also had a light-emitting diode (LED) near each
phototube and a laser fiber that could deposit laser light near the
center of the tank.  LED or laser light was introduced only during
special calibration runs (to be discussed in
Section~\ref{sec:calibproc}).

Each scintillator box is referred to by a unique four-character name
based on its location in MACRO\@.  For example, the seventh tank in the
Bottom (B) layer of the third supermodule is referred to as 3B07.  The
two ends of the box are referred to as the 0- and the 1-end, so a box
end is referred to by a name such as 3B07-1.  Each box is also
identified by a unique integer as detailed in Table~\ref{tab:erpnum}.

\begin{table}[htb]
  \leavevmode
  \begin{center}
    \begin{tabular}{|l||l|l|l|l|l|l|} \hline
      boxes   & SM1     & SM2     & SM3     & SM4     & SM5     & SM6    \\
      \hline \hline
      B01-B16 & 001-016 & 101-116 & 201-216 & 301-316 & 401-416 & 501-516 \\
      C01-C16 & 017-032 & 117-132 & 217-232 & 317-332 & 417-432 & 517-532 \\
      W01-W07 & 033-039 & 133-139 & 233-239 & 333-339 & 433-439 & 533-539 \\
      E01-E07 & 040-046 & 140-146 & 240-246 & 340-346 & 440-446 & 540-546 \\
      N01-N07 & 047-053 & \cdash  & \cdash  & \cdash  & \cdash  & \cdash  \\
      S01-S07 & \cdash  & \cdash  & \cdash  & \cdash  & \cdash  & 547-553 \\
      T01-T17 & 054-070 & 154-170 & 254-270 & 354-370 & 454-470 & 554-570 \\
      W08-W14 & 071-077 & 171-177 & 271-277 & 371-377 & 471-477 & 571-577 \\
      E08-E14 & 078-084 & 178-184 & 278-284 & 378-384 & 478-484 & 578-584 \\
      \hline
    \end{tabular}
  \end{center}
  \caption{\protect\label{tab:erpnum}Numbering of scintillator boxes.}
\end{table}

\subsection{Scintillator Electronics\protect\label{sec:erp}}

The phototube signals from each end of each scintillator box were
fanned out to several electronic subsystems.  The Energy
Reconstruction Processor (ERP) was a self-triggering system of ADCs
and TDCs for each box.  Other electronic subsystems, not relevant to
this analysis, included a gravitational collapse trigger, a
slow-moving monopole trigger, a two-face coincidence detector and a
waveform digitizing system.

There was a different ERP system for each supermodule of MACRO\@.  (For
ERP purposes, the North face and South face each constituted a
supermodule, so there were a total of eight ERP systems.)  For each
box in the supermodule, the ERP had two ADCs and two TDCs for each end
(a total of four ADCs and four TDCs per box).  One of the ADCs saw an
input that was attenuated by a factor of ten compared to the other;
the unattenuated ADC often saturated to its maximum count of 4095 on
muon-level phototube pulses.  All TDCs in a supermodule shared a
common stop.  One of the TDCs had a lower threshold to generate a TDC
start.  On most events the two TDCs started at almost the same time
but occasionally a small signal could start the low-threshold TDC
without starting the high-threshold TDC\@.  The eight measurements for
each box are summarized in Table~\ref{tab:erpsig}.

\begin{table}[htb]
  \leavevmode
  \begin{center}
    \begin{tabular}{|l||l|} \hline
      name & description \\ \hline \hline

      ADC0U & 0-end unattenuated ADC \\
      ADC1U & 1-end unattenuated ADC \\
      ADC0A & 0-end attenuated ADC \\
      ADC1A & 1-end attenuated ADC \\
      TDC0H & 0-end high-threshold TDC \\
      TDC1H & 1-end high-threshold TDC \\
      TDC0L & 0-end low-threshold TDC \\
      TDC1L & 1-end low-threshold TDC \\ \hline
    \end{tabular}
  \end{center}
  \caption{\protect\label{tab:erpsig}ERP signals for each box.}
\end{table}

ERP activity for a box began when a signal crossed a front-end
threshold (typically around 100~mV) and started the TDC for that box.
If both box ends crossed threshold within 200~ns, this coincidence was
called a minimum bias trigger (MBT).  When the MBT occurred, delayed
versions of the input signals were gated onto an ADC capacitor for
200~ns.  The common stop for all TDCs occurred about 400~ns after the
first MBT in the supermodule.  If no MBT occurred within 200~ns of a
TDC start, the event was ignored and the circuitry was cleared.

Even if an MBT occurred, the event could be rejected.  The circuitry
waited until charge integration was complete, and then estimated the
energy of the event based on the unattenuated ADC values at both ends.
(The estimate was done using a pre-stored lookup table, indexed by
6-bit non-linear flash ADCs that digitized the integrated charge from
each tankend.)  If the estimated energy exceeded a trigger threshold
the acquisition was triggered and the event was held to be read out.
If the energy was too low the event was rejected and the circuitry was
cleared.  Note that activity in a single box was sufficient to trigger
the acquisition.  The acquisition read out all boxes that had
sufficient energy to trigger.

There were actually two different energy trigger thresholds, sometimes
called Ehigh and Elow, and sometimes called muon and GC (gravitational
collapse) respectively.  For the events considered in this analysis, at
least one box had the more stringent (muon-level) trigger, but all
boxes in that supermodule that passed at least the GC level were read
out.

Channels in different supermodules had a different common stop.  If
a particle crossed a supermodule boundary, one cannot reconstruct the
timing using ERP information alone.  However, the stop signal for each
supermodule was sent to a CAMAC TDC called the interERP TDC\@.  By using
the interERP TDC one can relate the timing in different supermodules
and reconstruct the timing for the whole event.

\section{Streamer Tubes\protect\label{sec:st}}

A MACRO streamer tube consisted of a 12~m long wire at high voltage
(typically 2.5~kV) in a gas chamber of cross section 3~cm $\times$
3~cm formed by graphite-coated plastic at ground.  Primary ionization
from the passage of a high energy charged particle led to the
formation of a streamer which migrated to the wire, producing a pulse
of current on the wire.  All the streamer tube planes except those in
the lower vertical walls were also equipped outside the gas cell with
pickup strips at some angle to the wires ($26.5^\circ$ in horizontal
planes and $90^\circ$ in vertical planes).  The strips picked up an
oppositely-charged pulse by induction when a streamer formed under
them.  Wires and strips had a pitch of about 3~cm.

The wires and strips that resided in horizontal planes are called
central wires and strips, those on the East and West vertical walls
are called lateral wires and strips, and those on the North and South
vertical walls are called frontal wires (recall there are no strips on
the lower lateral or the frontal planes).

The streamer tube electronics used in this analysis provided simply
the digital information of which wires and strips had a pulse above
40~mV threshold within 10~$\mu$s of the trigger.  Offline software
uses this digital information to construct tracks in various views
(such as the central wire view, the central strip view, the lateral
wire view, etc.).  If one can unambiguously associate tracks in two or
more different views one can reconstruct the tracks in three
dimensions.

Specifically, to reconstruct a track in the central wire or strip
views, the standard MACRO tracking software (used throughout this
analysis) requires four or more collinear hits.  The lateral tracking
is more complicated.  In addition to the wire hits in the lateral
planes, which provide a hit in the (Y,Z) view, the tracking also pairs
central wire hits with central strip hits in the same plane to form
more (Y,Z) candidate coordinates.  To reconstruct a lateral track, the
software requires a total of at least 5 collinear hits, of which at
least 3 must be on lateral wires.  If there are 5 or more lateral
hits, no central hits are required.  In this analysis, frontal wires
and lateral strips are not considered for tracking.

Other electronics, not relevant to this analysis, recorded the digital
information of hits in a much-longer time window, and also recorded
(with limited precision) timing and charge information for each
streamer.

Further technical details on the scintillator and streamer tube
systems may be found in References~\cite{tech} and~\cite{tech01}.

\section{Calibrations\protect\label{sec:calibs}}

As will be detailed in Chapter~\ref{chap:meas}, the heart of the
experimental analysis is the recognition of upgoing muons using the
time of flight (TOF) in the scintillator system.  The scintillators
were intrinsically capable of subnanosecond timing resolution.
However, box-to-box variations and week-to-week variations made a
weekly calibration procedure critical to the success of the analysis.

\subsection{Timing Reconstructions\protect\label{sec:reconstructions}}

All timing reconstruction is done using the ERP TDCs for the hit
channels.  Within a single supermodule, all TDCs received a common
stop.  The time of phototube activity at a tankend is reconstructed
as

\[\calrec{TIME_i} = (TDC_i + \calrec{TWC})*\calcon{TICK} + \calcon{TOFF}\]

\noindent
where

\[\calrec{TWC} = \frac{\calcon{TWC1}}{\sqrt{ADC_i}} +
   \frac{\calcon{TWC2}}{ADC_i}\]

\noindent
is a timewalk correction which corrects for the fact that if two
pulses begin at the same time, the pulse of lower amplitude will cross
threshold later and thus start the TDC later (see
Figure~\ref{fig:timewalk}).  In the above, $i$ indicates the 0- or 1-
tankend, $TDC_i$ and $ADC_i$ are the measurements from ERP and
$\calcon{TICK}$, $\calcon{TOFF}$, $\calcon{TWC1}$ and $\calcon{TWC2}$
are calibration constants, determined separately for each channel of
each box on a weekly basis.

\picfig{fig:timewalk}{An illustration of timewalk.  A large pulse and
  a small pulse are superposed, beginning at the same time.  The small
  pulse crosses the TDC start threshold (indicated by the dotted line)
  later than the large pulse.}{\pawfig{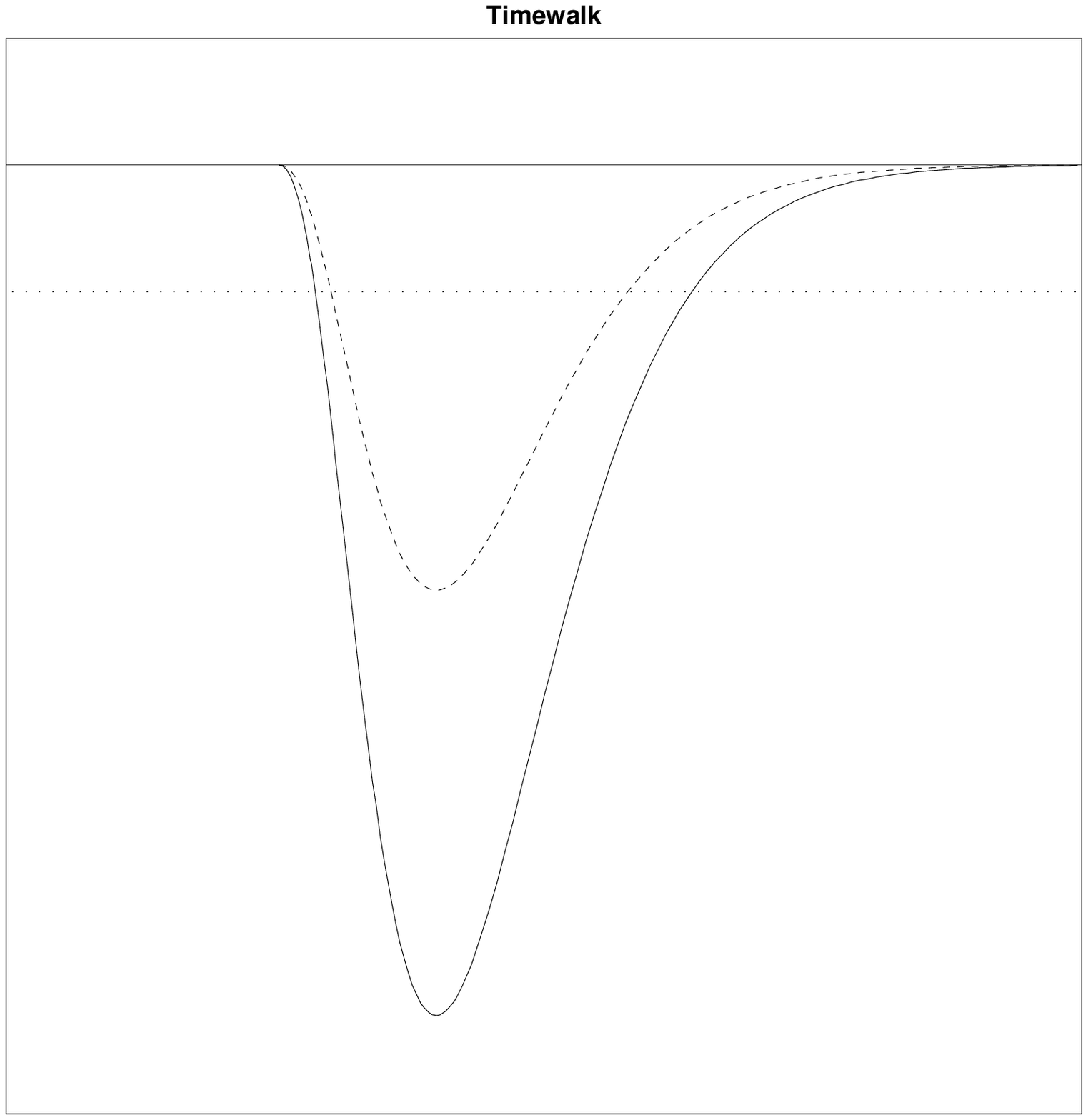}}

In calculating TOF, the time of light propagation within a
scintillator box cannot be neglected.  For a z-axis along the long
dimension of a box, with the origin at the center of the box, if a
particle passes through the box at location $z$,

\[ TIME_0 = T + (\frac{L}{2} + z)\frac{n}{c}\]
\[ TIME_1 = T + (\frac{L}{2} - z)\frac{n}{c}\]
\begin{equation} \protect\label{eqn:meantime}
  \calrec{T} = \frac{TIME_0 + TIME_1}{2} + const
\end{equation}
\[ \calrec{z} = \frac{TIME_0 - TIME_1}{2}\frac{c}{n}\]

\noindent
where $T$ is the time the light is produced, $L$ is the box length,
and $\frac{c}{n}$ is the speed of light in the oil.  So the difference
in meantimes of two boxes gives the time of flight between the two
boxes, independent of $z$.

Most events involve ERP hits in only one supermodule, and the above
``intraERP'' reconstructions are all that is needed to determine the
relative timing of all scintillator hits.  However, when scintillators
fired in more than one supermodule (and recall that for ERP purposes,
the North face and the South face are each a supermodule unto
themselves), an interERP adjustment must be made.  Two types of
information are available: each ERP system provides data about when
each box in that supermodule fired relative to the common stop signal
on that supermodule, and the interERP TDC system provides data about
when the common stop on one supermodule's ERP system occurred relative
to that on the other ERP system.  The latter information can be
reconstructed from the interERP TDCs using an interERP slope and an
interERP offset for each supermodule.  Putting the interERP and
intraERP information together, the reconstruction equation is

\vspace{-4ex}
\begin{eqnarray*}
  \calrec{\Delta T}&=&\calrec{TMEAN_a} - \calrec{TMEAN_b} +
     (IETDC_a*\calcon{IESLOPE_a}+\calcon{IEOFF_a}) - \\
  &&(IETDC_b*\calcon{IESLOPE_b}+\calcon{IEOFF_b})
\end{eqnarray*}

\noindent
where the IE (interERP) terms give the difference between the common
stops on the two different supermodules.  Here $\calrec{TMEAN}$ is the
mean time calculated using the intraERP constants (see
Equation~\ref{eqn:meantime}), $IETDC$ is the measured time in the
channel of the interERP TDC corresponding to the hit supermodule, and
$\calcon{IESLOPE}$ and $\calcon{IEOFF}$ are calibration constants.

In the case of more than two supermodules, or non-adjacent
supermodules, it may or may not be possible to adjust the event with
the interERP TDCs and calibrations that are available, as detailed in
the following paragraphs.

Every pair of supermodules was read out by a different computer; often
a pair of supermodules is referred to as a ``microvax,'' because three
microvax computers (one for each pair of supermodules) were used to
read out the equipment during runs.  In this nomenclature, Microvax~1
is SM1 and SM2, Microvax~2 is SM3 and SM4, and Microvax~3 is SM5 and
SM6.  Generally, if all the triggers in an event were in SM1 and SM2,
only Microvax1 would be read into the event record, etc.

Therefore, there was a different interERP TDC hardware module
physically located on each microvax; in an event only read out by
Microvax~1, the Microvax~1 interERP TDC would be read into the event
record.  With this TDC information, it is possible to reconstruct the
interERP timing among the North face, SM1 and SM2 (recall that, while
the North face is generally considered a part of SM1, it has a
separate ERP system and so an interERP adjustment is required to
reconcile ERP timing between the North face and other faces in SM1).
There was a sufficient number of N/SM1 and SM1/SM2 events each week to
calculate the necessary calibration constants, an interERP slope and
an interERP offset for each supermodule (including the North face, 6
parameters in all).  Similar considerations hold for Microvax~3 and
the SM5, SM6 and South face ERPs.

With the exception of interERP, Microvax~2 was primarily responsible
for reading out only SM3 and SM4; however, not only the common stop
signals from SM3 ERP and SM4 ERP, but also a copy of the common stop
signal from each of the other six ERP supermodules (N, SM1, SM2, SM5,
SM6 and S), was fed into its own channel on the Microvax~2 interERP
TDC\@.  There was a sufficient number of SM2/SM3, SM3/SM4 and SM4/SM5
events each week to calculate an interERP slope and an interERP offset
for the four supermodules 2-5, eight parameters in all.  Note that SM2
has one interERP slope and interERP offset for the hardware on
Microvax~1, and a different interERP slope and interERP offset for the
hardware on Microvax~2.  For a SM2/SM3 event, both Microvax~1 and
Microvax~2 triggered and read out their respective interERP TDCs, but
only the Microvax~2 interERP TDC is useful to adjust the timing
between SM2 and SM3 (because the Microvax~1 interERP TDC has no
information about when the SM3 common stop occurred).  With the
interERP slopes and offsets determined from SM2/SM3 events and SM3/SM4
events, it is also possible to adjust an event with hits only on, for
example, SM2 and SM4.

However, there were few events each week between N, SM1, SM6 or S and
either SM3 or SM4.  Thus, no calibrations are available for the
Microvax~2 interERP slopes and offsets for those supermodules.
Therefore, if an event occurred with ERPs fired in SM1 and SM4 only,
the event cannot be adjusted.  There is no information about SM4 in
the Microvax~1 interERP TDC, and the information present about SM1 in
the Microvax~2 interERP TDC is uncalibrated.

However, suppose SM2 also fired so there are hits in SM1, SM2 and SM4.
The relative timing of SM2 and SM4 is known from the Microvax~2
interERP TDC and calibrations, while the relative timing of SM1 and
SM2 is known from the Microvax~1 interERP TDC and calibrations.
Thus, it is possible to use SM2 as a bridge to synchronize calibrated
data from both the Microvax~1 interERP TDC and the Microvax~2 TDC:

\[ \mathrm{t_{SM1} - t_{SM4} = (t_{SM1})_{MVAX1} - (t_{SM2})_{MVAX1} +
  (t_{SM2})_{MVAX2} - (t_{SM4})_{MVAX2}} \]

\subsection{Calibration Procedure\protect\label{sec:calibproc}}

The calibration constants have been determined anew for every week the
detector was in operation.  For a given week, the inputs to the
calibration process are all the clean, well-tracked single muons
detected that week from the cosmic radiation, plus data from special
runs involving laser and LED that were performed at the beginning of
the week.  Actually, the special runs were not performed on all
supermodules every week, so some of the constants might remain static
for up to a few weeks at a time.

The special calibration runs were performed on one day a week,
designated as Calibration and Maintenance Day.  Any scheduled hardware
interventions (such as replacements and repairs of electronics or
phototubes) would occur first, and then the special calibration runs
would occur, including the newly-installed hardware.  The first normal
run after maintenance and calibration was used to label the set of
calibrations.  For example, 2~Aug~1994 was a Calibration and
Maintenance Day.  After maintenance and calibrations were finished,
the first normal run was Run~8035, beginning at 8:05~pm.  Normal runs,
with an average length of about 8~hours, continued for a week, until
Run~8057 was stopped at 8:47~am on 9~Aug to begin maintenance and
calibrations.  The special calibration runs on 2~Aug, along with the
muons collected in normal running from Run~8035 through Run~8057, are
used to produce the calibrations that are subsequently used to
reconstruct data in Runs~8035 through 8057.  This set of calibration
constants is referred to as Set~8035, or sometimes as Week~8035.

\subsubsection{LED Calibrations}

LED calibrations relied on a programmable pulse generator.  A digital
trigger output was fed into an ERP channel (the so-called ``LED fake
box'') while an analog output was fed to the LEDs in many boxes.  The
time delay between the trigger output and the analog output was
programmable.  (The pulser could also fake monopoles in special runs
not considered here.)  An LED run typically consisted of 16~events at
each of about 25 different delay settings ranging from 0 to 650~ns.
For each event the fake box was the first to trigger, while the timing
of other boxes relative to the fake box depended on the delay setting.
For each TDC, and for each delay setting, the average TDC value is
determined and plotted versus the programmed delay (see
Figure~\ref{fig:tdcSlope}); a linear fit is used to determine the
$\calcon{TICK}$ parameter.

\picfig{fig:tdcSlope}{Determination of the $\calcon{TICK}$ parameter.
  For a single box, for a single special LED run, TDC0H versus delay
  setting is shown.  Each dot is one event.  The boxed area is blown
  up in the inset, showing the typical spread in TDC values among the
  16~events at a single delay setting.  The parameter is determined by
  a fit to the linear portion of the graph.}{\pawfig{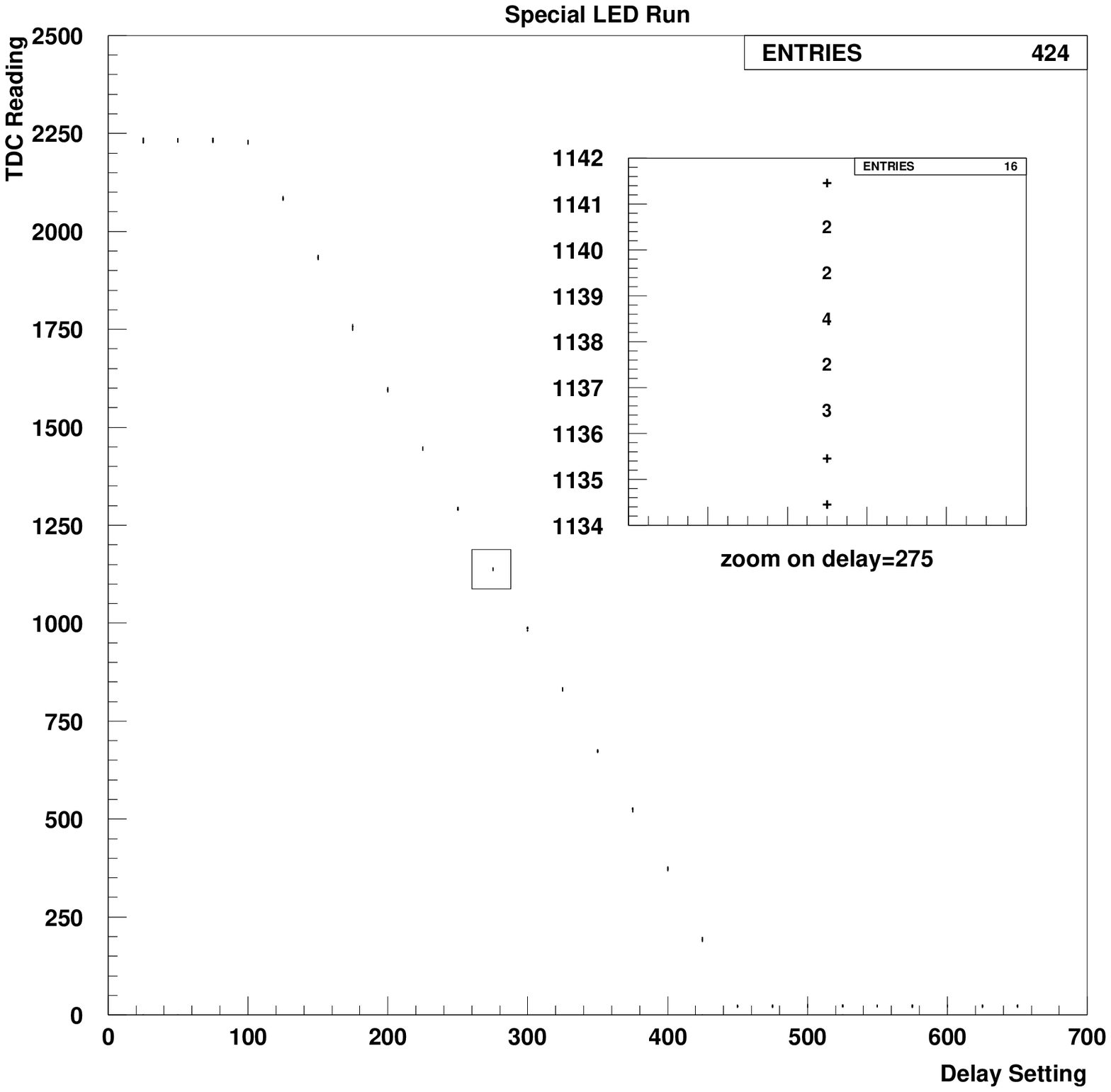}}

\subsubsection{Laser Calibrations}

One laser was fanned out through an optical splitter box and sent via
optical fibers to many scintillator boxes.  A computer-controlled
attenuator wheel was programmed to reduce the light before it reached
the splitter box.  Two stand-alone photomultiplier tubes measured the
light before and after the attenuator wheel respectively.  The
unattenuated PMT was discriminated and the digital signal was fed into
a channel of the ERP (the so-called ``laser fake box'' channel).

A laser run typically consisted of 50 pulses at each of about 20
attenuator settings.  In addition to energy calibrations (not used in
this analysis) the laser events are used to determine the timewalk
constants.  The fake box was the first ERP channel to trigger on each
event, and its timing was independent of the attenuator setting.
However, the timing of the other boxes relative to the fake box
depended on the amplitude of the pulse (due to timewalk) and thus
varied with the attenuator setting.  For each TDC channel, at each
attenuator setting the average ADC and average TDC are determined, and
the timewalk constants $\calcon{TWC1}$ and $\calcon{TWC2}$ are
determined to make the corrected TDC independent of attenuator
setting.

This procedure is underconstrained; there may be many choices for
$\calcon{TWC1}$ and $\calcon{TWC2}$ that give the right curvature.
The different choices may give different values for the corrected TDC
value, but that can be accommodated by the $\calcon{TOFF}$ offset
constant.  In practice, if the timewalk constants are determined anew
every time a laser calibration run is available, each of the
$\calcon{TWC1}$, $\calcon{TWC2}$ and $\calcon{TOFF}$ constants may
show considerable variation from week to week even when the apparatus
is fairly stable; the differences arise from random fluctuations in
the underconstrained timewalk determination.  Therefore, we chose to
update the timewalk constants very rarely.  Figure~\ref{fig:laser}
shows the performance of the timewalk correction in a laser run that
was taken a couple of years after the timewalk constants were
determined.

\picfig{fig:laser}{Determination of timewalk constants.  The data
  shown are for a single channel, for a single special laser run.  The
  lower set of dots gives the raw TDC values versus ADC values.  The
  upper set of dots gives the corrected TDC values using the
  parameters $\calcon{TWC1}=1163, \calcon{TWC2}=-1745$.  Each TDC
  count represents around 1/6~ns.}{\pawfig{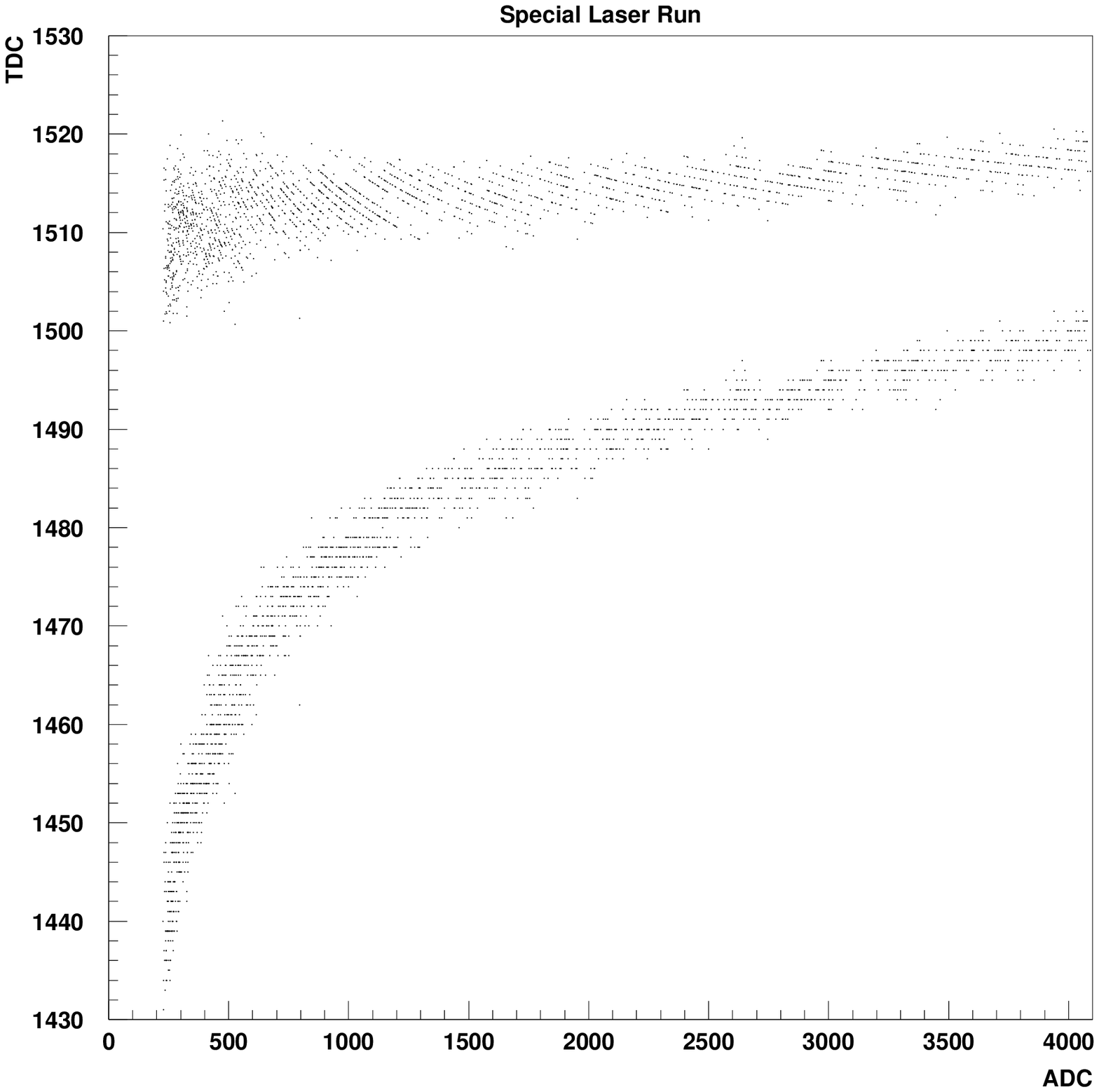}}
\subsubsection{Muon Calibrations}

During each week, clean, well-tracked single muons were collected.
From the streamer tube tracks, the position within hit tanks and the
flight path between tanks is determined.  $\calcon{TOFF}$ constants are
determined in a two-step procedure, making use of the $\calcon{TICK}$ and
$\calcon{TWC}$ constants already determined from the special runs.

First, the timing within a single tank is considered.  A linear fit is
made to a plot of reconstructed time difference between the two tank
ends versus the position determined from tracking (see
Figure~\ref{fig:muon1tank}).  The fit parameters give the effective
speed of light within the tank and the relative difference between the
$\calcon{TOFF}$s for the two ends of the tank.

\picfig{fig:muon1tank}{Timing within a single tank.  The data shown
  are for a single channel, for a week of muons.  Each dot is one
  event, with the x coordinate given by the point a streamer tube track
  intersects the 1200~cm tank; 0 is the center of the tank.  The y
  coordinate is the difference between the reconstructed time on the -0
  end of the tank and the -1 end, before the offset is corrected.  The
  reciprocal of the slope gives twice the effective speed of light in
  the tank, while the intercept shows the amount that must be added or
  subtracted to the $\calcon{TOFF}$ constant on one
  end.}{\pawfig{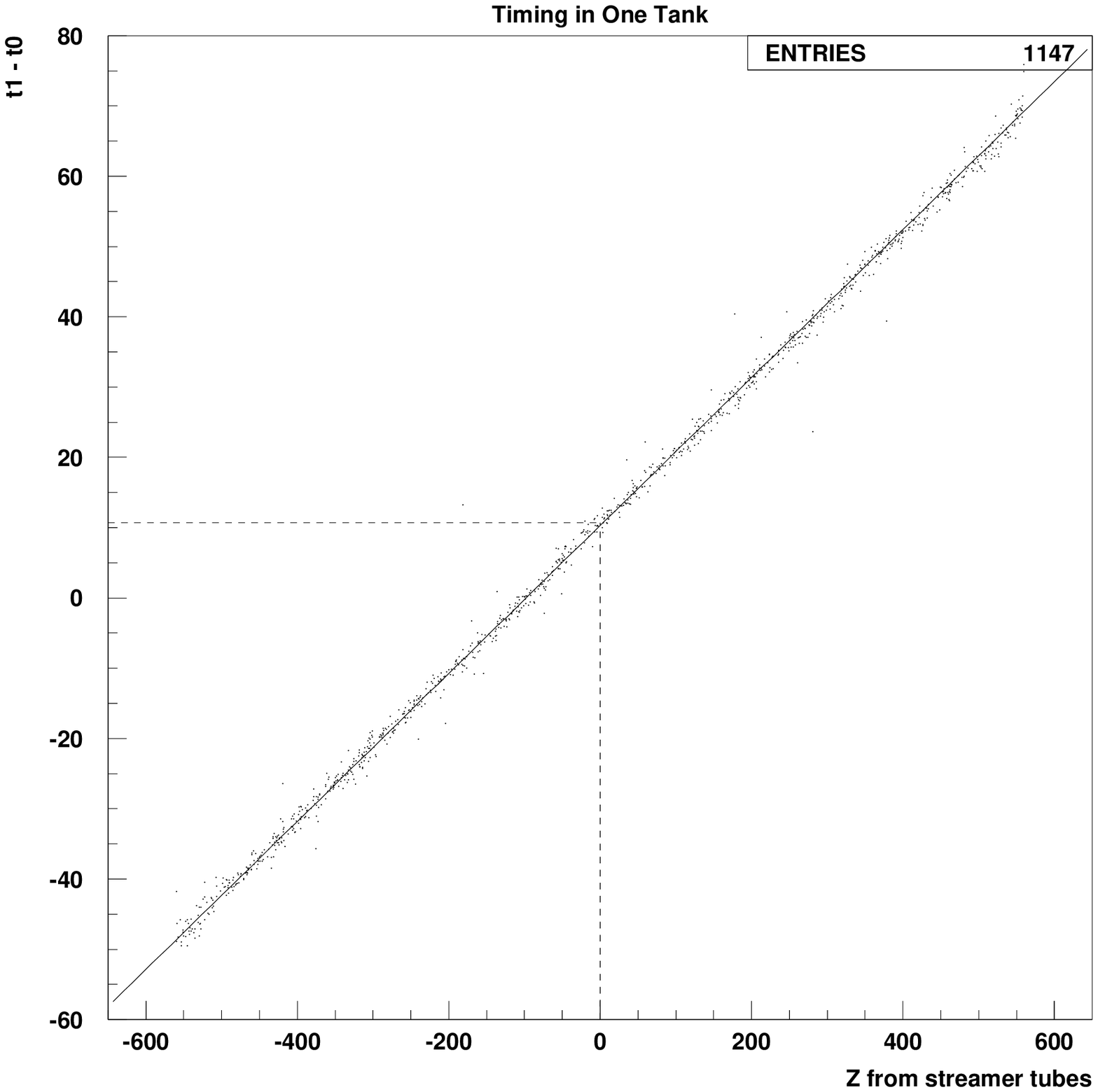}}

After this initial adjustment of $\calcon{TOFF}$, time of flight
between boxes is considered to determine the offsets of the various
boxes relative to one another.  An iterative procedure is used.  Using
an initial guess for the offsets (the results of the previous week's
fit, modified by the results of the single-tank offset adjustment),
the velocity of every particle is reconstructed.  If the mean computed
velocity of particles hitting a certain tank is not equal to $c$ (the
sample is dominated by downgoing relativistic muons), the offset
constant for that box is adjusted to reposition the mean to $c$.  This
will affect the calculated velocities of all events that hit that box,
and feed back to affect the mean velocity calculated for boxes in
other planes.  The iteration continues until the mean velocity for all
boxes is within a tolerance (typically 0.001) of $c$.

\subsubsection{InterERP Calibrations\protect\label{sec:ierpcal}}

After the intraERP constants are all determined using events that do
not cross supermodule boundaries, then interERP events from the weekly
sample of clean, well-tracked muons are used to determine the interERP
constants (a slope and an offset for each supermodule).  An
optimization is performed using the CERNLIB computer code
MINUIT~\cite{minuit} in which $\calcon{IESLOPE}$ and $\calcon{IEOFF}$
are allowed to freely vary to minimize the Mean Squared Error (that
is, to minimize the sum over all events of the square of the
difference between reconstructed and expected time-of-flight).

The North and South faces each had an independent ERP system.  As with
other supermodules, their $\calcon{TICK}$ and $\calcon{TWC}$ constants
can be determined using the laser and LED runs, and the relative
$\calcon{TOFF}$ difference between the two ends of each box can be
determined using muon tracks that intersect each box.  All of these
quantities are known before interERP calibrations begin.  However,
because these ERPs hosted only one plane, there are no multi-plane
intraERP events to use to determine the relative $\calcon{TOFF}$
constants between different boxes in the North or South face.
Therefore, $\calcon{TOFF}$ constants for each box in the North and
South faces are also variables to be determined by the interERP MINUIT
fit.

\chapter{Monitoring and Modeling the Detector \protect\label{chap:sim}}
\section{Muon Analysis\protect\label{sec:muonanal}}

With the exception of small effects (due to electron neutrino
interactions or nuclear recoil), this analysis is essentially an
analysis of muons in the MACRO detector.  Atmospheric muons (muons
produced in the atmosphere by cosmic ray showers) reached the detector
at the rate of about $10^4$ per day.  Neutrino-induced muons produced
outside the detector reached the detector about once every few days, while
the semi-contained neutrino-induced muons which are the subject of
this analysis were created once every few weeks.  Therefore, the
starting point for this analysis is a procedure for identifying and
characterizing muons in the detector.  Furthermore, the health of the
detector hardware can be monitored as a function of time via the muon
analysis applied to the copious atmospheric muons.

In addition to the measurement of semi-contained upgoing muons which
is the aim of this analysis, the analysis also uses studies of
downward throughgoing muons and downward stopping muons to monitor the
detector.  All of these analyses are extremely similar and, utilizing
a modular software environment, parts of the different analyses are
executed by the same code.  Here will be given a very brief overview of
the downward throughgoing analysis, necessary to understand the rest
of this chapter.  The semi-contained upgoing muon analysis will be
described in much more detail in Chapter~\ref{chap:meas}.

When a single event is analyzed, utilizing reconstructed streamer tube
tracks (see Section~\ref{sec:st}), information from at least two views
(wire, strip, or lateral) is combined to define a track in
three-dimensional space (a ``space track'').  Next, two or three
scintillator boxes that lie on the track or very near it and that
triggered in the event are identified and associated with the track.
If a space track cannot be created from the single-view tracks in the
event, or if there are not two scintillators that fired along a space
track, the event is rejected.  Otherwise, after a few more cuts to
ensure the quality of the data in the event, the time of flight
between two scintillator hits is determined using the timing
reconstructions (see Section~\ref{sec:reconstructions}), and the
pathlength between the two hits is determined by using the
reconstructed position along the box, along with a database of the
positions of all scintillator boxes.  (If there are more than two
boxes associated with the track, the two with the largest separation
are used to characterize the event.)  From these a velocity can be
determined which will be represented by the value
$\beta\,\equiv\,\frac{v}{c}$, with the convention that downgoing
particles have positive $\beta$ and upgoing particles have negative
$\beta$.  The vast majority of muons in MACRO have $\beta$ very near
unity.  A sample of downgoing muons can be created by selecting from those
events passing all cuts only the events with $\beta$ near +1.

Another quantity, {\bf tError}, can be defined as the difference
between the measured time of flight and the time of flight that would
be expected for a relativistic particle traversing the same
pathlength.  When the apparatus is well-functioning and
well-calibrated, {\bf tError} values cluster near zero.  Unlike
$\beta$, the experimental error on the measurement of {\bf tError} is
almost independent of pathlength.

Most detected muons, whether upward or downward, originate outside the
detector, and traverse the entire detector to stop somewhere on the
other side.  We may create a sample of muons that originate in the
detector (if they are upgoing) or stop in the detector (if they are
downgoing) by selecting events with incomplete tracks; that is, tracks
whose continuation strikes several streamer tube channels that did not
fire.  The exact algorithm will be detailed in
Chapter~\ref{chap:meas}.

Applying these analyses to all events in the dataset yields 24,000,000
downward muons (a rate of about ten per minute) of which 210,000
stopped (a rate of about one every ten minutes).  There are also about
690 externally-produced upward muons identified (about one every two
days), and a refined version of the analysis (described in
Chapter~\ref{chap:meas}) finds 76 identifiable semi-contained
neutrino-induced upward muons (about one every three weeks).

Figure~\ref{fig:upmuBeta} shows the distribution of the quantity
$1/\beta$ for all events passing a simple muon analysis.  (For fixed
pathlength, the error in $1/\beta$ is proportional to the error in the
measured time of flight; hence, the experimental data is often
presented as a distribution of $1/\beta$.)  Note that published MACRO
analyses on upward throughgoing muons use a more sophisticated
analysis which eliminates more mistimed events.

\picfig{fig:upmuBeta}{The distribution of $1/\beta$.  Downgoing events
  are at $\beta = +1$ and upgoing at $-1$.  (a)~is on a linear scale
  and (b)~is logarithmic.}{\twofig{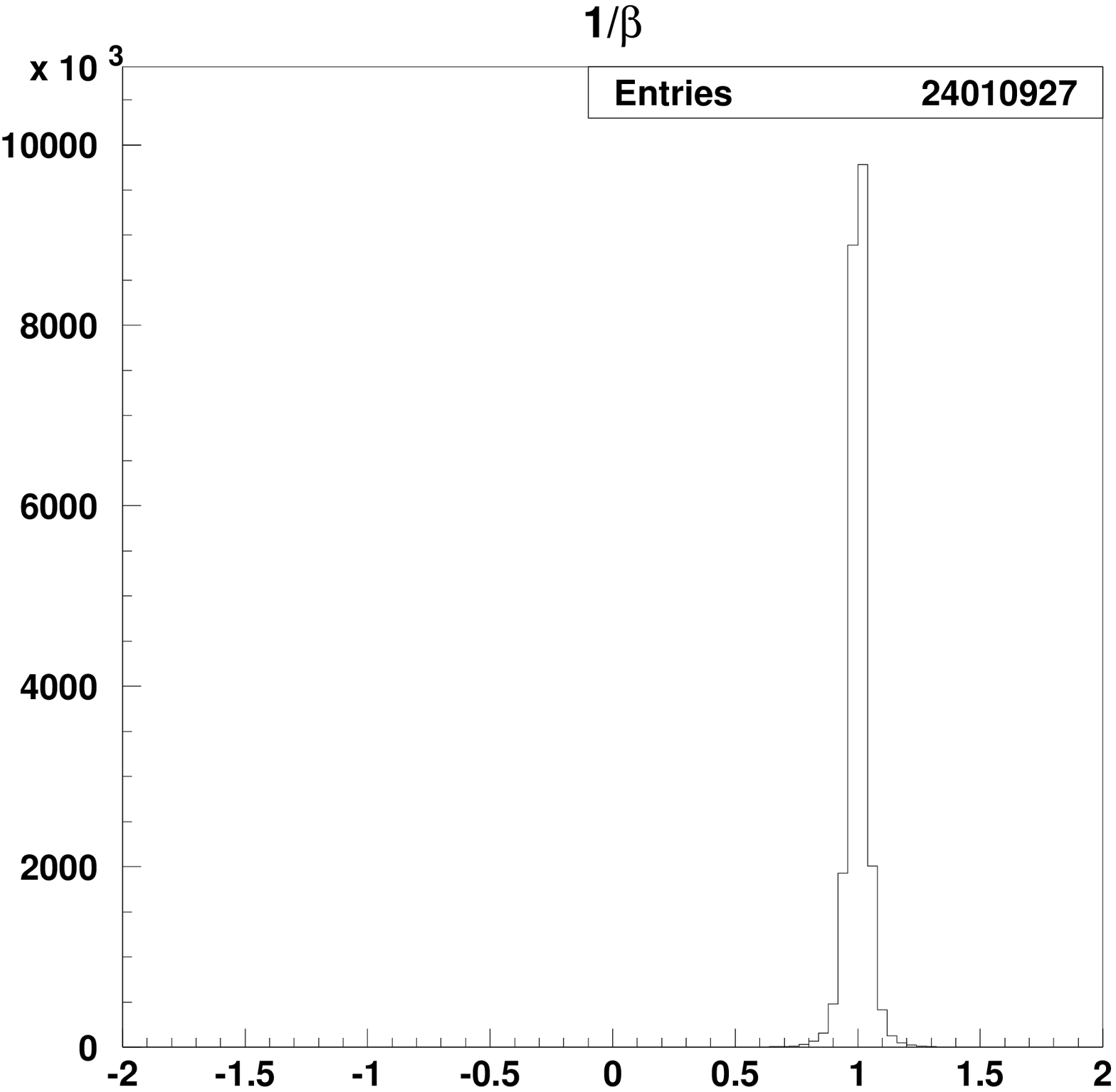}{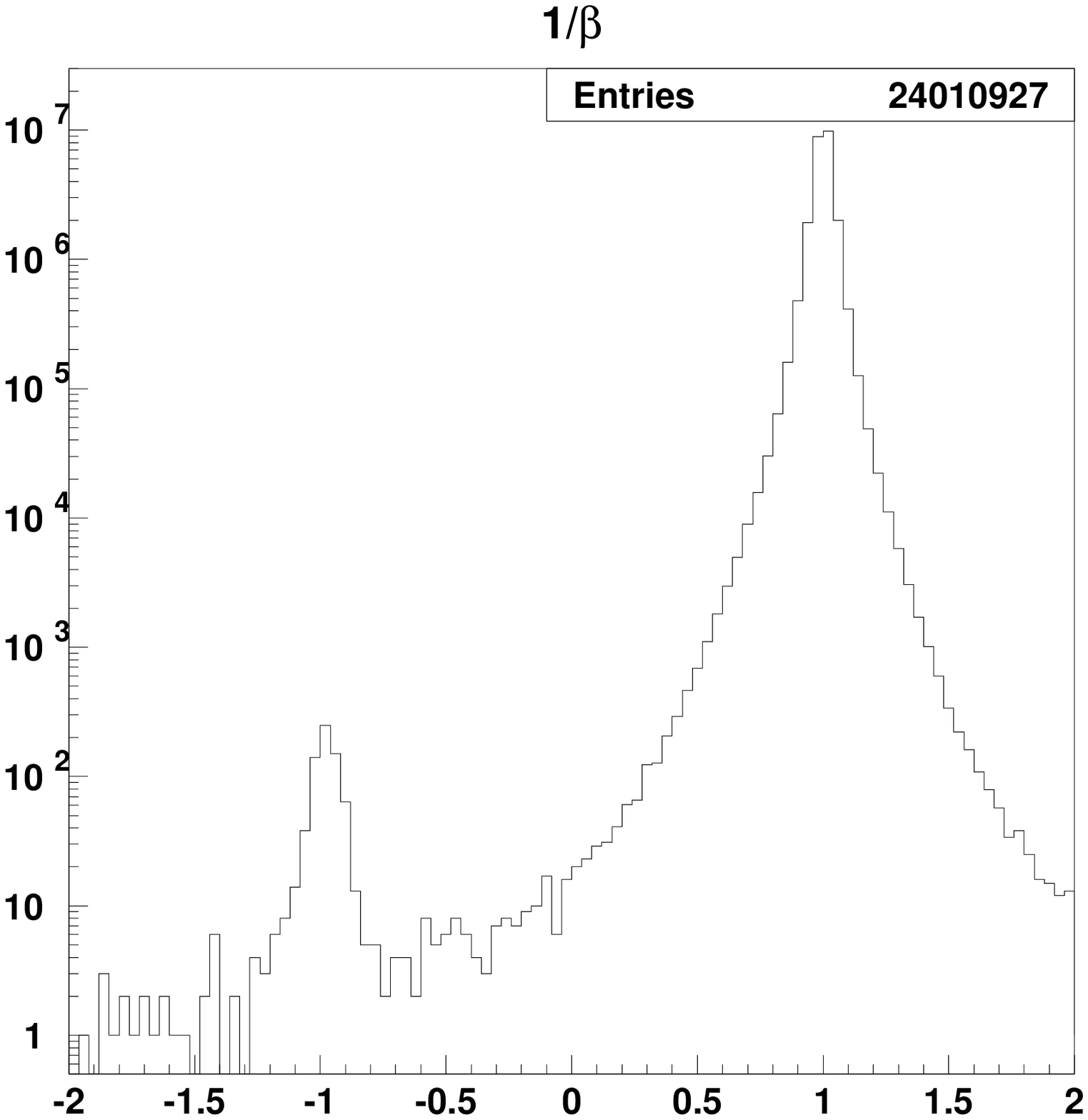}}

\section{Microcuts: Eliminating Bad Channels from the Analysis
  \protect\label{sec:microcuts}}

Using the calibrations generated as detailed in
Section~\ref{sec:calibs}, the timing can be reconstructed reliably in
most boxes most of the time.  However, due to hardware problems or
errors in the calibration software, results for a certain box may be
unreliable for a specific period of running time.  Because there were
more than $10^5$ potential background events (downgoing atmospheric
muons) for each signal event, it is necessary to eliminate any box
from the analysis if its timing was not reliable.  This causes a small
loss of acceptance for the analysis, but pays off with a big increase
in data quality.  These cuts are referred to as {\em microcuts}.

It makes sense to compute the microcuts on a weekly basis.  That is,
when reconstructing the data for each different week of running in the
dataset, a different list of boxes to be eliminated is in effect.  At
the lab, most hardware interventions took place on designated
``Calibration and Maintenance Days.''  The calibration constants are
produced on a weekly basis, to be in effect from one Calibration and
Maintenance Day to the next.  So problems due to bad calibration
constants or bad hardware often began and/or ended on Calibration and
Maintenance Day.

The microcuts are all determined empirically by looking at the real
data, in particular at the distribution of {\bf tError} (defined as the
difference between measured time-of-flight and expected time-of-flight
assuming downgoing relativistic particles) for events passing all cuts
of the standard muon analysis.  When a box was operating well and
is well-calibrated, {\bf tError} is distributed symmetrically about zero
with an RMS on the order of a nanosecond.  For each box, a decision is made
for each calibration period (usually one week) whether to apply a
microcut.  Figure~\ref{fig:scatter} shows the value of {\bf tError} for each
event for a single week of running, as a function of box number
(recall Table~\ref{tab:erpnum}).  
Figure~\ref{fig:meansig} shows the value of the mean and the RMS of
the {\bf tError} distribution for each box-week (i.e.,\ each of 476
boxes has one entry for each of 262 calibration periods, except for a
few boxes that were completely dead for some weeks).  A box is cut (by
a microcut) for any week in which the box's {\bf tError} distribution
has a mean more than $\pm\,3$~ns from zero, or an RMS of more than
2.7~ns.  Furthermore, there are weeks in which all the boxes have a
good mean and RMS, but in a particular supermodule there is an
unusually high number of ``outliers,'' events with {\bf tError} of
more than $\pm\,15$~ns.  Any week when a supermodule shows more than
ten events in the outlier region (after the microcuts for individual
bad boxes in that supermodule have already been applied), the entire
supermodule is cut for that week.  Figure~\ref{fig:microcuts} shows
the box-weeks that are cut by these criteria.  They are not randomly
distributed; out of 1004 box-weeks cut, 98\% of the cuts fall into one
of two categories:

\begin{figure}[p]
  \makebox[6in]{
    \resizebox{!}{3.5in}{\includefigure{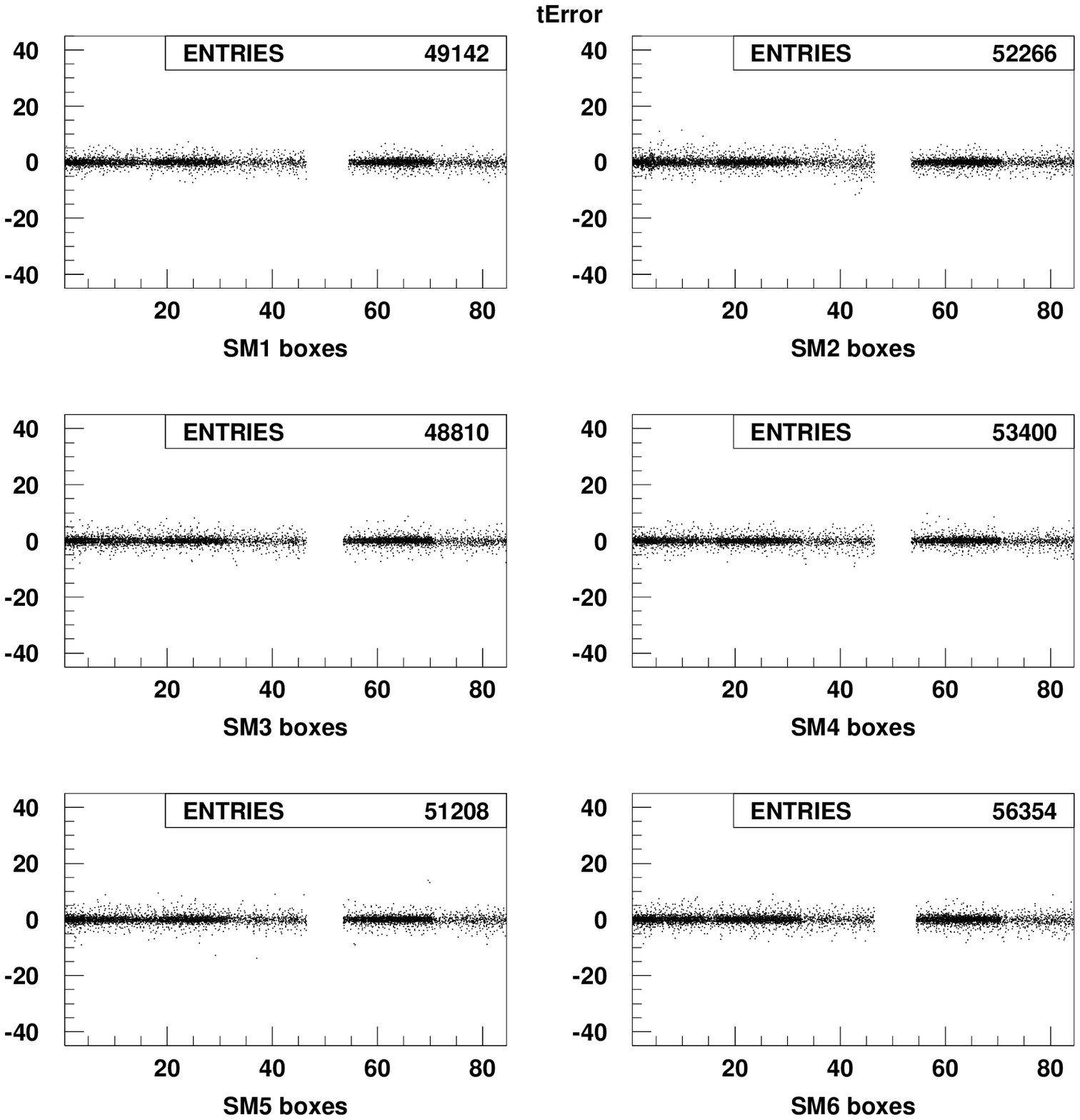}}
  }
  \caption{\protect\label{fig:scatter}
    Scatter plot of {\bf tError} for every box.  For every
  event, two dots are plotted: $+${\bf tError} for the entry box, and
  $-${\bf tError} for the exit box.  All of the events for a typical week are
  plotted.  \ignore{Problems are seen in box 62 of SM2 and boxes 78
    and 79 of SM5.}  The gaps from 47-53 are because the North face
  and South face boxes are not shown here; they are handled in the
  interERP microcuts.}
  \makebox[6in]{
    \resizebox{!}{3.5in}{\includefigure{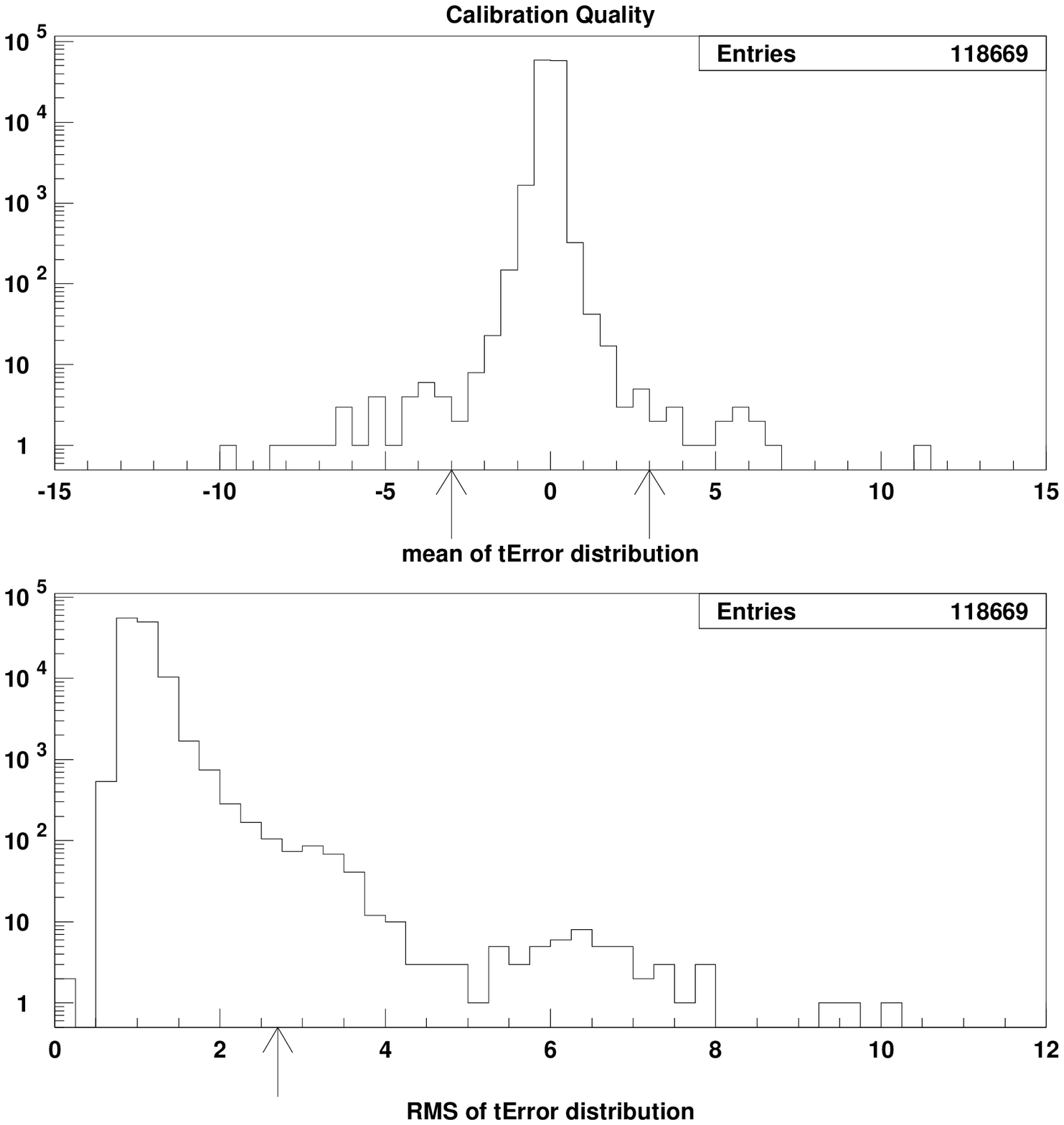}}
  }
  \caption{\protect\label{fig:meansig}
    Distribution of means and RMS of the {\bf tError}
  distribution, for all boxes and all weeks.  The arrows show the
  locations of the microcut criteria.  Note the logarithmic scale.}
\end{figure}
  
  \picfig{fig:microcuts}{Box-weeks eliminated by microcuts.  The
    vertical axis is the identification number of the box eliminated
    (see Table~\ref{tab:erpnum}).  The horizontal axis is the run
    number labeling the calibration week during which the box was
    eliminated.}  {\pawfig{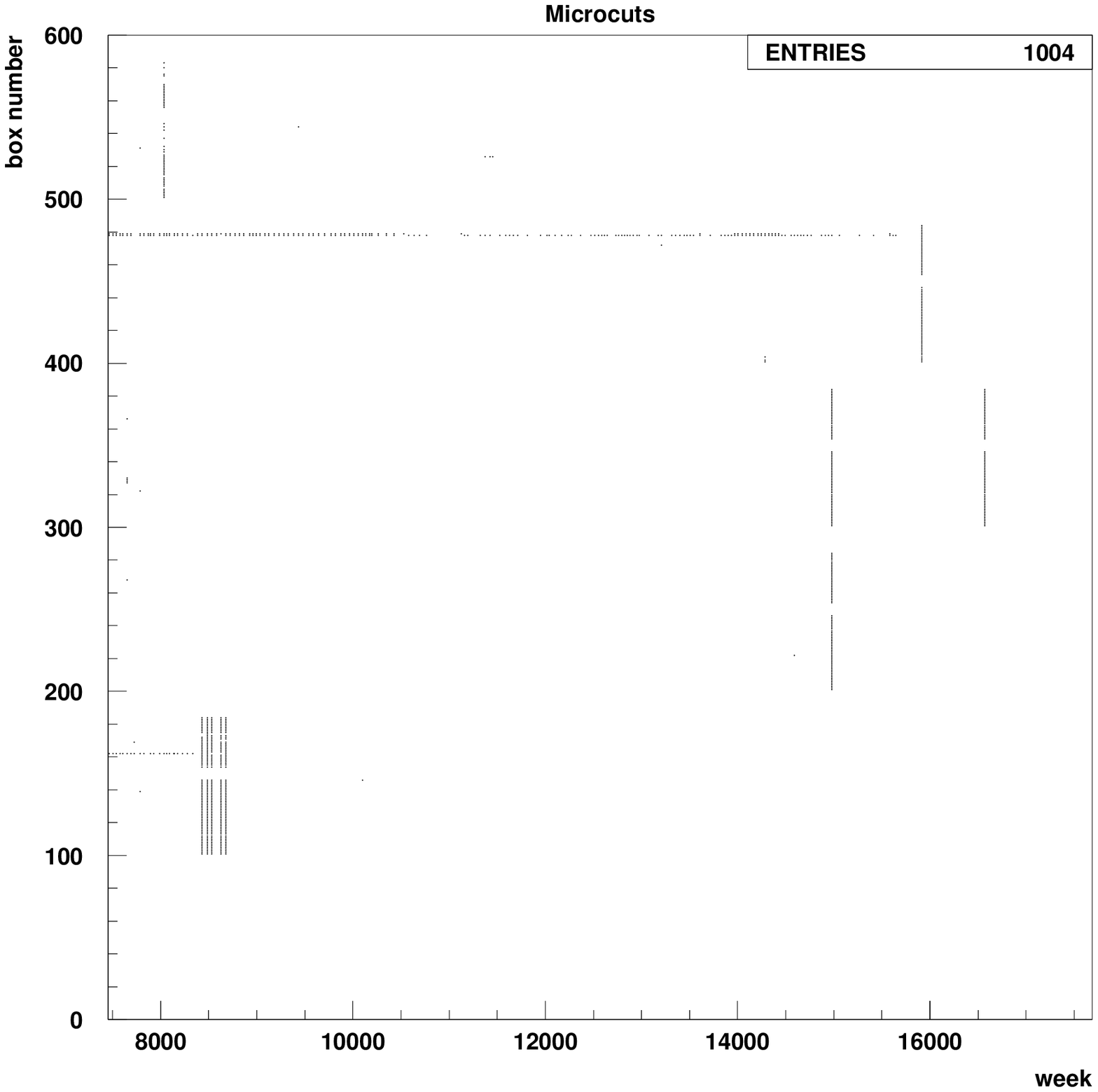}}
  
\begin{itemize}
  
\item Ten cases of entire supermodules cut.  In SM-weeks SM6-8035 and
  SM4-14978 almost every box had a bad mean and RMS\@.  In SM-weeks
  SM2-8430, SM2-8487, SM2-8534, SM2-8628, SM2-8677, SM3-14978,
  SM5-15194 and SM4-16571 the majority of the data was good (boxes had
  good mean and RMS) but there were an excessive number of outliers.
  
\item Three particular boxes had problems for extended periods of
  time.  Boxes 478 (5E08) and 479 (5E09) were bad throughout the first
  four years of the running period (actually, RMS was near the cut so
  the boxes are marked as bad for most weeks, but survive the cut for
  some weeks).  These boxes became normal after a hardware
  intervention in 1998.  Box 162 (2T09) was bad during the first
  several months of the running period.

\end{itemize}

Only 20 other box-weeks are cut.  All told, less than 0.1\% of all
box-weeks are cut.

Similarly, due to hardware or software failures, interERP events could
not be reliably reconstructed all the time, so additional microcuts
are applied to interERP events.  Because the North and South faces had
their own ERP systems, there are 7 interERP pairs of adjacent
``supermodules'' -- N/1, 1/2, 2/3, 3/4, 4/5, 5/6, and 6/S\@.
Distributions of {\bf tError} for all the events crossing a given pair
in a week are accumulated, and the pair is cut if the mean of the
distribution exceeds $\pm\,0.5$~ns, if the RMS of the distribution
exceeds 1.5~ns, or if there are ten or more outliers more than
$\pm\,15$~ns from zero.  See Figure~\ref{fig:ierpucuts}.

\picfig{fig:ierpucuts}{InterERP microcuts.  (a) shows the distribution
  of {\bf tError} for each of seven interERP pairs for a typical week's
  worth of data. (b) shows the distribution of mean and RMS
  for all pair-weeks.  The arrows show the locations of the microcut
  criteria.  (c) shows the pair-weeks that are cut by this microcut.}
  {
    \makebox[6in]{
      \resizebox{2.95in}{!}{\includefigure{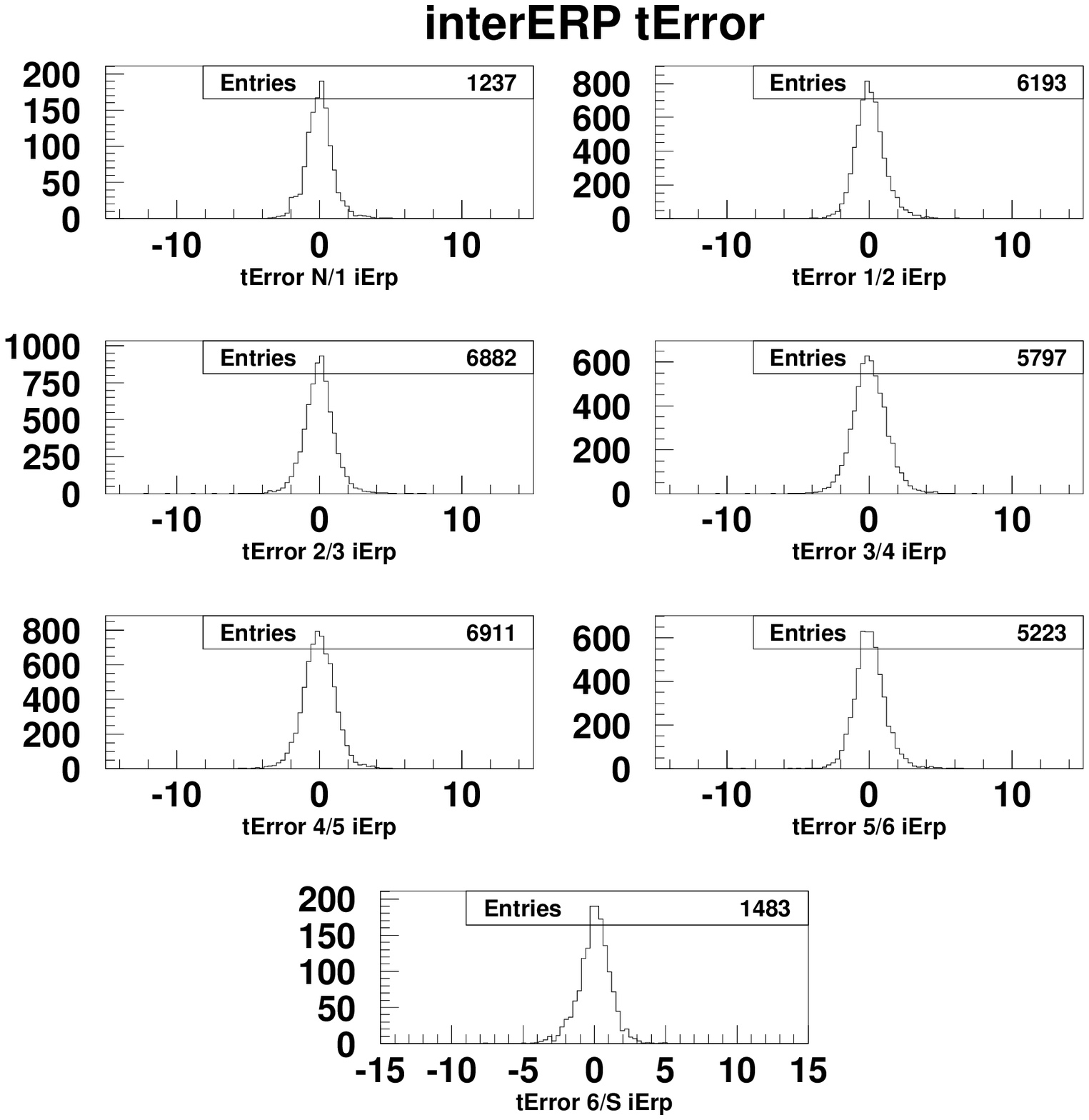}}
      \resizebox{2.95in}{!}{\includefigure{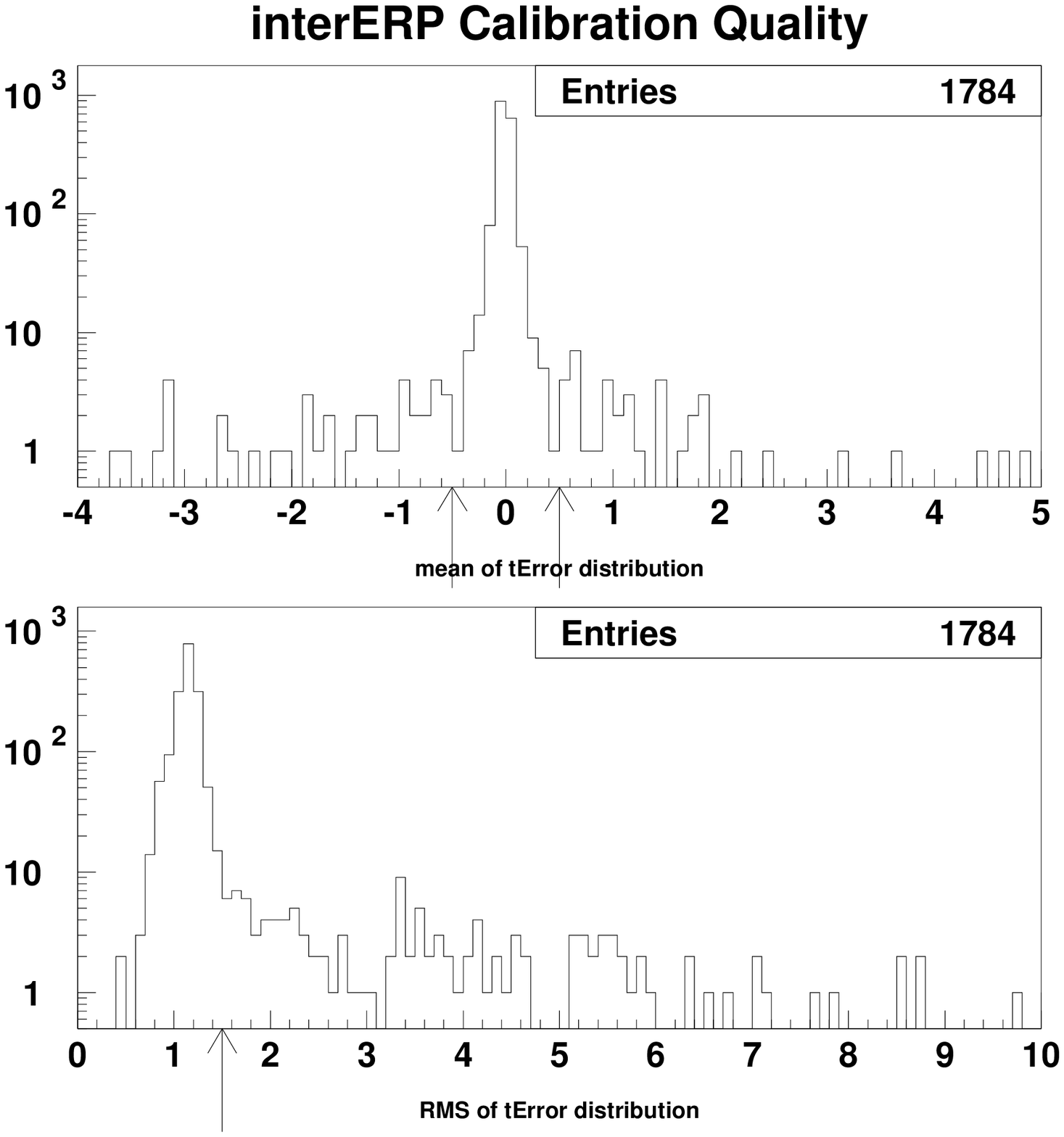}}}
    \makebox[6in]{\makebox[2.95in]{(a)} \makebox[2.95in]{(b)}}
    \makebox[6in]{\ } 
    \makebox[6in]{
        \resizebox{2.95in}{!}{\includefigure{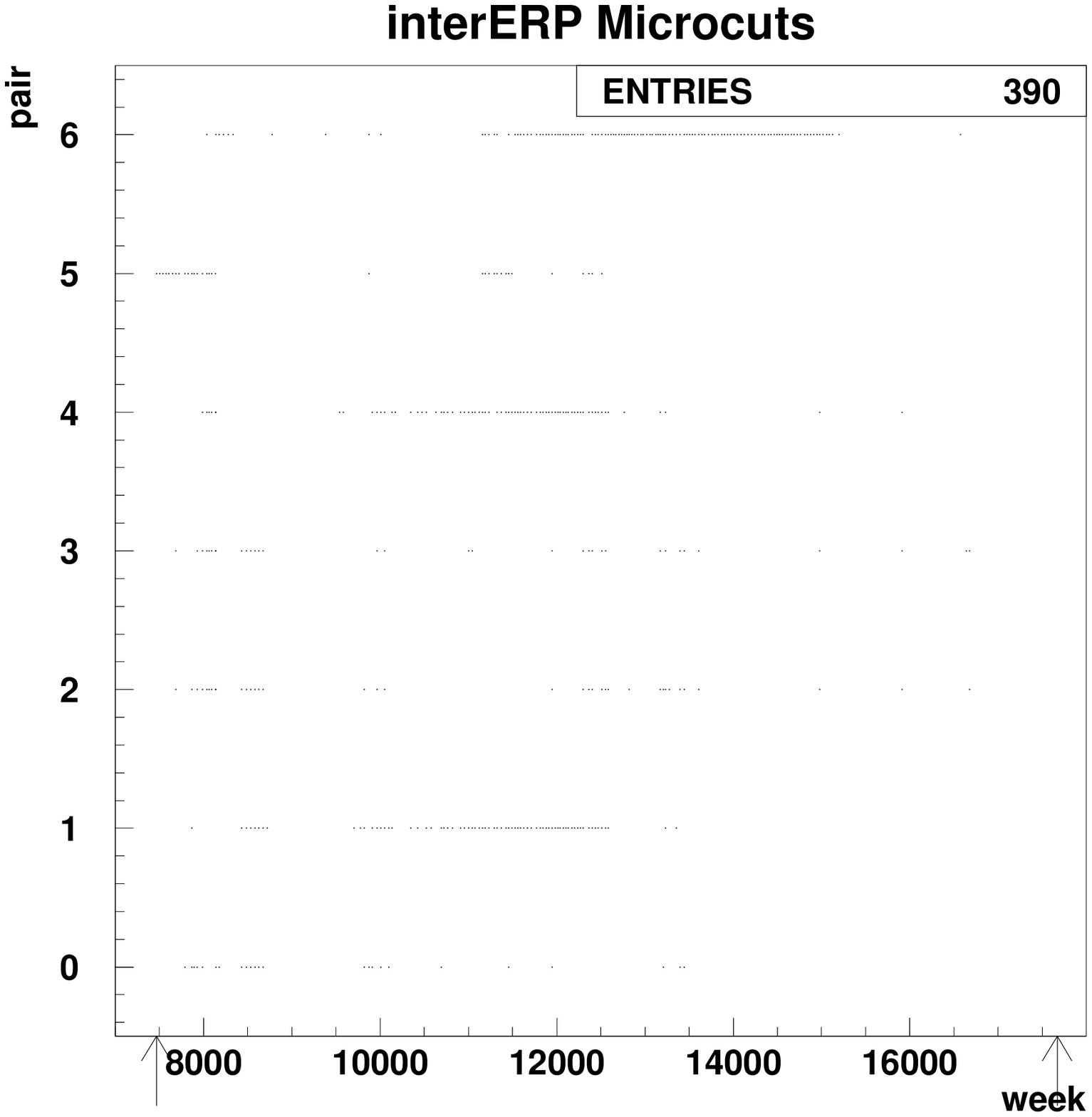}}
    }
    \makebox[3in]{\makebox[2.95in]{(c)}}
  }
  
Serious malfunctions of the interERP hardware were common.  The
interERP microcuts eliminate 390 pair-weeks, which is 21\% of all
possible pair-weeks.  All microcuts of both kinds reduce the number
of downgoing atmospheric muons reconstructed from the dataset for this
analysis by 0.5\%.

\section{Detecting Dead and Inefficient Channels\protect\label{sec:hwprobs}}

We will want to be able to calculate the event rate expected in the
detector under certain hypotheses about the neutrino physics.
To do so with the highest accuracy, we must be able to account for the
effect of imperfections in the detector -- channels that were dead or
inefficient for some period of time.

Around $10^5$ atmospheric (downward) muons traversed MACRO each week
and are fully reconstructed by the nominal muon analysis.  A
scintillation counter was hit about every 10-30~minutes on average
(depending on the type and location of the counter), and a streamer
tube channel (defined as a unit of eight wires, about 25~cm wide) was
hit about every 30-100~minutes.  These muons provide an excellent tool
for monitoring the efficiency of every part of the apparatus as a
function of time.

As with the microcuts, the efficiency is computed on a weekly basis.
The number of muons each week gives fairly good statistics for each
channel, and, as discussed in Section~\ref{sec:microcuts}, problems
due to bad calibration constants or bad hardware often began and/or
ended on Calibration and Maintenance Day.

Starting with events fully reconstructed by the standard muon
analysis, we may define a {\it hit} as an event in which a given
channel appears to be intersected by the muon track, and a {\it
  firing} as an event in which the channel was hit and it fired.
However, to understand the efficiency of a channel, it is not enough
to know just the ratio of firings to hits for a week.  For example, if
a channel were hit 100 times in a week and fired 50 times, there are a
couple of extreme possibilities.  The channel may have been 50\%
efficient all week; or it may have been completely dead for half the
week and fully efficient for half the week.  Also, it is worthwhile to
keep track of exactly when a channel was dead because several channels
may have been dead at the same time.  A supermodule that was
completely dead half the week and fully alive half the week would have
a different efficiency than one that had part of the SM dead for the
first half of the week, and the remaining part dead for the second
half of the week.

There was a constant probability per unit time of a muon hitting a
given type of channel, so the time between events is distributed
exponentially (see Figure~\ref{fig:mtbe}).  For each type of channel a
cut is defined that excludes less than 1\% of the exponential part of
the distribution.  If a channel did not fire for a period of time
exceeding the cut for that type of channel, the channel is considered
dead during that period.  Further statistical power is gained by
considering not only individual channels, but also groupings of
channels.  For example, if a ten minute period elapsed during which no
scintillator fired in an entire supermodule, we can be sure the
supermodule was dead for those ten minutes -- even though looking at
each channel individually, a ten minute gap would not be long enough
to be convincing evidence that the individual channel was dead.

\picfig{fig:mtbe}{Distribution of time between events, with
  exponential fit, for different types of channels for a typical week.
  The arrows show the location of cuts; whenever a channel did not
  fire for a period of time greater than the cut, it was considered
  dead for that period.  To the extent the distribution is truly
  exponential, the reciprocal of the fit slope is equal to the mean of
  the distribution, i.e.,\ the mean time between events.}{
  \makebox{\rule{0pt}{1in}\hspace{\textwidth}}
  \pawfig{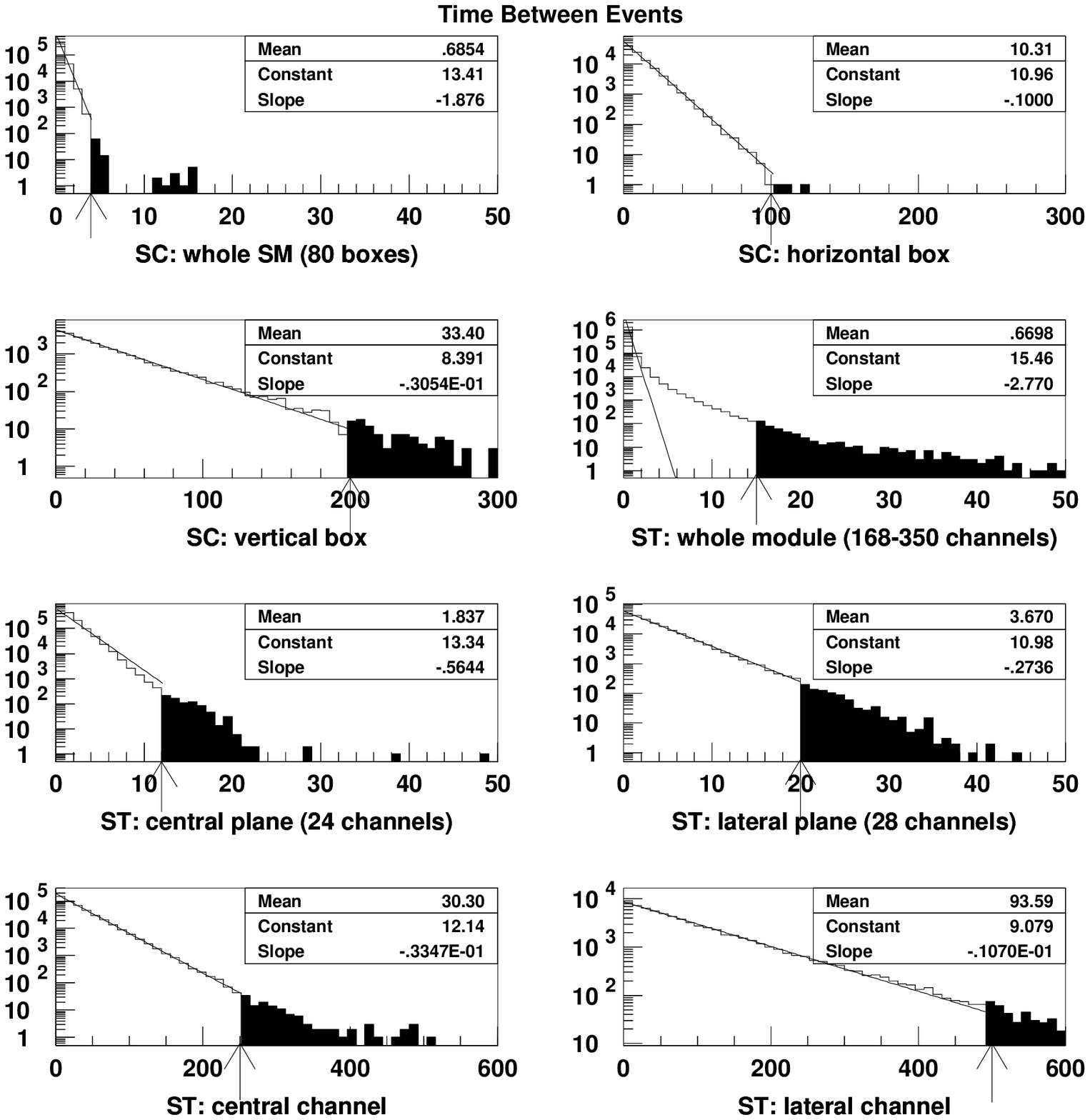}}

After all the dead periods for each channel are identified, the
efficiency (the ratio of firings to hits) is computed for each channel
summing over all the non-dead periods of the week.  For each channel a
database stores the beginning and ending time of each dead period, as
well as the efficiency for each week.

\section{Detector Simulation}

\subsection{Why the Detector Must Be Simulated}

We know (within the uncertainties to be detailed in
Chapter~\ref{chap:rate}) the flux of atmospheric neutrinos and the
differential cross sections for neutrino interactions with neutrons,
protons, and nuclei.  And we know how muons and other particles
created in neutrino interactions lose energy in matter and how our
detector recorded energy loss.  Therefore, it is possible in principle
to compute what the apparatus would measure.  However, all of these
processes are very complicated; what practical strategy can one follow
to calculate the prediction?

It is inherent within particle physics that we cannot predict exactly
what will happen when a high-energy charged particle traverses matter.
This is partly due to our ignorance -- we do not know exactly where
the molecules of matter are located or how close the high-energy
particle will approach them -- and partly due to the inherently
probabilistic nature of quantum mechanics.  While it is possible to
analytically calculate average quantities (for example, the average
energy loss per unit pathlength of a muon traversing Gran Sasso rock),
one can calculate efficiencies and acceptance of complex experimental
apparatus more accurately using an approach called Monte Carlo
simulation.

In the Monte Carlo approach, we use a computer program to simulate the
event of interest thousands or millions of times.  For example, we may
simulate 100,000 interactions of atmospheric neutrinos in the MACRO
apparatus.  For every high-energy particle produced by the
interaction, the program ``tracks'' the particle through the
apparatus.  This means it propagates the particle forward one
small step, then rolls the quantum mechanical dice to decide if it
undergoes a hard interaction, losing significant energy and changing
direction and possibly creating secondary particles, or if it merely
loses a little energy and continues on.  The program has a model of
the geometry of the detector so that it may track the particle from
one type of volume (say crushed rock) to another (say structural
iron).  If the particle is going through an active detector element
(in MACRO, that means liquid scintillator or streamer tube gas), the
energy loss is recorded, and at the end of the event the program
calculates what the detector would have recorded for that event.

For any one event, the results of the Monte Carlo are not reliable.
For example, it might happen that a real neutrino of a certain energy
and momentum interacts at a certain point in the real detector one
day, producing an upgoing muon that hits two scintillators and is
counted by the analysis software.  However, if one were to simulate
that interaction just once, it might happen that the upgoing muon
chooses to undergo a hard interaction, changes course, and exits the
detector through a crack.  It is only on average, if one simulates
tens of thousands of neutrinos of all energies and directions,
interacting at all points in the detector, that one obtains a reliable
estimate of the probability of detecting events of certain
characteristics.

There is an implicit requirement of using judgment to formulate
variables that are sufficiently generic to allow a meaningful
comparison (in physics terms) between the Monte Carlo prediction and
the physical measurement.  Nature has given us only about 100 events
in this analysis.  There is no way a Monte Carlo program could
predict, ``There will be one event that originated in the 5th plane of
the 4th Supermodule, hitting boxes 4C14 and 5W12; and one event
that....''  However, a Monte Carlo could accurately predict ``Only
20\% of the detected events will originate below the 5th plane.''

\subsection{Geant and GMACRO}

Geant 3.21~\cite{geant}, part of the CERN Programming Library, is not
so much a program as a set of facilities for creating Monte Carlo
programs.  It provides the following facilities (among others):

\begin{itemize}

\item {\bf Detector Geometry.} Geant provides subroutines for defining
  the detector geometry, which will be used by the tracking package.

\item {\bf Physics Processes.} Geant allows various processes that may
  occur as high-energy particles traverse matter to be turned on or
  off, and parameters governing the processes to be varied.
  
\item {\bf Tracking.} A subroutine call can introduce a primary
  particle at an interaction vertex, and Geant will subsequently track
  the particle.  As a result of the tracking, energy may be deposited
  in active detectors traversed by the primary particle, and secondary
  particles may be produced (which will subsequently also be tracked.)

\item{\bf Hits Management.} For detector volumes that are considered
  active volumes, Geant will keep track of all the energy deposited by
  all the primary and secondary particles.  Geant makes this
  information available after tracking is finished, when the detector
  response is being calculated.

\end{itemize}

GMACRO is an extension of Geant customized for the MACRO detector.
GMACRO includes calls to the Geant Geometry routines to define the
MACRO geometry, as well as event generators to randomly create initial
conditions for various classes of events (single atmospheric muons,
multiple muons, isotropic muon flux, atmospheric neutrino-induced
upgoing muons, etc.).  Relevant generators will be discussed in a
little more detail below.  GMACRO also contains code to calculate how
the actual MACRO detectors would respond to the energy loss determined
by Geant.  For the scintillators, GMACRO uses the time and amount of
energy loss to compute whether a box would create an ERP trigger, and
to compute the ADCs and TDCs that would be recorded in the ERP\@.  (It
does this by inverting the reconstruction equations from
Section~\ref{sec:reconstructions}, utilizing a special calibration
database filled with nominal values, identical for all boxes.)  In
streamer tubes, GMACRO probabilistically simulates the generation of
streamers and the charge induced on streamer tube wires; simulates the
charge induced on nearby strips; and determines which wires and strips
are above threshold.  Finally, GMACRO can write the output of each
event in a format almost identical to that written for real events in
the lab, so that both real and simulated data can be analyzed
by the same analysis software.  In addition to the simulated readout
hardware values, the event record for GMACRO events also contains a
record of ``Monte Carlo truth values,'' values known at the time the
event is simulated that cannot be determined perfectly from the
hardware.  For example, if the event is a simulated neutrino
interaction, the energy of the parent neutrino is recorded in the
event record.

The contained-neutrino interaction generator has a very central role
in this analysis, and will be described in some detail in
Chapters~\ref{chap:rate} and~\ref{chap:meas}.  Here is a brief description of two other
GMACRO event generators that play a smaller role in this analysis.

\subsubsection{The Atmospheric Muon Generator\protect\label{sec:atmugen}}

MACRO recorded the passage of several million atmospheric muons per
year (compared to just a few hundred neutrino-induced muons) so the
starting point to verify the accuracy of the simulation is to compare
simulated atmospheric muons with real data.  At the surface, the
initial muon intensity goes approximately as $\sec(\theta_{zen})$ where
$\theta_{zen}$ is the angle with respect to the vertical (the
so-called {\em zenith angle}) -- this approximation assumes a flat
atmosphere, and is good to within about $60^\circ$ of vertical.  The
energy distribution (for energies above 1~TeV, for which the muon has
a chance of penetrating down to Hall B) is approximately a power law,
going as $E^{-3.7}$, as derived from the power law distribution of
primary cosmic rays reaching earth and the scaling laws governing
shower evolution~\cite{gaisser}. (These scaling laws will be elucidated
a little further in Chapter~\ref{chap:rate}, in the context of
atmospheric neutrino production.)

The angular distribution of events reaching MACRO is determined not
only by this initial distribution at the surface but also by the shape
of the mountain.  Obviously in directions where the mountain cover is
thicker, a larger fraction of the initial atmospheric muons will range
out and not reach the detector.  An elevation map of the region around
MACRO was incorporated into the GMACRO atmospheric muon generator, and
iteratively corrected until now the mountain map is based on the
observed muon intensity from each direction.  In each direction, the
generator computes the initial distribution and then applies an energy
loss to that distribution according to the thickness of material.
Figure~\ref{fig:atmoang} compares the zenith and azimuth distributions
for real data and simulated atmospheric muons, for events passing the
standard muon analysis.  The figure suggests the simulation of
acceptance as a function of angle is accurate.  However, the energy
spectrum at the detector is not entirely accurate.  Standard GMACRO
uses a simple power-law spectrum of muons at the surface, but a more
accurate, and harder, surface spectrum given by Gaisser~\cite{gaisser}
is

\begin{equation} \protect\label{eqn:fancyspec}
  E_\mu^{-\gamma} \left\{ \frac{1}{1 + \frac{1.1E_\mu \cos\theta_{zen}}{115 GeV}}
      + \frac{0.054}{1 + \frac{1.1E_\mu \cos\theta_{zen}}{850 GeV}} \right\}
\end{equation}

\picfig{fig:atmoang}{Comparison of the zenith and azimuth
  distributions of real data and simulated atmospheric muons.  (a) All
  muons.  (b) Stopping muons.  Each graph shows the azimuth
  distribution for a slice of zenith angle values.  Real data is
  represented by the solid line, simulated data by the dotted
  line.}{\twofig{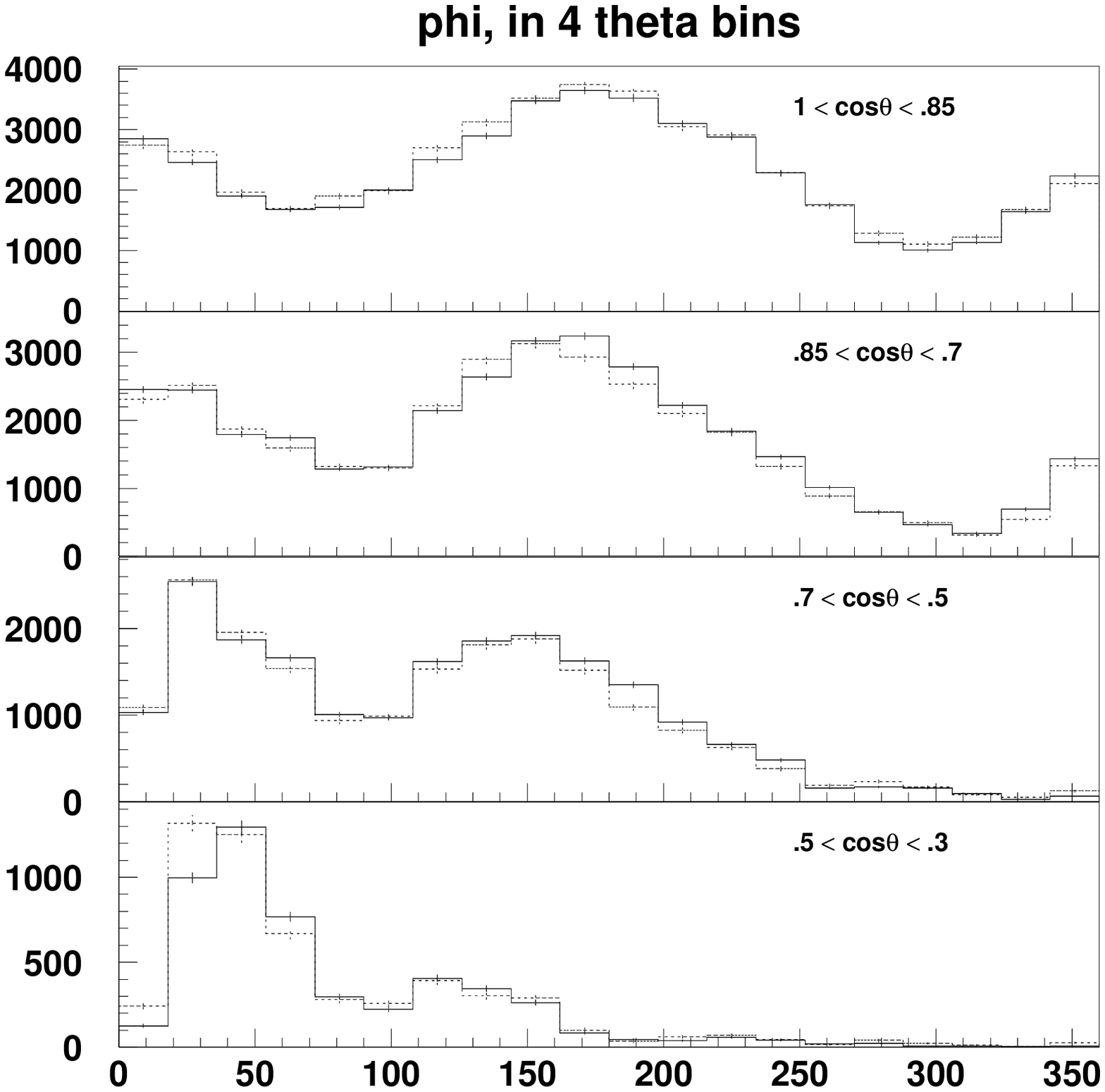}{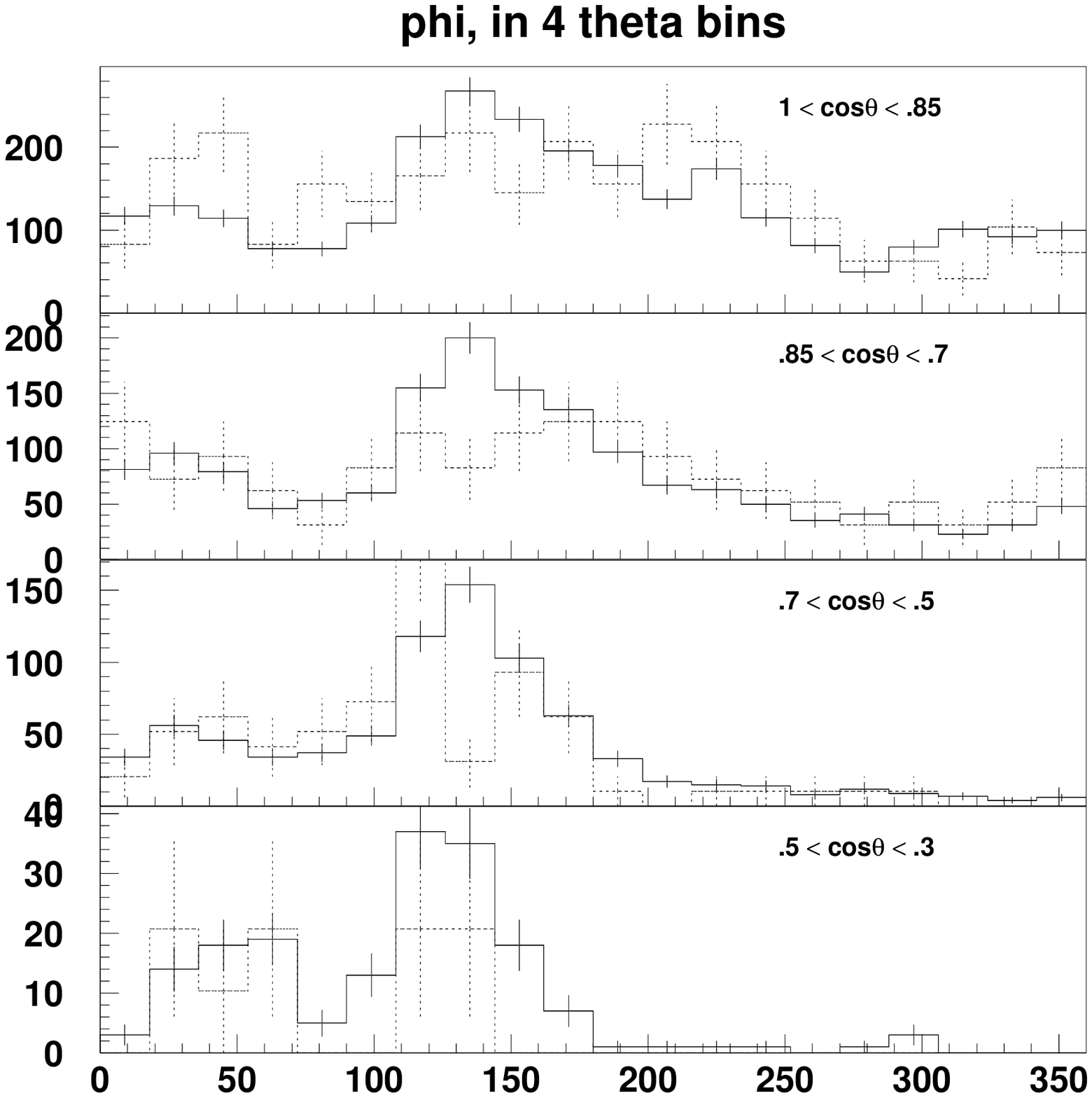}}

\noindent
where the first term arises from pion decay and the second from kaon
decay.  Muons need about 1~TeV at the surface to reach MACRO; at that
energy, the simple power law overestimates the vertical flux by 19\%,
while by $E_\mu \cos\theta_{zen}\, =\, 1000$~TeV the power law is
indistinguishable from Equation~\ref{eqn:fancyspec}.  Also, standard
GMACRO propagates muons from the surface to the detector using average
energy loss with no straggling.  Using standard GMACRO, the ratio of
identified stopping muons to muons passing the standard analysis
(mostly throughgoing) is 0.78\%, while in real data the ratio is
0.88\%.  Thus, the number of simulated stopping muons is too low by
11\%.  I modified GMACRO to implement Equation~\ref{eqn:fancyspec} at
the surface, and to propagate the muons to the detector
probabilistically using the survival probabilities calculated for Gran
Sasso rock by Bilokon, {\it et al.}~\cite{gsstraggle}.  This resulted
in a stopping fraction of 0.83\%, i.e.,\ 5\% fewer stopping muons in
the simulation than in the real data.  With only 527 simulated
stopping muons, the error due to counting statistics on the simulated
stopping fraction is 4.4\%, i.e.,\ the simulated stopping fraction is
$0.83\% \pm .04\%$.  In my judgment, the discrepancy in stopping rate
can be entirely accounted for by counting statistics and uncertainties
in the power law exponent and the Gran Sasso rock values of critical
energy and length scale for radiative loss.

\subsubsection{The Upgoing Muon Generator\protect\label{sec:upmugen}}

The upgoing muon generator used in this analysis, developed by MACRO
collaborator Colin Okada, provides more detail in the Monte Carlo
truth values than the standard generator distributed with GMACRO\@.  The
generator starts with a tabulated version of a published atmospheric
neutrino flux calculation (see Chapter~\ref{chap:rate} for a
discussion of what goes into a neutrino flux calculation).  Ignoring
any opening angle between the parent neutrino direction and the muon
direction, as well as fluctuations in the energy loss of muons
propagating through the rock toward the detector, we can
semi-analytically transform the neutrino flux into an upgoing muon
flux at the detector according to the equation

\[ \frac{d^3\Phi_\mu(E_\mu,E_\nu,\Omega)}{dE_\mu dE_\nu d\Omega} =
    \frac{\rho N_A}{\frac{dE}{dx}(E_\mu)} ~
    \frac{d^2\Phi_\nu(E_\nu,\Omega)}{dE_\nu d\Omega}
    \int_{E_\mu}^{E_\nu} dE_\mu^\circ 
         \frac{d\sigma (E_\nu,E_\mu^\circ)}{dE_\mu^\circ}
\]

\noindent
where $\Phi_\mu(E_\mu,E_\nu,\Omega) $ is the flux of muons with energy
$E_\mu$ derived from a parent neutrino with energy $E_\nu$ from
direction $\Omega$; $\rho$ is the density of rock in $g/cm^3$, $N_A$
is Avagadro's number, $\frac{dE}{dx}$ is the muon energy loss per unit
length, which is a function of muon energy; $\Phi_\nu(E_\nu,\Omega)$
is the flux of neutrinos; $\frac{d\sigma
  (E_\nu,E_\mu^\circ)}{dE_\mu^\circ}$ is the cross section for a
neutrino of energy $E_\nu$ to undergo a charged current interaction in
the rock and produce a muon of energy $E_\mu^\circ$.  In this
scenario, the muon is produced with energy $E_\mu^\circ$ some distance
from the point where the flux is measured, by which time the muon has
lost energy to $E_\mu$.

The neutrino cross section is computed by integrating the naive parton
model cross section for neutrino-nucleon interactions over the parton
distribution functions GRV94-LO~\cite{GRV94}.  The average muon energy
loss in rock surrounding the detector is taken from the computation by
Lohmann, {\em et al.}~\cite{lohmann} (and recall energy loss
fluctuations are ignored).

The generator produces muons at the detector according to the
calculated flux.  The event record stores the parent neutrino energy,
the zenith angle of the parent neutrino with respect to vertical at
the detector, the initial muon energy (before energy loss while
propagating from the interaction point to the detector), the distance
the muon has traveled to reach the detector, and the muon energy at
the detector.
\subsection{Tuning the Monte Carlo\protect\label{sec:mctune}}

Both Geant and GMACRO provide a number of discrete and continuous
parameters which must be chosen carefully to provide an accurate
simulation without wasting too much computer time calculating
irrelevant processes.  The parameters used in this analysis are as
follows:

{\tt
  \hskip 1in TRAK  1 \hfill

  \hskip 1in HITS  1 \hfill

  \hskip 1in DIGI  1 \hfill

  \hskip 1in RAWD  1 \hfill

  \hskip 1in NSMOD 6\hfill

  \hskip 1in NHMOD 2\hfill

  \hskip 1in MULS  1\hfill

  \hskip 1in LOSS  1\hfill

  \hskip 1in ANNI  1\hfill

  \hskip 1in BREM  1\hfill

  \hskip 1in COMP  1\hfill

  \hskip 1in DRAY  1\hfill

  \hskip 1in PAIR  1\hfill

  \hskip 1in MUNU  1\hfill

  \hskip 1in RAYL  0\hfill

  \hskip 1in DCAY  1\hfill

  \hskip 1in PFIS  2\hfill

  \hskip 1in PHOT  1\hfill

  \hskip 1in HADR  5\hfill

  \hskip 1in CUTS 10*0.0005 \hfill

  \hskip 1in SETS    1='CPLS' 'LPLS' 'FPLS' 'CPLT' 'LPLT' 'FPLT'\hfill

  \hskip 1in SETS    7='HTPT' 'TPT1' 'TPT2'\hfill

  \hskip 1in SETS    10='HTPS' 'TPS1' 'TPS2'\hfill

  \hskip 1in SETS    13='SQTP'\hfill

  \hskip 1in c WNOI 1\hfill

  \hskip 1in c SNOI 1\hfill

  \hskip 1in TIME 0. 0. 0\hfill

  \hskip 1in TRHW 120*6.87\hfill

  \hskip 1in TRHS   1=2.55 3*3.15 2.55 3*3.0 2.55 3.0\hfill

  \hskip 1.5in {\textmd{\it  \small(Same TRHS parameters repeated for all 
  twelve modules)}}

  \hskip 1in ELAT 0.85\hfill

  \hskip 1in TRWT 6.5\hfill

  \hskip 1in TRST 2.65\hfill

  \hskip 1in TRLT 9.5\hfill

  \hskip 1in TROT 200.\hfill

  \hskip 1in BACK 'BOTH'\hfill

  \hskip 1in ERPP 1=1 2=10 3=5 4=12 5=07\hfill

  \hskip 1in ERPP 6=0.55 7=0.55 8=1.4 9=.7 11=1\hfill

  \hskip 1in PNRM 2.\hfill
}

One would have to read the Geant and GMACRO documentation to know the
exact meaning of each card, but some of the more important choices are
described here.

All relevant physics processes are turned on except Rayleigh
scattering.  All types of particles are tracked until they have lost
energy and have less than 0.5~MeV remaining (as controlled by the CUTS
parameters).

The propagation of low energy neutrons in the detector requires some
special attention.  GHEISHA~\cite{gheisha}, the default hadron tracking
code in Geant, is only valid down to neutron energies of about 20~MeV\@.
However, this is not really adequate.  Even a very low energy neutron
may propagate up to a few meters through the rock in the interior of
the detector.  And if a neutron reaches a scintillator tank, even if
its kinetic energy is less than an MeV, it may be captured by a carbon
nucleus with the release of several MeV -- enough to fire the ERP
trigger.  Thus, it is necessary to track neutrons below 20~MeV\@.
However, Figure~\ref{fig:neutron_bounce} shows that GHEISHA gives
implausibly long and complex tracks when asked to track neutrons as
they lose energy all the way down to 0.5~MeV\@.  To improve the
situation, an alternative hadron tracking code purported to track
low-energy hadrons more accurately, MICAP~\cite{micap}, was
incorporated into Geant and selected by setting the HADR parameter to
5.  As seen in the figure, it gives plausible results for the
low-energy neutrons.

\picfig{fig:neutron_bounce}{Simulation of low energy neutrons.  The
  upper figures show, in addition to MACRO channels that fired, the
  starting and ending points of secondary particles (dashed lines).
  The lower figures add indicators of all GMACRO elements through
  which a particle passed.  (a)~shows the improper simulation of the
  GHEISHA hadron code. (b)~shows the same event, this time using the
  MICAP hadron interaction code.}
{\twofig{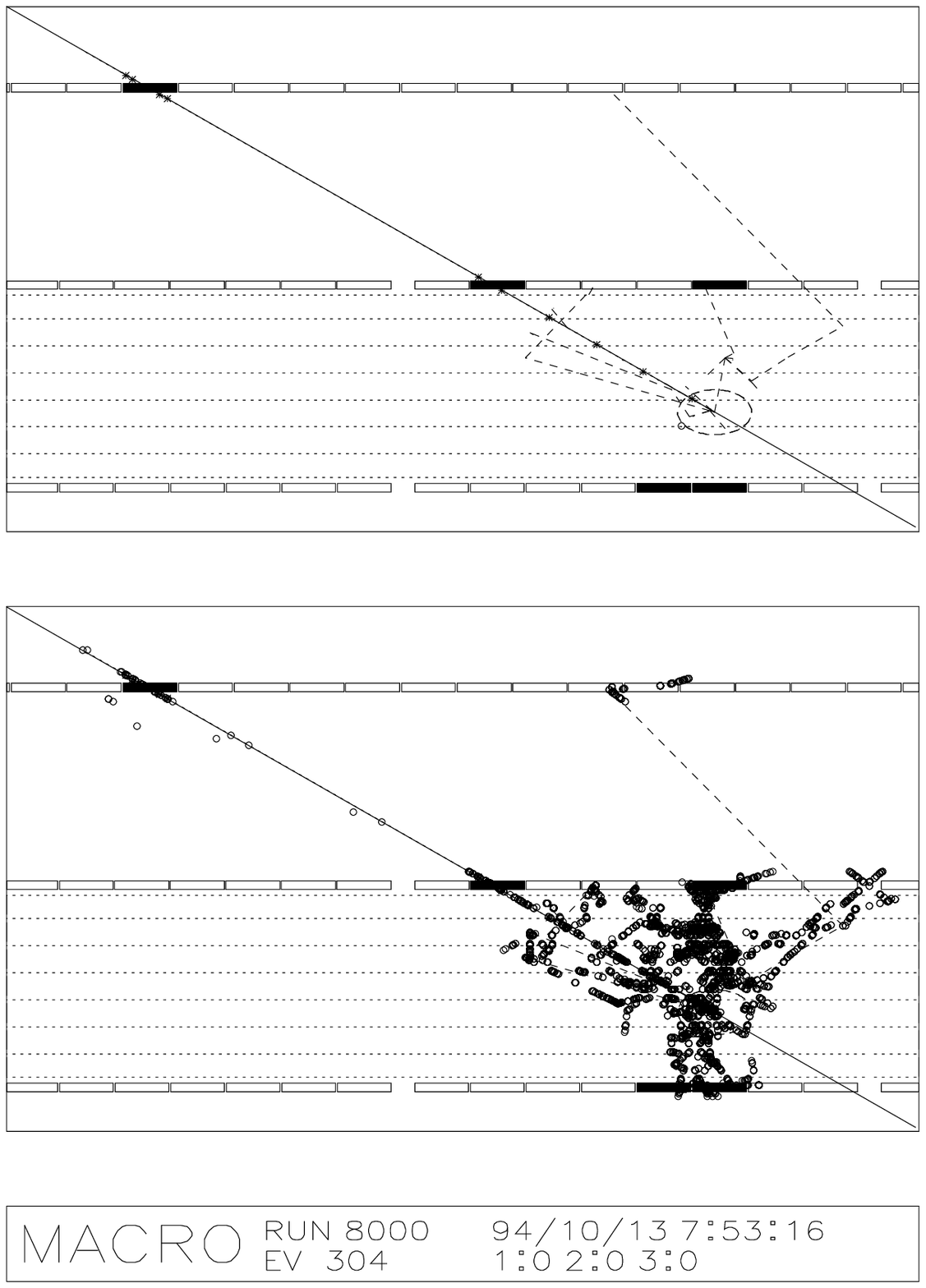}{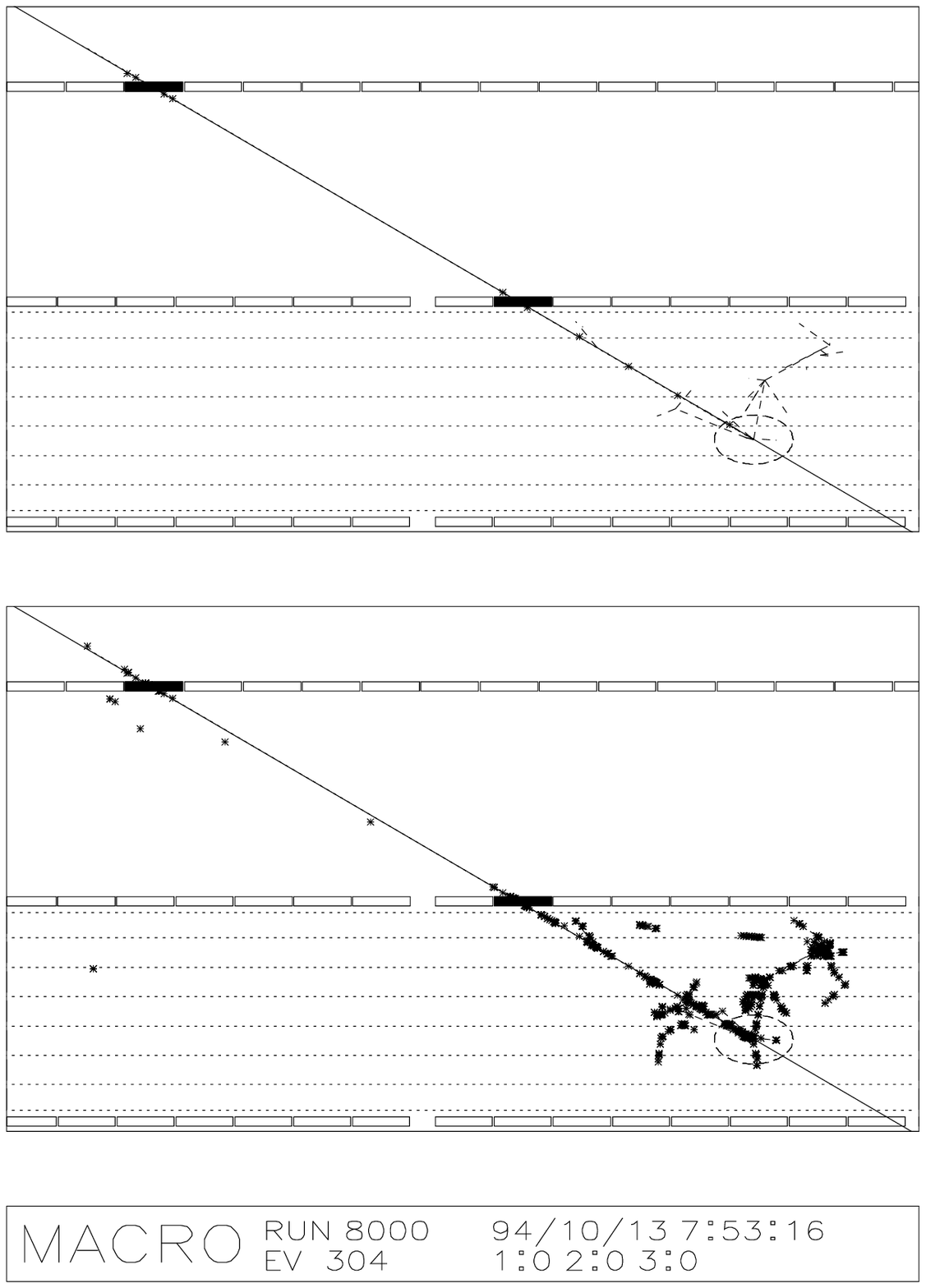}}

Measurements made by MACRO are not perfect.  The timing and position
derived from the high-threshold TDCs do not agree perfectly with the
low-threshold TDCs, nor does the position derived from ADCs agree
perfectly with position derived from TDCs.  Sometimes these problems
arise because more than one particle hits the same box, at different
places and different times (this is especially likely in
contained-vertex neutrino interactions).  If only one particle hits
the box, GMACRO could give almost perfectly-consistent results in all
channels provided that certain non-ideal behaviors were not simulated.

GMACRO models errors in timing as due to two sources -- a gaussian
error due to details of how the light propagates from its point of
generation to the tank end where it is detected, and a gaussian error
in the accuracy of time digitization in the ERP\@.  The first error
results in both TDCH and TDCL at a given tank end making the same
(slightly-incorrect) measurement, while the latter introduces
differences between the measurements of TDCH and TDCL\@.  The default
value of the gaussian error in the digitization (ERPP 8=0.7 ticks)
results in a distribution of time(TDCH)-time(TDCL) that is too narrow;
a value of 1.4 ticks reproduces the data well (see
Figure~\ref{fig:timeHL}a).  The error for position derived from TDCs
is dominated by the uncorrelated variation in time of propagation to
the two different tank ends.  For real data, our best estimate of
where the particle really went through the tank is derived from
streamer tube tracks, so ERPP 6 and 7 (for horizontal and vertical
tanks respectively) are tuned to match the Monte Carlo and real-data
distributions of the distance from a track to a scintillator hit as
determined by scintillator timing.  The default value of 0.7~ns gives
too large a mismeasurement; the value of 0.55~ns fits the data well
(see Figure~\ref{fig:timeHL}b).

\picfig{fig:timeHL}{(a)~Distribution for real and simulated events of
  the difference between time determined from high-threshold TDCs and
  time determined from low-threshold TDCs.  (b)~Distribution for real
  and simulated events of the square of the distance from the location
  of a scintillator hit, as determined by scintillator timing, to the
  associated streamer tube track.}{\twofig{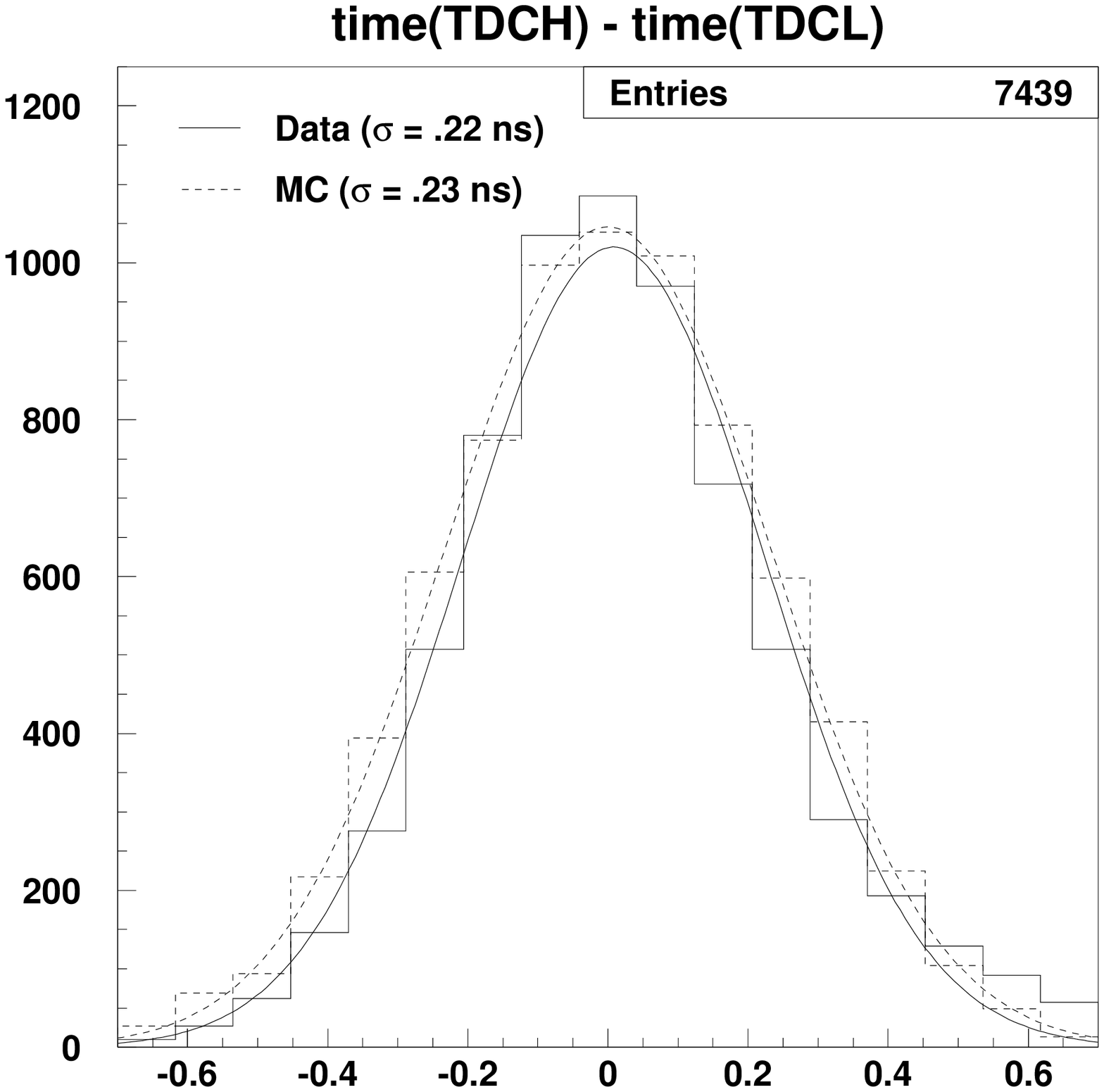}{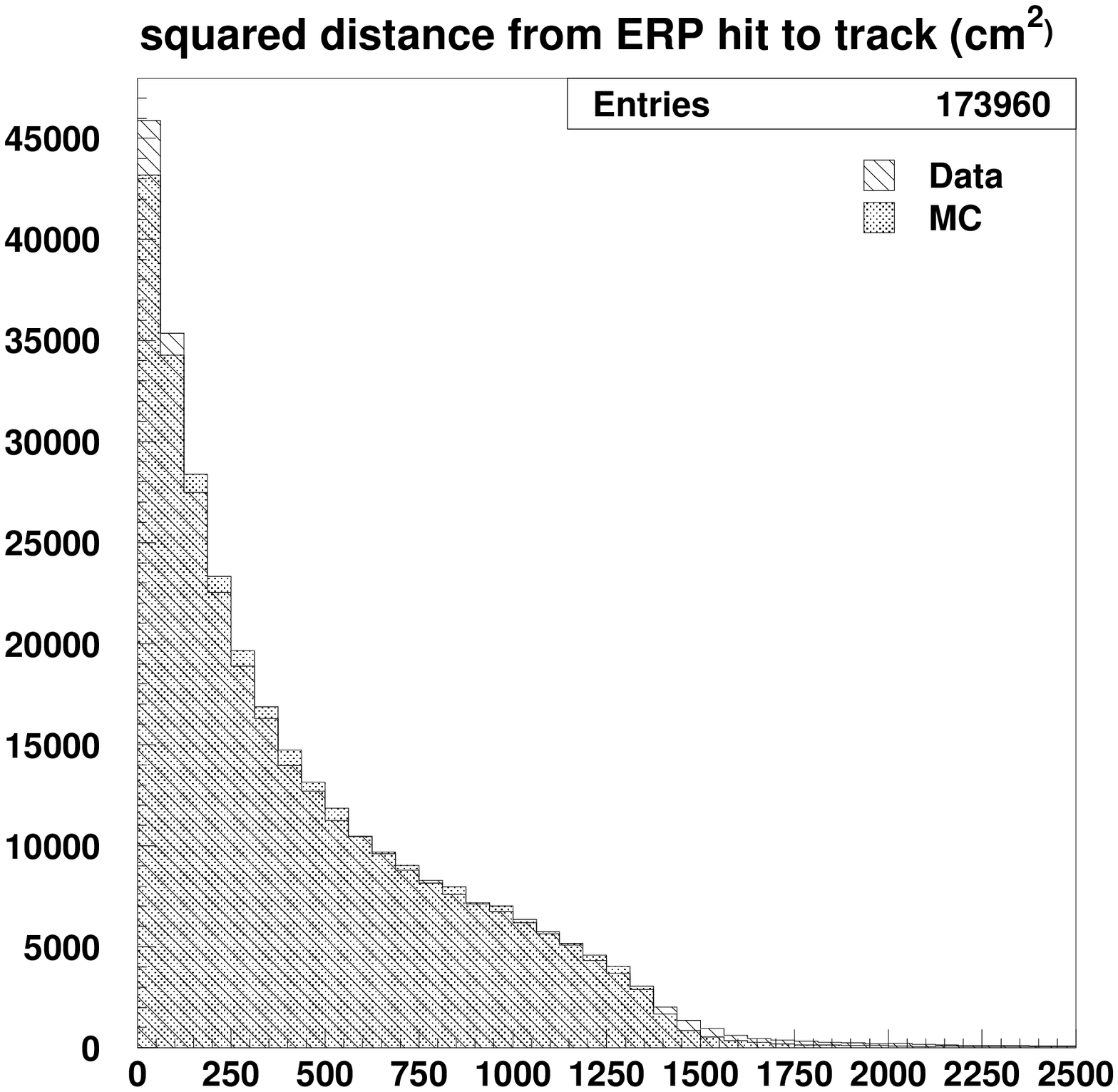}}

For energy deposition at a given point along the tank, the average
number of photoelectrons created at the tank end is proportional (over
the energy range of interest) to the energy deposition, where the
constant of proportionality is a function of position along the tank
(called the response function).  To infer the position of an event
from ADCs, one searches for a position along the tank where the
response function gives the measured ratio of light reaching the two
tank ends.  The error in this measurement is dominated by fluctuations
of photoelectron statistics.  For both construction of simulated
events and reconstruction of real or simulated data, the product of
the response function times the energy deposition is referred to as
``photoelectrons.''  However, there is ambiguity in the absolute
normalization because a multiplicative factor applied to the response
function could be compensated by the reciprocal applied to the gain
constant.  In GMACRO, normally the quantity called photoelectrons is
treated as if it were the true number of photoelectrons, and Poisson
fluctuations are taken about that number (ERPP 11=1 turns on the
Poisson fluctuations).  Comparing GMACRO events to real events, the
inaccuracy in position reconstructed from ADC is too great.
Therefore, I added code to GMACRO to shift the number of
photoelectrons by a constant for purposes of fluctuation; i.e.,

\[ P_{GMACRO}(n|\mathrm{mean} = \overline{n}) =
   P_{Pois}(xn|\mathrm{mean} = x\overline{n}) \]

\noindent
where the value of the normalization factor $x$ is set by the new
control card PNRM\@.  Setting PNRM to 2. gives a reasonable agreement
between Monte Carlo and data for the distribution of the difference
between position determined by ADC and position determined by TDC (see
Figure~\ref{fig:adcz}).

\picfig{fig:adcz}{Distribution for real and simulated events of the
  difference between position determined from TDCs and position
  determined from ADCs.}{\pawfig{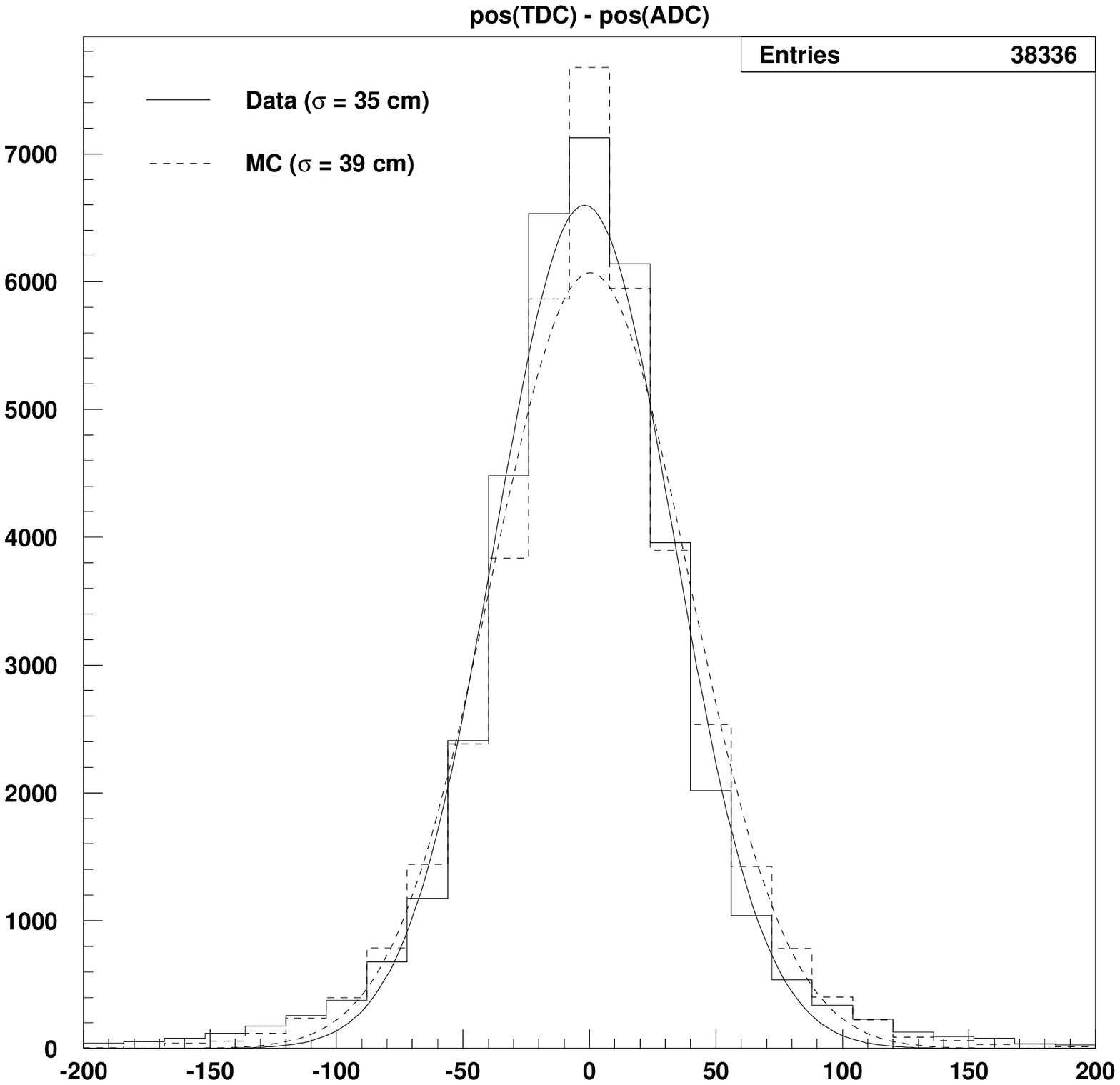}}

In real data the collaboration attempted to tune the ERP lookup tables
for each box (defined in Section~\ref{sec:erp}) so that the box had a
constant energy threshold for triggering on energy deposition anywhere
along the long axis of the box.  The threshold was low enough that any
muon that passed through a box from top to bottom or side to side
would be sure to trigger.  However, for corner clippers the particle
might or might not trigger depending on the effective trigger threshold.
In GMACRO, the energy threshold is perfect so if the energy deposited
in the box exceeds the threshold, the box triggers.  The threshold is
a settable parameter.

In Section~\ref{sec:hwprobs} a procedure to determine the efficiency
of each scintillator or streamer tube channel was described.  The same
procedure can be applied to calculate channel efficiencies for
simulated data (although in simulated data there are no dead periods).
The Geant/GMACRO parameters used in this analysis are chosen to match
the simulated to nominal real efficiencies.  For real data, a given
type of channel tended to have a nominal efficiency most weeks but on
some weeks the efficiency could drop on many or all channels
(presumably due to problems with high voltage, readout electronics, or
streamer gas).  
Real data from the week labeled by calibration set 17668 had most
channels of all types near nominal, except for strips.  The strip
efficiencies were quite sensitive to streamer gas composition, and for
most weeks a lot of strip channels had efficiencies 10-20\% lower than
the best channels.  Calibration set 13936 was one of just a few weeks
with a narrower distribution at high efficiencies.  So in what
follows, set 17668 is used as the real data for comparison with Monte
Carlo efficiencies for scintillators and streamer tube wires, and set
13936 is used for strips.

Even in GMACRO, the calculated efficiency is not 100\%, nor even
exactly the same for all channels of a given type.  Inefficiency can
occur when particles hit the box but have a very short pathlength in
the box and do not deposit enough energy to trigger; or because of
tracking errors due to hard scattering, hits due to secondary
particles, or quantization error in the streamer tubes, all of which
are simulated.  The magnitude of all these effects varies depending on
the location of the box.  In tuning the Monte Carlo, the goal is to
have the distribution of simulated efficiencies match the distribution
of measured real efficiencies.

For horizontal boxes of scintillator, the default values of 5~MeV for
Elow triggers and 10~MeV for Ehigh triggers (ERPP 2=10 3=5) works
well.  Figure~\ref{fig:effScintHor} shows good agreement of trigger
efficiency for horizontal scintillator boxes in the real and simulated
data.  In these comparisons, each entry in the histogram is the
efficiency for one channel.

\picfig{fig:effScintHor}{Efficiency (as defined in the text) of
  horizontal scintillators for real and simulated events.}
{\pawfig{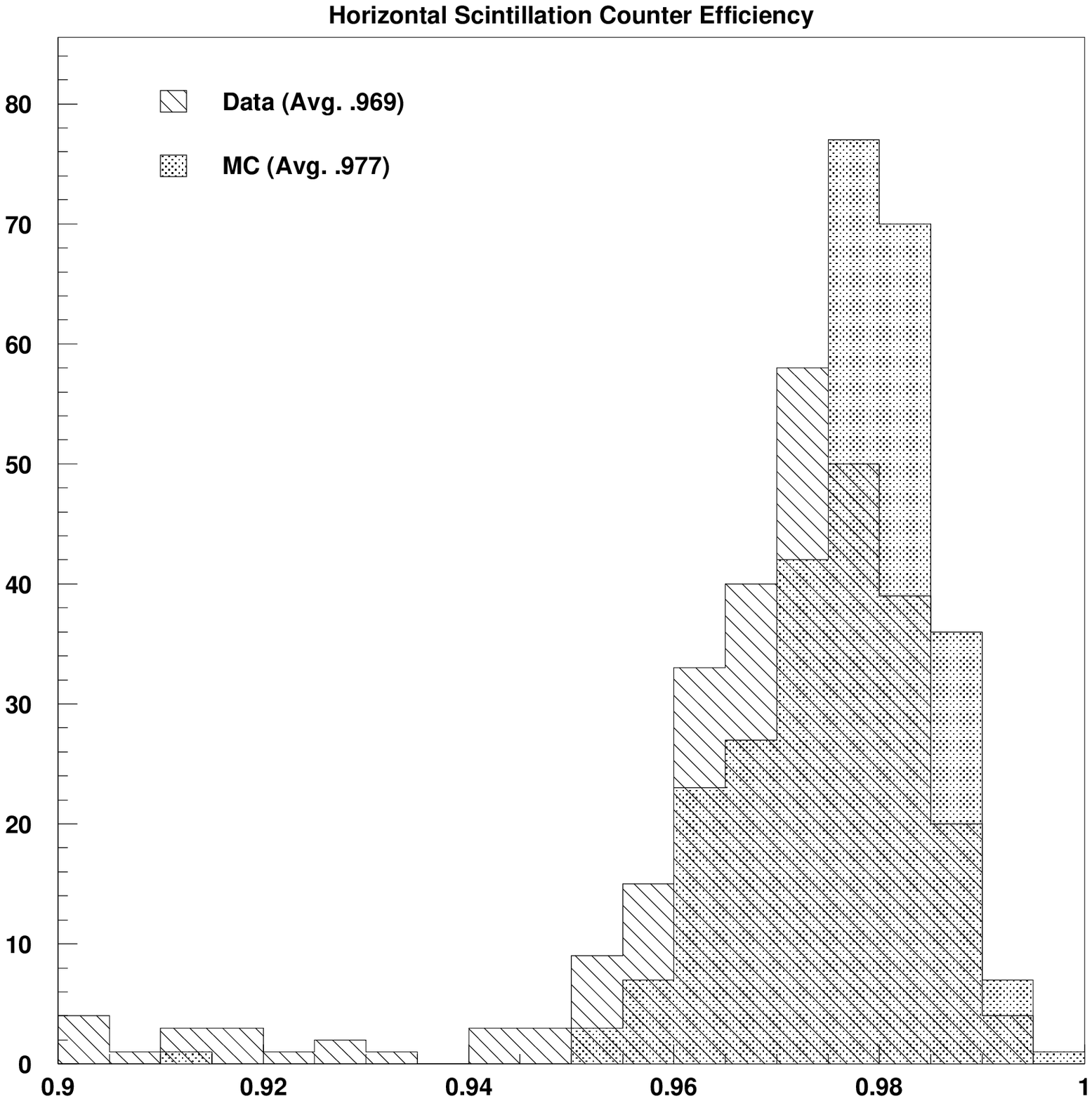}}
  
For the vertical scintillators, the default trigger thresholds of 5
and 10~MeV for simulated events give measured efficiencies that are
too high.  There were gaps between scintillation tanks in the vertical
layers so slight tracking errors may cause discrepancies, in which a
track that appears to hit a box corresponds to a particle that missed
the box, or vice versa.  It may be that the discrepancy in measured
efficiencies is not really due to the trigger thresholds, but due to
the probability to have such minor tracking errors.  On the other
hand, in physics analyses, events that appear to hit the box but
really miss it are partially compensated for by events that appear to
miss the box by a small amount but really hit it.  The latter will be
counted as good events in muon analysis, but they will not affect the
measured efficiencies.  (The efficiency calculation limits itself to
events that appear to hit the box, and asks how many of those events
caused the box to trigger.)  Apparently, minor tracking errors of this
type are more likely in real data than in simulated data: for real
data, in about 1.3\% of events involving a fired vertical box, the
track misses the box (in such cases the pathlength is coded as -1 in
the histogram in Figure~\ref{fig:effScintVert}(b)) while for simulated
data the number is just 0.8\%.  Therefore, the vertical box energy
thresholds have been tuned (ERPP 4=12 5=7) to make the measured
efficiencies a little high, so the combination of efficiency for
tracks that appear to hit the box plus the efficiency for nearby
tracks that appear to miss the box are comparable between Monte Carlo
and data.

\picfig{fig:effScintVert}{(a)~Efficiency of vertical scintillation counters
  for real and simulated events.  (b)~Distribution of pathlength
  inside vertical boxes.  Events in the artificial bin at -1 appeared
  to miss the box entirely.}
{\twofig{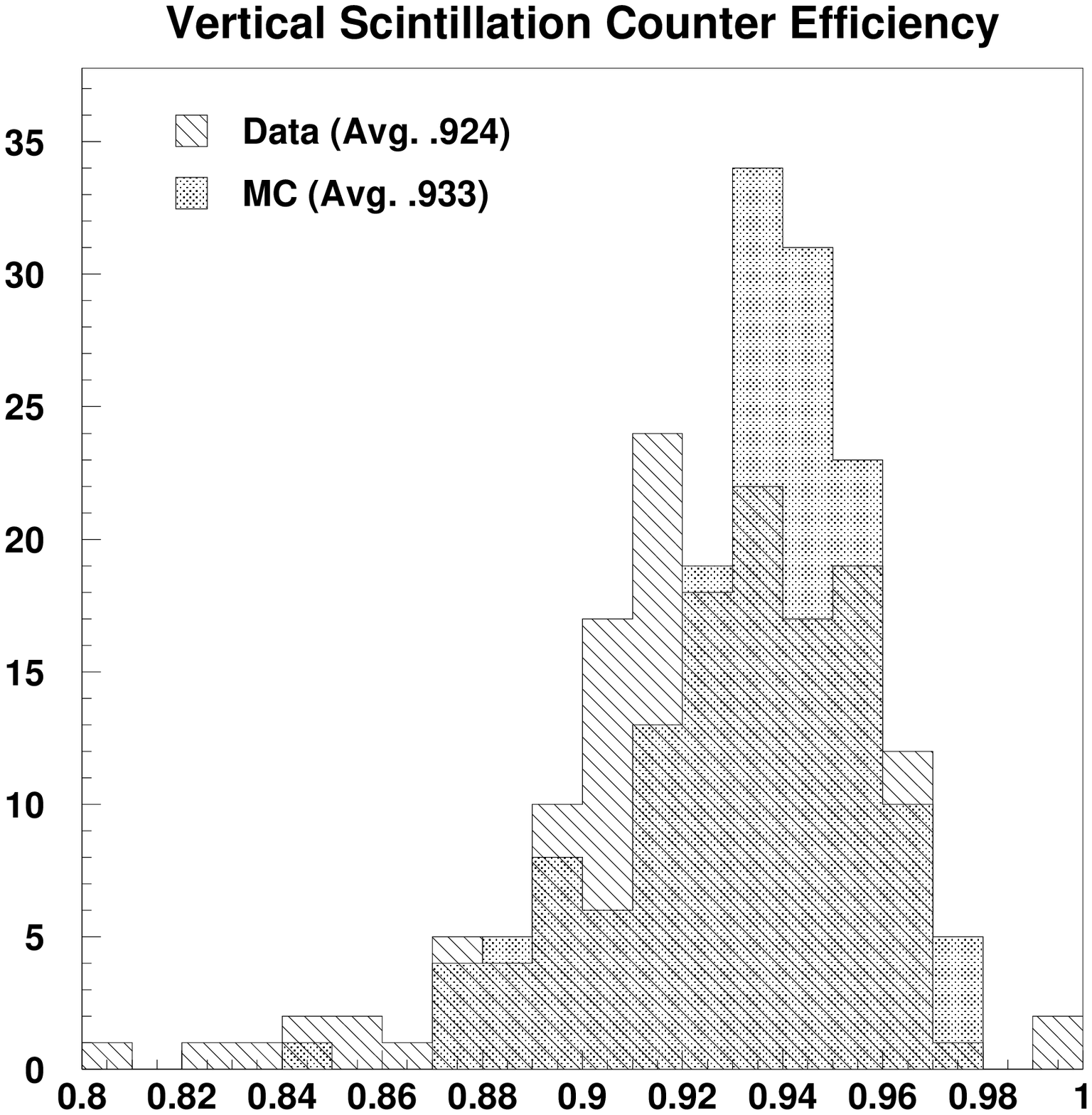}{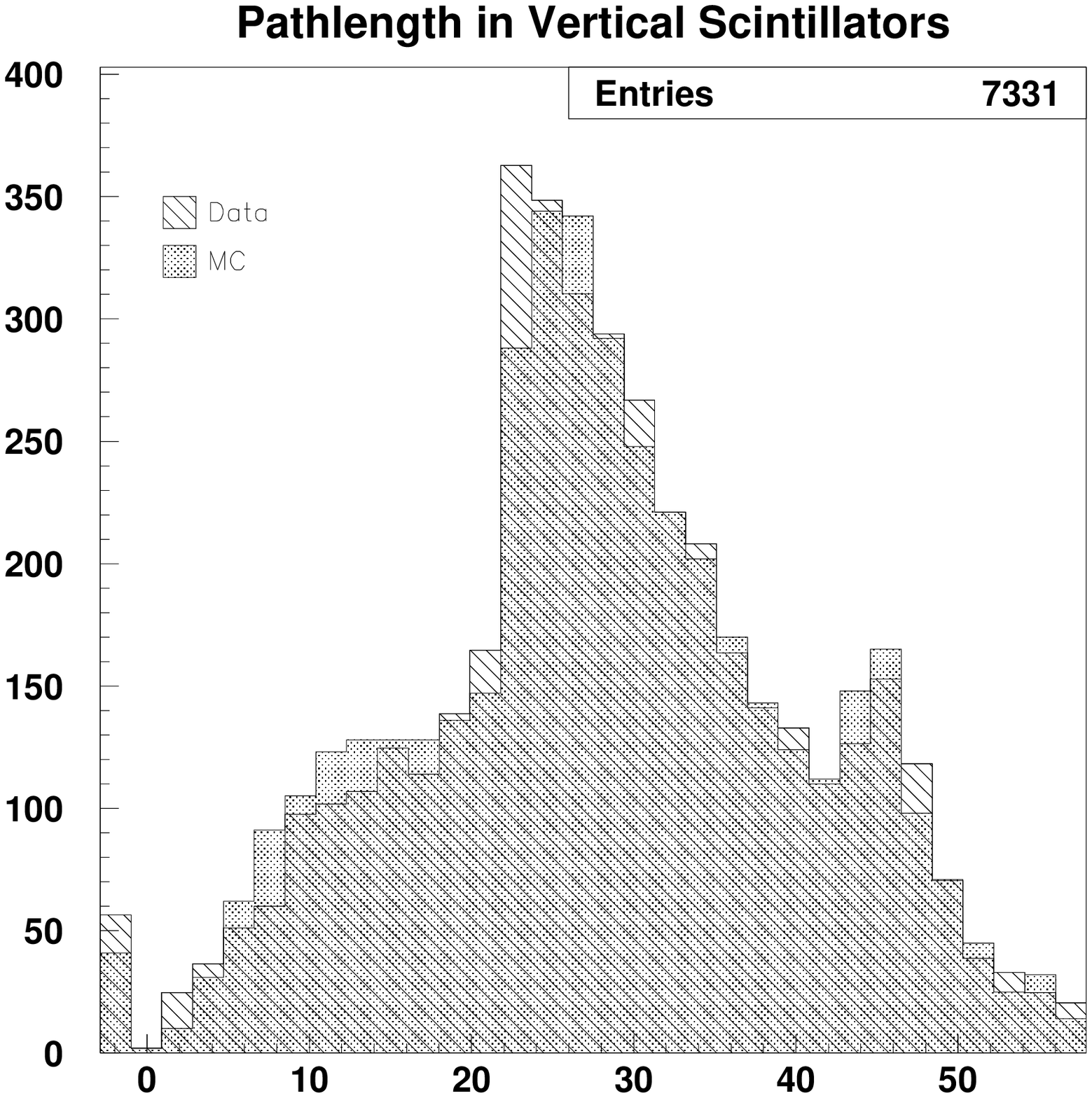}}

Each time a particle is tracked through a streamer tube, GMACRO
calculates the charge induced on the streamer tube wire, as well as
the charge induced on nearby strips.  This calculation does not use
user-settable parameters.  However, the threshold charge above which
the Streamer Tube Acquisition System (STAS) considers the channel to
have fired is parametrized.  Thus, physical inaccuracies in the charge
calculation can largely be compensated by choosing the STAS thresholds
to reproduce the measured efficiencies and hit width distributions.

For central streamer tube wires, a threshold of TRHW=6.87 (arbitrary
units) had been determined some years earlier by MACRO collaborators
who tuned the Monte Carlo parameters.  Figure~\ref{fig:effWire} shows
good agreement between simulated and real efficiencies.  For the
attico planes, there were not any previous efforts to tune the
parameter TRWT.  The default value of 7.0 gives low efficiencies,
while 6.5 gives good agreement.  (Figure~\ref{fig:effWire} is summed
over lower and attico planes but each alone gives good agreement
between Monte Carlo and data.)

\picfig{fig:effWire}{(a)~Efficiency of central streamer tube wires for
  real and simulated events. (b)~Distribution of number of adjacent
  wires fired in a plane.}{\twofig{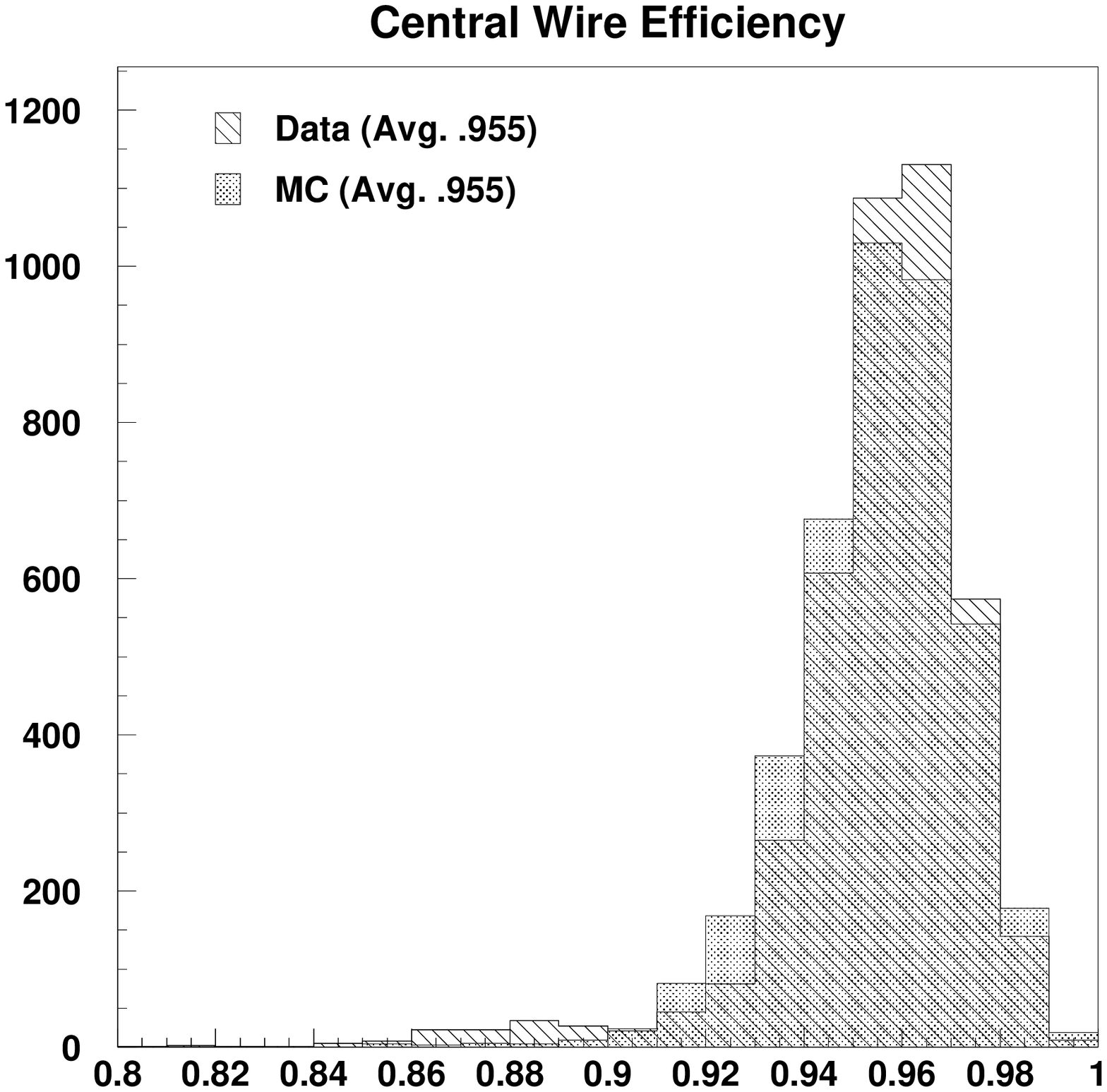}{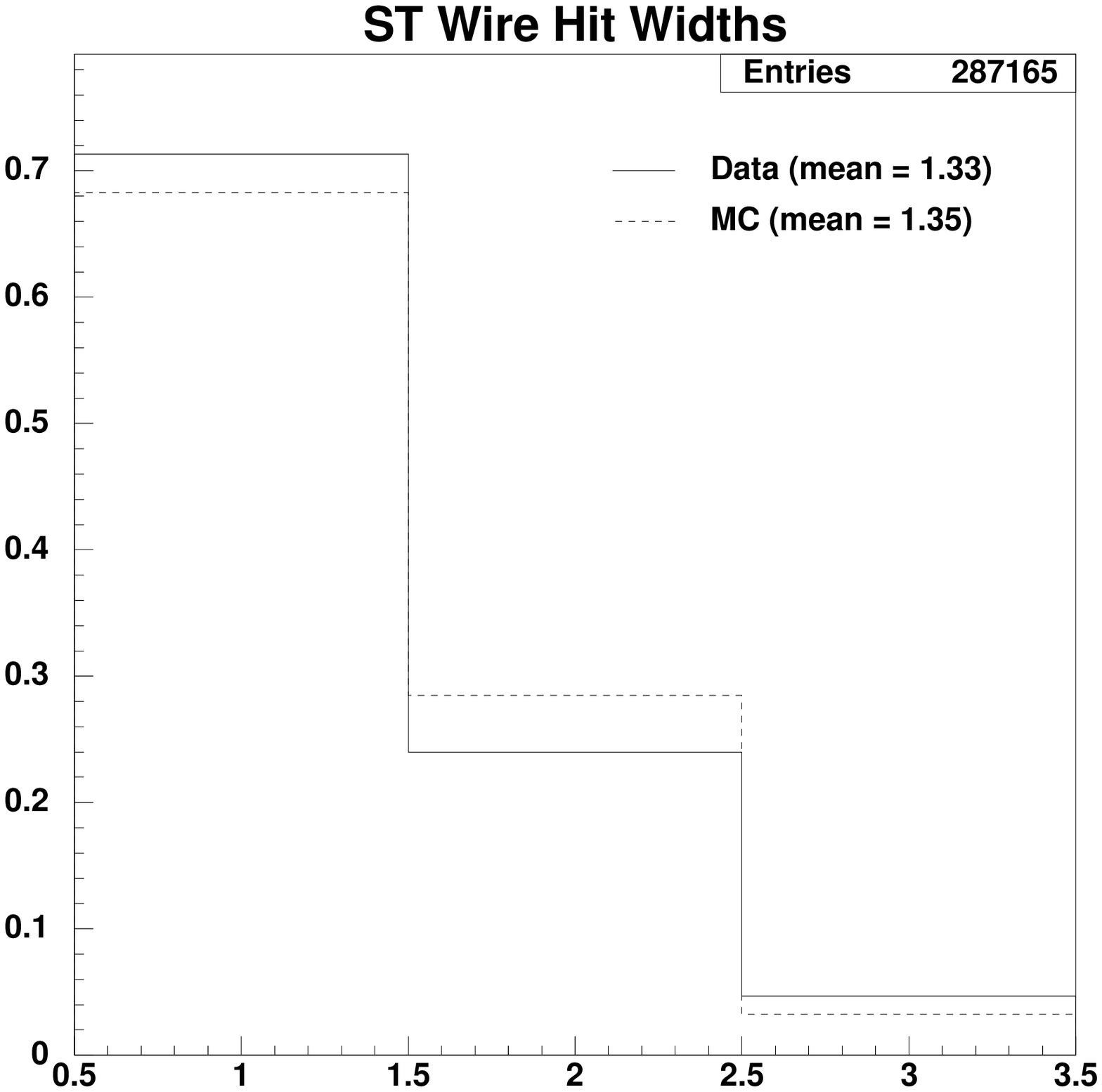}}

The situation is more complicated for central streamer tube strips.
Streamer tubes in the first, fifth and ninth streamer tube planes bore
a considerable weight (of scintillator tanks or track etch detector)
and the plastic tubes deformed, with the result that the strips were
$\sim$ 1~mm closer to the wires in these planes than in other planes.
Thus, a larger charge tended to be induced on these strips.  This
effect is not directly simulated in GMACRO\@.  However, one can apply a
lower STAS threshold to these planes in GMACRO (using the TRHS
parameter) and achieve an equivalent effect -- an increase in the
probability the STAS considers the channel fired.  Again, values that
had been determined earlier by MACRO collaborators (specifically, 2.55
for planes 1, 5, and 9; 3.15 for planes 2-4; and 3.0 for planes 6-8
and 10) give a good match between data and Monte Carlo.  For attico
horizontal strips, the default value of TRST=3.8 works well.
(Figure~\ref{fig:effStrip}.)

\picfig{fig:effStrip}{(a)~Efficiency of central streamer tube strips
  for real and simulated events, plotted separately for each plane.
  The datapoints give the mean and RMS of the efficiencies for all
  channels in that plane.  Note the elevated efficiency in planes 1, 5
  and 9.  (b)~Number of adjacent strips fired in a plane of central
  streamer tube strips.  In both figures, the statistical uncertainty
  on the mean is a tiny fraction of the RMS, and the vertical separation
  between data and Monte Carlo is statistically significant.  The
  horizontal separation is artificial, for clarity of presentation.}
{\twofig{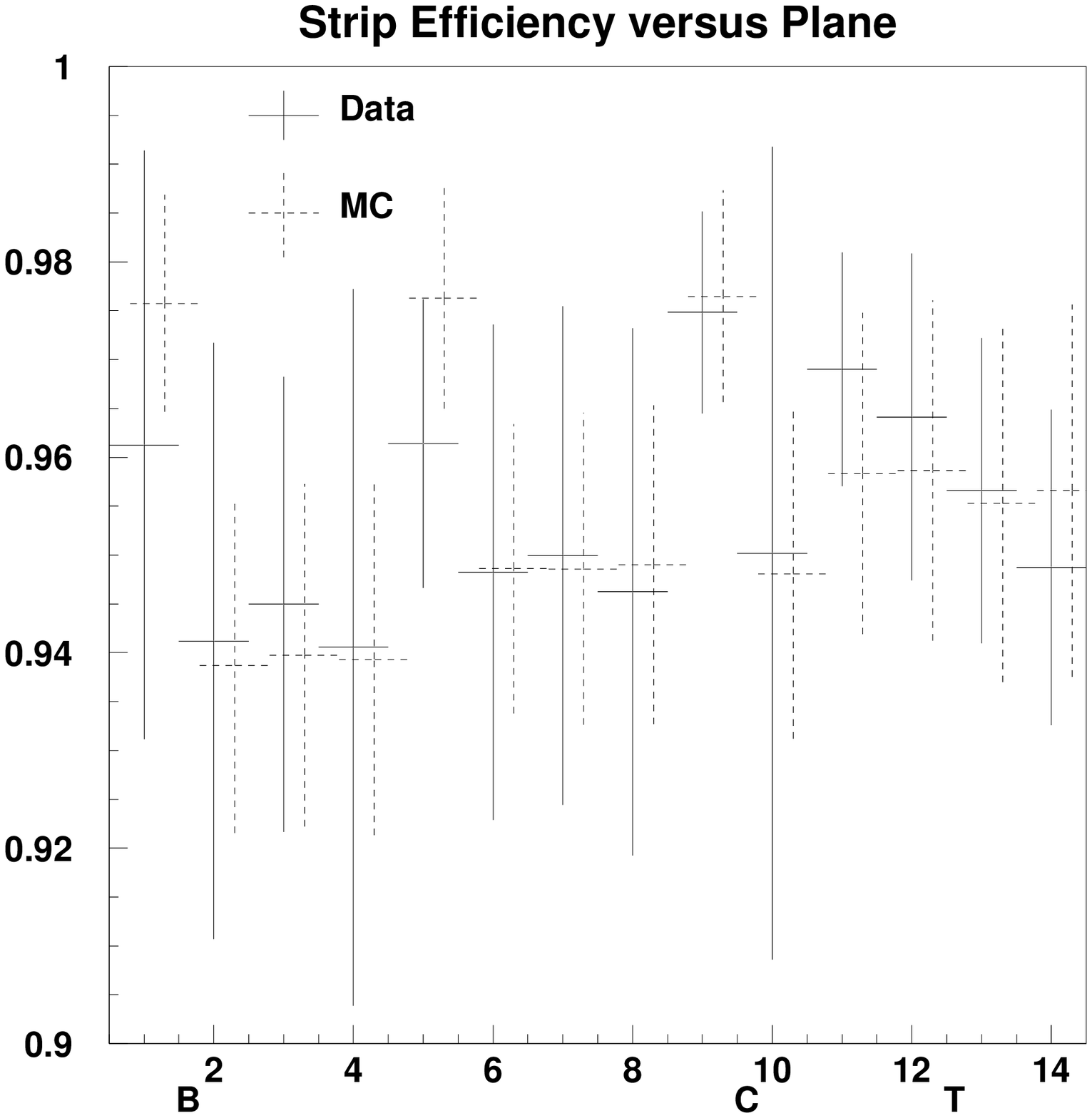}{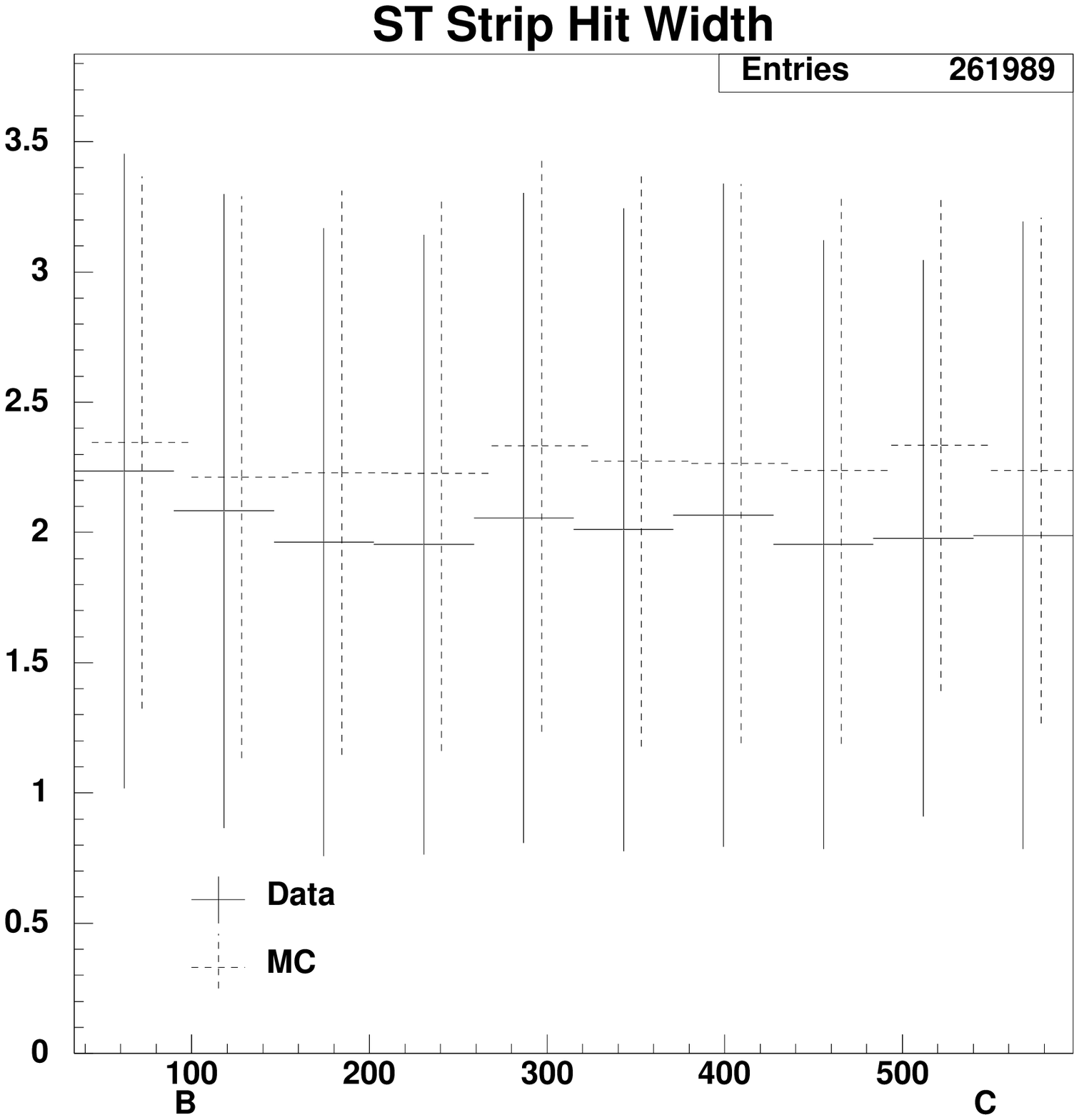}}

For the lateral and frontal planes of streamer tube wires in the lower
detector, the default parameters result in measured efficiencies
greater than in real data.  Perhaps this is not really due to an
inaccurate simulation of streamer tube efficiencies, but rather
reflects a greater accuracy of reconstructed tracking parameters in
simulated events.  Thus, if (in a real event) the reconstructed track
points into a lateral streamer tube channel, the real particle may
have missed that channel.  However, in the simulated data the
probability of such a misreconstruction is lower.  The only parameter
available to tune the lower lateral streamer tube efficiencies is
``ELAT.''  Setting ELAT to 0.85 means that any time the tracking
software thinks a lateral streamer tube should fire, it is only put in
the acquisition as having fired 85\% of the time.  However,
atmospheric muons typically strike two adjacent cells when passing
through most lateral planes, and our definition of efficiency
considers the hit efficient if {\em either} of the cells fire.  Thus,
it would be necessary to reduce ELAT to ridiculously low values to
cause the measured efficiencies of simulated events to change
significantly.  The lowest value of ELAT I was comfortable with, 0.85,
results in measured efficiencies of simulated events still a percent
or two too high.  For the attico, the standard GMACRO code does not
provide a parameter to vary the efficiency of the attico lateral
planes independently of the lower planes.  However, it was not
possible to set the parameter to match the efficiencies of both kinds
of streamer tubes.  Therefore, I coded a new threshold parameter,
TRLT, into GMACRO to control the attico lateral planes separately from
the lower detector.  Setting TRLT to 9.5 gives a good match of the
efficiencies.  (Figure~\ref{fig:effLat}.)

\picfig{fig:effLat}{(a)~Efficiency of lateral streamer tube wires for
  real and simulated events. (b)~Hit width distribution.  The upper
  figure is for attico, and the lower for the lower part of the
  detector.}{\twofig{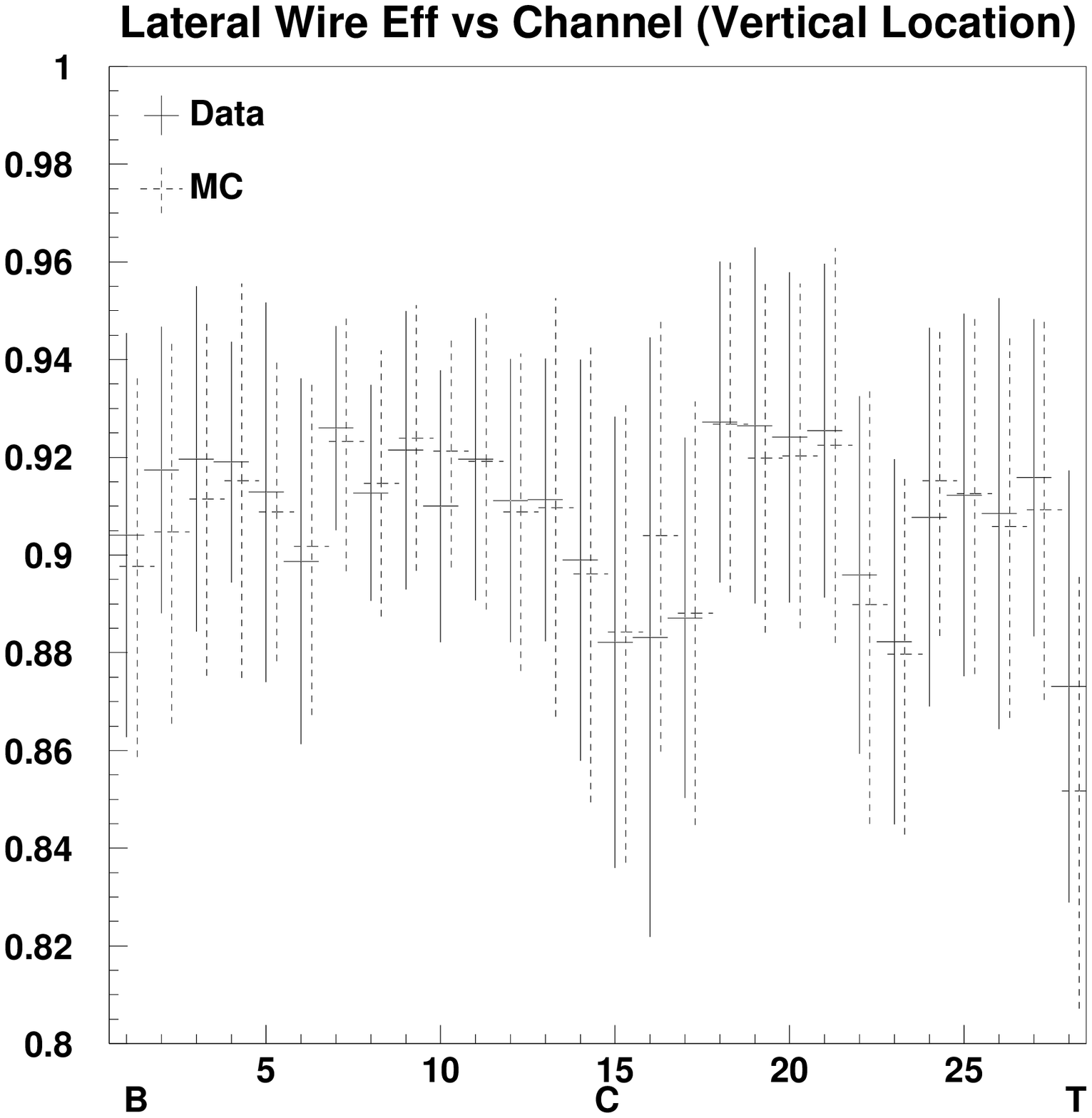}{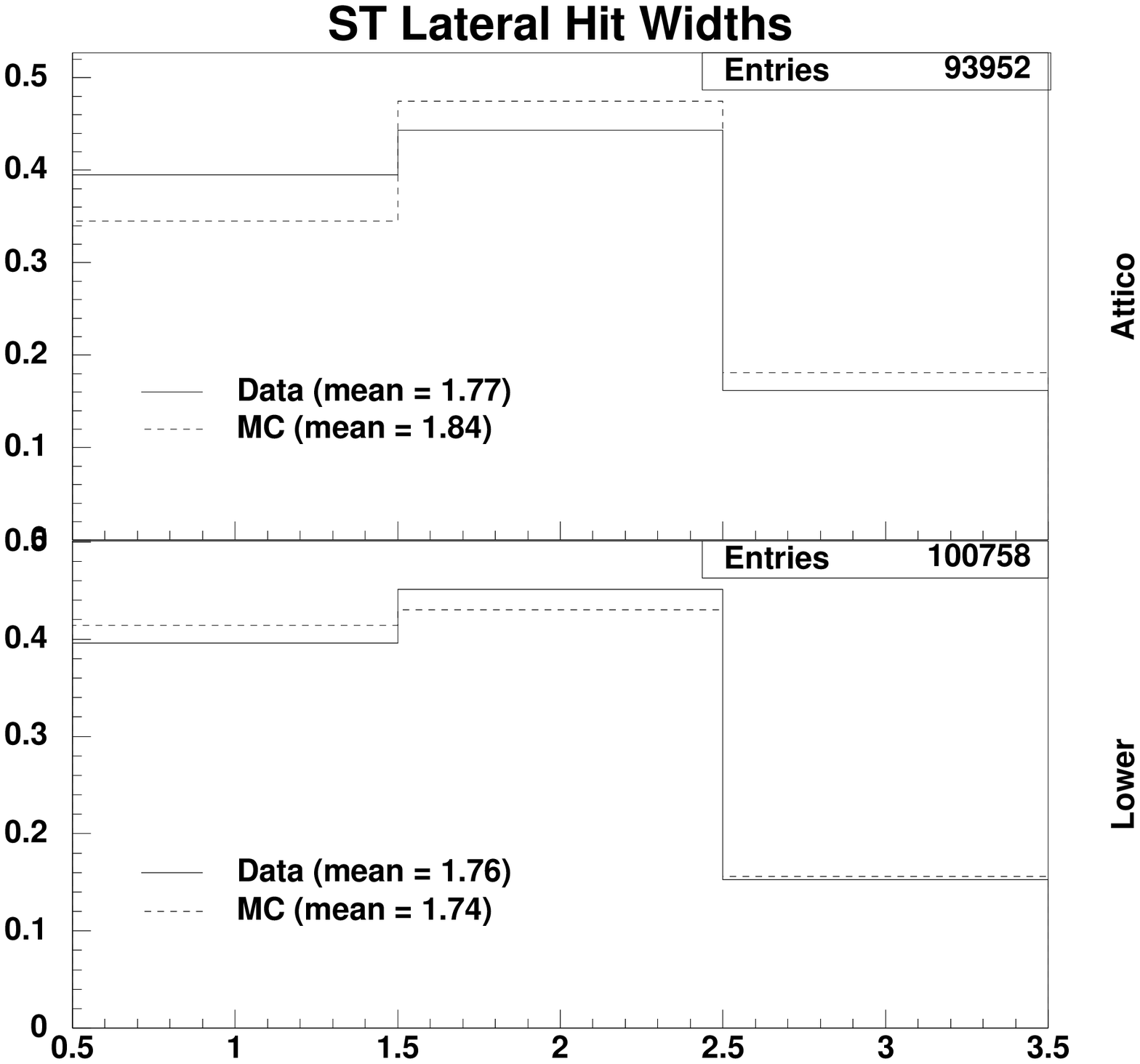}}

Lateral strips are not used in this analysis, so their performance has
not been tuned.

\subsection{Simulating Problems with the Real
  Hardware\protect\label{sec:liveAdj}}

With the nominal Monte Carlo, we can make predictions of what would
have been observed if the detector operation had always been nominal.
However, during the data-taking there were a number of hardware
problems that degraded the detector performance, some minor (for
example an ERP channel that must be excluded from offline analysis
because it gives unreliable timing for a week or two) and some major
(for example an entire supermodule that was excluded from the
acquisition for a few days because of a bad high voltage power
supply). To draw conclusions about the neutrino flux based on the
observed rate of neutrino interactions, the effect of the observed
hardware irregularities must be estimated.

For MACRO high energy neutrino analyses, in which the neutrino-induced
muon traverses the entire detector, the topology of neutrino events is
similar enough to the topology of the copious downgoing atmospheric
muons that we may use a clever technique called {\em equivalent
  livetime}.  In this approach, for a period of running when the
detector is known to have been operating well in all channels, the
rate of downgoing muons is calculated.  Then for the entire dataset,
the equivalent livetime is defined as the total number of downgoing
muons observed divided by the nominal rate.  The idea is that if there
were a period of time, say 100~hours, during which detector problems
were severe enough to cause the loss of 5\% of the downgoing muons (so
the rate during that period was 95\% of the nominal rate), that would
be equivalent to having run for 95~hours at nominal efficiency.

Around 800 downgoing muons stop in the detector each week, giving a
topology similar to the semi-contained upgoing neutrino interactions
with which this analysis is concerned.  We might hope to define an
equivalent livetime procedure using the rate of downward stopping
muons to monitor detector efficiency.  However, on reflection we must
abandon this approach.  First of all, detector problems could cause
the rate to go either up or down.  For example, widespread detector
problems could cause events to be not detected at all and reduce the
measured rate.  On the other hand, streamer tube inefficiency could
cause an event that really had throughgoing topology to appear as if
it were semi-contained.  Thus, the more numerous throughgoing events
could bleed over and infect the apparent semi-contained rate.  Because
the ratio of downward throughgoing muons to downward stopping muons
(about 100:1) is much greater than the ratio of (neutrino-induced)
upward throughgoing muons to upward semi-contained events (about
10:1), measuring an increased rate of apparent stopping muons does not
give a good estimate of the expected increase in the rate of apparent
semi-contained neutrino events.  Furthermore, the distribution of
which streamer tube planes are most important is different for
downgoing stopping muons than for semi-contained neutrino events.  For
these reasons, the equivalent livetime approach can give only a rough
idea of the effect of hardware problems but cannot give a good
quantitative estimate.

Therefore, instead of relying on the equivalent livetime approach, the
detailed measurement of dead periods and inefficiencies in the real data
(described in Section~\ref{sec:hwprobs}) are applied to the simulated
data.  For each week of real data-taking, a database has been created
listing for each channel the dead periods, as well as the efficiency
during the non-dead periods.  To simulate the detector during a given
week, first, events are generated using the nominal Monte Carlo.
Then, for each event, a time during the week is picked at random and
the event record is altered by removing channels that were dead and
probabilistically removing channels that were inefficient at the
chosen time.  After channels are eliminated, then the event is
analyzed.  If any microcuts are in effect for that week, they are
applied by the analysis software.

\subsubsection{How Well Does It Work?}

Starting with a Monte Carlo sample of 100,000 downgoing atmospheric
muons that reach the detector, if no detector irregularities are
simulated, 63,340 muons pass all cuts of the muon analysis.  Then a
series of jobs analyze all 100,000 input events while simulating
conditions (dead periods, inefficiency, and microcuts for each
channel) of each analysis week (one job for each week).  About 1/3 of
the weeks give more than 60,000 reconstructed muons; the greatest is
61,759.  Thus, even during the very best weeks, the simulation
indicates that non-ideal performance of the detector (most
significantly the computer deadtime) caused about a 2.5\% reduction in
efficiency, from 63,340 events detected to 61,759.

Also for each week, the number of real events that reconstructed as
good muons, divided by the number of minutes of data-taking that week,
yields a rate.  Half the weeks show a rate between 10.5 and 11.3 muons
per minute.  If one or more supermodules was turned off for long
periods during the week, the rate is much lower.  With about 100,000
real muons reconstructed each week, the statistical error on the
measured rate is around 0.3\%.

The atmospheric muon generator in GMACRO has no concept of time or
rate; it produces a flux with the proper angular and energy
distributions, but does not say anything about the predicted rate.
The Monte Carlo prediction may be normalized using the weeks in which
the effect of detector problems was least; that is, the weeks in which
the measured rate was greatest.  In a scatter plot of number of
simulated muons passing all cuts versus rate in the real data
(Figure~\ref{fig:hwprobFixNorm}), the 5 weeks with the highest real
rates (all within 0.6\% of each other) (the weeks were 15647, 16017,
16049, 16080, and 16184) also have some of the highest numbers of
simulated muons (all within 0.5\% of each other).  The average of the
rates (11.19) and the average of the number of simulated muons
(61,679) for these 5 weeks is used to determine the livetime of the
Monte Carlo sample as $\frac{61,679\ \mu}{11.19\ \mu/min} = 5510$ min.
With this assumption, the number of simulated muons reconstructed for
each week can be used to predict the real data rate for that week.
This choice of how to normalize the Monte Carlo is obviously quite
arbitrary; if instead of using the 5 weeks with highest rate, we chose
a different number of weeks between 1 and 9, the calculated livetime
of the simulation would differ by no more than 0.2\% from the original
choice.

\picfig{fig:hwprobFixNorm}{Weeks with the highest measured rates.
  Each point represents one week; the measured muon rate in real data
  is on the x-axis while the y-axis gives the number of simulated
  muons that reconstruct when the conditions for that week are
  simulated.  The five weeks in the upper right are averaged to set
  the normalization of the Monte Carlo.}
{\pawfig{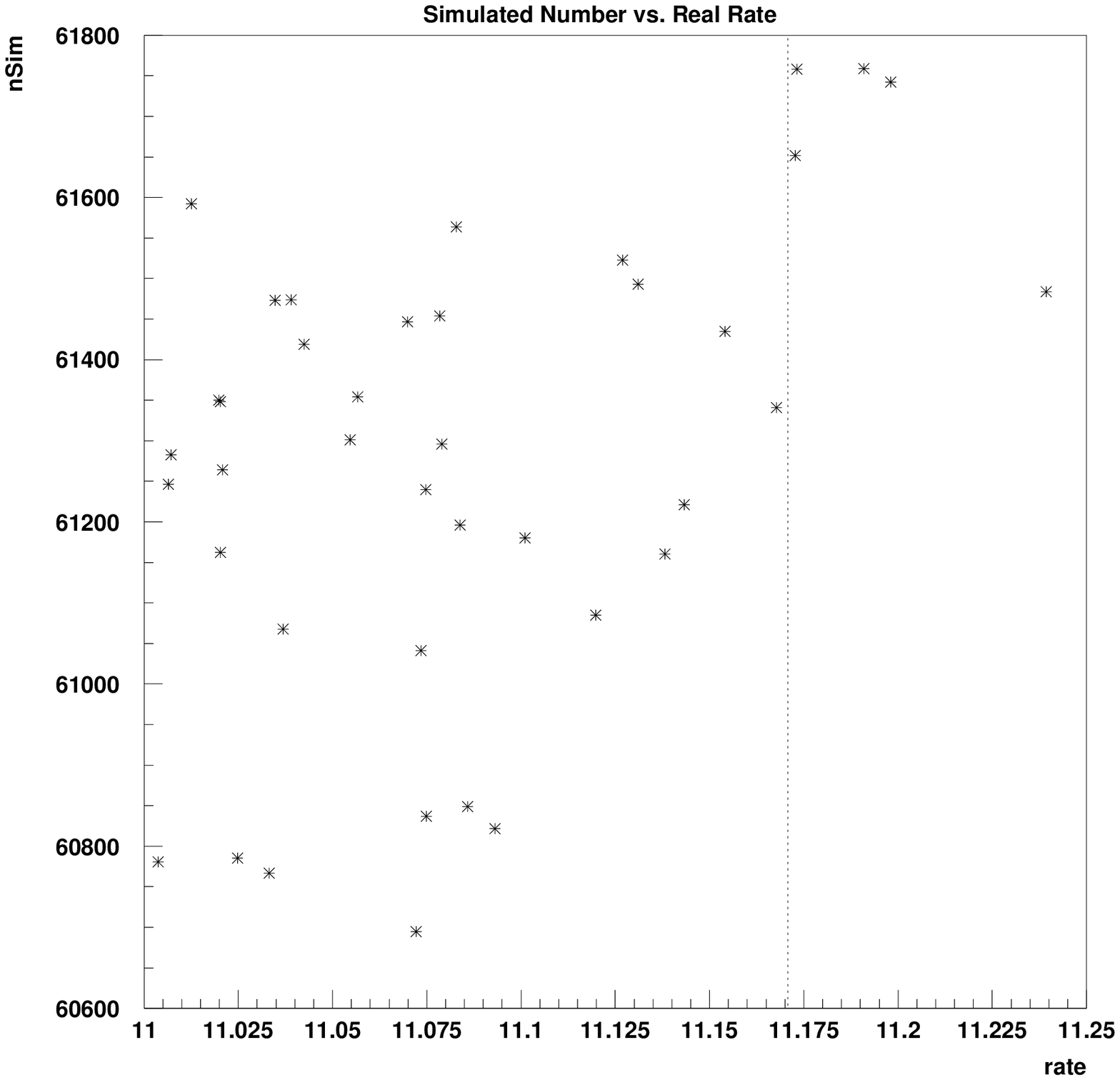}}

The simulation predicts the measured rates fairly well
(Figure~\ref{fig:hwprobRate}).  Most weeks have rates near the nominal
$\sim$11/min, and the simulation predicts that.  Some weeks have
significant irregularities (for example, two supermodules may have
been turned off during the entire week), and the simulation does a
good job of predicting the rate during those weeks as well.  The error
(predicted rate minus measured rate) appears like a random variable
with variance (according to a gaussian fit to the error histogram) of
2\%.  Actually, this error is too large to be accounted for by
counting statistics in the number of muons.  Apparently the simulation
algorithm, which is not exact, sometimes overestimates and sometimes
underestimates the effect of irregularities but does so in a random
way.  Summing over all 262 analysis weeks, the errors tend to cancel.
In fact, if we multiply the predicted rate for each week times the
real livetime for that week, the simulation predicts the total number
of muons in the entire dataset to within 0.3\% (23,790,000 predicted
versus 23,726,256 observed).

\picfig{fig:hwprobRate}{Comparison of predicted rate (y-axis) to
  measured rate for atmospheric muons (mostly throughgoing).  Each
  point represents one week.  The dashed line at $45^\circ$ is not a
  fit, but shows the ideal locus, predicted = measured.  The inset
  shows the distribution of fractional differences, (predicted -
  measured)/measured.}
{\resizebox{6in}{!}{\includefigure{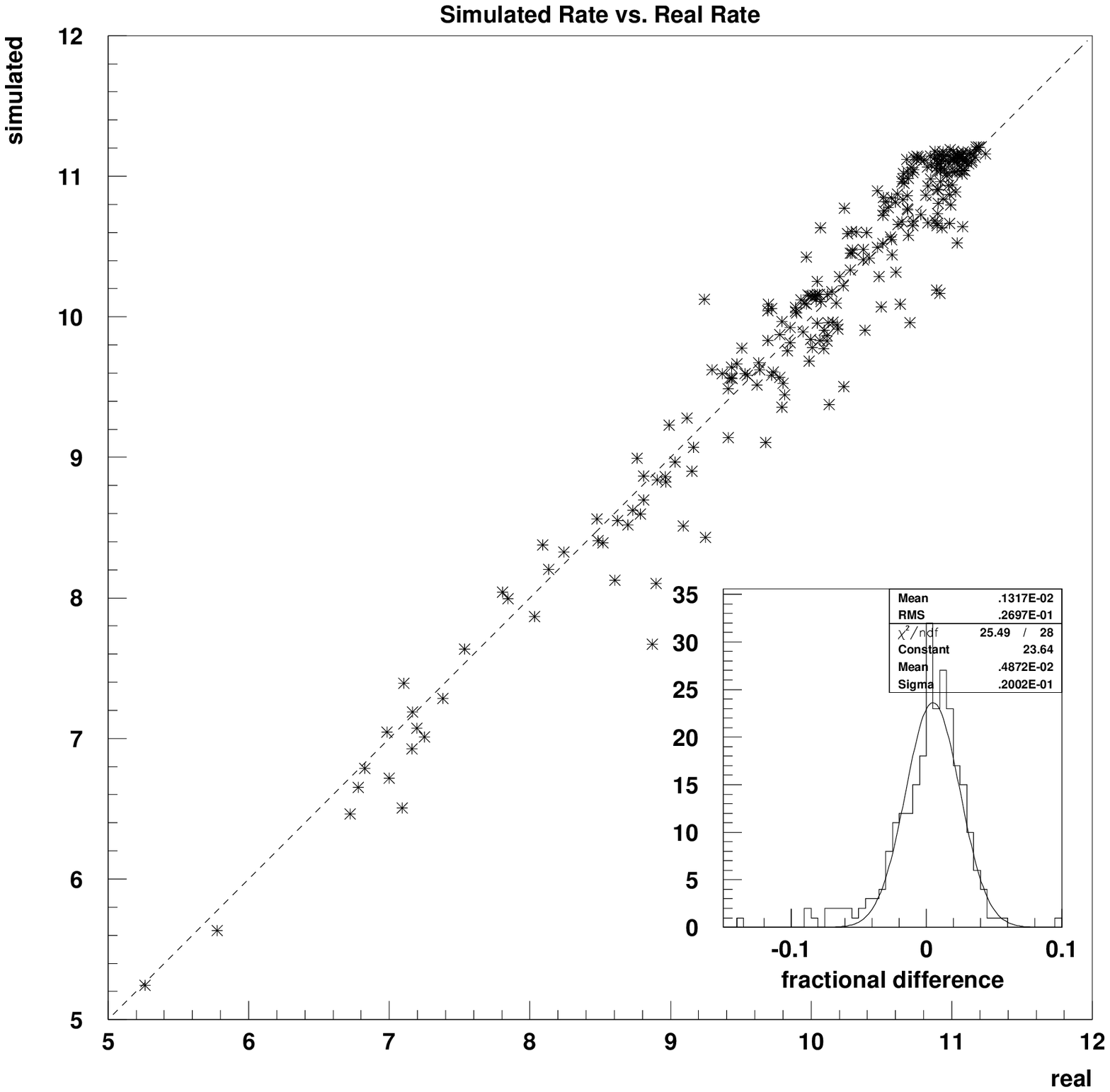}}}

Another way to frame the same problem is to ask the simulation to predict
the inefficiency in any given week due to detector irregularities.
(Here the term inefficiency is used very loosely, to refer to the
difference between the muon rate for the week and the nominal rate of
11.19.)  Figure~\ref{fig:hwprobInef} shows the results, which again are
fairly good.

\picfig{fig:hwprobInef}{Comparison of predicted inefficiency (y-axis)
  to measured inefficiency.}
{\resizebox{6in}{!}{\includefigure{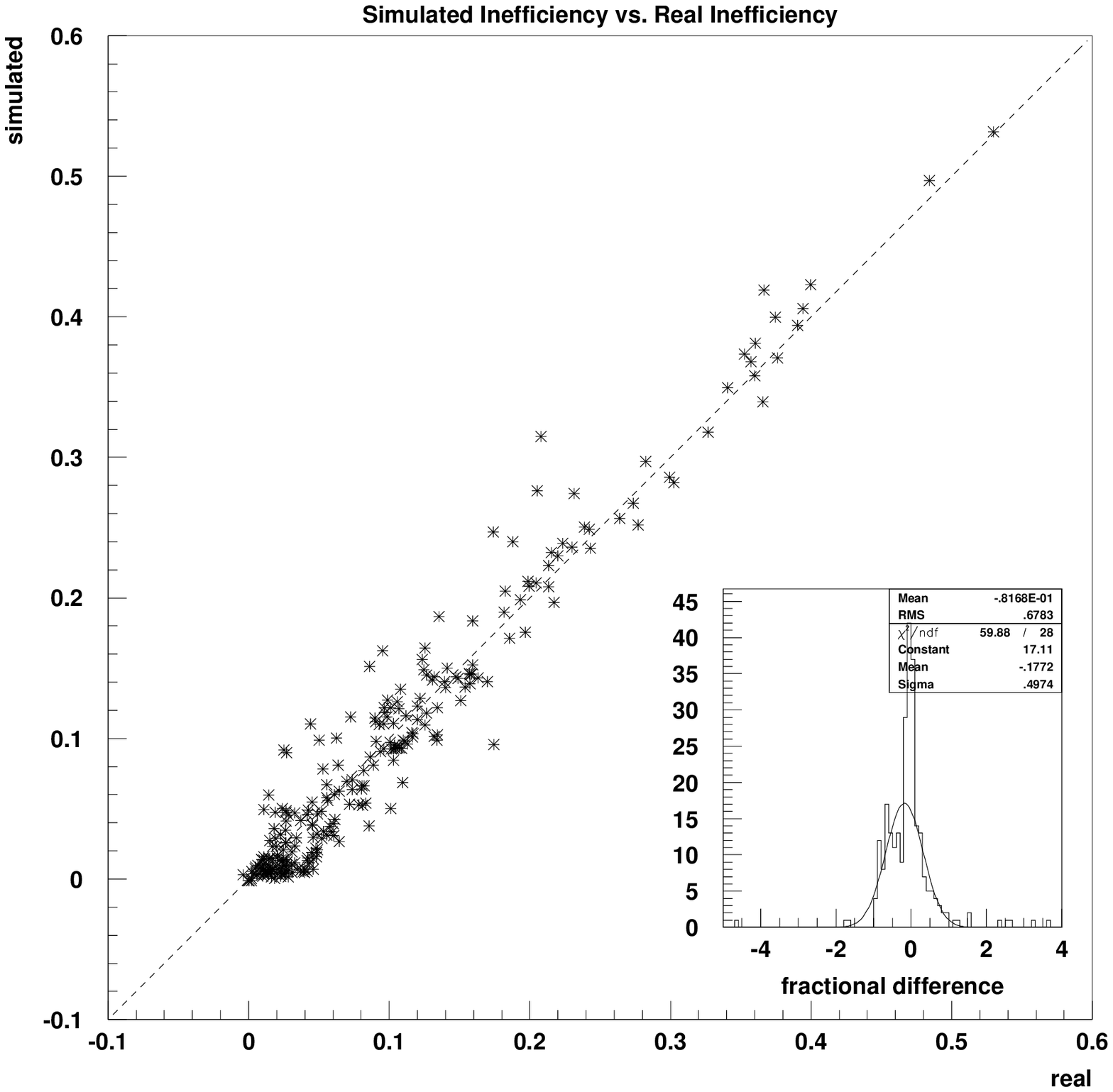}}}

Summed over all weeks, the simulation predicts there would have been
26,970,000 muons if there had been no hardware problems (assuming the
normalization of the simulation livetime above is correct).  The
prediction is reduced by 11.8\% taking hardware problems into account,
while the observed number is 12.0\% less than the no-problem
prediction.  As before, all the week-by-week errors canceled
and the prediction of inefficiency summed over all weeks is accurate.

The task of predicting the efficiency for throughgoing muons is easier
than the task of predicting the efficiency for the semi-contained
events with which this thesis is concerned.  In
Chapter~\ref{chap:meas}, criteria will be defined for muons that
appear to either originate (upgoing) or stop (downgoing) in the
detector.  Compared to throughgoing muons, the rate of such muons is
more sensitive to exactly how many channels have problems when and
where.  Less than one upgoing neutrino-induced event of this topology
occurred per week; however, between 600 and 1000 downward stopping
atmospheric muons were recorded most weeks.  If the simulation can
accurately predict the rate of stopping muons week by week, that would
give us some confidence that the efficiency it predicts for the
neutrino analysis is reliable.

Of the 100,000 simulated downgoing muons (the same sample used above),
527 are identified as stopping muons when hardware problems are not
simulated.  Simulating hardware problems, half the weeks show 480 or
more simulated stopping muons.  As discussed in
Section~\ref{sec:atmugen}, the simulation does not get the muon
spectrum underground exactly right, and underpredicts the rate of
stopping muons by 5.0\%.  We could use the above calculation, that the
livetime of the simulation is 5510~min, and convert the
number of stopping simulated muons to a prediction of the real rate of
stopping muons.  However, that would just recover the fact that the
overall normalization of the stopping fraction is off by 5.0\%.
Instead, let us re-normalize the simulation to get a simulation
livetime for stopping muons.

Using the same weeks identified above that have high rates of
throughgoing muons (ostensibly, these are the weeks that had the least
significant hardware problems), the measured rates average to 0.0966
stopping muons per minute, and all 5 weeks are within 4\% of this
average.  (The error due to counting statistics is greater when
considering the stopping muon problem, on the order of 3.5\% (800
events) for the real data and 4.5\% (500 events) for the simulated
data.)  The number of simulated stopping muons averages to 514.6, and
all 5 weeks are within 2.6\% of this average.  That suggests a
simulated livetime of 5326~minutes.  (Using a different number of
weeks between 1 and 9 to set the normalization yields a livetime
differing by no more than 1\% from the original choice.)  Using this
equivalent livetime, summing over all 262 weeks the simulation
predicts there would have been 227,800 stopping muons if there were no
hardware problems, or 204,700 taking the hardware problems into
account.  The measured number is 207,831.  The predicted total is off
by about 1.5\%.  To put the same result another way, the predicted
inefficiency due to hardware problems is 10.1\% while the measurement
(difference between number of muons expected if all weeks had nominal
performance and the number of muons actually observed) is 8.8\%.

\picfig{fig:hwprobStopRate}{Comparison of predicted rate of stopping
  muons (y-axis) to measured rate.}
{\resizebox{6in}{!}{\includefigure{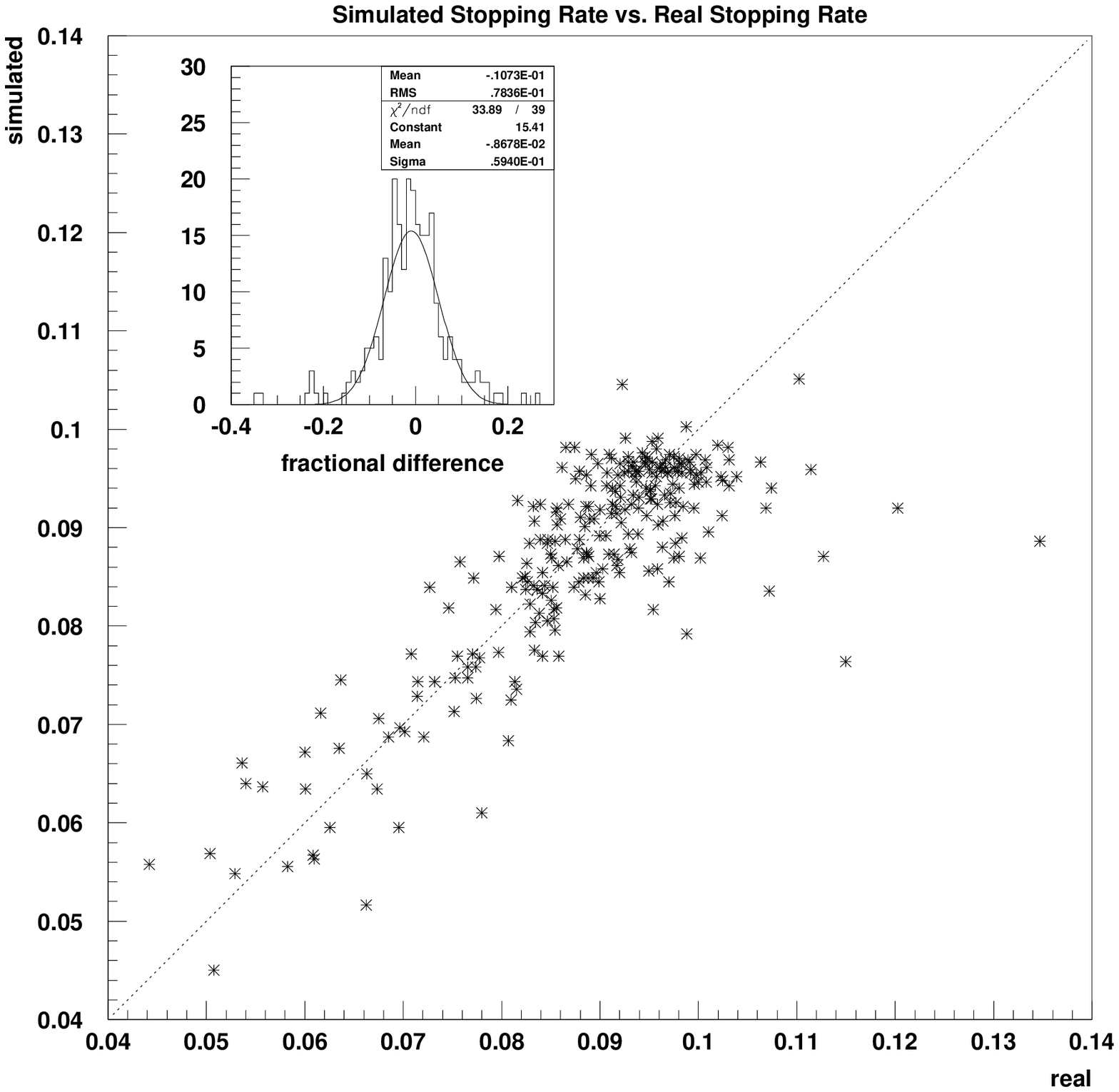}}}

In conclusion, over the course of the run, the procedure for
estimating the effect of hardware problems appears to be accurate to
around 1.5\% for a set of events quite similar to the neutrino-induced
muons with which this thesis is concerned.

\chapter{Predicting the Atmospheric Neutrino Interaction Rate
  \protect\label{chap:rate}}

To search for evidence of neutrino oscillations, we must have a good
understanding of the expected event rate in the absence of
oscillations, and how it changes as a function of the oscillation
parameters.  For the experimental signatures considered in this
analysis, the neutrino energies must be hundreds of MeV or higher
(i.e.,\  above $\numu$ charged current threshold).  In this regime,
almost all neutrinos are atmospheric neutrinos, 
i.e.,\ neutrinos from hadronic showers that develop when high energy
cosmic rays hit the atmosphere.  One can use knowledge of the primary
cosmic ray flux and hadronic interaction models to make predictions
about the flux of atmospheric neutrinos.  The flux may be folded with
a charged current cross section as the basis of event rate
calculations.

This chapter is organized as follows:  The first section is about
calculating the neutrino flux.  It starts with some general
principles, then explains some of the subtleties involved in making
accurate calculations.  The Bartol flux, a calculation used to make
predictions in this analysis, is discussed in a little more detail.
The second section is about the neutrino cross section.  First, a
number of calculational approaches are described.  Then two particular
calculations are detailed: the Lipari model, upon which predictions in
this analysis are based, and NEUGEN, an alternative cross section
calculation that has been used as an independent check on the
predictions.

\section{Atmospheric Neutrino Flux\protect\label{sec:flux}}

When primary cosmic ray particles (typically protons) reach the
atmosphere of earth, hadronic interactions will occur (the atmosphere
is about eight nuclear interaction lengths).  Atmospheric neutrinos
typically descend from mesons (pions or kaons) in the hadronic shower,
through one of the processes

\vspace{1.5ex}
\begin{minipage}{2in}
  \begin{tabbing}
    $p + X \rightarrow \, $\=$ \pi^+ \, X$ \\
                           \>\hskip 0.5ex $\hookrightarrow \, $\=$\mu^+ \, \numu$ \\
                           \>                     \>\hskip 0.5ex $\hookrightarrow \,
                                            e^+ \, \numubar \, \nue$
  \end{tabbing}
\end{minipage}
\hskip 1in
\begin{minipage}{2in}
  \begin{tabbing}
   $p + X \rightarrow \, $\=$\pi^- \, X$ \\
                          \>\hskip 0.5ex $\hookrightarrow  \, $\=$\mu^- \, \numubar$ \\
                          \>                \>\hskip 0.5ex $\hookrightarrow \,
                                           e^- \, \numu \, \nuebar$
  \end{tabbing}
\end{minipage}
\vspace{2ex}

\noindent
where the charged $\pi$ may be replaced with a charged K, which decays
to $\mu$ about 70\% of the time, or a $K^0_L$, which decays leptonically
66\% of the time.

To first order, a flux calculation requires only a knowledge of the
flux of primary cosmic rays as a function of energy, the density
profile of the atmosphere, and a model of hadronic interactions (as
well as leptonic interactions, which are well-known).  However, if
more precision is desired, a number of subtleties must be considered.

Flux calculations have passed through several generations.  The first
calculations were made in the early 1960s, within the first decade after
reactor neutrinos had been detected; and the first observations of
atmospheric neutrinos were reported in 1965~\cite{kgf,reinesAtmo}.  A
revival of calculations occurred in the 1980s as atmospheric neutrino
interactions constituted a background in searches for nucleon decay deep
underground.  When some observations showed a discrepancy from the predicted
ratio of $\nue$ to $\numu$, attention turned to the possibility of neutrino
oscillations.  Experimenters demanded greater precision, and progress was
made steadily from 1987 until about 1996.  The flux calculation used in this
analysis dates from 1996.  The definitive observation of oscillations by
Super-Kamiokande in 1998 reinvigorated the field.  Since then, many
technical improvements have been demonstrated; unfortunately at this time no
complete calculation has been performed taking the new findings into
account.  The computational intensity of the calculations is scaling along
with the computational resources of the authors.

\subsection{Transport Equations}

The simplest models are analytic (or semi-analytic) solutions of 
{\it transport equations}.  This approach does not simulate individual
interactions but rather treats the flux of a particular particle
species as a time-independent field which evolves as a function of
energy and depth in the atmosphere.  The transport equation approach
dates back at least to the 1941 work of Rossi and Greisen on
electromagnetic cascades in the atmosphere~\cite{rossi}, and has been
successfully extended to consider all the particles discovered in
cosmic ray cascades since then.  To be tractable, the equations must
ignore any transverse momentum imparted to secondary particles; the
showers are taken to evolve collinearly with the primary particle's
momentum.  

If we ignore the production of nucleons by non-nucleonic secondary
hadrons, and ignore continuous ionization loss, the transport equation
for nucleons is then

\[ \derivative{\Phi_N(E,X)}{X} = - \frac{\Phi_N(E,X)}{\lambda_N(E)} +
   \int_E^\infty \differential{E'} \frac{\Phi_N(E',X)}{\lambda_N(E')}
                 \frac{F_{NN}(E,E')}{E}
\]

\noindent
Here $\Phi_N(E,X)$ is the flux, differential in E, of nucleons that
have traversed a distance X (measured in $\gcmsq$) through the
atmosphere.  On the right-hand side of the equal sign, the first term
is a sink in which nucleons of energy $E$ interact and go to a lower
energy (possibly producing mesons at the same time), while the second
term is a source in which nucleons of higher energy $E'$ interact to
produce nucleons of energy $E$ (and possibly some mesons).
$\lambda_N(E)$ is the {\em interaction length} of nucleons in air, so
that the differential probability of a hard interaction while
traversing $\differential{X}$ is ${\differential{X}}/\lambda_N$.  At
typical cosmic ray energies (1~TeV), $\lambda_N$ is about 80~$\gcmsq$.
$F_{NN}(E,E')$ is a dimensionless quantity, defined such that $1/E\ 
F_{NN}(E,E') \differential{E}$ gives the probability that, when a
nucleon of energy $E'$ undergoes a hard scattering process, a nucleon
of energy $E$ results.

For unstable hadronic secondaries ($\pi$,K),

\[ \derivative{\Phi_{i=\pi,K}(E_i,X)}{X} = 
    - \left (\frac{1}{\lambda_i} +  \frac{1}{d_i} \right ) \Phi_i(E_i,X)
    + \sum_j \int \differential{E_j} \frac{\Phi_j(E_j,X)}{\lambda_j}
    \frac{F_{ji}(E_i,E_j)}{E_i}  \]

\noindent                
where $i$ and $j$ indicate particle species, which in the source term
includes nucleons.  The $d_i$ term is a sink due to particle decay.
The decay length must be given in units of matter traversed (i.e.,\ 
$\gcmsq$); therefore, one must adjust the naive decay length to account
for the density of air at that point, as well as the Lorentz dilation
of the particle lifetime.  One finds that

\[ d_i = \rho \gamma \tau_i \]

\noindent
where $\rho$ is the local density of air, $\gamma$ is the Lorentz
factor and $\tau_i$ is the proper lifetime of the particle.  $\rho$
depends on the altitude rather than the slant depth, so the angular
dependence enters explicitly.  The atmospheric profile is to a good
approximation exponential, and we can derive

\[ d_i = \frac{ E \tau_i X \cos \theta_{zen}}{m_i h_0} = 
   \frac{E}{\epsilon_i} X \cos \theta_{zen} \]

\noindent
where $h_0$ is the scale height of the exponential density profile, a few
kilometers, and the definition of $\epsilon_i$ is obvious.  A long-lived
particle has small $\epsilon$.  Numerically, $\epsilon_{K^\pm}$~=~850~GeV,
$\epsilon_{\pi^\pm}$~=~115~GeV, and $\epsilon_{\mu^\pm}$~=~1~GeV\@.  For
energies much greater than $\epsilon$, the particle will probably interact
or range out before it decays.  For energies much less than $\epsilon$, the
particle will probably decay before interacting.

Muons arise from the decay of pions rather than from interacting hadrons,
and we write (neglecting muon interaction and ionization loss),

\[ \derivative{\Phi_{\mu}(E_\mu,X)}{X} = 
    - \frac{\Phi_\mu(E_\mu,X)}{d_i}
    + \int \differential{E_\pi}
    \frac{\Phi_\pi\,(E_\pi,X)\,\epsilon_\pi}{E_\pi\,X\,\cos\theta_{zen}} ~
    \frac{F_{\pi\mu}(E_\mu,E_\pi)}{E_\mu}  \]

\noindent
where the source term is based on $\derivative{\Phi_\pi}{X})_{decay}$.
For electrons and neutrinos, similar transport equations apply.  The
$e$ and $\nue$ equations have no interaction or decay term, and a
source term due to muon decay.  The $\numu$ equation has no
interaction or decay term, and source terms due to both pion decay and
muon decay.

The neutrino probabilities in the muon decay term $F_{\mu\nu}$ must not be
calculated by averaging over angle in the muon's rest frame.  The muon
is produced by a weak interaction fully polarized in the pion rest
frame, and after boosting back into the lab frame the muons retain
some net polarization along the direction of flight.  If the muon
decays, it tends to throw the electron neutrino forward and the muon
neutrino backward, increasing the $\nue/\numu$ ratio at a given
energy.

The various transport equations are coupled, and can in principle be
solved, given the initial conditions that $\Phi_N(E,0)$ is the input
primary spectrum and $\Phi_i(E,0)$ is zero for all other species.  In
practice, analytic solutions can be found if the input spectrum is a
power law in E, $\Phi_N(E,0)~=~E^{-\gamma}$ (approximately true), and
if the cross sections are assumed to scale, so that

\[ \lambda_i(E_i) = \lambda_i \]
\[F_{ji}(E_i,E_j) = F_{ji}(\frac{E_i}{E_j},1) \]

\noindent
independent of energy (also approximately true over most of the
relevant energy range).

Gaisser~\cite{gaisser} gives some exact and some approximate results
from solving the transport equations with a certain model for the
hadronic interactions.  Qualitatively, for nucleons the shape of the
power law spectrum remains unchanged while the magnitude of the flux
attenuates exponentially through the atmosphere with attenuation
length 120~$\gcmsq$ (the vertical column of the atmosphere is about
1~Atm/$g_{grav}~\approx~1000~\gcmsq$).  The attenuation length is
greater than the interaction length for nucleons because it includes
the effect of regeneration due to scattering of higher-energy
nucleons.  Pions and kaons have the same energy spectrum as nucleons.
Their fluxes initially rise as they are produced by plentiful
nucleons, peaking around $100\,\sim\,200~\gcmsq$; but after the
nucleon source attenuates, the meson sink due to decay or interaction
becomes more important and the meson fluxes attenuate, eventually
exponentially.  The lepton fluxes are quite energy-dependent, as they
depend on the relative sizes of the meson interaction and decay terms.
The probability of a high-energy pion decaying before it interacts
goes as $1/E$, so at high energy the muon flux goes as $E^{-\gamma -
  1}$.  At lower energy all the pions decay and the muons, like the
hadrons, go as $E^{-\gamma}$.  For muons of low enough energy that
they cannot reach the earth, the daughter neutrinos follow the energy
dependence of the muons.  At low energy the fraction of muons from
kaon parents is 8\%, increasing to 27\% asymptotically (well above
1~TeV).  This is because kaons are so short-lived; at higher energies
for which the pion usually interacts the kaon continues to decay.  Due
solely to kinematics, the fraction of muon neutrinos due to kaons is
greater than the fraction of muons due to kaons, and becomes greater
than 50\% around 85~GeV~\cite{gaisser01}.

For high energy muons and neutrinos there is an explicit zenith angle
dependence, making horizontal leptons more numerous than vertical.
This arises from the meson decay term.  For a pion produced from a
nucleon at fixed X, a slanting track will be higher in the atmosphere
than a vertical track, and the less-dense atmosphere reduces the
interaction term relative to the decay term, making muon production
more likely.  For an exponential atmosphere, and approximating the
atmosphere as flat rather than curved above the detector, the
resulting dependence is exactly $1/\cos\theta_{zen}$.  This
approximation works well for zenith angles up to $60^\circ$, and is
borne out by measurements of atmospheric muons deep underground.  For
more horizontal leptons the formula must be altered, but it remains
true that the flux increases monotonically with zenith angle.

At high energy, another zenith angle dependence enters implicitly.  A
2.5~GeV muon has $\gamma\,c\,\tau\,=\,17$~km and many will reach the
earth and be stopped before they decay to give high-energy neutrinos.
For a given zenith angle, the neutrino flux underground is that
calculated at the value of X corresponding to the surface of the
earth.  For slanted tracks, X at the surface is greater (the tracks
have seen more $\gcmsq$ of atmosphere), the muons have had more chance
to decay, and the neutrino flux is increased.  This effect also goes
as $1/\cos\theta_{zen}$ near vertical, and increases the
horizontal flux.  For the lowest energy showers, all the pions decay
and all the muons decay so neither of these effects is operative, and
the neutrino flux is isotropic.

Above 1~GeV in the lab, the cross section for neutrino interactions on
nucleons is around $10^{-38}~\mathrm{cm}^2~* \frac{E}{1 \,
\mathrm{GeV}}$, and a trip through the earth encounters on the order
of $10^9~\gcmsq$, or $10^{33} \,
\frac{\mathrm{nucleons}}{\mathrm{cm}^2}$.  So for neutrino energies
less than a TeV the neutrino beam is essentially unattenuated, while
the probability of interaction in the earth is of order unity for
neutrinos above 100~TeV\@.  Under the approximations used in this
transport equation approach, the flux is up-down symmetric below a
TeV, because a detector sees an identical column of atmosphere at
$\theta_{zen}$ and $(\pi \,-\, \theta_{zen})$.  These assumptions also
yield an azimuthally-symmetric flux.  The most likely primary energy
to have produced a neutrino of energy $E_\nu$ is around $10\,E_\nu$.

\subsection{More Precise Calculations}

The simple transport equations above give a good qualitative
understanding of the neutrino flux underground; however, they are
inadequate for quantitative work because a number of important effects
have been neglected entirely, and others approximated.  Some clever
people have made heroic efforts to incorporate many effects into the
semi-analytic calculation through {\em ad hoc} mechanisms; but the
most reliable calculations resort to Monte Carlo techniques that
simulate individual showers and/or individual trajectories.  Monte
Carlo programs can easily incorporate effects that were ignored in the
semi-analytic calculations, such as muon ionization loss and
non-scaling cross sections.

Modern experiments have based their analyses on either of two Monte
Carlo calculations: Super-Kamiokande uses the 1995 work of Honda {\em
  et al.}~\cite{hkkm}, referred to as HKKM, while Soudan-2 and MACRO
(including this analysis) refer to the Bartol group's
high-energy~\cite{bartolHigh} and low-energy~\cite{bartolLow}
calculations of 1995 and 1996, collectively referred to as Bartol.
Both HKKM and Bartol are ``one-dimensional'' calculations which ignore
transverse momentum of secondaries.  The most interesting new
calculation to appear on the scene since 1996 is that of Battistoni
{\em et al.}~\cite{fluka3D} which includes a new hadronic interaction
code (FLUKA) as well as adding in some of the three-dimensional
effects ignored by the other two calculations.  However, this
calculation is still under development and is not yet ready to be used
as the main prediction for serious experimental analyses.

Discussion follows on several points that must be considered by the
authors of calculations.  Generally, the choices made by HKKM and
Bartol will be indicated.  Afterward, the Bartol calculation (which
is used to make predictions for this analysis) is summarized.

\subsubsection{Primary Cosmic Ray Flux\protect\label{sec:priflux}}

At the time Bartol and HKKM were written there was controversy about
the overall normalization of the primary flux above 3~GeV, with the
measurement of Webber~79~\cite{webber} more than 50\% higher than
Leap~87~\cite{leap} around 10~GeV and above.  HKKM based its
calculation on a flux that split the difference, while Bartol used a
flux just a few percent above the Leap~87 result.  In the 1990s, a
number of new measurements were made which all generally fall within
$\pm 10\%$ of the Leap~87 measurement (see
Figure~\ref{fig:primary}).  Future calculations will probably
disregard Weber~79 and fit the remaining data.

\picfig{fig:primary}{Measurements of the primary flux of cosmic ray protons.
  Note that the flux below 10~GeV depends on solar epoch;
  IMAX~\protect\cite{imax} and MASS2~\protect\cite{mass2} made their
  measurement near solar maximum while LEAP~\protect\cite{leap},
  CAPRICE~\protect\cite{caprice} and BESS~\protect\cite{bess} were near
  minimum.  Also shown are the flux models used by the Bartol and HKKM
  calculations.  (Figure is due to Lipari~\protect\cite{lipariVenice}.)}
{\resizebox{6in}{!}{\includefigure{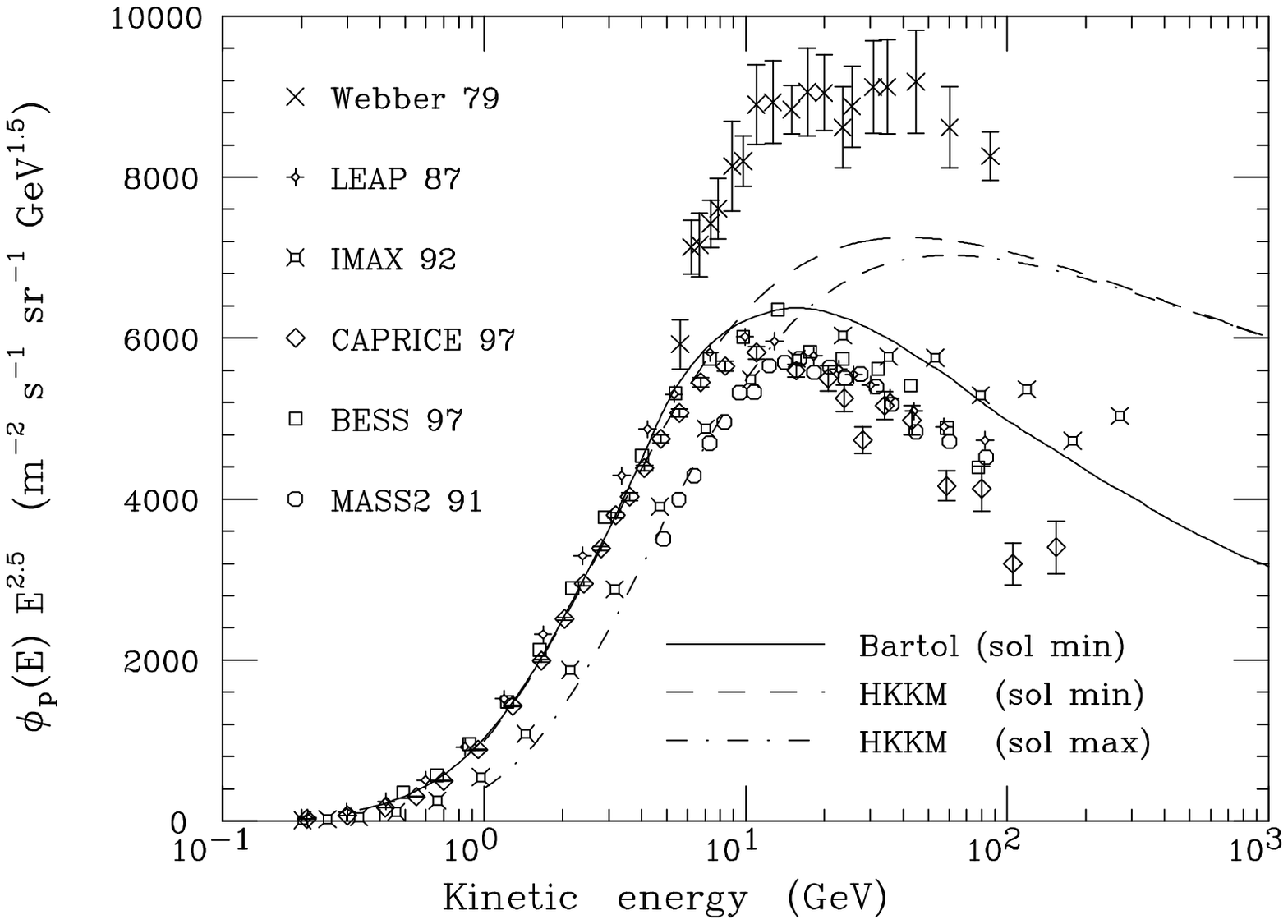}}}

The composition of the primary cosmic rays is also somewhat important,
because neutrons produce a few more negative pions and protons produce
more positive.  This is important for the experiments because the
cross sections of neutrinos and antineutrinos differ.

At energies below a few GeV, the primary flux varies with the 11-year
solar cycle.  When the sun is most active and the solar wind is
strongest, it carries a magnetic field that excludes the lowest-energy
primaries from the solar system.  Bartol and HKKM both use good models
of the solar modulation.

\subsubsection{Nucleon-Nucleus Interactions}

Most of the primaries that interact are protons, and heavier primaries
can be considered as a superposition of independent nucleons; thus the
relevant primary reaction is nucleon-nucleus where the nucleus is a
component of air.  A hadronic interaction model determines several
quantities, which are in general energy-dependent: the multiplicity of
produced mesons, the energy spectrum of produced mesons, the
$\pi^+/\pi^-$ ratio and the $\pi/K$ ratio.  All of these will directly
impact the neutrino fluxes.  Note that the approximation that cross
sections scale, necessary to find an analytic solution, has been
abandoned in the Monte Carlo approach in favor of energy-dependent
functions.

The multiplicity and spectrum of pions in proton-nucleus interactions
is well-measured only for those pions that carry off a large fraction
of the primary nucleon's energy (e.g.,\ $>\approx\,20\%$ for protons
on beryllium at 20~GeV).  For softer pions, the extrapolation is
entirely based on the model of nuclear effects used in the
calculation, with HKKM using the Lund color string-based model
FRITIOF~\cite{fritiof}, and Bartol using their own TARGET model, based
on a parametrization~\cite{emulsion} of pseudorapidity distributions
in nucleon-nucleus reactions.

\subsubsection{Atmospheric Muon Fluxes}

Muon fluxes are necessarily calculated as an intermediate step in the
calculation of neutrino fluxes, so the model can be checked by
comparing it with experimental muon data.  The calculated flux of
high-energy muons that reach the ground can be compared with reliable
high-statistics measurements of muons at ground-level.  However, at
lower energy muons decay in flight so measurements must be made in the
atmosphere.  There have been several recent measurements of
atmospheric muons as a function of height, taken during the ascent of
balloon experiments.  In the flux calculations, a wrong primary flux
can be compensated by a wrong lepton yield per primary particle to
yield a correct muon flux.  Indeed HKKM and Bartol have differences of
order tens of percent in their primary fluxes and interaction models
in compensating directions, so that both models match the muon data at
ground level, and predict very similar neutrino fluxes.  Their
agreement with higher altitude muons is less impressive.

The Monte Carlo approach can add in two terms that were ignored in the
analytic calculation: ionization loss of propagating muons in the atmosphere
(which would amount to about 2~GeV for a muon traversing the entire
atmosphere) and hard interactions of muons in air.

\subsubsection{The Earth's Magnetic Field}

The geomagnetic field plays a significant role in modifying the
neutrino flux, especially at neutrino energies below about 1~GeV\@.  The
most important effect is the suppression of low-energy primaries.  As
charged particles approach the earth they begin to bend in the earth's
magnetic field, with the degree of bending determined by the
particle's rigidity, which is the ratio of momentum to charge.  The
lowest-rigidity particles will be bent away so that they will never
reach the earth.

For calculation, it is more fruitful to think of the problem
backwards.  Is it possible for a primary to approach from a certain
angle at a certain point in longitude and latitude?  By running the
equations of motion backward it is possible to determine if the
particle originated from infinity.  If the particle's past trajectory
is trapped in the earth's magnetic field, or if the past trajectory
passed through the earth, the trajectory is not a possible source of
primaries.  We will refer to such trajectories as ``forbidden,'' and
all other trajectories as ``allowed.''  

The problem was solved analytically by St\"{o}rmer~\cite{stormer} in
1930 for a perfect dipole field, and neglecting the shadowing effect
of the solid earth.  If the dipole strength is $M$, at a given
magnetic latitude $\lambda_M$, a positive particle is allowed to
approach from position $(r,\theta_{zen},\phi)$ (where $\theta_{zen}$
is the local zenith angle, $r$ is the radius measured from the
dipole's center, and $\phi$ is a local azimuthal angle measured
clockwise from local magnetic north) if its rigidity is greater than
the St\"{o}rmer rigidity,

\[ R_S^+(\lambda_M,r,\theta_{zen},\phi) =
  \left( \frac{M}{2\,r^2} \right) 
  \left\{ \frac{\cos^4 \lambda_M}{[1\,+\,(1\,-\,\cos^3\lambda_M\,
        \sin\theta_{zen}\,\sin\phi)^{1/2}]^2} \right\}
\]

\noindent
Note that the rigidity cutoff is greatest at the magnetic equator and
vanishes at the poles (where a particle can ride in along a magnetic
field line).  The zenith angle dependence shows that in the $\sin\phi
> 0$ hemisphere, horizontal particles have a higher cutoff than
vertical; while the opposite holds in the other hemisphere.  Finally,
the cutoff is greatest for particles arriving from the east and moving
to the west (see Figure~\ref{fig:eastwest}).  This last result breaks
the azimuthal symmetry of the neutrino flux, and is known as the
east-west effect.  It was measured in charged particles in
1933~\cite{mexicoCity,mexicoCity2}.  Super-Kamiokande recently
made the first observation of the effect in
neutrinos~\cite{SKEastWest}.  For the earth, the maximum St\"{o}rmer
rigidity is for a horizontal particle traveling east at the magnetic
equator, $(R_S)_{max} = M_\oplus/2\,r_\oplus^2$ = 60~GV\@.  (The unit GV
is 1~GeV divided by the charge of one electron, so 60~GV would
correspond to a 60~GeV proton or a 120~GeV alpha, etc.)  For most
locations on the earth and most directions of approach, the cutoff is
substantially less.

\picfig{fig:eastwest}{The origin of the east-west effect.  The circle
  is a slice of earth perpendicular to the magnetic axis.  Three
  trajectories are shown approaching the same point on earth, for
  positively-charged particles with the same rigidity but different
  origins.  The west-going trajectory is forbidden, so the west-going
  flux is reduced.  Likewise, the rigidity cutoff for the vertical
  trajectory would be more than that for the east-going
  trajectory.}{\resizebox{3.5in}{!}{\includefigure{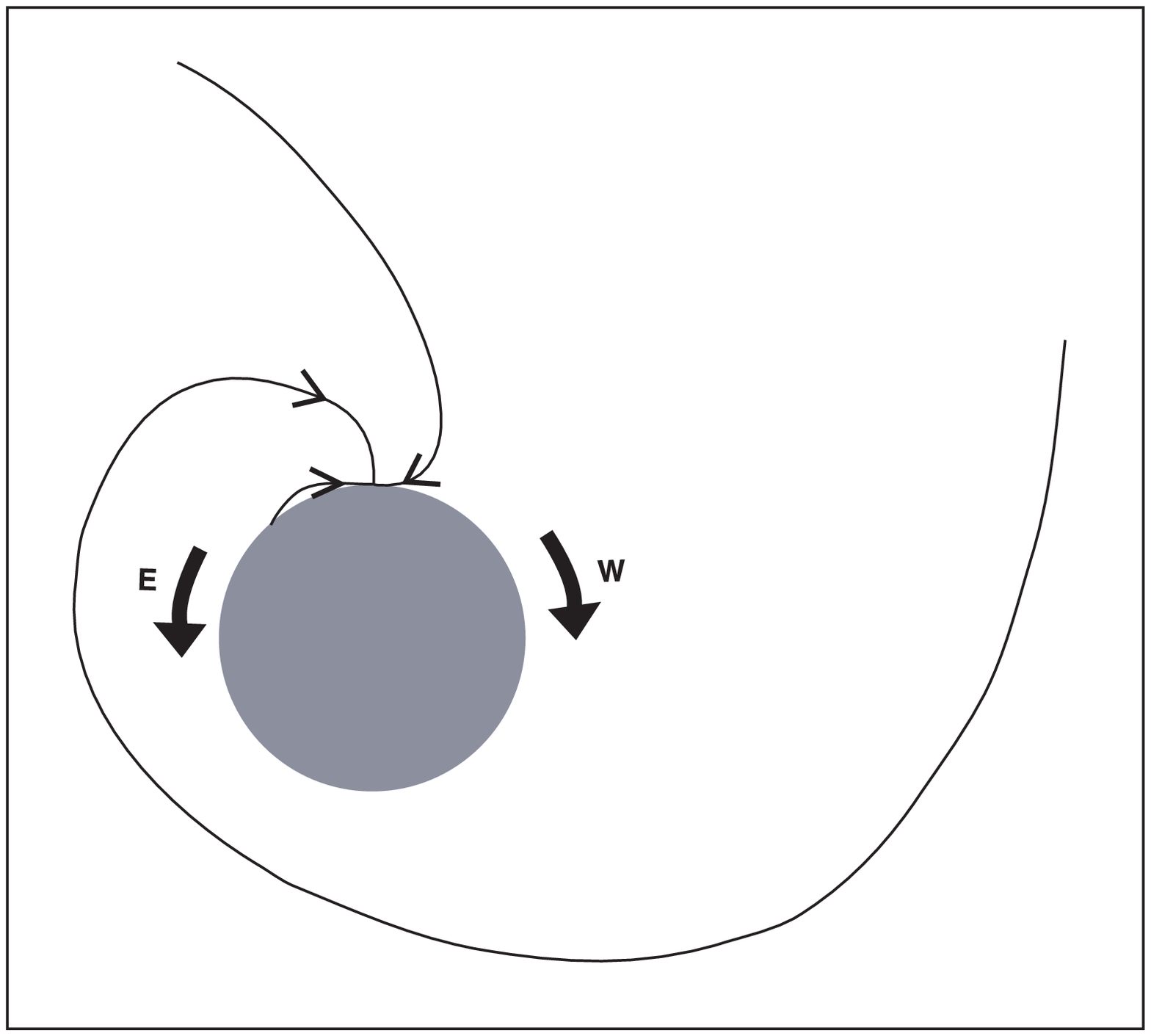}}}

Geomagnetic effects also break the up-down symmetry of the neutrino
flux.  Consider a detector at low magnetic latitude, where the cutoff
is larger than average.  All the downgoing neutrinos originate within
tens of kilometers, and have the same cutoff, so the flux of downgoing
neutrinos will be lower than average.  However, the upgoing neutrinos
come from all over the earth and their flux will be close to average.
Thus, that detector will see fewer downgoing neutrinos than upgoing.
For detectors at high latitude, the asymmetry goes the other way.
However, at neutrino energies above a few GeV, the neutrinos come from
primaries that are above cutoff and the up-down symmetry is restored.

A multipole expansion of the earth's magnetic field shows it is 90\%
dipole; however, both Bartol and HKKM forgo the perfect dipole model
and calculate the trajectories in a detailed model of the actual
geomagnetic field, the International Geomagnetic Reference
Field~\cite{IGRF}.

By Liouville's Theorem, the flux remains constant along a trajectory in a
magnetic field.  Therefore, the geomagnetic cutoff can be implemented in a
Monte Carlo calculation by simply throwing an isotropic flux of primaries
and then eliminating those that come from forbidden trajectories.

All modern calculations consider the effect of the geomagnetic field
on primaries, but so far no detailed calculations have included
geomagnetic effects on secondaries.  Muons travel long distances and
can be affected by the earth's field.  Lipari has recently
shown~\cite{lipariEastWest} that near horizontal, the east-west effect
is enhanced for decay products of $\mu^+$ (i.e.,\ $\nue$ and
$\numubar$) and depressed for the daughters of $\mu^-$ (i.e.,\ $\numu$
and $\nuebar$).  Due to the different cross sections for neutrinos and
antineutrinos, this affects the zenith angle distribution.  It is most
significant for neutrinos below 1~GeV\@.

\subsubsection{Transverse Momentum of Secondaries}

Probably the most important limitation of HKKM and Bartol is that they
suppress the three-dimensional kinematics of the interactions in the
atmosphere.  This results in an enormous simplification of the
computation.  One can initially calculate the distribution of neutrino
yield from nucleons in the atmosphere of various energies, independent
of angle.  Then one can generate events choosing only primaries that
point at the detector.  If the trajectory of the primary is allowed
(in the sense defined in the previous section), the neutrino yield is
sampled from the pre-calculated distribution.  If the trajectory is
forbidden, it makes no contribution to the neutrino flux.

It was long assumed that any effects due to the non-collinearity of
real shower development would average to zero, or at least something
negligible.  For high-energy neutrinos, where the opening angle in
each reaction or decay is small, this is approximately true.  However,
for low-energy neutrinos which can have wide opening angles, it
becomes possible for trajectories to migrate preferentially out of a
high-density region of phase space into a lower-density region and the
distributions can be greatly affected.  For example, the FLUKA-3D
calculation has shown that the near-horizontal flux of neutrinos below
1~GeV is greatly enhanced by including transverse momentum with wide
opening angles, and Lipari has given a qualitative
explanation~\cite{lipari3D}.

Including the effect of the geomagnetic field on secondary muons (discussed
above) is a second example of three-dimensional effects ignored by HKKM and
Bartol.

\subsection{The Bartol Flux\protect\label{sec:bartol}}

This analysis uses the Bartol flux for acceptance studies and to make
predictions of no-oscillations and oscillated measurements in MACRO\@.  The
flux tables were computed including geomagnetic effects for the Gran Sasso
location by Todor Stanev, using the procedures described in Reference
\cite{bartolLow} for neutrino energies a GeV and below, and Reference
\cite{bartolHigh} for energies above 1~GeV\@.

For energies below 10~GeV, the calculation was done separately for
solar minimum and solar maximum epochs.  When running the code, a
solar midcycle choice is also available, which just returns the
average of the solar minimum and solar maximum fluxes.  Also,
geomagnetic effects on primaries were only considered for neutrino
energies below 10~GeV\@.  Both of these effects are negligible for
neutrinos above 10~GeV\@.  So above 10~GeV the flux is up-down
symmetric, while for energies below 10~GeV the fluxes are different
for upgoing and downgoing particles.  While an east-west effect would
have appeared in the original Monte Carlo events on which the tables
are based, the tables were calculated integrating over azimuthal
angle.

The final tables depend on three indices -- neutrino type ($\numu,
\nue, \numubar, \nuebar$), neutrino energy, and zenith angle.  For a
given neutrino type, the flux at a given energy and zenith angle is
found by finding three energy bins surrounding the desired energy and
interpolating across angle bins for each to find the flux at the given
angle, and then interpolating across these energies to find the flux
at the given energy and angle.  The data is binned in 63 energy bins,
spaced logarithmically from 50~MeV to 100~TeV, and in 12 angular bins
for $\cos\theta_{zen}$ = (1.00, 0.75, 0.50, 0.25, 0.15, 0.05, -0.05,
-0.15, -0.25, -0.50, -0.75, -1.00).  The flux is assumed to be the
same in all azimuthal directions.

Figure~\ref{fig:flux} shows some results of the flux calculation.

\picfig{fig:flux}{The Bartol flux, calculated using geomagnetic
  effects at Gran Sasso.  As a function of zenith angle, the flux is
  shown in several energy bands.  The relative normalization shown is
  not always accurate, having been adjusted to cleanly separate each
  band in the figure.  The open symbols are the upgoing flux reflected
  across the horizontal to give a sense of the up-down asymmetry.
  Less than 1\% of detectable events in this analysis come from
  neutrinos below 0.4~GeV\@.}{\pawfig{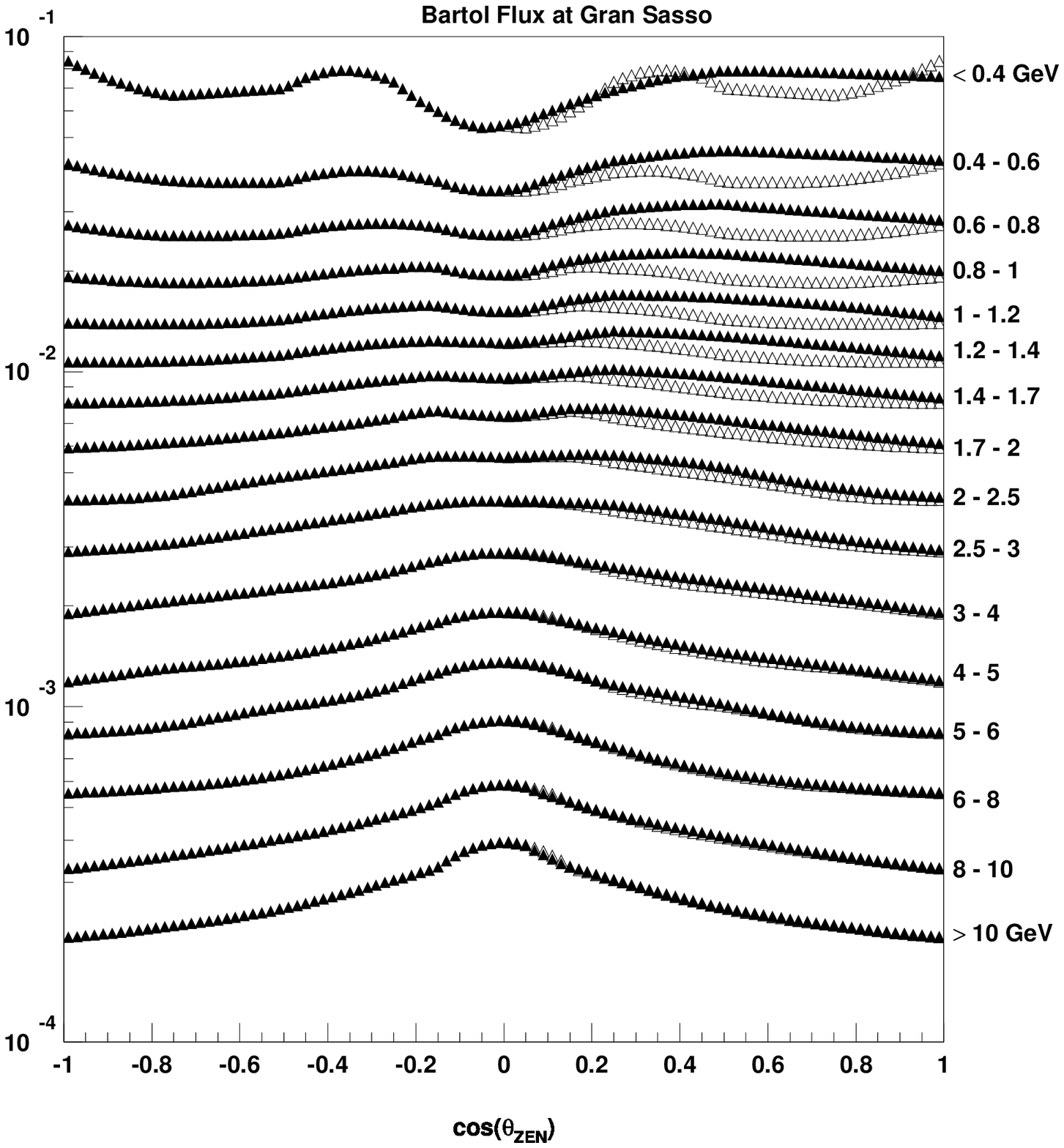}}

\section{Neutrino Cross Section\protect\label{sec:nuSigma}}

After the flux is calculated, the other major ingredient of a neutrino
interaction rate calculation is the neutrino cross section.  This
section introduces the basic concepts, and explains a number of models
and approximations that have been used.  Finally, two particular
calculations are described -- the Lipari model, on which the
predictions for this analysis are based, and the NEUGEN model which
has been used as an independent check.

Most atmospheric neutrinos pass through MACRO without depositing any
measurable energy.  It is only the rare neutrino that undergoes a hard
interaction that produces a detectable event.  (A 1~GeV neutrino
traversing MACRO from top to bottom has an interaction probability of
a few per trillion.)  The present work measures primarily muons
resulting from the charged current interaction of muon neutrinos or
antineutrinos.  Neutral current and electron neutrino interactions
must also be considered as a lesser source.

The weak cross section for a neutrino to scatter off a point particle
of mass $M$ goes as $s$, the Mandelstam variable, which is $M^2 \,+\,
2\,M\,p$\raisebox{-0.7ex}{\textsc{lab}}.  In ordinary matter the
number of electrons is of the order of the number of nucleons but the
cross section is around three orders of magnitude lower due to the low
electron mass; so we can neglect neutrino-electron scattering and
concentrate on neutrino-nucleon scattering.

Different conceptual models are used to discuss neutrino-nucleon
interactions in different regimes.  We can discuss all of them with a
common set of kinematic variables, illustrated in
Figure~\ref{fig:nukine}.  In the figure, quantities represented as
$\fourvec{x}$ are four-vectors.

\picfig{fig:nukine}{Neutrino interaction kinematics.}
{\makebox[3.33in]{\resizebox{3.2in}{!}{\includefigure{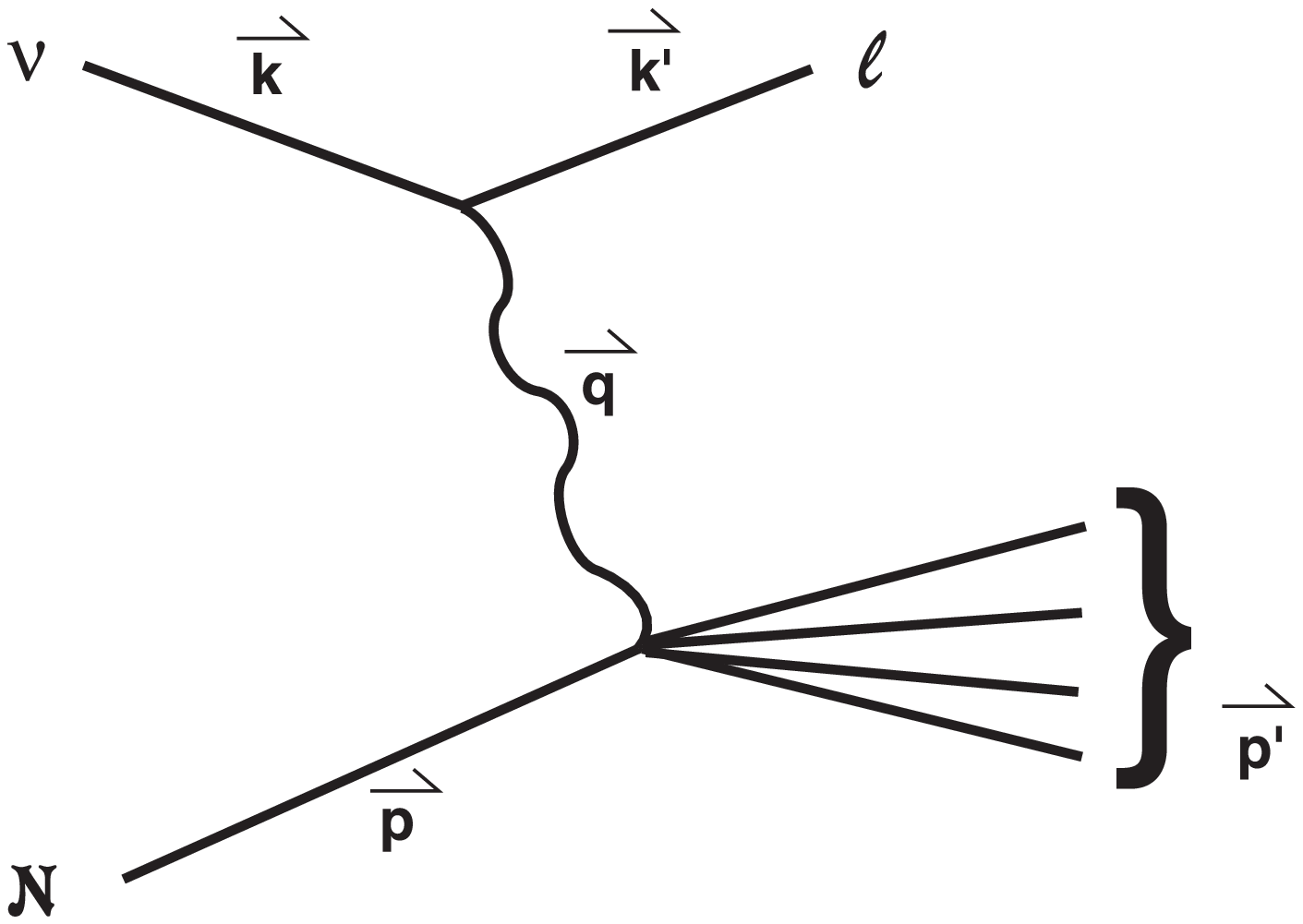}}}
  \raisebox{18ex}{
      \framebox{
        \parbox{1in}{
          \begin{tabbing}
            $Q^2 \, $\=$ = -(q^2)$ \kill
            $\fourvec{q} $\>$ = \, \fourvec{k} - \fourvec{k'}$ \\
            \\
            $Q^2 $\>$ = -(\fourvec{q}^2)$ \\
            \\
            $W   $\>$ = \, \fourvec{p'}^2$ \\
            \\
            $y   $\>$=\fourvec{p}\cdot\fourvec{q}/\fourvec{p}\cdot\fourvec{k}$\\
                  \>$\stackrel{lab}{=} (E_\nu - E_l)/E_\nu$ \\
            \\
            $x   $\>$ = Q^2/(2 \fourvec{p}\cdot\fourvec{q}) $ \\
                  \>$\stackrel{lab}{=} Q^2/(2 M_N (E_\nu - E_l))$
          \end{tabbing}
        }
      }
  }
}

At high $Q^2$, the neutrino is interacting with the partons within a
nucleon, and the formalism of Deep Inelastic Scattering (DIS) is
applicable.  The standard model gives exact formulas for the coupling
of neutrinos to partons.  Published phenomenological {\em parton
  distribution functions} give the probabilities of partons carrying a
given fraction of the nucleon momentum.  Parton distribution functions
have been determined using a variety of experimental probes --
electron-nucleon scattering, muon-nucleon scattering, neutrino-nucleon
scattering and nucleon-nucleon scattering.  If the parton model were
perfect, all of these probes would give the same result.  In reality,
there are small but significant differences, and not all structure
functions reproduce neutrino scattering data well.

The lepton kinematics can be fixed in inelastic events by choosing two
independent scalar variables.  Most Monte Carlo calculations choose
$x$ and $y$.  $y$ is chosen according to $\derivative{\sigma}{y}$
integrated over all $x$, and after $y$ is fixed $x$ is chosen with the
aid of the parton distribution functions.

The choice of $x$ and $y$ gives the lepton kinematics unambiguously;
however, the hadronic situation is confused.  After the interaction,
there is a struck parton plus spectator partons at rest in the lab.
Some phenomenological hadronization prescription must be applied to
produce the observable hadrons.

At smaller $Q^2$, the neutrino ``sees'' the nucleon as a coherent
whole, and perturbative calculations are not applicable.  Most of the
non-perturbative portion of the cross section is included in two
cases: quasi-elastic scattering ($\nu \, n \rightarrow l^- \, p$ or
$\overline\nu \, p \rightarrow l^+ \, n$) and single pion production
(QEL and 1PI, respectively).  Note that while most of the cross
section of high energy (multi-GeV) neutrinos is deep inelastic, even
very high energy neutrinos may sometimes participate in a low $Q^2$
event, so the non-perturbative cross section applies at all energies.

In the quasi-elastic case, the final hadronic mass is predetermined and
one variable (perhaps $Q^2$) suffices to specify the kinematics.
Due to Lorentz invariance and other symmetries, the uncertainty in
quasi-elastic scattering may be confined to a few form factors, as in
the classic treatment of Llewellyn Smith~\cite{lsmith}.  The hadronic
current is written in terms of scalar, pseudoscalar, vector, axial
vector and tensor form factors.  The scalar and tensor factors are
assumed zero due to the non-observation of ``second-class currents.''
The pseudoscalar factor contributes proportional to $M_l/M_N$ and may
be ignored.  The vector form factor may be related to well-determined
electromagnetic form factors via the ``Conserved Vector Current''
hypothesis.  Only the axial vector form factor is undetermined.  Its
value at $Q^2 = 0$ is known from nuclear beta decay, and in neutrino
scattering experiments it has been determined that the $Q^2$
dependence is consistent with a dipole form,

\[ F_A(Q^2) = \frac{-1.23}{(1 - Q^2/M_A^2)^2} \]

\noindent
with $M_A$ around 1~GeV\@.

Because the neutrino usually reacts with a bound nucleon rather than a
free nucleon, the naive cross section is modified by nuclear effects.
The most important are the Fermi motion of nucleons confined to the
nucleus, and Pauli blocking in which some low-energy-transfer events
are forbidden because the scattered nucleon is excluded from the
energy level to which it would otherwise go.

The situation for single pion production is more complex.  The
extra degrees of freedom make a simple form factor approach
unworkable.  Fogli and Nardulli~\cite{fogli} have calculated
differential cross sections in good agreement with available data
using a number of restrictive assumptions, applying some form factors
known from quasi-elastic scattering, and summing a number of effective
Feynman diagrams using exact propagators of all relevant resonances.
They also include a non-resonant background and calculate interference
terms with the resonances.  Rein and Segahl~\cite{rein} have taken a
different approach, modeling the nucleon as three point particles with
harmonic oscillator potentials between each pair using a hamiltonian
suggested by Feynman, Kislinger and Ravndal~\cite{FKR}.  Each excited
state of the nucleon can be related to a real baryonic resonance by
comparing quantum numbers, and the matrix elements for a weak charged
current interaction to excite the ground-state nucleon to a particular
resonance can be calculated exactly.

Because the hadronization process in DIS is not straightforward,
there is no clear boundary between DIS and resonant production.  If
one takes account of single pion production completely in the
exclusive channel, any events in the inclusive (DIS) channel that
result in single-pion production may be double counting that part of
the cross section.  Two approaches have been taken to this problem.

One may omit the calculation of the exclusive channels (quasi-elastic
or resonant scattering).  By considering higher order QCD effects in
extrapolating structure functions to small $x$ (where experimental
measurements do not extend), one may hope to be effectively
compensating for non-perturbative effects.  In the 1990s, due to the
electron-proton collider HERA, structure functions became
available which were valid to lower values of x than any earlier
structure functions.

Other models do a full calculation in both the inclusive (DIS)
and exclusive channels, but apply some prescription for subtracting a
fraction of one or both cross sections to eliminate the
double-counting.

\subsection{The Lipari Cross Section Model\protect\label{sec:lipari}}

The final results in this analysis are compared with the theoretical
prediction incorporating a cross section model due to
Lipari\footnote{Lipari is a MACRO collaborator, but the code is freely
  available.}.  The main feature of the Lipari model is that it
incorporates the no-double-counting prescription of Lipari, Lusignolo
and Sartogo~\cite{LLS} which calculates three cross sections:

\[ \sigma_{\mathrm{TOT}} ~=~ \sigma_{\mathrm{QEL}} ~+~
  \sigma_{\mathrm{1PI}} ~+~ \sigma_{\mathrm{DIS}} \]

\noindent
where the 1PI cross section is restricted to final state hadronic mass
$\sqrt{W}$ (see Figure~\ref{fig:nukine}) of less than 1.4~GeV and the
DIS cross section to above 1.4~GeV\@.  This eliminates the low $Q^2$
events, which are dominated by the exclusive channels, from being
counted as part of the DIS cross section.

This model reproduces well the total cross section data at high
energy, as well as the Brookhaven data at sub-GeV energies (see
Figure~\ref{fig:lipariData}).  However, experimental data exists in
contradiction to the Brookhaven data.  Furthermore, where this model
predicts an enhancement of the antineutrino cross section at sub-GeV
energies, existing data seem to show a decrease.

\picfig{fig:lipariData}{Comparison of the $\numu$ charged current cross
  section with selected experimental data.  Figure is reproduced from
  Reference~\protect\cite{LLS}.}
  {\resizebox{6in}{!}{\includefigure{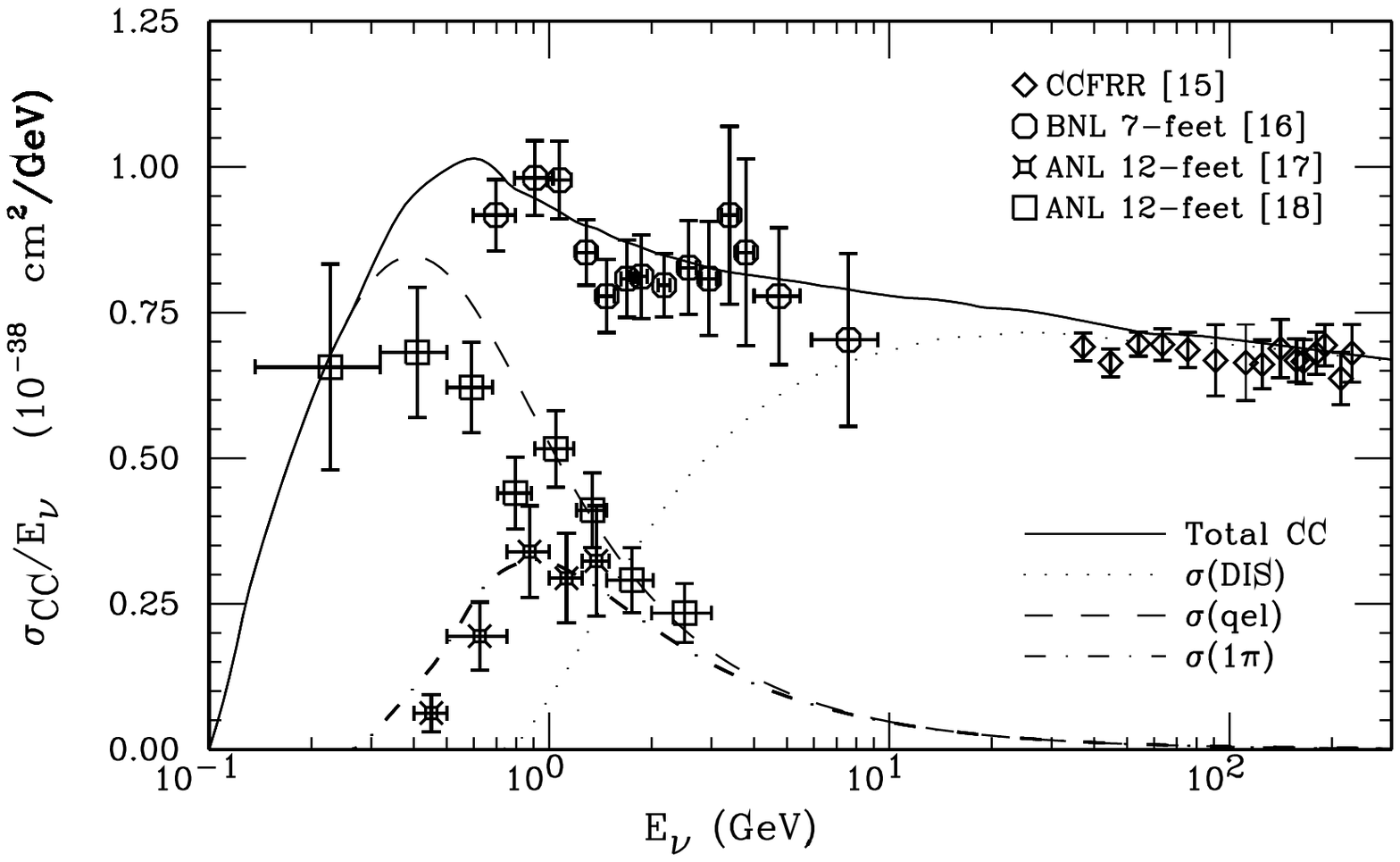}}}
  
For the quasi-elastic cross section, Lipari uses the equations of
Llewellyn Smith described above, with axial mass $M_A$ = 1~GeV\@.
Neutral current elastic scattering is omitted from the model.

Lipari calculates the one-pion cross section according to

\[ \sigma_{\mathrm{1PI}}(Q^2) = \int_0^{W_{max}} dW\ \sigma_{FN}(Q^2,W) \]

\noindent
where $\sigma_{FN}$ is the cross section of Fogli and Nardulli
(described above) and $W_{max}$ is \mbox{$(1.4\ \mathrm{GeV})^2$};
i.e.,\ his one-pion total cross section at some value of $Q^2$ is the
total cross section of Fogli and Nardulli with the portion resulting
in a final hadronic mass over 1.4~GeV subtracted out.  However, to
simplify the calculation, the exact differential cross section of
Fogli and Nardulli is not used; instead the process is assumed to pass
through the $\Delta$ baryonic resonance, with the resonant mass
sampled from the $\Delta$ Breit-Wigner distribution ($M_\Delta$ =
1.2~GeV and $\Delta M_\Delta$ = 120~MeV) and a simple phase space
decay to a nucleon and one pion.  Note that any final states with a
hadronic mass less than 1.4~GeV and more than one pion cannot be
produced in this model.  Finally, note that the lepton spectrum is
exactly that of Fogli and Nardulli; only the hadronic state is
affected by the approximation.

In the code provided by Lipari, the deep inelastic cross section is
calculated to leading order, and any leading-order parton distribution
functions may be used, via the CERNLIB code PDFLIB~\cite{pdflib}.  We
choose the GRV94-LO set~\cite{GRV94}.  (The authors have published a
minor update, GRV98-LO~\cite{GRV98}.  The new version changes total
cross section by less than 1/2\% and does not noticeably change the
lepton kinematic distributions.)  Immediately after the charged
current interaction, the hadronic system consists of a struck parton
and two spectator partons, which are handed to the Lund string
fragmentation model~\cite{jetset} to produce the final hadronic state.
Beginning with a quark and a diquark, Lund will correctly produce an
energetic leading hadron in the lab frame containing the struck quark,
and will conserve baryon number.

\subsection{The NEUGEN Cross Section Model\protect\label{sec:neugen}}

As a consistency check, the prediction of the Lipari model has been
compared with a prediction incorporating a completely independent
cross section model, NEUGEN, which was originally developed for the
Soudan experiment by Giles Barr~\cite{barr} and subsequently extended
by Hugh Gallagher~\cite{gallagher} and Geoff Pearce.  It is now
maintained by the MINOS collaboration.

This model also computes separate cross sections for quasi-elastic,
resonant, and deep inelastic scattering processes.  The prescription
to avoid double counting is less sophisticated than Lipari; a fixed
fraction of the low-multiplicity part of the deep inelastic cross
section is suppressed by hand.

The quasi-elastic calculation is essentially identical to that of
Lipari, with an axial mass $M_A$ = 1.032~GeV\@.

The resonant cross section implements the model of Rein and Segahl
(described above), considering 17 different resonances of the baryon.

The neutrino-parton cross section for the deep inelastic process is
computed using any available parton distribution functions from
PDFLIB; we again choose GRV94-LO\@.  The hadronic final state is
considered as a coherent whole, with a mass and momentum given by the
kinematic variables $x$ and $y$.  A final hadronic multiplicity is
picked based on the mass according to the KNO distribution~\cite{KNO},
and the various kaons, pions and nucleon are assigned momentum
probabilistically according to available phase space.  This leads to
inaccurate momenta distributions for high $Q^2$ events, with no
energetic leading hadron.

\chapter{Measurement of Neutrino Interactions in the
Detector\protect\label{chap:meas}}

\section{Topologies of Neutrino-Induced Muons in MACRO}

Neutrino-induced muons detectable by MACRO can exhibit any of several
topologies (see Figure~\ref{fig:sctop}).

\picfig{fig:sctop} {Topologies of neutrino-induced muons in MACRO\@.
  The sense (upgoing or downgoing) can be determined only for those
  tracks with two or more scintillator hits (represented by white
  stars).  (a) Throughgoing Upward Muons -- ANALYZED; (b) Throughgoing
  Downward Muons -- NOT ANALYZED; (c) Stopping Downward Muons -- NOT
  ANALYZED; (d) Contained-Vertex Upward Muons -- ANALYZED IN THIS
  WORK; (e) Contained-Vertex Downward Muons -- ANALYZED; (f) Stopping
  Upward Muons -- ANALYZED; (g) Fully-Contained Muons -- NOT
  ANALYZED.}{\includefigure{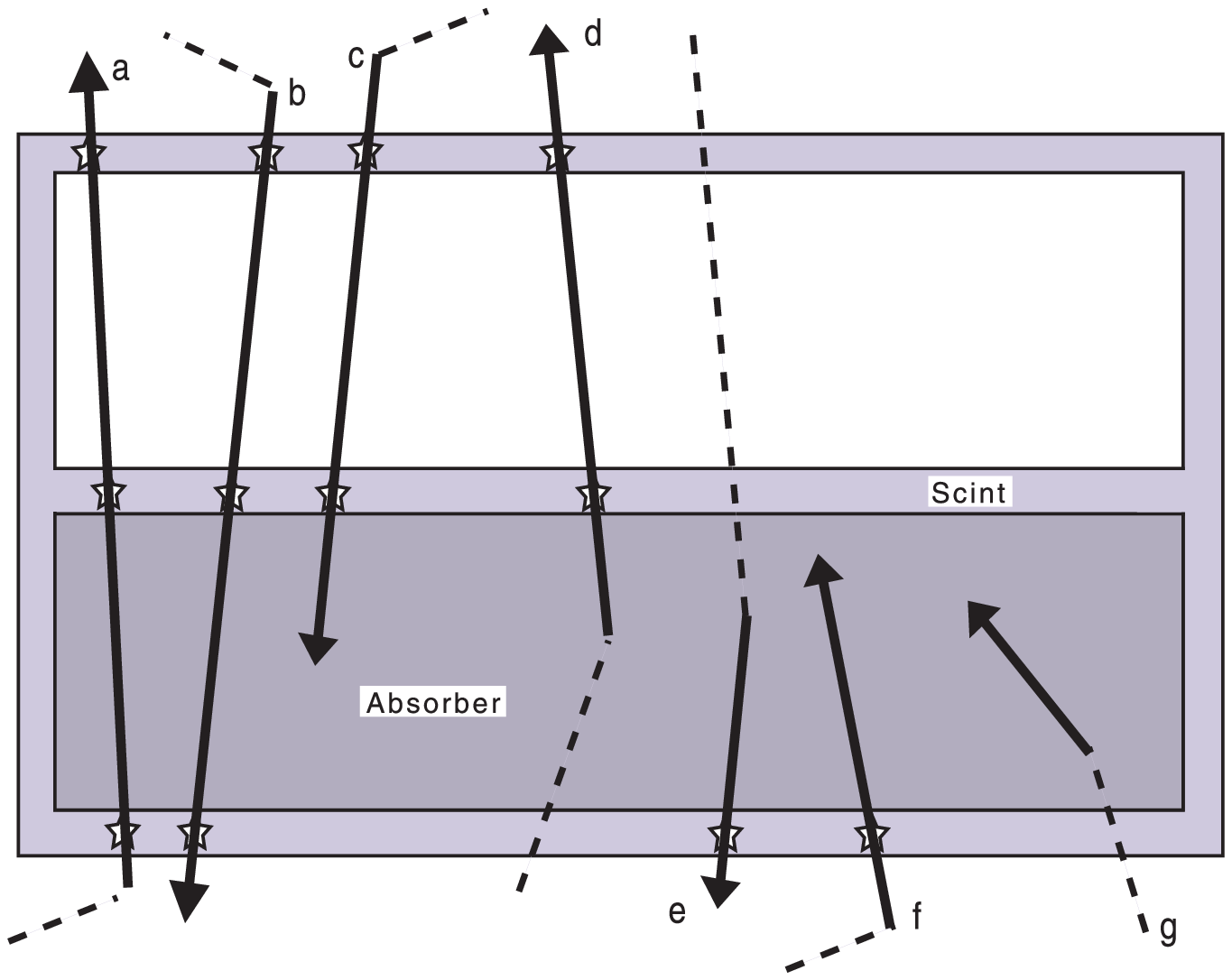}}

Category (a), Throughgoing Upward Muons, are the most numerous
detectable neutrino-induced events in MACRO\@.  MACRO has been
publishing results on this topology since 1994.  This thesis does not
update results on this topology.

Categories (b) (Throughgoing Downward Muons) and (c) (Stopping
Downward Muons) are unfortunately indistinguishable in MACRO from the
overwhelming background of atmospheric muons (muons produced in the
atmosphere by cosmic ray showers) and MACRO cannot discern anything
about neutrinos from these topologies.

Category (d) (Contained-Vertex Upward Muons) is the topic of this
analysis.  In this topology, the direction (upward or downward) must
be determined to distinguish category (d) from stopping atmospheric
muons which have the same topology as (c).  In MACRO the direction is
determined by scintillator timing between two layers (by observing
whether the lower box was hit before or after the upper box).
Therefore, this analysis requires the muon to hit the Center layer and
a higher tank.

Categories (e) (Contained-Vertex Downward Muons) and (f) (Stopping
Upward Muons) are indistinguishable from each other in MACRO\@. The
direction (upward or downward) cannot be determined because the muon
does not hit two scintillator layers.  However, all the events of these
categories are neutrino-induced, and it is possible to make neutrino
measurements based on this topology.  MACRO has published results on
this topology~\cite{lowenu}.  This analysis does not update results on
this topology, although it does rely on updated results in
Chapter~\ref{chap:interpretation}, Interpretation.

Finally, category (g) (Fully-Contained Muons) could perhaps be
measured.  However, MACRO is too granular to make this measurement
well, and no attempt has been made at its analysis.

For the three topologies that have been analyzed ((a), (d) and
(e)+(f)), Figure~\ref{fig:entop} gives the energy distribution (from a
Monte Carlo calculation assuming $\numu$ disappearance with maximal
mixing and $\dmm = 3.2~\times~10^{-2}~\mathrm{eV^2}$) of the parent
neutrinos 
giving rise to detectable events, based on the cuts made in the
analyses identifying the events.

\picfig{fig:entop}{Energy distribution of parent neutrinos
  giving rise to detectable events in MACRO\@.}{\pawfig{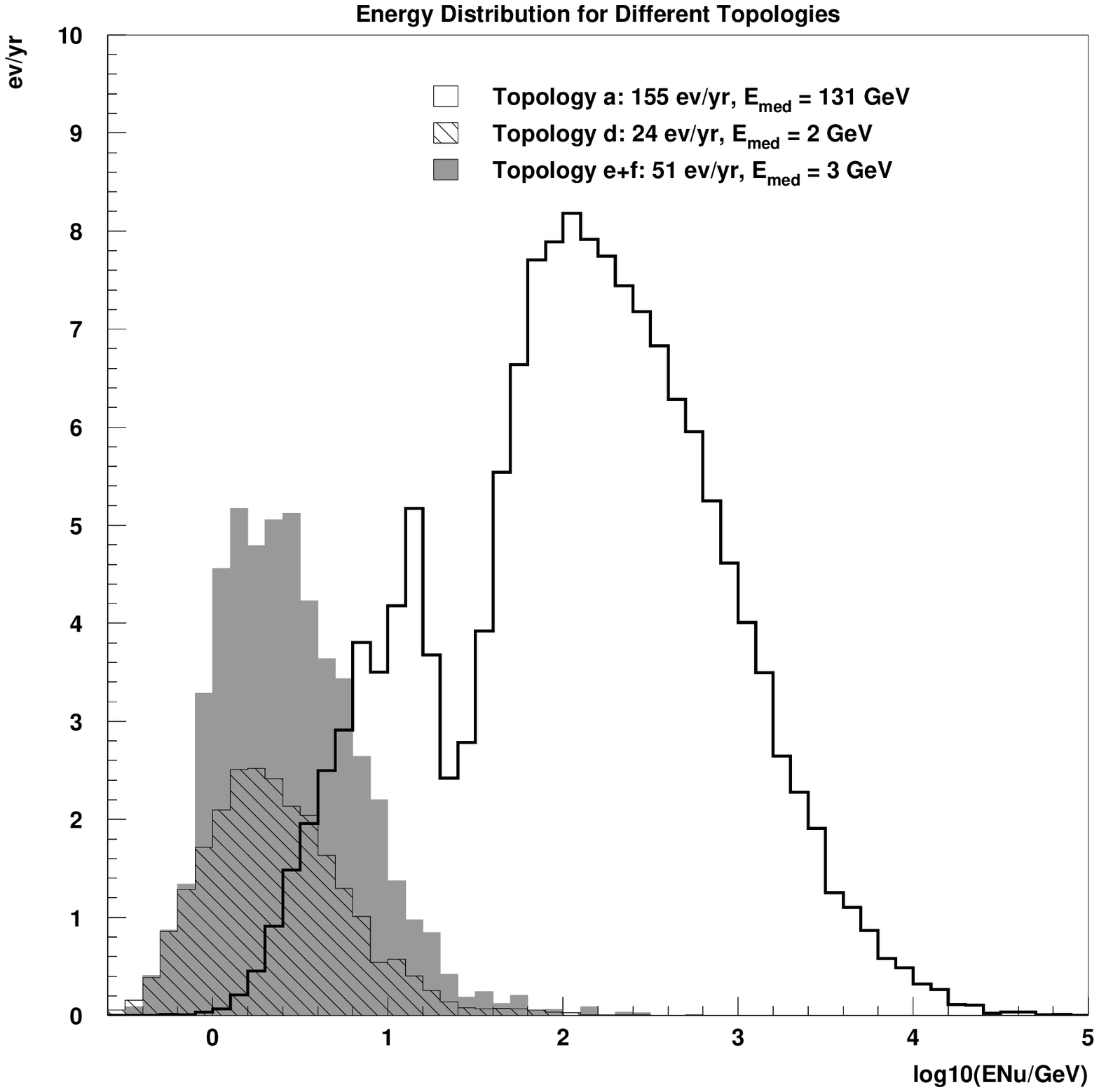}}

\subsection{Topology Criteria for This Analysis}

About five kilotonnes, 89\% of the mass of MACRO, resided below the
tenth layer of streamer tubes (which rested on the Center layer of
scintillation tanks).  Most of the mass was in the crushed rock
absorber, the iron trays containing the absorber, and the structural
iron.  This region of MACRO will be referred to as ``lower MACRO\@.''
The neutrino interactions in the detector follow the mass distribution
(approximately -- the interaction rate also depends on the Z/A ratio,
which varied between 0.47 (iron) and 0.51 (oil) in the various
materials making up MACRO).  This analysis limits itself to neutrino
interactions in lower MACRO which produce an upgoing muon that fires a
scintillator box in the Center layer as well as a higher scintillator
box.

In overview, the analysis proceeds by identifying a candidate streamer
tube space track associated with fired scintillator boxes in the C
layer and the attico.  Next, a containment cut is applied to ensure
that the track begins in lower MACRO (or ends in lower MACRO if it is
a downgoing stopping muon).  The event is then examined and rejected
if it has any characteristic that casts the accuracy of the timing in
doubt.  Finally, those events with time of flight consistent with an
upgoing muon are selected as the event sample.

\section{The Dataset}

\subsection{Data Flow and Analysis Software\protect\label{sec:runhandling}}

Fundamentally, we want to use the streamer tubes to identify the track
of a penetrating particle in the detector, and use the scintillator
timing to determine if the particle was upgoing (which means it was
almost certainly neutrino-induced) or downgoing (probably an
atmospheric muon).  Practically, we want to write a computer program
that utilizes all the information recorded in MACRO events to identify
muons of various classes.

In the Gran Sasso lab, whenever electronics (e.g.,\ the ERP) monitoring
the apparatus identified a ``trigger'' condition (e.g.,\ ERP muon
trigger), the status of various streamer tube and scintillator
electronics were read by a computer and stored on disk as one
``event.''  A ``run'' is a collection of consecutive events; a run
typically lasted a few hours and contained about $10^4$ events (many
events caused by other triggers had no muons), although it was
possible for a run to be much shorter.  All of the events from one run
are kept in one file on disk.

The file containing all the events in a run may be converted to new
formats or archived to tape.  The dataset on which this analysis is
based is composed of runs that had irrelevant information deleted and
were stored in a compact format called the Muon Analysis Data Summary
Tape (MUADST).  MUADSTs were created on a VAX/VMS computer using
unformatted FORTRAN WRITE statements.  MUADST files arrived at Caltech
by many different routes.  Many were transferred to tape in a raw
format and read on the Caltech High Energy Physics IBM RS/6000
computers utilizing a machine translation algorithm.  Some runs were
converted by MACRO collaborator Colin Okada to a transportable format
called DSTZ and brought to Caltech on tape.  Still other runs were
transferred over the internet from Gran Sasso computers in raw MUADST
format.

\begin{figure}[bth]
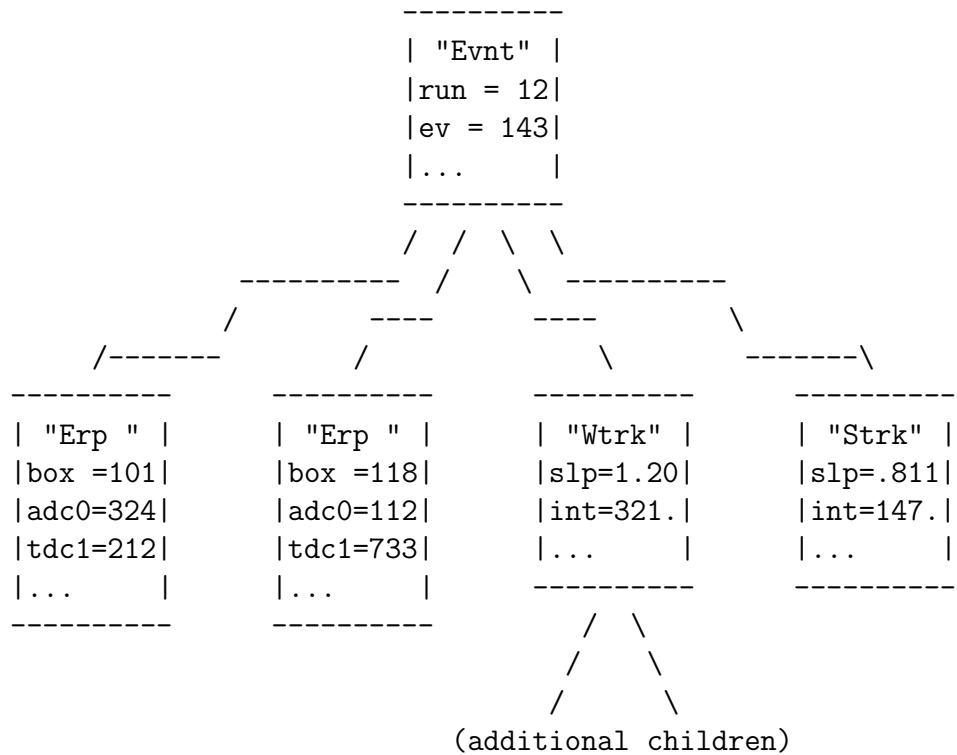

  \begin{verbatim}

                             ----------
                             | "Evnt" |
                             |run = 12|
                             |ev = 143|
                             |...     |
                             ----------
                             /  /  \  \
                   ----------  /    \  ----------
                  /        ----      ----        \
          /-------        /              \        -------\
     ----------      ----------      ----------      ----------
     | "Erp " |      | "Erp " |      | "Wtrk" |      | "Strk" |
     |box =101|      |box =118|      |slp=1.20|      |slp=.811|
     |adc0=324|      |adc0=112|      |int=321.|      |int=147.|
     |tdc1=212|      |tdc1=733|      |...     |      |...     |
     |...     |      |...     |      ----------      ----------
     ----------      ----------         /  \
                                       /    \
                                      /      \
                                (additional children)
  \end{verbatim}
  \caption{\protect\label{fig:tree}An example of a  FARFALLA event tree.}
\end{figure}

Regardless of the format in which received, all runs were converted to
a common format called Fisica Analisi Rete Facendo Altrimenti
Letargico Lavoro Allegro (FARFALLA)\footnote{This is very bad Italian,
  which literally translates Physics Analysis Network Making Otherwise
  Dull Work Exciting and Gay.  The acronym, FARFALLA, is the Italian
  word for butterfly.}~\cite{farfalla}.  FARFALLA is a \cpp class
library (developed by me and MACRO collaborator Chris Walter) for
creating and using custom data summary files (DSTs).  FARFALLA stores
data in a tree structure in which small nuggets of related information
are placed into a ``node'' on the tree.  For example, all of the ERP
information (raw values of ADCs and TDCs, as well as reconstructed
values such as energy and time) for a single hit scintillator box
resides in one node.  In the FARFALLA DST files on disk or tape, all
of the nodes with information from the same event are collected
together into one event tree (see Figure~\ref{fig:tree}).  A file
consists of many trees (i.e.,\ many events).  Walter and I created a
software distribution called MACROFAR which contains FARFALLA node
definitions for several different kinds of information that would
commonly be used in DSTs for MACRO\@.  Many of these nodes contain
reconstructed data, including ERP timing reconstructions and
reconstructed streamer tube tracks.  The FARFALLA DSTs created for
this analysis contain only nodes defined in MACROFAR, enumerated in
Figure~\ref{fig:dstnodes}.

\begin{figure}[htb]
  \begin{center}
    \framebox{\hskip 0.25in
      \parbox{2in}{\bf
        triggerNode\break
        timeNode\break
        erpNode\break
        interErpNode\break
        wireTrackNode\break
        stripTrackNode\break
        lateralTrackNode\break
        otherWireHitsNode\break
        otherStripHitsNode\break
        otherLateralHitsNode\break
        otherFrontalHitsNode\break
        otherLstripHitsNode
        }
      }
  \end{center}
  \caption{\protect\label{fig:dstnodes}
    Nodes in the DSTs forming the dataset for this analysis.}
\end{figure}

Analysis of the FARFALLA DSTs takes place in a software environment
called Data Analysis SHell (DASH)~\cite{dash}, developed by MACRO
collaborator Ed Kearns.  When integrated with FARFALLA, DASH reads one
event tree from disk into memory, and then calls several user-written
``modules'' to operate on the tree in memory.  Some modules might
create new nodes and add them to the tree, which will be visible on
the tree to subsequent modules.  Some modules act as ``filters,''
meaning they make a decision whether processing will continue 
(i.e.,\ whether subsequent modules will be called) or the event will be
rejected without further processing.  After one event is either
processed fully or rejected, the next event is read in and processed.

Some of the DASH modules used in this analysis utilize the services of
Tracks And Boxes Object-Oriented (TABOO), a class library I wrote that
can take information from MACROFAR nodes and do such things as the
following:

\begin{itemize}
\item Combine streamer tube tracks in different views to form a track
  in three-dimensional space.

\item Given a space track, create a list of scintillator boxes
  intersected by the track.

\item Given a space track, create a list of streamer tube channels
  intersected by the track.

\end{itemize}

\subsection{Real Data}

The dataset for this analysis extends from Run 7473, 29 Apr 1994, to
Run 17701, 02 Jun 1999, a calendar time of 5.1~years.

All runs with run numbers less than 80,000 in the MUADST dataset for
this period were converted to FARFALLA\@.  At the time each run was
converted, several variables were checked to ensure the quality of the
data.  I made a subjective decision for each run whether it appeared
to be a calibration run (a run with many firings of the calibration
LEDs or laser), or if it appeared to have hardware problems so severe
they could cause excessive deadtime or other inefficiencies (the
latter are referred to as ``bogus'' runs).  Calibration and bogus runs
are listed in Figures~\ref{fig:calruns}-\ref{fig:bogusruns} and were
eliminated from the dataset.

\begin{figure}[p]
  \begin{center}
    \framebox[6in]{
      \parbox{5.5in}{\bf\small
        7489, 7490, 7491, 7492, 7493, 7494, 7495, 7496, 7526, 7527,
        7529, 7532, 7534, 7535, 7571, 7572, 7573, 7574, 7575, 7577,
        7595, 7596, 7597, 7598, 7599, 7600, 7601, 7602, 7603, 7604,
        7640, 7641, 7642, 7643, 7644, 7645, 7646, 7647, 7648, 7649,
        7650, 7661, 7680, 7681, 7682, 7683, 7684, 7685, 7687, 7688,
        7713, 7714, 7715, 7716, 7717, 7718, 7719, 7720, 7721, 7722,
        7766, 7767, 7768, 7769, 7770, 7775, 7776, 7815, 7816, 7817,
        7818, 7819, 7820, 7852, 7853, 7854, 7855, 7856, 7857, 7881,
        7882, 7883, 7884, 7886, 7887, 7888, 7889, 7890, 7891, 7892,
        7919, 7920, 7921, 7922, 7923, 7924, 7925, 7926, 7927, 7978,
        7979, 7980, 7981, 7982, 7983, 7984, 7985, 7986, 8022, 8023,
        8024, 8025, 8026, 8027, 8058, 8059, 8060, 8061, 8062, 8063,
        8085, 8086, 8087, 8088, 8089, 8090, 8091, 8092, 8128, 8129,
        8130, 8131, 8132, 8133, 8134, 8135, 8161, 8163, 8164, 8165,
        8166, 8167, 8168, 8169, 8170, 8171, 8172, 8173, 8174, 8175,
        8199, 8200, 8201, 8202, 8203, 8204, 8205, 8206, 8207, 8208,
        8218, 8219, 8220, 8257, 8258, 8259, 8260, 8261, 8262, 8263,
        8264, 8265, 8266, 8267, 8268, 8269, 8270, 8271, 8272, 8273,
        8274, 8315, 8316, 8317, 8318, 8319, 8320, 8321, 8322, 8323,
        8324, 8325, 8374, 8375, 8376, 8377, 8378, 8379, 8380, 8381,
        8415, 8416, 8417, 8418, 8419, 8420, 8421, 8422, 8423, 8424,
        8425, 8426, 8470, 8472, 8473, 8476, 8477, 8478, 8479, 8480,
        8481, 8482, 8522, 8523, 8524, 8525, 8526, 8527, 8544, 8545,
        8546, 8547, 8670, 8671, 8672, 8673, 8674, 8675, 8676, 8711,
        8712, 8713, 8714, 8715, 8716, 8717, 8718, 8719, 8767, 8768,
        8769, 8770, 8771, 8800, 8801, 8803, 8808, 8809, 8811, 8812,
        8814, 8854, 8855, 8856, 8857, 8858, 8859, 8913, 8914, 8915,
        8958, 8959, 8960, 8961, 8962, 9067, 9068, 9069, 9070, 9071,
        9072, 9073, 9074, 9075, 9076, 9115, 9116, 9117, 9165, 9166,
        9167, 9168, 9169, 9170, 9171, 9172, 9173, 9174, 9175, 9176,
        9222, 9223, 9224, 9225, 9226, 9227, 9274, 9275, 9276, 9277,
        9278, 9321, 9322, 9323, 9324, 9325, 9360, 9361, 9362, 9404,
        9405, 9406, 9407, 9408, 9409, 9473, 9474, 9475, 9534, 9535,
        9536, 9537, 9538, 9577, 9578, 9579, 9580, 9581, 9635, 9636,
        9637, 9638, 9639, 9640, 9641, 9642, 9643, 9644, 9701, 9702,
        9703, 9704, 9705, 9813, 9814, 9815, 9864, 9865, 9866, 9910,
        9911, 9913, 10136, 10137, 10138, 10263, 10264, 10469, 10470,
        10520, 10521, 10572, 10573, 10575, 10576, 10631, 10632, 10633,
        10717, 10723, 10724, 10725, 10758, 10759, 10800, 10801, 10802,
        10803, 10804, 10805, 10806, 10809, 10811, 10815, 10816, 10858,
        10860, 10862, 10863, 10864, 10865, 10900, 10903, 10904, 10906,
        10908, 10909, 10945, 10946, 10951, 10994, 10996, 10997, 10998,
        10999, 11035, 11036, 11037, 11038, 11039, 11040, 11041, 11042,
        11043, 11072, 11117, 11625, 11626, 11627, 11628, 11695, 11696,
        11697, 11698, 11699, 11700, 11701, 11702, 11703, 11704, 11705,
        11706, 11707, 11708, 11709, 11710, 11711, 11712, 11713, 11753,
        11754, 11755, 11756, 11757, 11758, 11759, 11760, 11761, 11762,
        11763, 11764, 11765, 11766, 11767, 11768, 11769, 11770, 11771,
        11772, 12507, 12508, 12509, 12510, 12512, 12513, 12546, 12547,
        12549, 12550, 13184, 13186, 13187, 13599, 13637
        }
      }
  \end{center}
  \caption{\protect\label{fig:calruns}Calibration runs eliminated
    from the dataset.}
\end{figure}

\begin{figure}[htb]
  \begin{center}
    \framebox[6in]{ \parbox{5.5in}{\bf\small 7701, 7702, 7735, 7736,
        7737, 7758, 7759, 7786, 7787, 7811, 7821, 7822, 7823, 7969,
        7970, 7971, 7972, 7973, 7974, 7976, 7987, 8007, 8008, 8009,
        8010, 8011, 8055, 8056, 8072, 8105, 8302, 8343, 8369, 8370,
        8371, 8391, 8392, 8393, 8504, 8543, 8566, 8575, 8577, 8633,
        8732, 8828, 8884, 8894, 8922, 8923, 9119, 9120, 9349, 9352,
        9364, 9412, 9413, 9414, 9415, 9416, 9427, 9433, 9446, 9456,
        9457, 9520, 9594, 9631, 9649, 9650, 9651, 9669, 9670, 9679,
        9709, 9711, 9713, 9714, 9715, 9716, 9717, 9720, 9721, 9730,
        9780, 9781, 9782, 9783, 9784, 10594, 10609, 10760, 10761,
        10828, 10829, 10969, 10970, 10971, 10972, 10973, 10974, 10975,
        11073, 11198, 11199, 11368, 11369, 11414, 11415, 11416, 11417,
        11449, 11483, 11484, 11573, 11799, 11801, 11802, 11809, 11812,
        11814, 11815, 11816, 11839, 11840, 11841, 11842, 11843, 11844,
        11861, 12094, 12193, 12275, 12276, 12283, 12284, 12288, 12309,
        12847, 12879, 12977, 12978, 12979, 12980, 12981, 12982, 12983,
        12984, 12985, 12986, 12987, 13028, 13049, 13051, 13600, 13601,
        13602, 13604, 13605, 13606, 13629, 13630, 13638, 13639, 13640,
        13642, 13671, 13702, 13718, 13720, 13723, 13725, 13726, 13727,
        13728, 13990, 14007, 14015, 14059, 14075, 14114, 14173, 14210,
        14250, 14283, 14381, 14382, 14383, 14384, 14385, 14729, 14753,
        14754, 14758, 14759, 14760, 14761, 14762, 14764, 14765, 14766,
        14767, 14769, 14770, 14773, 14774, 14775, 14776, 14777, 14882,
        14883, 14907, 14908, 14909, 14910, 14923, 15342, 15348, 15349,
        15350, 15351, 15579, 15696, 15745, 15813, 15837, 15838, 15850,
        15851, 15877, 15897, 15898, 15964, 15965, 16016, 16122, 16123,
        16259, 16260, 16265, 16266, 16269, 16303, 16304, 16306, 16307,
        16400, 16418, 16490, 16585, 16585, 16587, 16641, 16642, 16782,
        16789, 16842, 16843, 17143, 17144, 17148, 17149, 17150, 17151,
        17152, 17153, 17233, 17345, 17346
        }
      }
  \end{center}
  \caption{\protect\label{fig:bogusruns}Bogus runs eliminated
    from the dataset.}
\end{figure}

As described in Section~\ref{sec:hwprobs}, a database was created with
the dead periods and weekly efficiency for each scintillator and
streamer tube channel.  Actually, in creating the database it was
found necessary to eliminate a handful of runs that had inconsistent
times in the event records.  (The time used is that written to the
event record by the acquisition computer's clock, which the MACRO
collaboration calls VAX time.  The time available from the LNGS atomic
clock, when it was working properly, is in principle far more accurate
than VAX time; unfortunately, it was also far more prone to
malfunctions.)  There was nothing really wrong with these runs except
that the VAX time irregularity made it difficult to calculate time
between events for the efficiency database.  The runs that were
eliminated at this stage are shown in Figure~\ref{fig:badtimeruns}.

\begin{figure}[htb]
  \begin{center}
    \framebox[6in]{
      \parbox{5.5in}{\bf\small
        7724, 8722, 8882, 8907, 8933, 8943, 9094, 9110, 9140, 9195,
        9737, 9744, 10071, 10121, 10401, 10402, 10454, 10455, 10480,
        11231, 12051, 12816, 13203, 13213, 13991, 14609, 15036, 16426,
        17288
        }
      }
  \end{center}
  \caption{\protect\label{fig:badtimeruns}Runs eliminated from the
    dataset due to inconsistent VAX times.}
\end{figure}

Remaining are 8362 runs, with 66.4 million events including 23.7
million downgoing events passing the standard muon analysis.  Summing
the total livetime yields 4.5~years.  The difference of 0.6~years
between the livetime and the calendar time of 5.1~years is presumably
due to scheduled downtime for maintenance and calibrations, short
periods of downtime due to acquisition crashes, long periods of
downtime due to major equipment failures, and the effect of
eliminating runs from the dataset.

\subsection{Simulated Data}

The Monte Carlo dataset was generated by running GMACRO, using the
GMACRO parameters determined by comparison with downgoing atmospheric
muons (Section~\ref{sec:mctune}), with the neutrino flux of the Bartol
group (Section~\ref{sec:bartol}) at the Gran Sasso location, and the
neutrino event generator of Lipari (Section~\ref{sec:lipari}).  Three
aspects of the marriage of the detector-independent flux and cross
section calculations to the GMACRO detector simulation are discussed
below: adjusting the flux to the solar cycle conditions in effect
during the data-taking, computing the total interaction rate of
atmospheric neutrinos in MACRO, and placing the vertex of the
interaction in MACRO\@.

\subsubsection{The Solar Cycle\protect\label{sec:solcyc}}

As mentioned in Section~\ref{sec:priflux}, the neutrino flux at the
lowest energies is dependent on the solar cycle.
Figure~\ref{fig:sunspots} shows an actual measurement~\cite{intersol}
of solar activity during the period of data-taking for this analysis.
Using the divisions shown in the figure, we may model the period from
the beginning of data-taking (April 1994) to April 1998 as solar
minimum, and from April 1998 to the end of the dataset (June 1999) as
solar midcycle.  The first period has 77\% of the minutes of
data-taking, so 77\% of the the Monte Carlo livetime is generated
using the Bartol solar minimum flux, and the remainder using the
Bartol solar midcycle flux.  Because the interaction rate is slightly
higher in the solar minimum epoch, 77.9\% of the generated events are
from that epoch to correspond to 77.0\% of the generated livetime.

\picfig{fig:sunspots}{Solar activity, as defined and measured in
  Reference~\protect\cite{intersol}, as a function of time.  The long
  arrow shows the period of this analysis modeled as solar minimum,
  and the shorter arrow is solar
  midcycle.}{\resizebox{6in}{!}{\includefigure{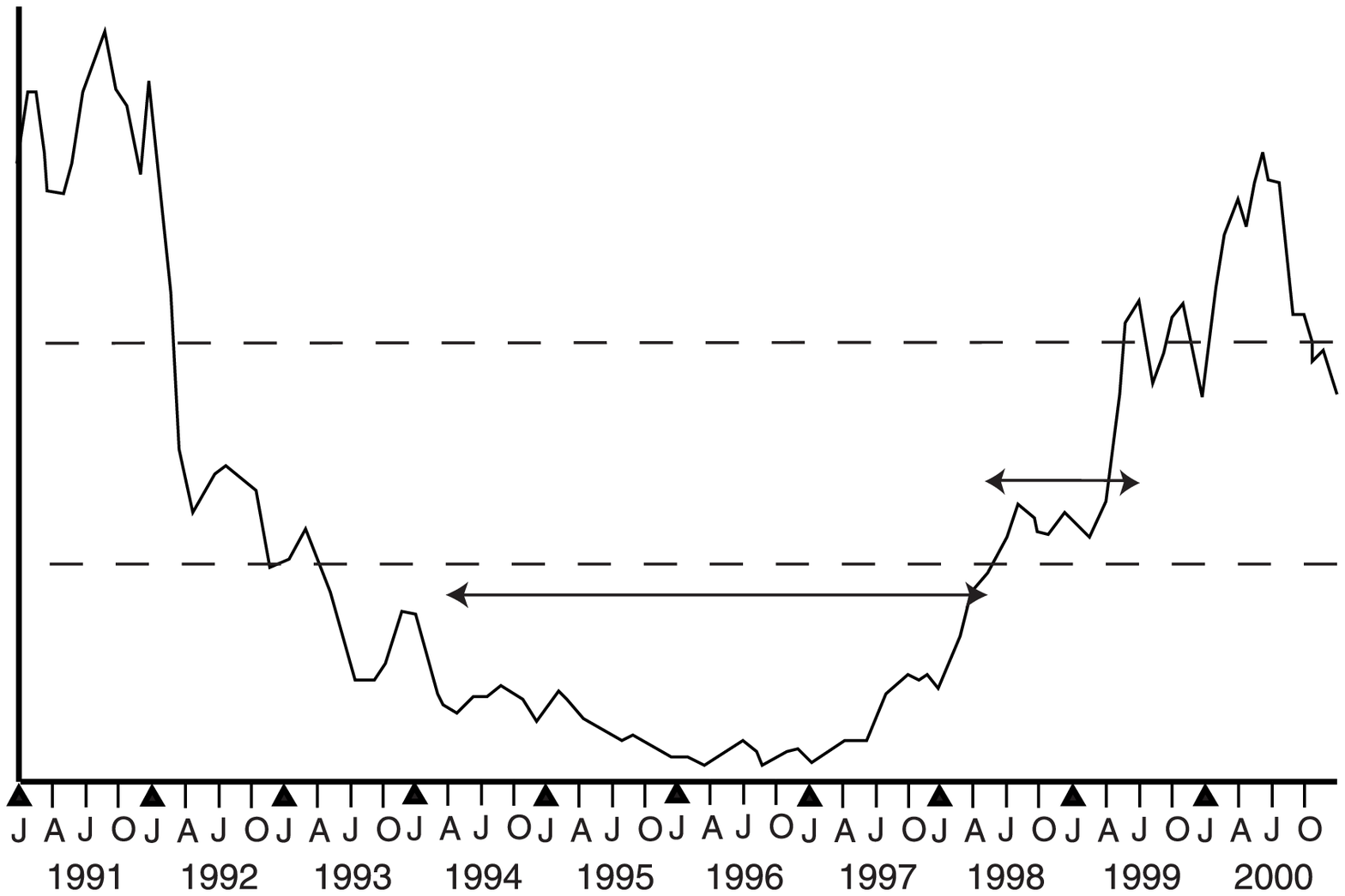}}}

\subsubsection{Interaction Rate and Mass of MACRO}

The subject of this analysis is neutrino events with vertices in the
detector.  To utilize the results of an event generator, consisting of
a flux and a cross section model, the event vertex must be placed
randomly in the detector according to the probability of a neutrino
interaction at each point.  The cross section on electrons is
negligible, so the interaction probability at a point depends on the
density of protons and neutrons at that point.  The cross section on
neutrons is 75\% higher than that on protons, so the Z/A ratio at the
point is also relevant.  If Z/A were constant across all the materials
in MACRO, the interaction probability would be proportional to the
mass density at each point.  The detector consisted of a number of
different materials -- liquid scintillator, plastic scintillator tanks
and streamer tubes, crushed rock absorber, and various shapes of
steel, including structural I-beams and containers for the crushed
rock.  Z/A ranges from 0.466 in iron to 0.506 in liquid scintillator.

During construction the composition of MACRO was monitored closely --
not with a mass calculation in mind but with energy loss calculations
in mind -- and the observations were incorporated into GMACRO, the
detector simulation software.  Scrutinizing the mass information
implicit in the GMACRO geometry, I saw where uncertainties could
significantly affect the result, and verified or updated the
parameters, and even modified the GMACRO geometry slightly to conform
with known masses of components.  Then, the detector geometry as
represented in GMACRO was used as the basis of both the event rate
calculation and the random placement of event vertices.

At the end of initialization, the event generator considers the chosen
flux and the calculated cross sections, and reports the interaction
rate per neutron and per proton.  (For Bartol solar minimum flux and
Lipari cross section, the rates are $0.20\times 10^{-30}$
interactions per proton per second, and $0.36\times 10^{-30}$
interactions per neutron per second.)  A generation volume enclosing
the entire MACRO detector has been defined, and a million random points
distributed uniformly in the generation volume are sampled.  The Geant
volume tree may be walked to determine in what material the point
lies; however, because the MACRO geometry in GMACRO is defined with
some overlapping volumes (for example, the volume containing
structural iron between the two modules of Supermodule 1 overlaps with
the volume containing all other Supermodule 1 materials), care has to
be taken to be sure Geant does not report the point is in air when it
is really in structural iron or absorber.  After sampling, the
proportion of the generation volume occupied by each material is
known.  For each material, knowing the volume it occupies, its
density, and its Z/A ratio, we may calculate the number of neutrons
and protons, and thence the interaction rate, in that material.
Summing over all materials gives the total interaction rate in MACRO\@.

Because the signature of neutrino-induced semi-contained muons
requires an upgoing muon to hit the center scintillator layer and a
higher layer, almost all events detected by the analysis are generated
in the lower part of MACRO\@.  Table~\ref{tab:mass} shows the mass, as
calculated using GMACRO, for the lower half of MACRO\@.

\begin{table}[htb]
  \leavevmode
  \begin{center}
    \begin{tabular}{|r|l|l|l|} \hline
      Medium   & Volume ($\mathrm{m}^3$) & 
                 Density ($\mathrm{g}/\mathrm{cm}^3$) &
                 Mass (tonnes)    \\
      \hline\hline
      Crushed Rock Absorber & 1490   & 2.17                  & 3240     \\
      Iron                  & ~111   & 7.87                  & ~876     \\
      Scintillator          & ~398   & 0.870                 & ~346     \\
      Plastic               & ~135   & 1.63                  & ~220     \\
      Mineral Oil           & ~~20.6 & 0.870                 & ~~18.0   \\
      Air                   & 7410   & $1.2 \times 10^{-3}$  & ~~~8.93  \\
      Streamer Tube Gas     & ~296   & $1.0 \times 10^{-3}$  & ~~~0.296 \\
      \hline
      TOTAL                 &        &                       & 4709     \\
      \hline
    \end{tabular}
    \caption{\protect\label{tab:mass} Masses of components of lower MACRO\@.}
  \end{center}
\end{table}

The above calculations all take place during program initialization.
Later, during the event loop, the generated neutrino interactions must
be placed in MACRO according to the interaction probability at each
point.  The generator uses a Monte Carlo rejection algorithm: the
interaction vertex is placed at a random point in the generation
volume with uniform probability, Geant then walks the volume tree to
determine in what material the point exists, and then the choice of
location is accepted with probability equal to the ratio of the
interaction probability in that material to the maximum interaction
probability in any material (iron), taking the density and Z/A into
account.  If the vertex is not accepted, another is generated,
repeating until one is accepted.

\subsubsection{The Generated Dataset}

The generator utilizes neutrinos from all directions (downgoing as
well as upgoing) with interaction vertices in a generation volume that
contains the entire detector (attico as well as lower MACRO).  The
generator calculates the event rate at solar minimum, taking into
account the mass and Z/A ratio of the actual material in the
generation volume, as 886 events/year for $\numu\ +\ \numubar$, and
452 events/year for $\nue\ +\ \nuebar$ (not considering oscillations).
For solar midcycle, the interaction rates are about 4\% lower.  The
generated dataset consists of 100,000 $\numu$ and $\numubar$ events
(77,527 with the solar minimum spectrum and 22,473 at solar midcycle)
and 50,344 $\nue$ and $\nuebar$ events (39,516 minimum and 10,828
midcycle), both corresponding to 88~years at solar minimum and
25~years at solar midcycle, a total of 113~years.

Two other generators are used to check for backgrounds.  Simulated
downgoing atmospheric muons are checked to see if they pass all the
cuts and appear to be semi-contained neutrino events.  However, the
100,000 generated events correspond to a simulated livetime of
less than a half week.

Upgoing muons due to neutrino interactions below the detector are used
to estimate the rate of throughgoing upward muons that inadvertently
pass the containment cut, thus appearing to be semi-contained.  The
100,000 generated events correspond to a livetime of 143.5~years.

\section{Analysis Chain}
\subsection{Preparing the Data for Analysis}

First a DASH module called \dashmodule{badFilt} applies the microcuts
for scintillator quality (see Section~\ref{sec:microcuts}); if an ERP
box that fired in the event is marked as bad for the week in which the
event occurs, that box is deleted from the event record so subsequent
analysis behaves as if the box did not fire.  Likewise, if boxes on
two supermodules fired and the interERP reconstructions for that pair
of boxes are marked as bad for that week, interERP TDC information is
deleted from the event record, in such a way that it will be
impossible to reconstruct the timing.  In this way, even if equipment
marked for a microcut fired during the event, the event may still be
accepted if the remaining equipment is adequate to mark the event
unambiguously as neutrino-induced.

Next, all ERP reconstructions (see Section~\ref{sec:reconstructions})
are computed and stored in the event record by the DASH module
\dashmodule{recal}.  Then, if ERP channels fired on more than one
supermodule, an attempt is made to adjust the timing in all hit ERPs
so that their relative timing is correct, utilizing the information in
the interERP TDCs.  The DASH module \dashmodule{ierpAdj} uses the
procedures detailed in Section~\ref{sec:reconstructions} to adjust all
hit boxes in an event to be mutually consistent, if possible.  If it
is not possible (either because the combination of hit supermodules is
intrinsically unadjustable, or because of missing or bad data
(underflow or overflow) in the interERP TDCs), the event is rejected
at this stage.

By this point in the analysis chain, any event not rejected contains
only good boxes with the relative timing reconstructed for all
scintillator hits.  In the real data, 97.6\% of all events survive to
this stage; most of those cut are interERP events in which the
interERP pair is marked for a microcut (see
Section~\ref{sec:microcuts}).

\subsection{Identifying the Best Candidate Track}

Now that data has been prepared, to search for neutrino-induced muons
the first step is to use reconstructed streamer tube tracks to
identify candidate space tracks.  As in the standard muon analysis
(described in Section~\ref{sec:muonanal}), any event that does not
have a viable candidate track is rejected and not processed further.
It would be possible to use the generic track finder from the standard
muon analysis, and then reject the event if the candidate track chosen
does not have the proper topology (recall that this analysis requires
scintillator hits in the Center layer and a higher layer).  However,
to increase efficiency slightly, a customized track finding module,
called \dashmodule{scTrack}, was developed.  \dashmodule{scTrack} only
chooses candidates with the proper topology.  Thus, in an event in
which one candidate track (perhaps due to an upgoing neutrino-induced
muon) has the right topology and one candidate track (perhaps due to a
recoiling pion) has the wrong topology, we can be assured that the
track with the right topology will be chosen and the event will not be
rejected at this stage.

\dashmodule{scTrack} begins by iterating over every possible pair of
reconstructed streamer tube tracks in different views -- that is, wire
and strip views, wire and lateral views, or strip and lateral views.
For each pair, it computes the space track in three dimensions, and
then identifies any fired scintillator boxes that are near the track.
The position of the scintillator hit is determined by using the
scintillator timing to find the coordinate along the box (as described
in Section~\ref{sec:reconstructions}), and the coordinates of the
center of the box in the other two dimensions.  The shortest distance
from the scintillator hit to the space track is computed, and if it is
within 75~cm, the hit is considered associated with the track.
Figure~\ref{fig:SC-ST} illustrates these concepts.  If the space track
can be associated with a scintillator hit in the Center layer and
another scintillator hit in the attico, the track is considered a
candidate.  If more than one hit in the Center layer could be
associated, the one nearest the track is chosen; likewise the closest
attico hit is chosen.  Thus, a candidate consists of one space track,
one Center hit and one attico hit.  Because the wire and strip
tracking require four hits to reconstruct a track, this algorithm
cannot find muons that traversed fewer than four of the horizontal
streamer tube planes.

\picfig{fig:SC-ST}{A scintillator hit and a space track.  The
  coordinates of the hit are taken to be the center of the box in the
  two short dimensions, and given by the scintillator timing along the
  long axis of the box.  The solid black arrows indicate the
  coordinates of the hit.  The dashed line is a nearby space track.
  The white arrow shows the distance from the hit to the space track,
  at the point of the track's closest approach to the
  hit.}{\resizebox{5in}{!}{\includefigure{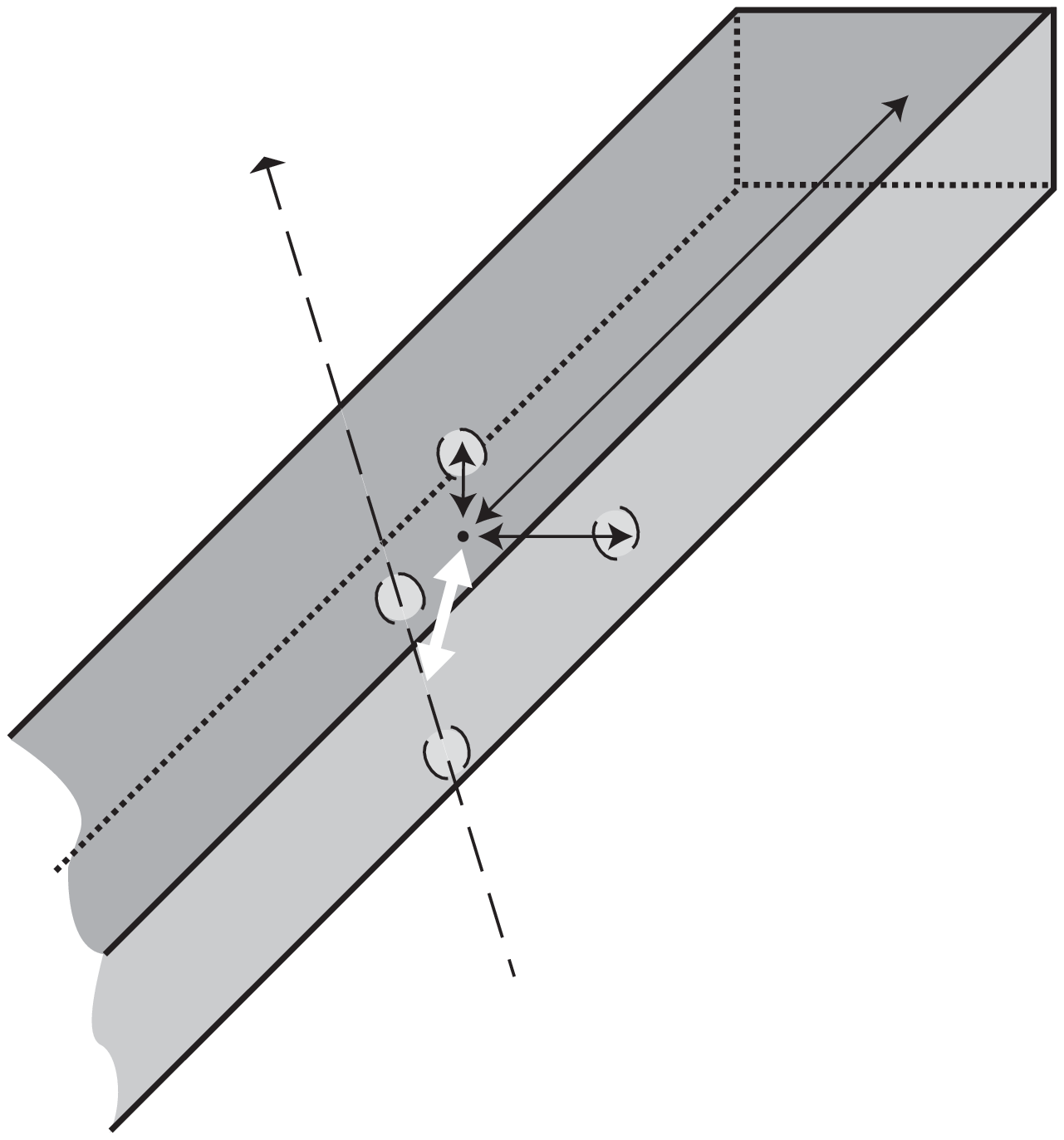}}}

If no candidate space track is found by iterating over pairs of
tracks, \dashmodule{scTrack} attempts to invoke an algorithm called
``combined tracking'' (or, more informally, ``Spurio tracking,'' after
the name of the MACRO collaborator who developed the algorithm).  The
combined tracking algorithm uses a single lateral track, in the Y-Z
view.  Recall (see Section~\ref{sec:st}) that lateral tracks may
contain, in addition to hits in the lateral view, combined wire-strip
hits that lie along the track in the Y-Z view.  Combined tracking
utilizes the X and D information from these central hits to
reconstruct a track in three dimensions that passes through all the
hits.  Combined tracking can be successful for muons that crossed as
few as two horizontal planes (plus some lateral planes), so it is more
efficient for near-horizontal muons than the standard tracking.

For most events in the dataset, whether downgoing atmospheric muons or
neutrino-induced muons or something else, there is at most one
candidate track.  However, if more than one candidate exists,
\dashmodule{scTrack} will choose the ``best'' track to be processed as
the candidate by the remaining modules in the analysis chain.  If
there are one or more ``quality'' tracks (defined as a track for which
the average of the squared distance from the two scintillator hits to
the track is less than $(50~\mathrm{cm})^2$), the quality track whose
time of flight between the two scintillator hits is closest to that
expected for an upgoing relativistic particle is chosen.  In this way,
if there are upgoing and downgoing candidates, an upgoing candidate
will be chosen.  If no quality tracks exist, the candidate is chosen
that has the lowest average of the squared distance to the track.
Information about the chosen candidate is placed on the FARFALLA tree,
and is available to all subsequent modules.

In the real data (which includes triggers for other physics processes
in addition to downgoing muons), a candidate track is identified for
about 1/3 of all events.  For simulated downgoing muons, 55\% have the
right topology to pass this requirement.  For simulated neutrino
interactions in the detector (including both $\numu$ and $\nue$, and
both charged current and neutral current interactions), only 6.4\%
have a candidate track.  Most of the rest range out in the absorber
without hitting any scintillators, or they go the wrong direction to
hit the Center layer and an attico layer.
\subsection{Containment}

At this point the remaining events are generally muons that traversed
at least part of the lower detector, the C layer, and a higher
scintillator tank.  However, the vast majority of these events are
throughgoing events.  It is necessary to identify a subset of the
remaining events that appeared to have their track start (or, in the
case of downward muons, end) within the interior of MACRO\@.  This is
determined by requiring that the candidate space track hit several
channels that did not fire, all below the lowest channel that did
fire.  This eliminates particles that entered or exited via a lower
face of the detector, as well as particles that entered or exited
through a known crack in the detector.

A DASH module, \dashmodule{scCrack}, was developed to identify such
events.  For the candidate space track, a list is generated of the
streamer tube channels (defined as a collection of eight wires, about
25~cm wide) and scintillation counters intersected by the track, and
it is determined if the hit channels fired in the event.  If a channel
that is hit but did not fire is in the database of dead channels for
the moment the event occurred (see Section~\ref{sec:hwprobs}), the hit
is deleted from the list and the algorithm proceeds as if the track
did not hit that channel.  The lowest fired channel defines the bottom
of the track.  A hit channel that did not fire may be below the track
or within the track.

\dashmodule{scCrack} then implements the logic defining the
containment cut.  The logic was developed primarily by looking at
stopping downward muons in the real data.  It became clear that
lateral streamer tube channels are unreliable for a containment cut.
There were many gaps between streamer tube channels in the lateral
walls, and if the actual path of a muon differed slightly from the
reconstructed space track, it often happens that the software
considers a channel hit when in fact it was slightly missed.
Therefore, \dashmodule{scCrack} considers a fired lateral channel on
the track to be evidence that the vertex was not contained, but an
unfired lateral channel is not considered to be evidence that the
vertex was contained.  Also, while it is sensible to insist that, to
consider a vertex contained, a hit scintillator counter below the
track did not fire, in practice it is seen that adding this
requirement eliminates very few events that are not eliminated by
considering streamer tube information alone.  In fact, in the final
sample of selected neutrino events, there are no fired scintillators
below the track.  Therefore the containment logic does not consider
the scintillator counter, so that certain uncertainties in the
simulation (for example, for slow neutrons that trigger a scintillator
far below the vertex) cannot affect the Monte Carlo rate prediction.

Therefore, the final containment logic is quite simple: that two live
horizontal streamer tube channels below the track be hit but not
fired.  Note that if the track extends downward through a lateral
wall, and a channel in the lateral wall did fire, the event is not
considered contained by this definition because any missing horizontal
channels are not below the track.  See Figure~\ref{fig:contain} for
examples of events passing or not passing the containment cut.

\picfig{fig:contain}{How the containment cut works.  The dashed line
  shows the space track, filled circles represent hit channels that
  fired, and open circles represent hit channels that did not fire.
  (a) and (b) pass the containment cut because two central channels
  below the track were hit but did not fire. (c) does not pass the
  containment cut.  Although several lateral channels below the track
  were hit but did not fire, lateral channels are not considered
  reliable enough to provide confidence that this event is really
  contained.  (d) does not pass the containment cut.  Although the
  bottom two hit central channels did not fire, they are not below the
  track due to the lateral channel that did
  fire.}{\scalebox{0.9}{\includefigure{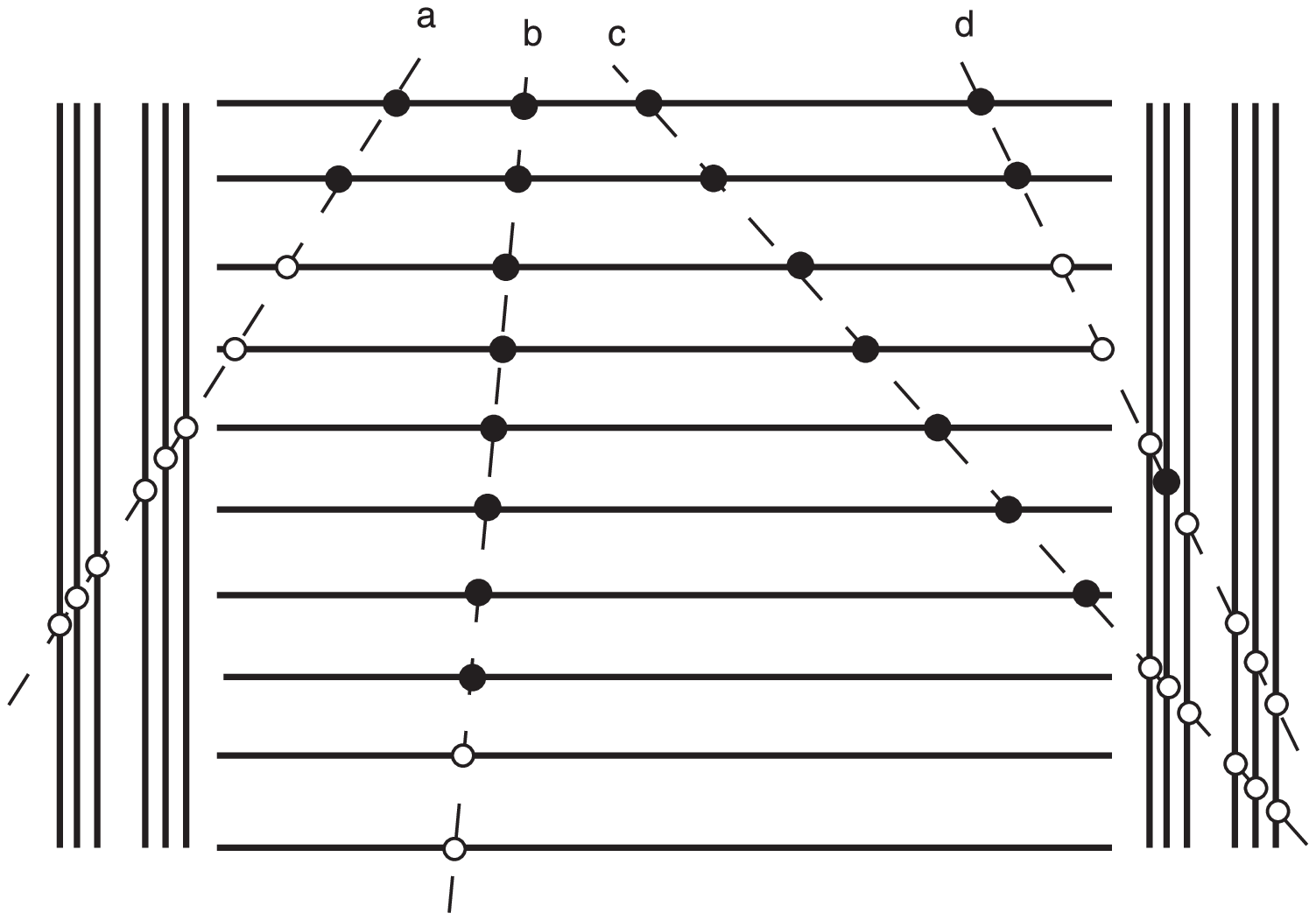}}}

Note that true neutrino interactions near a wall or a crack may not
pass the containment cut (nor may stopping muons that stop near a wall
or a crack).  This cut is similar in spirit to a ``fiducial volume''
cut in other experiments, although the cut is defined by topology
rather than volume.  The simulation should predict the rate of
interactions passing the cut, as well as true signal events that do
not pass the cut.

In the real data (mostly downward throughgoing muons), only 1.4\% of
events with a candidate track pass the containment cut.  For simulated
neutrino interactions, 85.7\% pass.

For each event selected at the end of the analysis, certain streamer
tubes that failed to fire led to the event being declared
semi-contained.  Let us call these ``critical channels.''  A final
check shows that the critical channels were indeed alive and
functioning.  Figure~\ref{fig:checkCrit} shows, for each critical
channel in each event in the final sample, the last time the channel
fired before the event, and the first time the channel fired after the
event.  The distribution has the expected exponential shape, and there
is no evidence that any of the tubes were experiencing any problems
firing.

\picfig{fig:checkCrit}{Timing in critical channels.  The entries
  at negative time show how many minutes before the event the channel
  last fired, while the positive entries show how many minutes after
  the event the channel next fired.}{\pawfig{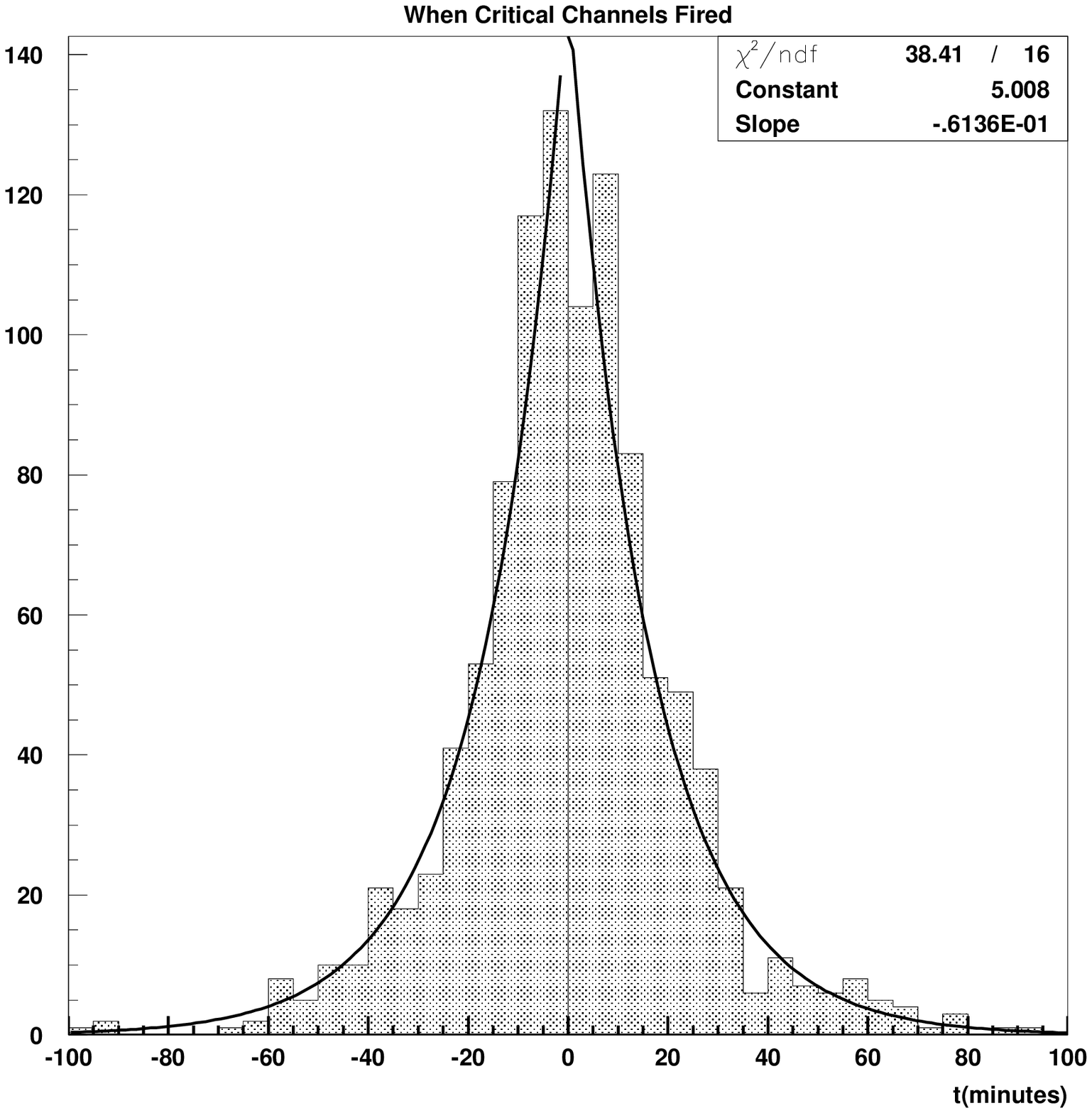}}

\subsection{First Background-Reducing Cuts}

The vast majority of candidates identified by the above procedure are
correctly measured.  However, there are hundreds of thousands of
downgoing atmospheric muons for each neutrino-induced muon identified
by this analysis, so even a tiny fraction of mismeasured events could
overwhelm the signal.  Therefore, it is necessary to apply a number of
cuts to reject any events that may mismeasure the direction of the
particle.

The cuts applied at this point can choose either upgoing or downgoing
particles.  The first cut requires that the values recorded by the
TDCs of the boxes on the track are within the normal valid range --
between 1000 and 4050 counts.  The second cut requires the time
reconstructed by high-threshold and low-threshold TDCs agree within
10~ns for each of the two boxes.  Only a fraction of a percent of real
events fail these two cuts.  The third cut requires the two
scintillator hits to be at least 300~cm apart, so a mistake of 20~ns
would be required to turn a downgoing relativistic particle into
upgoing.  This cut eliminates over 10\% of all candidates, but it also
eliminates the majority of mistimed events at negative $\beta$.
Figure~\ref{fig:mandCuts} shows the effect of these cuts on the
distribution of $1/\beta$.  (Recall that for fixed pathlength, the
error in $1/\beta$ is proportional to the error in the measured time
of flight.)  After these cuts, 275,606 events remain, 436 of them
upgoing, and a clear signal peak is emerging at $\beta~=~-1$.
However, there is still obviously a large background from mistimed
events.  The precision of the measurement can be improved by making
additional efforts to reduce background.

\picfig{fig:mandCuts}{The effect of the first background-reducing
  cuts.  The top graph shows the $1/\beta$ distribution of all
  semi-contained events before cuts.  For each of the three cuts, the
  left column shows events eliminated by the cut and the right column
  shows events surviving the cut.  The three cuts are, from top to
  bottom, Valid TDCs, TDCH-TDCL Agreement, and
  Pathlength.}{\resizebox{6in}{!}{\includefigure{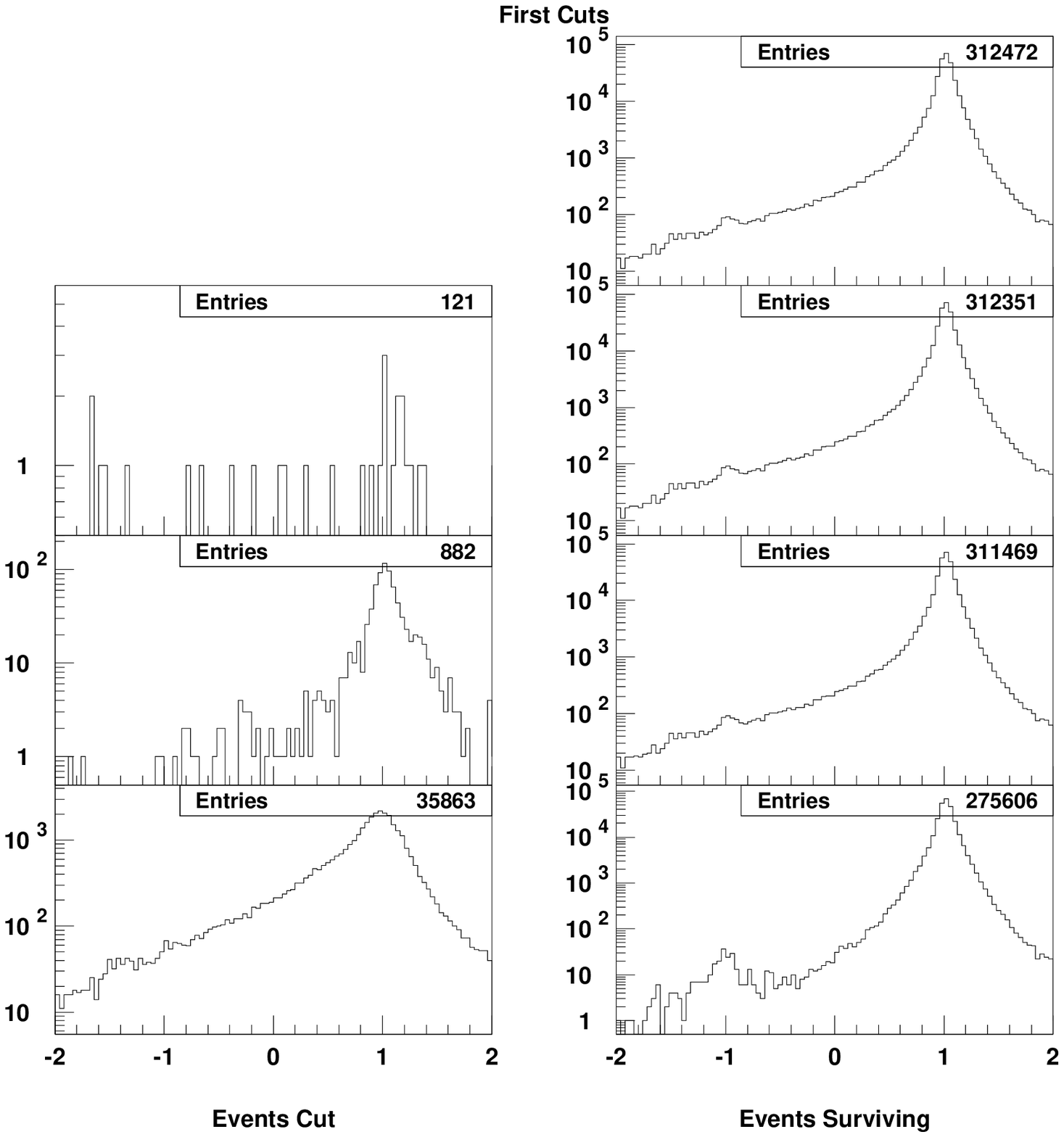}}}

\subsection{Further Background Suppression Methods}

Seven different additional cuts, which will be referred to as
``optional cuts,'' have been developed and tested.  All the previous
cuts were capable of being passed by either downgoing or upgoing
particles; however, much of the remaining background at negative
$\beta$ is initiated by downgoing particles.  A downgoing particle may
create backscattered upgoing particles in the detector by photonuclear
interactions or other processes; a bundle of 2 or more muons that
originated from the same high energy primary cosmic ray may arrive at
different times in such a way as to mimic upgoing timing; or particles
that initiate large electromagnetic showers in or near the detector
may cause several particles to enter one box at different times and
places, making the timing and position reconstructions unreliable.
Therefore, many of the following cuts will reject an event if there is
any indication of a downgoing particle; thus, the analysis is no
longer up-down symmetric.

EARLIER FACE CUT: Generally one expects the earliest scintillator hit
on the candidate track to be the earliest hit in the event; if not, it
could be because one or more boxes are mistimed.  However, often
nearby hits in the same face as the earliest hit on the track are a
few nanoseconds earlier.  This cut rejects the event if there is an earlier
scintillator hit (off the track) in a different face than the earliest
hit on the track.  This cut is up-down symmetric.  It incurs some
inefficiency for real neutrino events because a secondary may hit a
lower box before the muon reaches the Center layer.

CLUSTER CONSISTENCY CUT: Usually a group of scintillator hits near
each other in space are within a few nanoseconds of each other in
time; if not, it could be because one or more boxes are mistimed.
This cut rejects the event if there is a scintillator hit within
250~cm of a hit on the candidate track which differs by more than
10~ns in time.  This cut is up-down symmetric.  It incurs some
inefficiency for real neutrino events because a slow particle from the
vertex or a secondary particle created elsewhere may hit a nearby box
at late times.

ERP MULTIPLICITY CUT: Large showering events, which often have
inaccurate timing in some boxes, usually cause triggers in several ERP
boxes.  This cut rejects the event if there are four or more fired
scintillator boxes not intercepted by the track.  Although only two
boxes are officially associated with the candidate track, any other
boxes hit by the track are not counted toward the number of hits off
the track.  This cut is up-down symmetric.  It incurs some
inefficiency for real neutrino events because high energy neutrino
events can have a high multiplicity of secondary particles moving in
different directions.

NO DOWNGOING TRACK CUT: Sometimes a downgoing muon in the detector
creates an upgoing secondary.  This cut checks all candidate tracks
other than the chosen candidate, and rejects the event if any
candidate has associated scintillator hits with downgoing time of
flight measurements.

LEAST SQUARES FLOW CUT: This cut determines parameters $a$ and $b$ for
the equation $t~=~az~+~b$ by minimizing the squared error over the
set of all $(z,t)$ coordinates, one coordinate for every ERP hit.
This cut rejects the event if $a$ is negative, which indicates the
general flow of the event is downgoing.

MAJORITY OF PAIRS UPGOING CUT: For every pair of ERP hits, if they are
at different heights, it is determined if the pair timing is upgoing
or downgoing.  This cut rejects the event if half or more of the pairs
are downgoing.

ALL PAIRS UPGOING CUT: This cut rejects the event if even one pair of
ERP hits has a non-tachyonic downgoing time of flight.  However, it is
less strict than the previous cut in determining the pair to be
downgoing: it will not count a pair as downgoing if the time
difference is small, or if it is less than the amount of time needed
to connect the two hits at the speed of light.  Thus, for events with
ambiguous timing, it is possible to pass this cut and fail the
Majority of Pairs Upgoing Cut.  It incurs some inefficiency for real
neutrino events because events with an upgoing muon may have other
downgoing secondaries.

Figure~\ref{fig:optCutsBoth} shows the effect of each
cut individually, as well as the effect of requiring events to pass
all seven optional cuts.

\picfig{fig:optCutsBoth}{The effect of each cut individually on the 436
  upgoing candidates; the cuts are not cumulative in these graphs.
  The first seven rows are, from top to bottom, the optional cuts in
  the order described in the text. The bottom pair shows the effect of
  requiring events to pass all seven optional
  cuts.}{{\includefigure{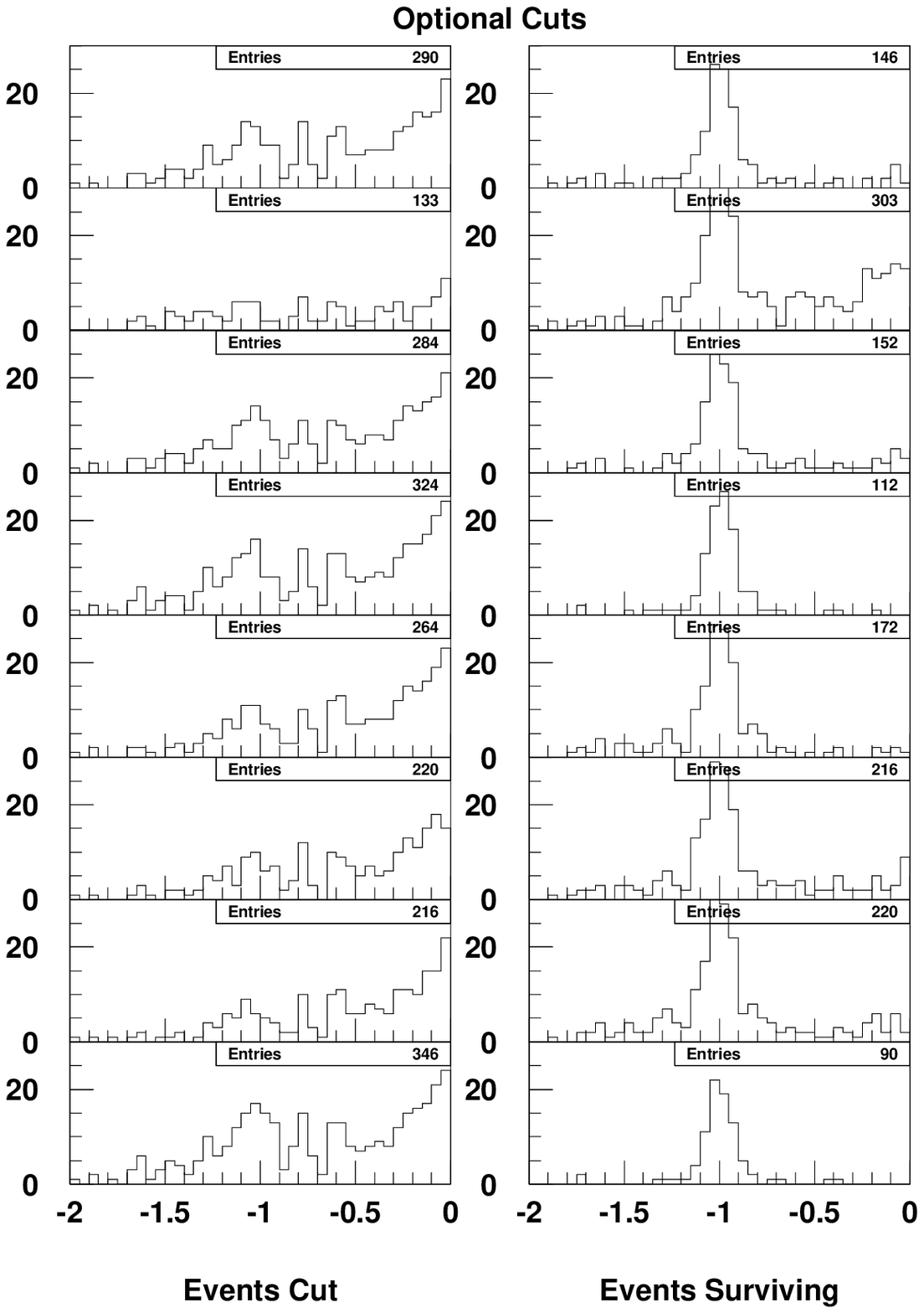}}}



An investigation was made to discern if any particular combination of
optional cuts could optimize signal and background in such a way as to
reduce the statistical uncertainty of the neutrino flux measurement.
Two kinds of background must be considered.  First, there are events
for which the time of flight is mismeasured, either due to transient
hardware failure or unusual conditions in the scintillator box.  For
these events, there is no preference for $\beta\,=\,-1$.  One can
estimate the background in the $\beta\,=\,-1$ peak by looking at the
distribution of events off-peak.  Figure~\ref{fig:betab4opt} shows the
distribution of $1/\beta$ for all events passing all the non-optional
cuts with none of the optional cuts applied.  It is clear that there
is a significant signal near $\beta\,=\,-1$, but with a large
background due to mistimed events.  The distribution of off-peak
events is skewed in the direction of $\beta\,=\,+1$ which suggests
they are the (badly-measured) tail of the downgoing muons
distribution.  If we define the signal region to be
$-1.2\,<\,1/\beta\,<\,-0.8$ (chosen by looking at Monte Carlo events,
as described in the next section), we may estimate the background in
the signal region utilizing a fit to the background events outside the
signal region.  In the figure, a linear fit is used which has no
motivation other than phenomenological.

\picfig{fig:betab4opt}{$1/\beta$ distribution for events passing all
  the non-optional cuts.  (a)~shows upgoing and downgoing events on a
  logarithmic scale while (b)~shows only upgoing events on a linear
  scale.  The linear fit uses the off-peak regions to estimate the
  background in the signal region.}{\twofig{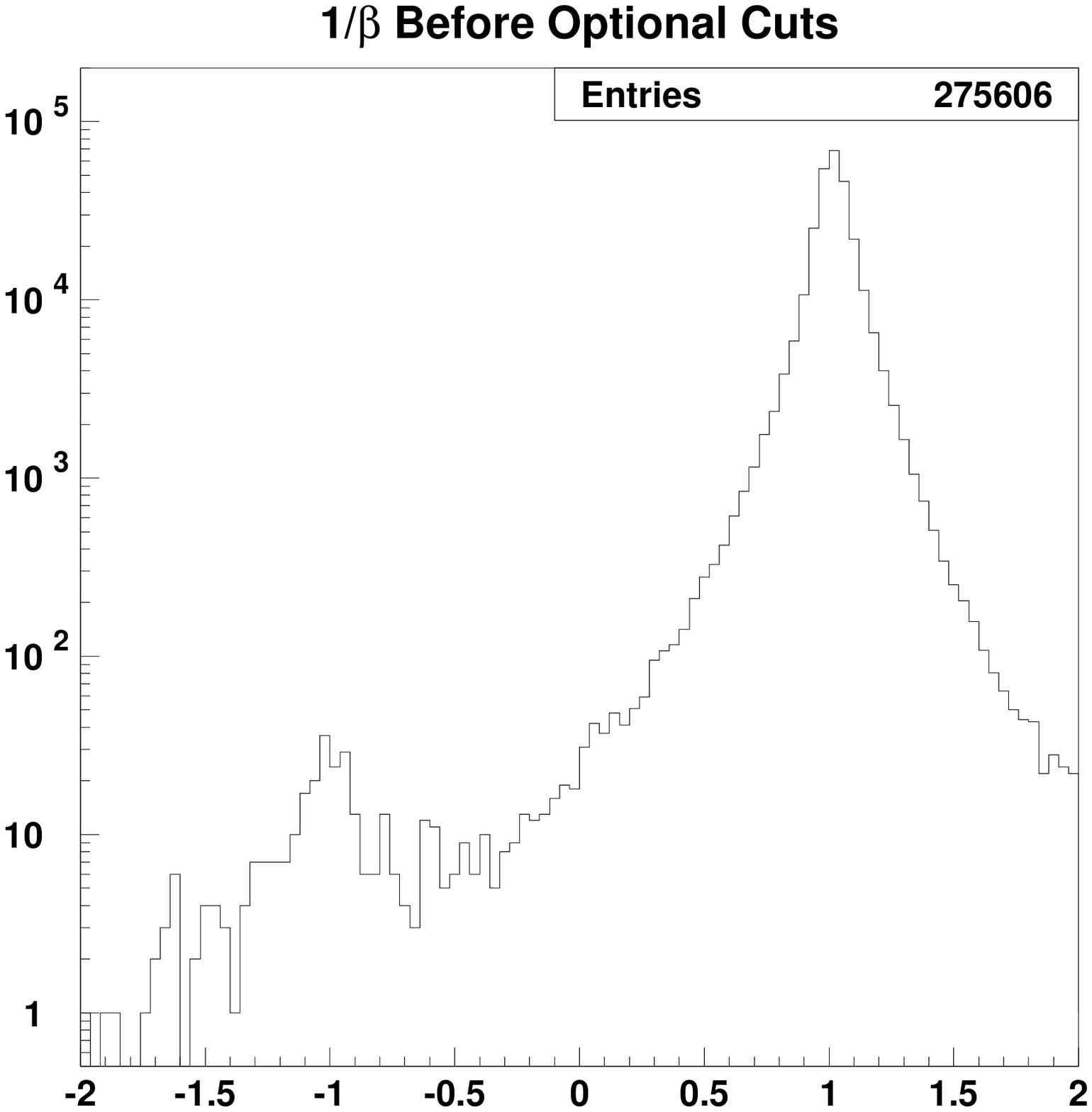}{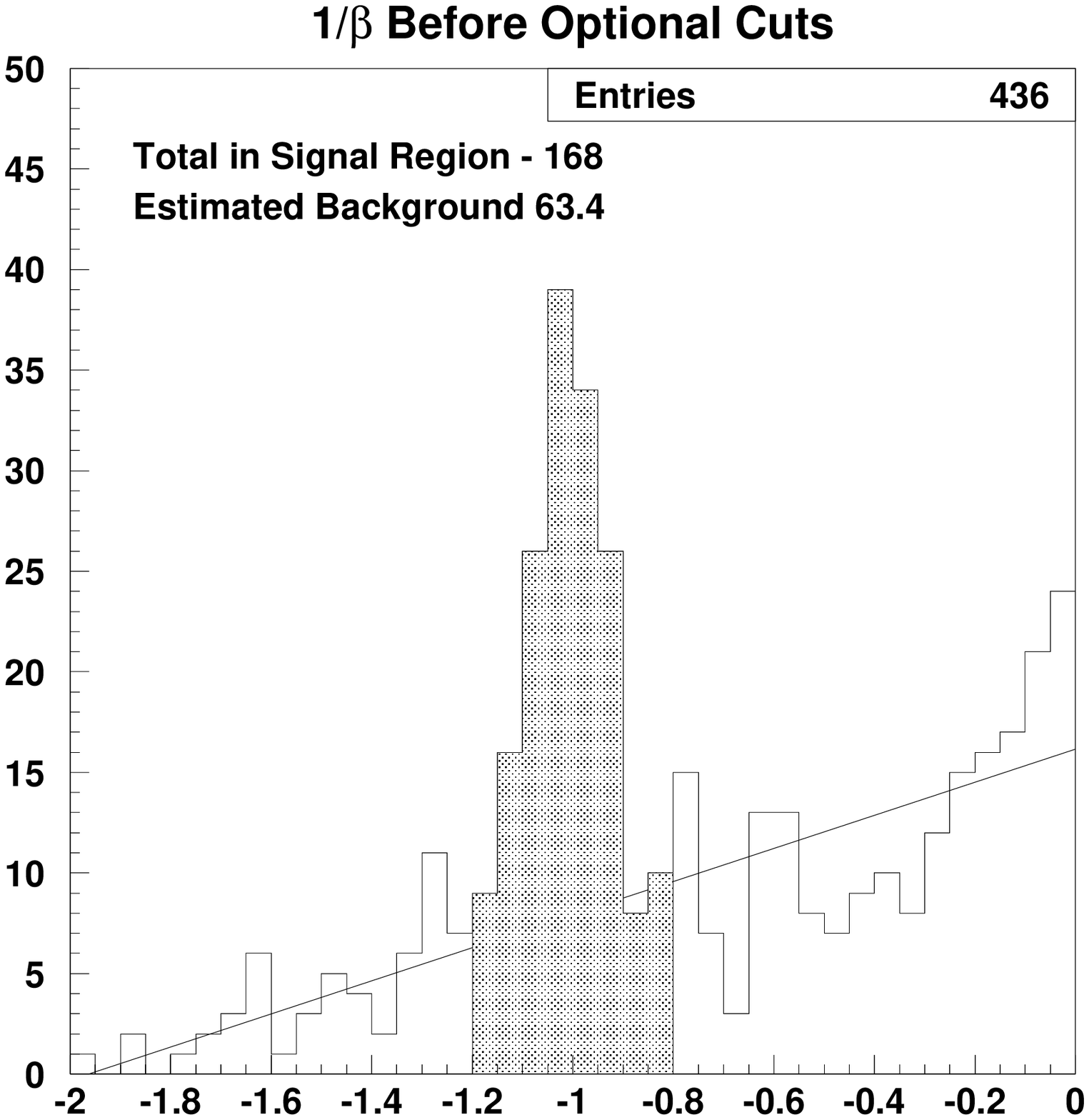}}

For Poisson-distributed signal and background processes, if we
estimate the signal $\tilde{S}$ to be the total events T (in the
signal region) minus the expected average $\overline{B}$ of the
background distribution, the fractional error ($\sigma/\mu$) in the
determination of signal is $\sqrt{T}/(T~-~\overline{B})$.  If we
restrict ourselves to insisting the events pass some definite
combination of cuts, there are 128 possible combinations of the 7
cuts, ranging from none of the cuts, to just the first cut, through
various combinations, finally to applying all 7.  We may apply the
linear background fit to the $1/\beta$ distribution of events passing
each combination of cuts and compute the fractional precision with
which the non-mistimed $\beta\,=\,-1$ signal may be determined.
(Actually, the pure signal events in the Monte Carlo have some
probability to produce a measured $\beta$ outside the signal region,
so the livetime-adjusted number of Monte Carlo events in the sidebands
is subtracted from the real data events in the sidebands before doing
the linear fit.)  When this prescription is applied, the fractional
error ranges over a surprisingly small range, from 10.1\% up to
11.6\%., as the estimated signal (which is legitimately different
depending on the cuts) ranges from 76 events up to 111 events.  Thus,
neither a loose set of cuts (which provides more signal) nor a tight
set of cuts (which produces less mistimed background) enjoys much
statistical advantage over the other.

Then, a second type of background must be considered -- true upgoing
relativistic particles which are not neutrino-induced (which, in
MACRO, are therefore almost certainly induced by downgoing muons).
This background cannot be estimated by looking outside the $\beta\,=\,
-1$ peak because it truly has $\beta\,=\,-1$.  As stricter cuts are
applied, the true signal will be decreased, and the Monte Carlo
sample, which consists of 100\% signal events, will be increasingly
rejected.  Ideally, if all backgrounds had been eliminated by earlier
processing and the Monte Carlo correctly simulated the chance of
signal passing each cut, the estimated signal in real data would be
proportional to the number of events passing the cuts in Monte Carlo
for whatever set of cuts is chosen.  Figure~\ref{fig:sVsMC} shows a
scatter plot, for each of the 128 combinations of optional cuts, of
the data signal, plotted versus the predicted signal from the Monte
Carlo.  It is seen that the ratio tends to be larger for looser cuts
(i.e.,\ cuts that produce a larger predicted signal).  This suggests
that the loose cuts are retaining some upgoing background in addition
to signal.

\picfig{fig:sVsMC}{Scatterplot of the different possible combinations
  of 7 optional cuts.  Each point represents one possible set of cuts
  (for example, one point represents all events passing cuts 1, 3, and
  7).  For each of 128 possible combinations, the estimated
  background-subtracted signal seen in the real data is plotted
  against the predicted signal from the simulation (arbitrarily
  scaled).  Poisson errors on the data are shown on the first and
  last points.  The combination of cuts used in this analysis, in which
  all 7 optional cuts are applied, is circled.}{\pawfig{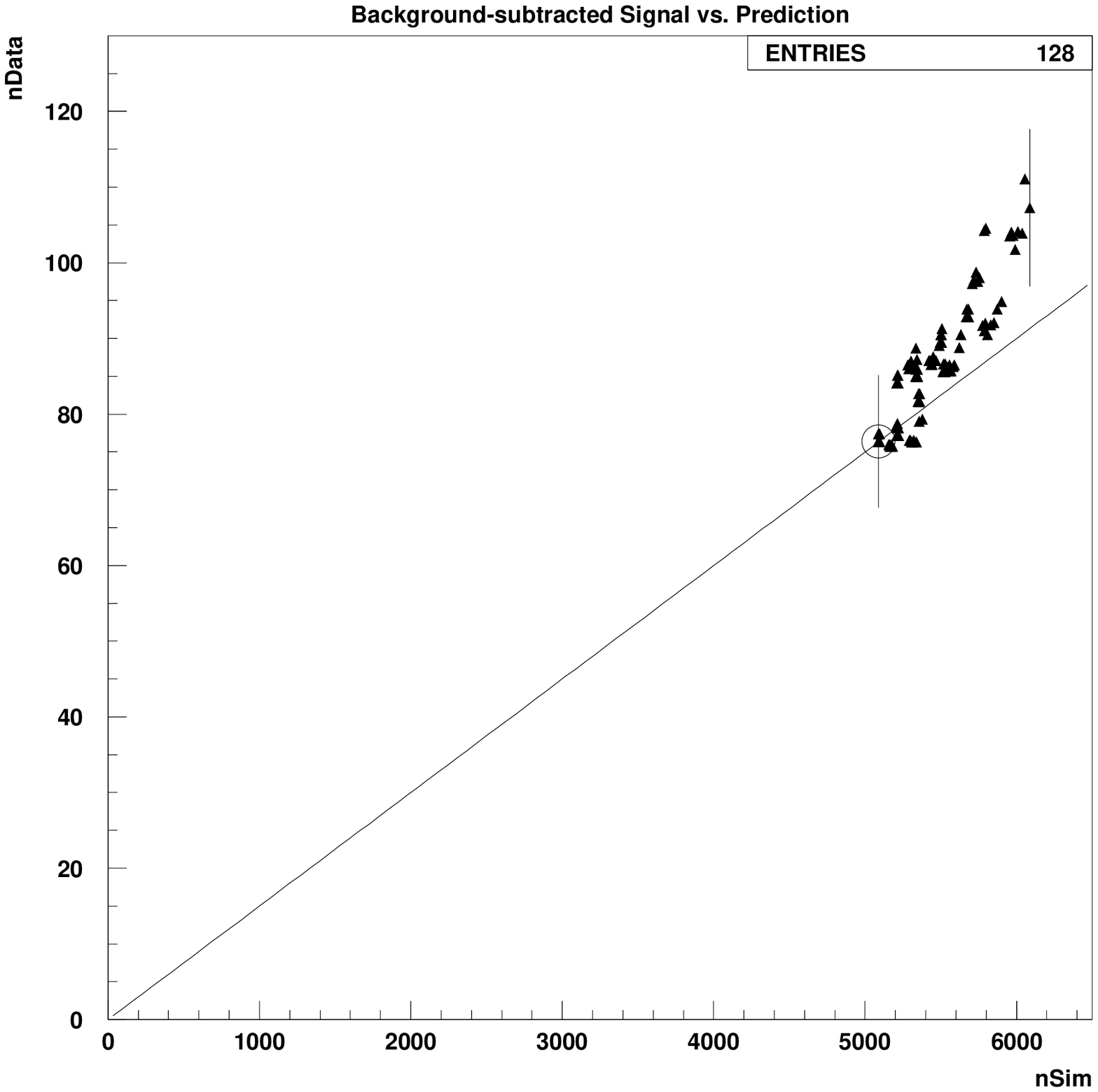}}

Therefore, it appears that to eliminate as much of the $\beta\,=\,-1$
background as possible, it is prudent to utilize all the optional
cuts.  This will also reduce signal, but it was shown above that
because the mistimed background is also reduced, we pay a small price
in statistical error by cutting hard.

\section{Results}

\subsection{The Selected Events}

Figure~\ref{fig:betab4opt} above shows the $1/\beta$ distribution for
events passing all non-optional cuts, including the containment cut,
but no optional cuts.  The vast majority are downward events
($\beta\,=\,+1$) stopping in the detector, while a few events are
badly mismeasured, and a signal of neutrino-induced events at
$\beta\,=\,-1$ is clearly visible.  Figure~\ref{fig:finalBeta} shows
that the mistimed background is cleaned up in the negative $\beta$
sector by the application of all seven optional cuts.  Within the
signal range of $-1.2\,<\,1/\beta\,<\,-0.8$, 77 events are found.

\picfig{fig:finalBeta}{The final $1/\beta$ distribution.  The signal
  region, containing 77 events, is shaded.}{\pawfig{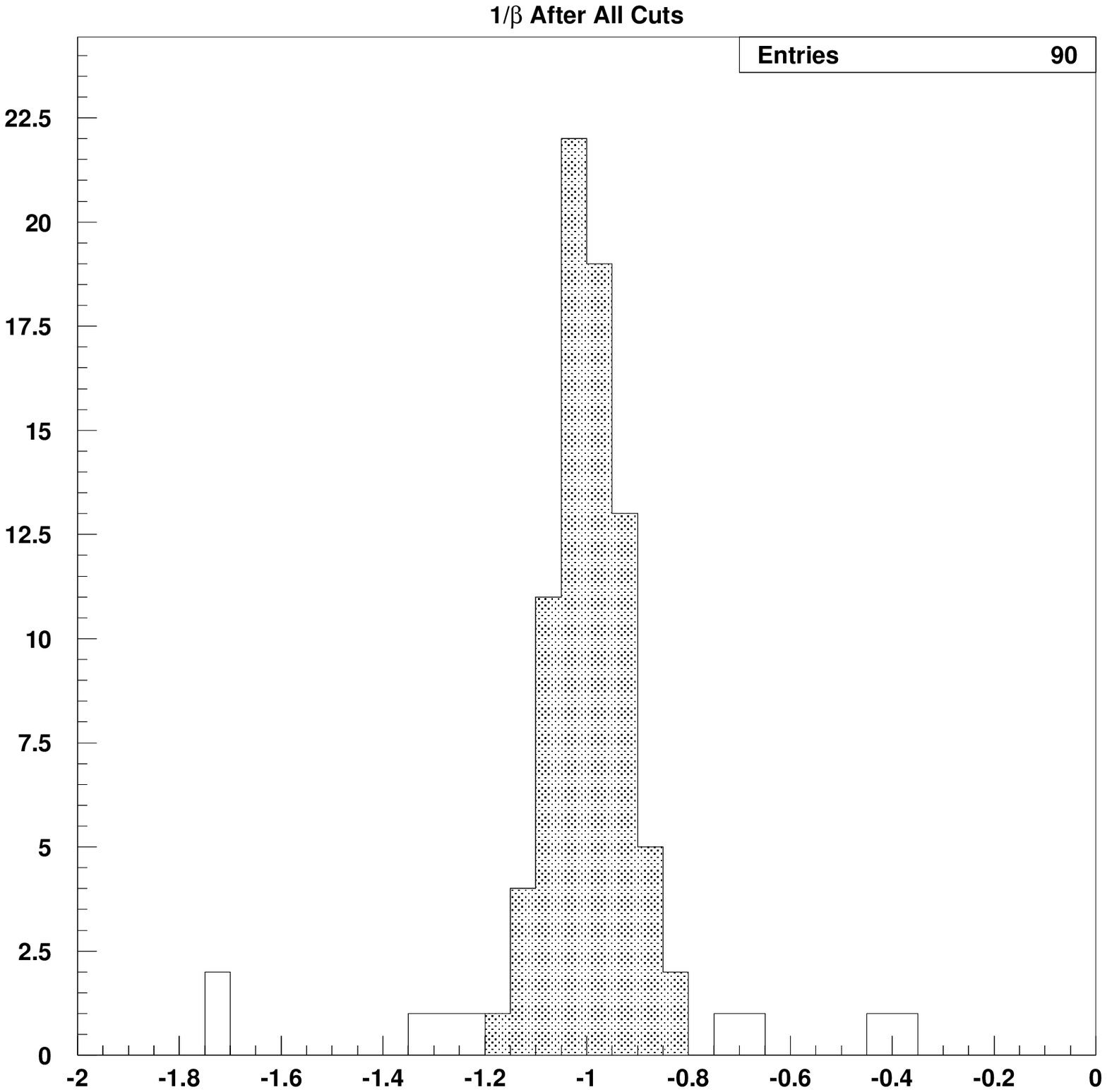}}
\subsection{The Monte Carlo Prediction}

Following the same analysis path as real data, the simulated data for
113 years of nominal detector operation yielded 5460 events passing
all cuts including all optional cuts (neglecting the effect of
neutrino oscillations).

For each real data run, we could define the livetime for that run as
the time from the first event recorded in the run until the last
event.  Summing this time over all the runs in the dataset yields
2,346,699~minutes (4.5~yr).  However, not all of these minutes are
equivalent.  During some of them MACRO was not in its nominal
configuration.  For example, sometimes one or more entire supermodules
were turned off.  In the earliest runs in the dataset, the attico
streamer tubes were not all installed and working.  Other times, even
if all the supermodules were in acquisition, some small parts of the
detector were dead or not fully efficient.  Often one scintillator
tank would be turned off due to hardware problems, sometimes for weeks
or months at a time.  At times the streamer tube gas mixture was not
well-controlled, and the strip efficiency declined precipitously.  For
all of these reasons, it is necessary for the Monte Carlo calculation
of the expectation to have a way of accounting for the effects of
MACRO not being always in a full, nominally-functioning configuration.

To incorporate the effects of the known dead channels and inefficiencies
into the Monte Carlo calculation, the database of channel dead periods
and channel efficiencies (see Section~\ref{sec:hwprobs}) is used.  The
entire dataset of simulated events is passed through 262 different
simulations, one for each week analyzed.  For each Monte Carlo event
to be analyzed, a clock time is chosen randomly, between the time the
first event of the week occurred and the time of the last event of the
week.  If the time chosen is not during a run in the dataset (that is,
it is between runs), the time is rejected and another chosen.  Any
channels that were dead at the chosen time are eliminated from the
event record before analysis proceeds, and any channels that were not
dead at that minute are probabilistically eliminated from the event
record according to their efficiencies stored in the database.  Next,
a decision is made probabilistically to eliminate all channels in a
microvax according to the measured average computer deadtime stored in
the database for that microvax for that week.  The algorithm assumes
maximum correlation between the microvax dead periods; that is, if the
event involves channels from two microvaxes and one microvax is chosen
to be dead, the other microvax will also be chosen to be dead as often
as possible, consistent with the difference of deadtimes between the
two microvaxes.  Finally, any channel that would have been eliminated
from the real data at that time by a microcut (see Section
\ref{sec:microcuts}) is eliminated from the simulated data as well.
After this process, the event is analyzed based on the channels that
have not been eliminated.  The resulting event may or may not pass all
cuts.

For each Monte Carlo event, if it passes all cuts in the simulation
for a given week, it is recorded to have a weight for that week equal
to the number of minutes of actual data-taking that week, divided by
the number of simulated minutes in the Monte Carlo generation.  If it
does not pass the simulation, its weight for the week is zero.
Summing the weight for the event over all 262 weeks gives its total
weight.  If the detector had been nominal at all times, the same
events would pass each week and the summed weight for all events would
be equal to the ratio between the total real livetime and the length
of time in the simulated generation.  The Monte Carlo prediction for
the number of events in the bins of any histogram would be the number
of Monte Carlo events in the bin, multiplied by the ratio of the real
livetime to the simulated generation time.  The effect of the recorded
non-ideal performance of the real detector is indicated by the
difference between the actual summed weight for each Monte Carlo event
and the ideal summed weight.  The Monte Carlo prediction for the bins
of any histogram is the sum of the weights of all simulated events in
that bin.  Figure~\ref{fig:simBeta} shows, for example, the $1/\beta$
distribution for the real and ideal detectors.  The simulated effect
of the non-ideal performance is a reduction of 15.0\% in the predicted
event rate in the signal region $-1.2\,<\ 1/\beta\,<\,-0.8$.

\picfig{fig:simBeta}{The predicted $1/\beta$ distribution (neglecting
  oscillations) for the ideal detector (solid line) and for the
  non-ideal detector as simulated.}{\pawfig{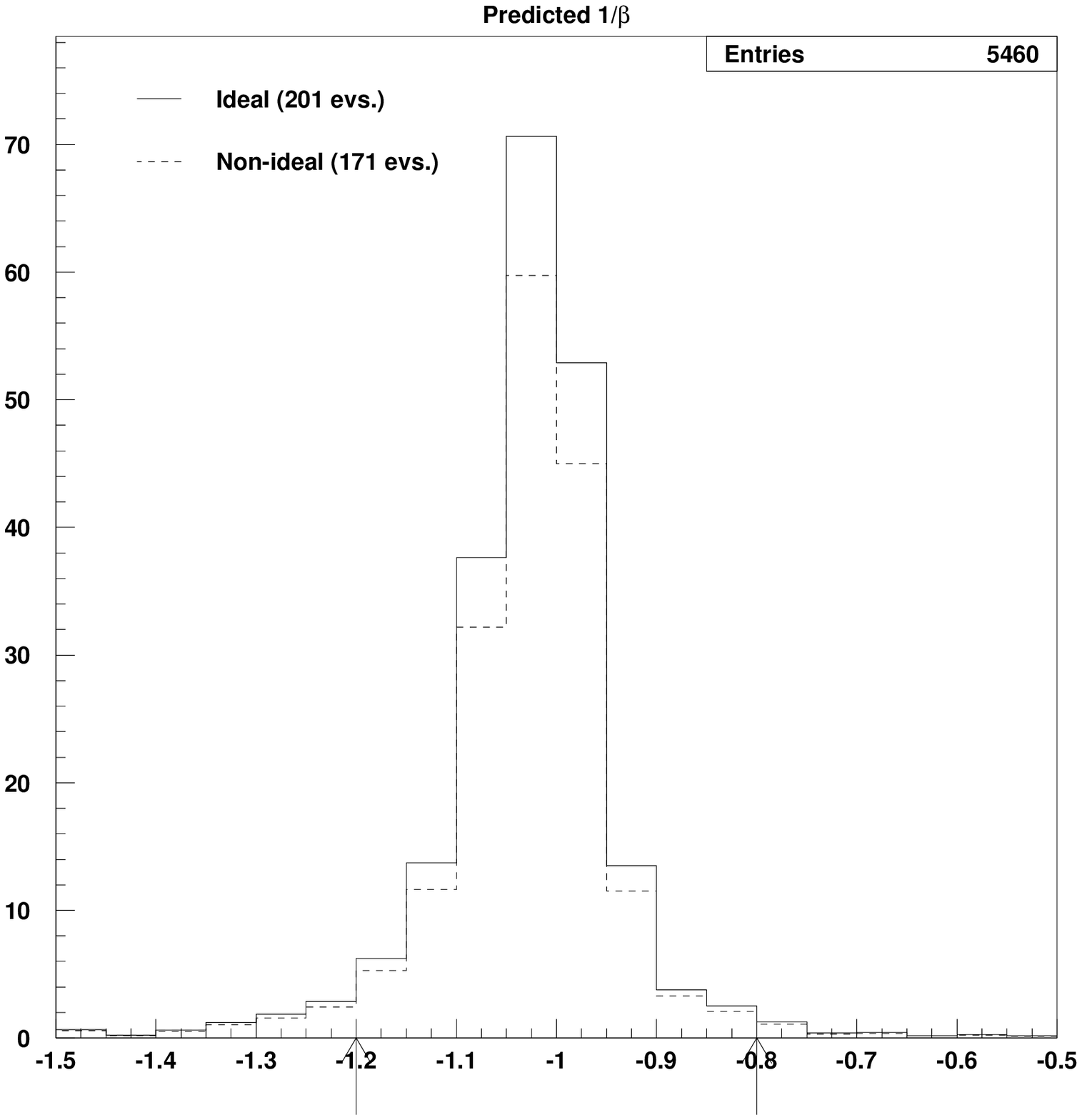}}

For the non-ideal detector, Table~\ref{tab:mcProps} shows several
properties of the simulated events passing all cuts, both assuming no
oscillations and assuming oscillations at a test point of $\sstt = 1,
\dmm = 3.2~\times~10^{-2}~\mathrm{eV^2}$.  The question of what this
analysis has to say about oscillation parameters will be taken up in
Chapter~\ref{chap:interpretation}.

\begin{table}[htb]
  \leavevmode
  \begin{center}
    \begin{tabular}{|c|r|r|}
      \hline
                             & No Oscillations & Oscillations \\
      \hline \hline
        Total                & 170.8           &  96.3 \\
      \hline
        $\numu$              & 108.2           &  57.6 \\
        $\numubar$           &  49.6           &  25.7 \\
        $\nue$               &  10.3           &  10.3 \\
        $\nuebar$            &   2.7           &   2.7 \\
      \hline
        $\numu+\numubar$ CC  & 154.9           &  80.4 \\
        $\numu+\numubar$ NC  &   2.9           &   2.9 \\
        $\nue+\nuebar$ CC    &  11.9           &  11.9 \\
        $\nue+\nuebar$ NC    &   1.1           &   1.1 \\
      \hline
        p                    &  67.8           &  37.8 \\
        n                    & 102.9           &  58.5 \\
      \hline
        Absorber             & 122.8           &  68.0 \\
        Iron                 &  31.8           &  18.1 \\
        Scintillator         &  10.9           &   7.0 \\
      \hline
    \end{tabular}
  \end{center}
  \caption{\protect\label{tab:mcProps}Properties of simulated events
         passing the analysis.  The first four rows show the
         particle interacting, the next four, the type of interaction,
         the next two, whether the interaction was with a proton or a
         neutron, and the last three, in what material in MACRO the
         interaction occurred.  Several materials in MACRO are not
         listed.}
\end{table}

\subsection{Comparison of the Measurement and the Prediction}

Figure~\ref{fig:mcData} shows a comparison of the measurement and the
prediction, with and without oscillations, before applying any
corrections or statistical and systematic errors.  It is immediately
evident that the data favor the hypothesis of oscillations.  The
significance of this assertion will be explored in the next chapter,
Interpretation.

\picfig{fig:mcData}{The selected real data events, plotted as a
  function of two variables, and compared to the Monte Carlo
  prediction, with and without oscillations.  (a)~$\cos(\theta_{zen})$
  with -1 representing vertical upwards and 0 horizontal.
  (b)~Pathlength (in centimeters) in lower MACRO, which roughly sets a
  minimum on the penetrating power of the initial
  muon.}{\twofig{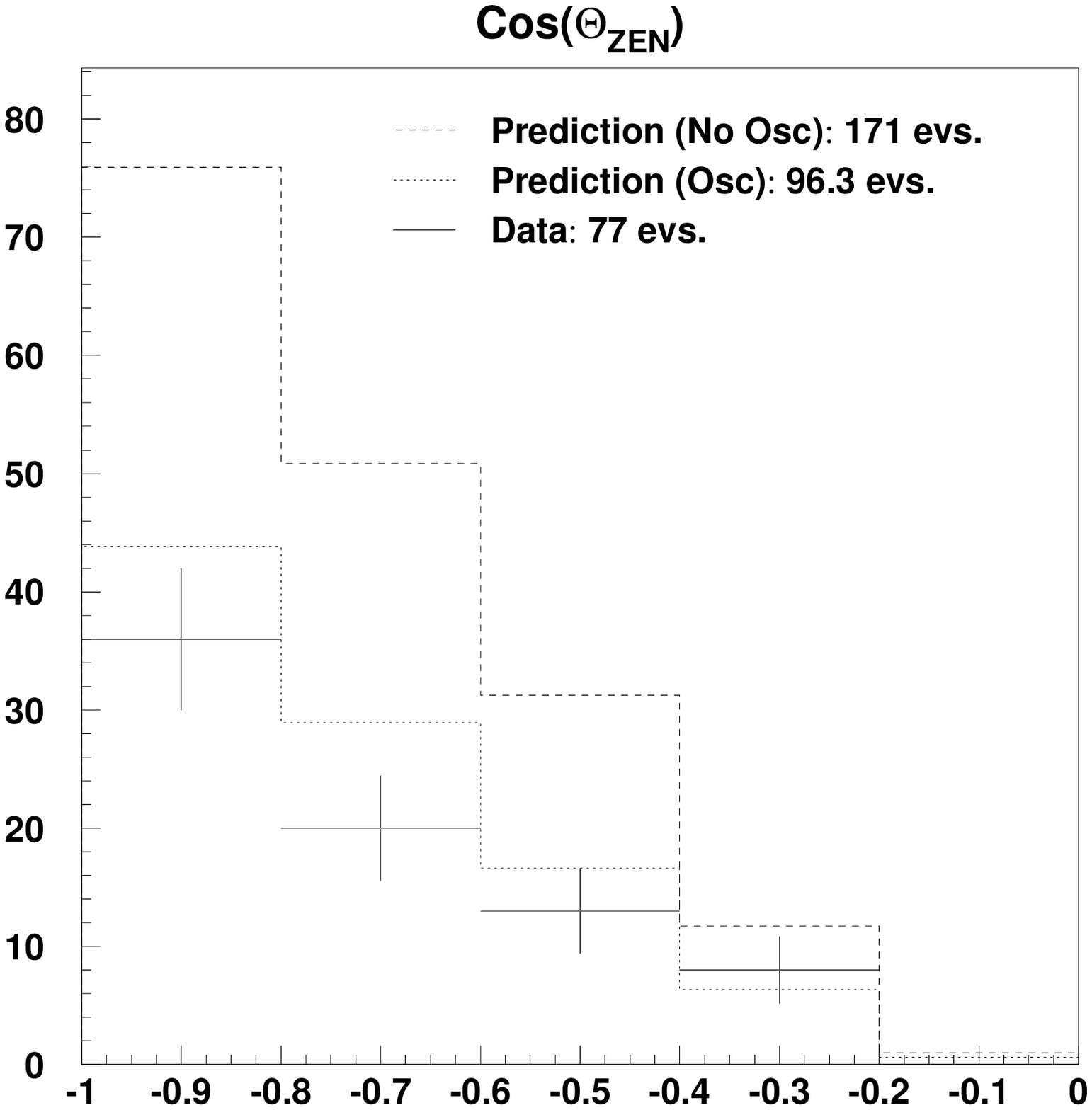}{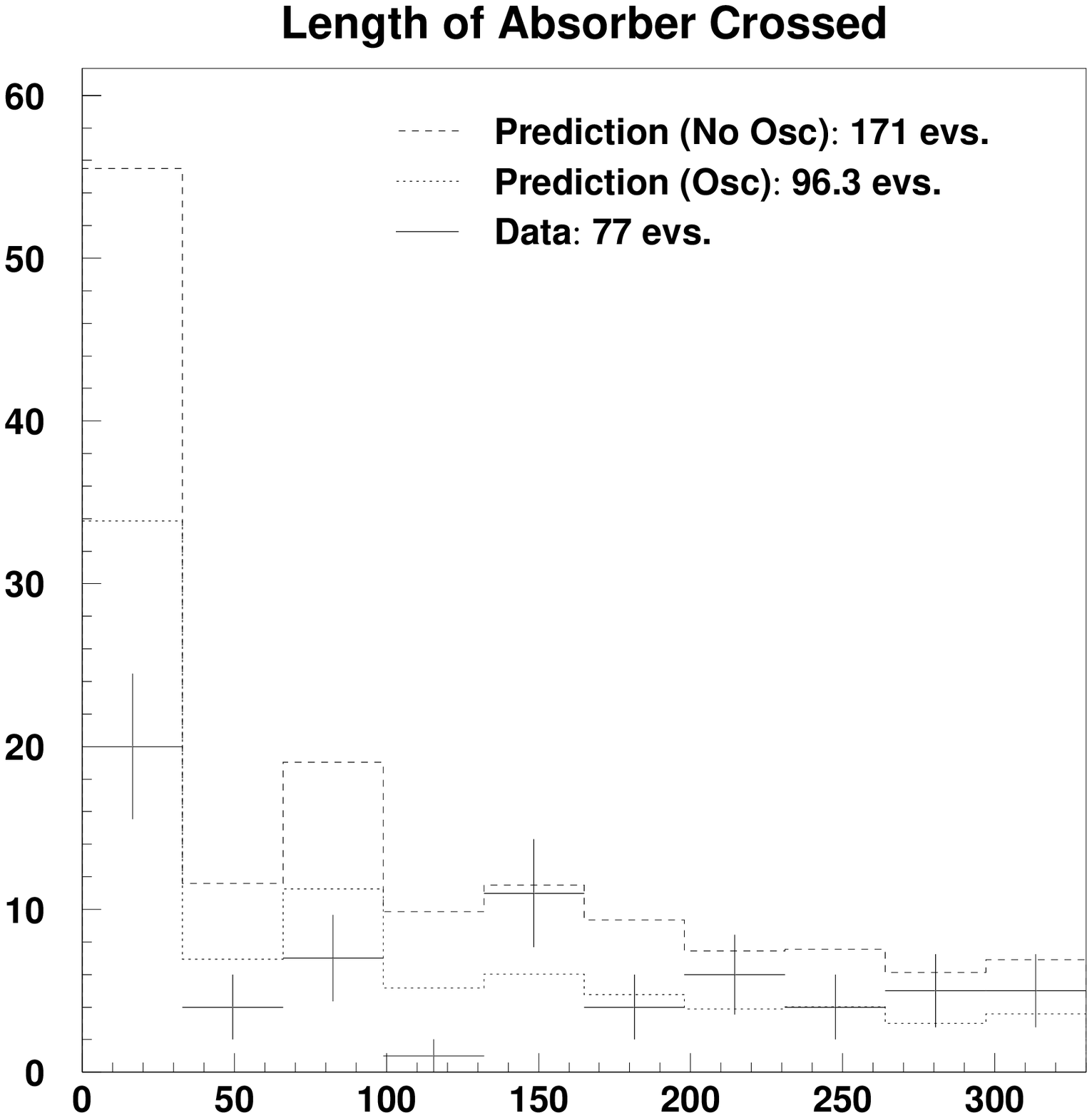}}


\chapter{Interpretation\protect\label{chap:interpretation}}

\section{Corrections and Estimates of Uncertainty in the Measurement}

\subsection{Externally-produced Neutrino-induced Muon Background}

CORRECTION and UNCERTAINTY

If the containment cut is not perfect, upward throughgoing
(neutrino-induced) muons produced outside the detector may appear to
be semi-contained.  A Monte Carlo dataset of 100,000~events, with an
equivalent livetime of 143.5~yr, has been created using the upgoing muon
generator described in Section~\ref{sec:upmugen}.  The generator is
based on the flux of upgoing muons due to atmospheric neutrinos that
would exist if there were no oscillations.  After putting the data
through the simulation of detector deadtime and inefficiency, the
events passing the semi-contained analysis sum to a prediction of
1.77~events for the livetime of this analysis.  Adjusting the flux for
oscillations, the prediction is 0.97~events.  (Here, and throughout
the rest of this chapter, data assuming oscillations are computed at a
test point of $\numu$ disappearance oscillations with $\sstt = 1; \dmm
= 3.2~\times~10^{-3}~\mathrm{eV^2}$.)  This is to be subtracted from
the measured events.  The correction is 1-2\% of the total number of
events, a statistical error is applied for fluctuations in the
background, and a systematic uncertainty of 1\% of the total number of
events is applied to account for any inaccuracy in this procedure.

\subsection{Mistimed Background}

UNCERTAINTY only

Even in the Monte Carlo, not all neutrino-induced events are measured
to have $\beta\,=\,-1$ (see Figure~\ref{fig:mcBeta}a).  In the
$|\beta|\,<\,1$ region ($|1/\beta|\,>\,1$) there are events in which
the muon (or perhaps some other charged particle from the interaction
that creates a track and is chosen by the analysis to characterize the
event) is actually moving at less than the speed of light, as well as
events that are mistimed due to multiple particles hitting a single
tank at different places.  At $|\beta|\,>\,1$, only the latter class
exists.  In the real data, 9~events exist in the regions
$-2\,<\,1/\beta\,<-1.2$ and $-0.8\,<\,1/\beta\,<0$ which is consistent
with the Monte Carlo prediction of 7.9~events (see
Figure~\ref{fig:mcBeta}b).  Therefore, there is no empirical evidence
for additional mistimed background in this analysis, in which all
optional cuts are applied; therefore, no subtraction is made.  If half
the events in the sidebands are really due to mistimed background, a
linear fit to them suggests about 1~event in the signal region, so an
uncertainty of 1~event is assessed due to mistimed background.  This
uncertainty is asymmetrical; it can only cause the measurement to be
higher than the true value.

\picfig{fig:mcBeta}{Even ``signal'' events from the simulation
  sometimes have measured $\beta$ outside the signal region.
  (a)~shows the $1/\beta$ distribution from the simulation with fine
  binning.  (b)~compares the sidebands of the data (solid line) and
  the simulation (dashed line) at a coarser
  binning.}{\twofig{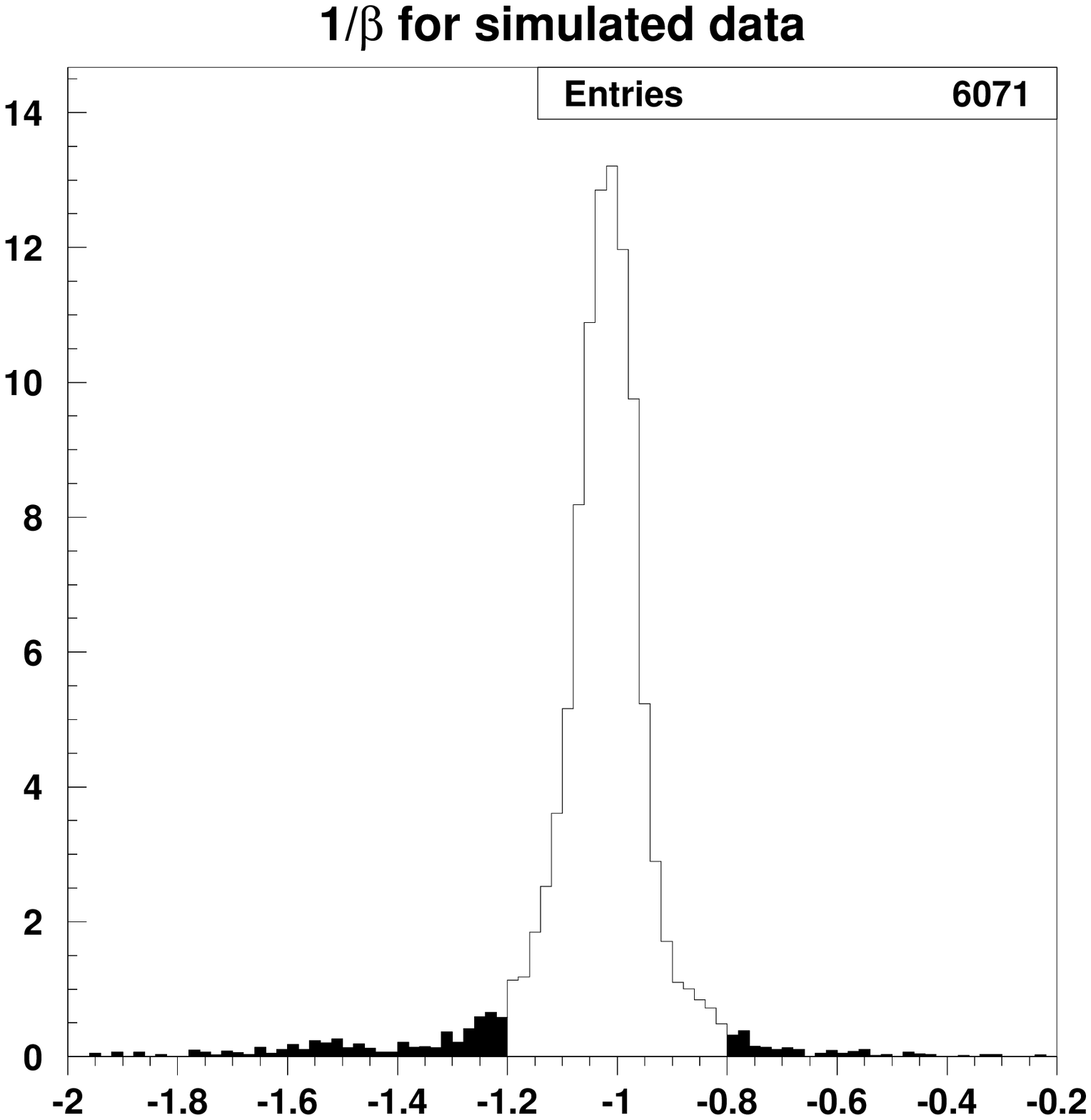}{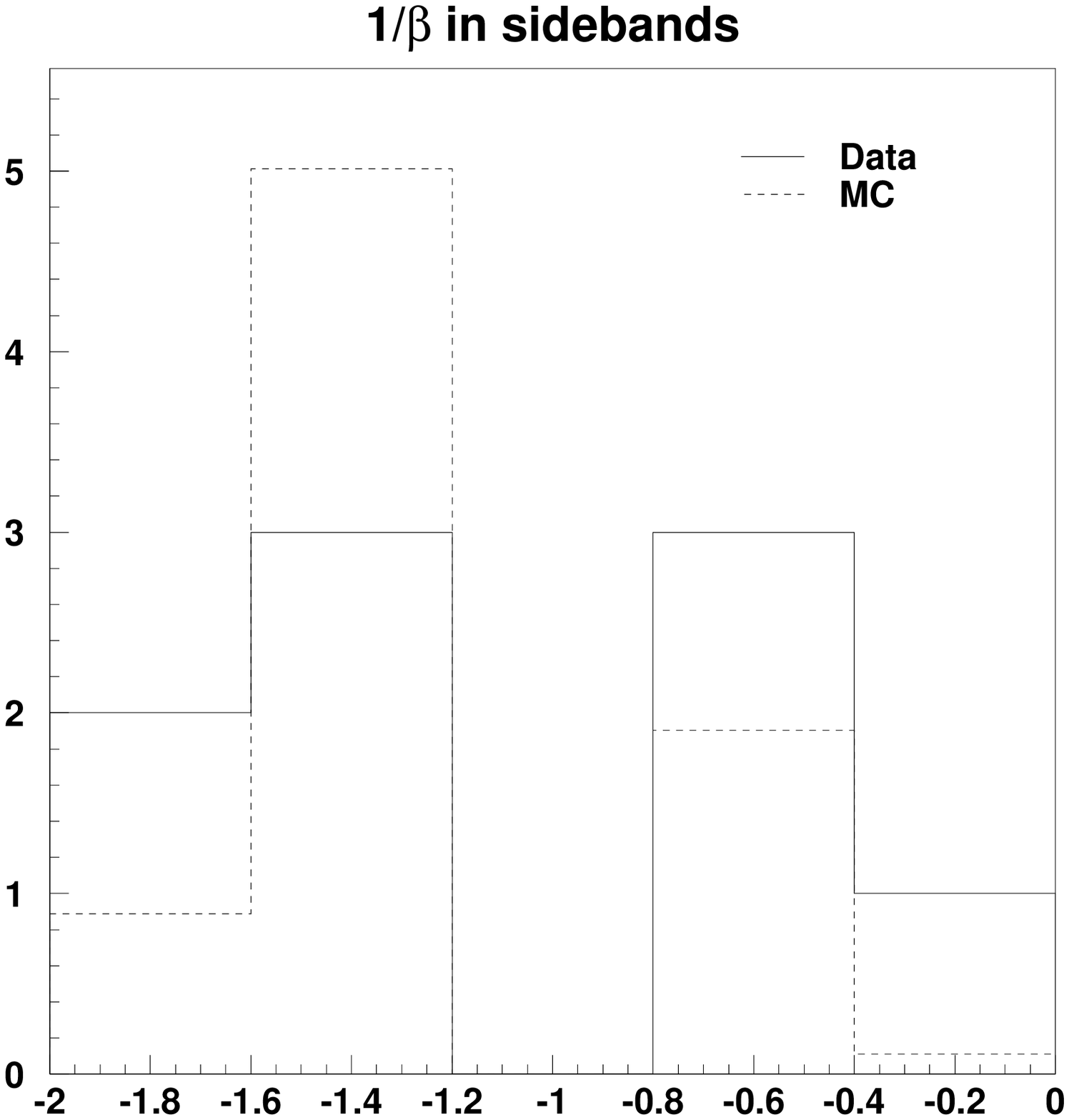}}

\subsection{Remaining $\beta\,=\,-1$ Background}

UNCERTAINTY only

This analysis attempts to minimize the backgrounds induced by
downgoing muons by implementing all of the optional cuts.
Figure~\ref{fig:sVsMC} in Chapter~\ref{chap:meas} showed that the less
restrictive combinations of cuts appear to have more $\beta\,=\,-1$
background than the all-optional-cuts combination used in this
analysis; however, that does not rule out the possibility that
background events remain in the all-optional-cuts combination.  A hand
scan of events revealed that the sample remaining after applying all
optional cuts appeared considerably cleaner than the events allowed by
the less restrictive combinations.  Because there is no clear evidence
of background, no subtraction is made.  A systematic uncertainty of 3
events is assigned.  This uncertainty is asymmetrical; it can only
cause the measurement to be higher than the true value.

As a check, 100,000 simulated downgoing atmospheric muons were passed
through the upgoing semi-contained analysis.  All failed to pass the
analysis, which is approximately to say none generated an upgoing
particle in the detector while escaping detection themselves.  This
represents a simulated livetime of under one week, however.  No
meaningful lower limit can be set on the background based on this
simulation, but at least it gives no evidence of background.
\subsection{The Nominal Detector Simulation}

UNCERTAINTY only

An inaccurate detector simulation would change the prediction but
would have no effect on what is measured.  However, because
discrepancies due to the simulation are an experimental rather than a
theoretical issue, it has been the custom in MACRO to incorporate
systematic uncertainty due to the simulation into the experimental
error.

This analysis requires extreme reliance on the details of the Monte
Carlo.  The great majority of neutrino-induced particles in the
detector range out without leaving a trace or hit just a couple of
streamer tubes.  Out of over 150,000 generated neutrino interactions,
fewer than a thousand create a track and hit two scintillators; the
rest do not even pass the preliminary stages of this analysis.  Thus,
we are relying on the simulation's estimate of the occurrence of a
relatively rare process.  Furthermore, specifying the initial
kinematics of an event is not enough to determine if the event will be
detectable.  To demonstrate this, 10,000 neutrino interactions were
generated, determining the initial kinematics of particles emerging
from the interaction.  Then, each of the 10,000 interactions was
tracked in GMACRO four times, the only difference being the random
number seed to the Monte Carlo.  Each run of GMACRO rolled the dice to
determine how the particles interacted in the detector, computed the
response of MACRO, and wrote a run file of 10,000 events to disk.
Utilizing a preliminary version of this analysis, 741 of the 10,000
neutrino interactions produced an event detected by the analysis in at
least one of the GMACRO runs, but only 249 of them were detected in
all four runs of GMACRO\@.  Each of the four runs produced 483~$\pm$~14
detectable events.  This means that around half of all detectable
events in the Monte Carlo had kinematics so susceptible to change that
they would not be detectable if the Monte Carlo made different
choices.  (See Figure~\ref{fig:burn}.)

\picfig{fig:burn}{The same neutrino interaction, tracked four times by
  GMACRO\@.  The initial particles are a 730~MeV $\mu^+$ moving up and
  to the right, and a 390~MeV neutron moving down and to the left.
  Only event (b) passes the analysis and is detected.  This is
  an extreme example of differences in tracking, not typical.}
{
  \makebox[5in]{
    \resizebox{2.5in}{!}{\includefigure{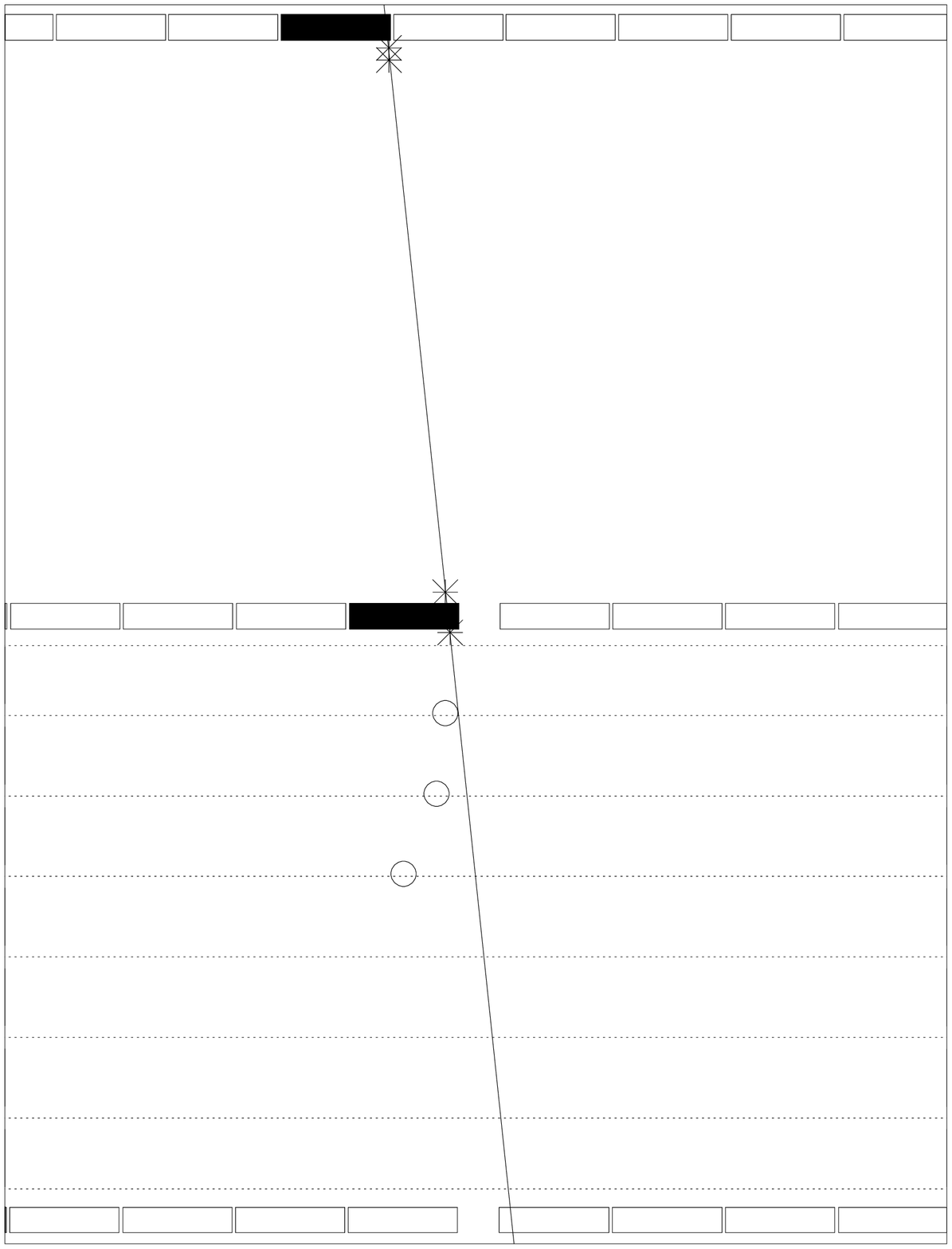}}
    \resizebox{2.5in}{!}{\includefigure{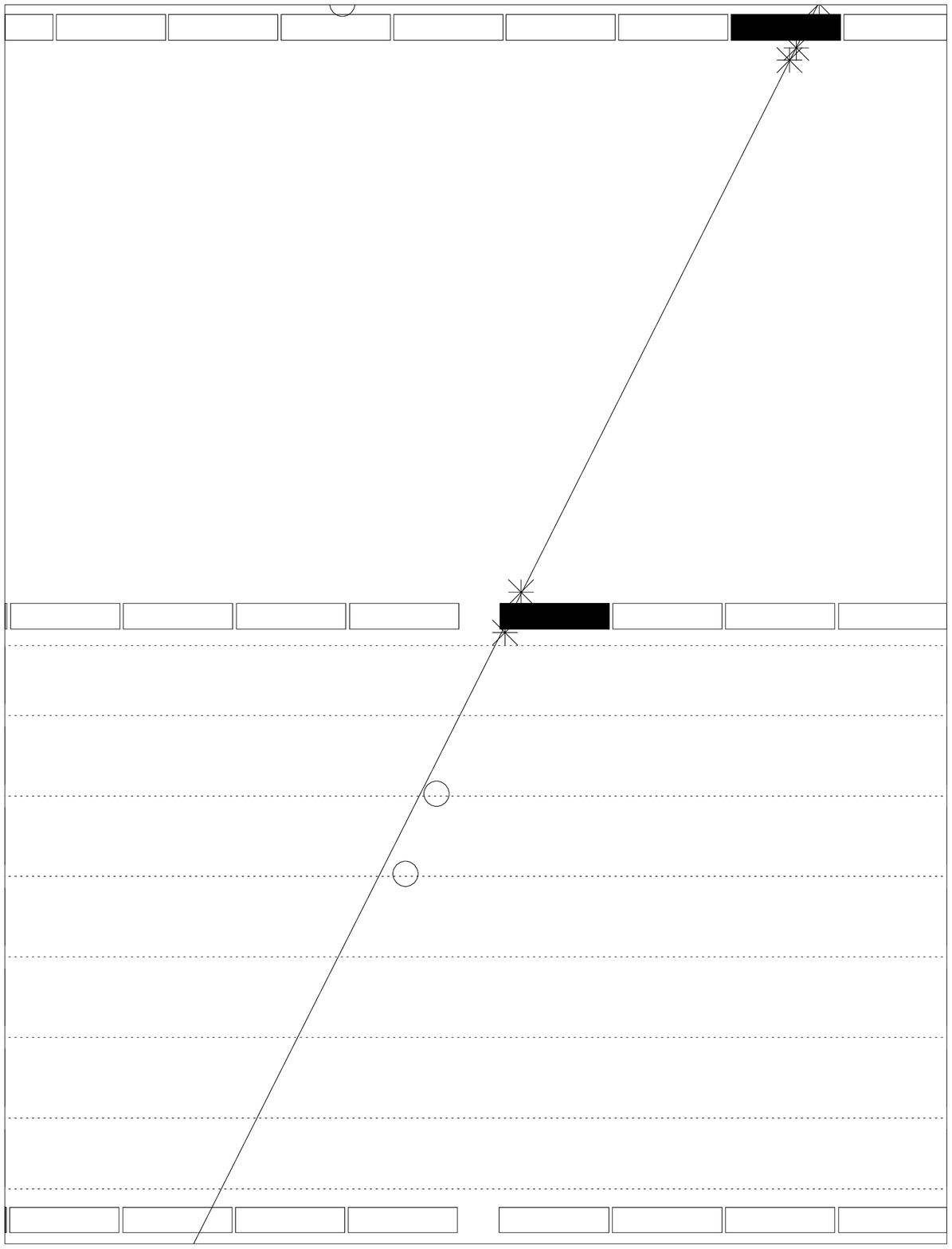}}}
  \makebox[5in]{\makebox[2.5in]{(a)} \makebox[2.5in]{(b)}}
  \makebox[5in]{
    \resizebox{2.5in}{!}{\includefigure{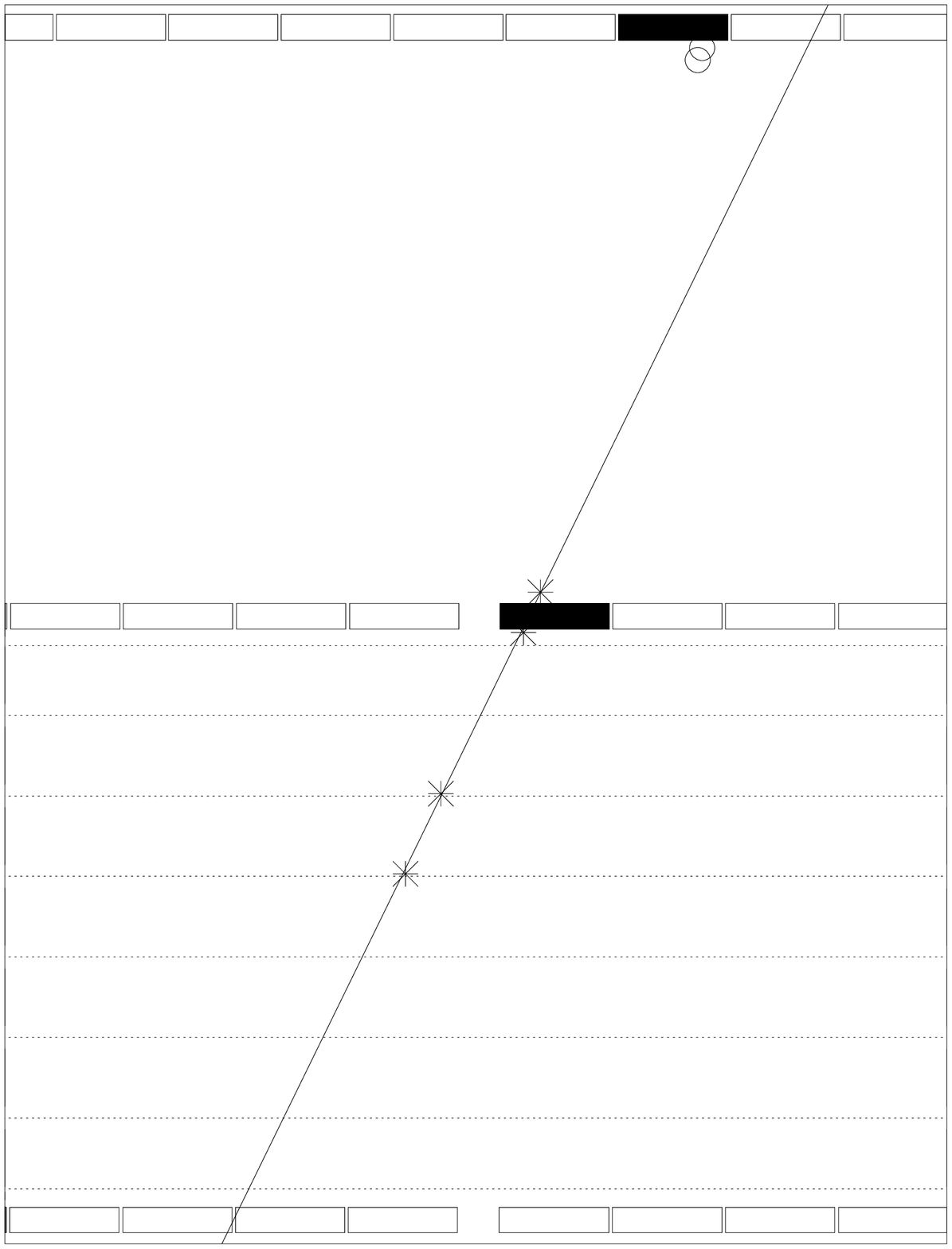}}
    \resizebox{2.5in}{!}{\includefigure{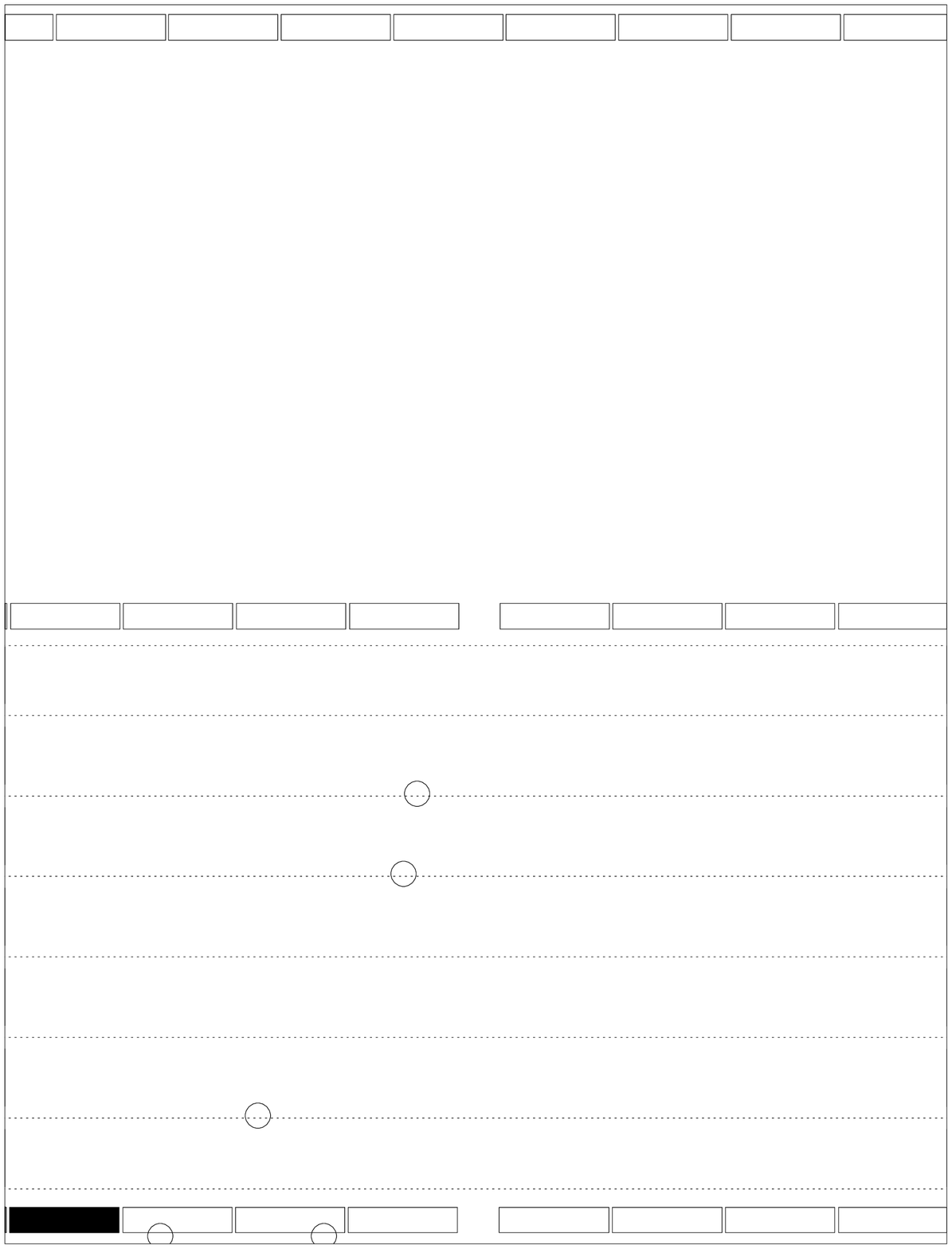}}}
  \makebox[5in]{\makebox[2.5in]{(c)} \makebox[2.5in]{(d)}}
}

It is difficult to estimate an uncertainty due to the simulation.
GMACRO is very well-certified to perform well on the type of event
MACRO sees most often -- single relativistic muons (see the plots in
Section~\ref{sec:mctune}).  However, the events on which this analysis
is based are sometimes much more complicated, involving multiple
charged and uncharged particles.  There is little empirical evidence
that GMACRO produces an accurate response in these events.  However,
many of the elements comprising GMACRO -- the materials, geometry and
physics processes -- are well-understood and certified in their own
right.  GMACRO's estimate of the timing and energy measurements when
multiple particles hit a single tank is not known to be good; in fact,
it is known that the GMACRO algorithm is an oversimplification of the
true case.  It is suspected that this is the chief source of
discrepancy between the true and simulated response, particularly for
high energy neutrino interactions with many secondary particles.

Evidence of inaccuracy in the simulation comes from an attempted
analysis of a higher-energy subset of semi-contained neutrino
interactions, which loosened the strict requirements on tracking from
the analysis presented in this work, and attempted to remain efficient
for multiple-particle events.  The higher-energy analysis relied
primarily on scintillator information and looked for events with a
generally-upward flow of particle timing.  The analysis had to be
abandoned because of inconsistencies between the classes of events
seen in the data and in the simulation.  It should be noted, however,
that the events in the higher-energy analysis were inherently more
difficult to simulate and to reconstruct due to the presence of
multiple particles, higher-energy particles, and more electrons and
photons.

Several of the major parameters within GMACRO have been varied to get a
sense of the variation.  Raising the mass by 5\% increases the
prediction by 3.8\%.  Changing the hadronic interaction model to the
version of FLUKA incorporated in Geant 3.21 decreases the prediction
by 5.5\%.  Lowering the ERP trigger thresholds in the simulation by
2~MeV decreases the prediction by 1.4\%.  Raising all the Geant CUTS
parameters from 0.5~MeV to 1~MeV (so that particles are tracked as
they lose energy only until they reach 1~MeV) increases the prediction
by 0.7\%.  Setting all streamer tube trigger thresholds to their
default values, rather than the values determined by tuning, decreases
the prediction by 4.0\%.  I believe these tests overstate the
uncertainty due to the simulation, because there are good reasons to
believe the parameters used in the prediction are more accurate than
the alternate parameters tested here.  All the variations mentioned in
this paragraph add in quadrature to 7.9\%.  Since this is probably an
overestimate, a systematic uncertainty of 6\% is assigned to the
simulation.

\subsection{The Simulation of Non-ideal Detector Performance}

The accuracy of the simulation of dead periods and inefficiency was
discussed at length in Section~\ref{sec:liveAdj}.  A conservative
uncertainty of 2\% is assessed for inaccuracies in this simulation.

\section{Estimates of Uncertainty in the Theoretical Prediction}
The systematic uncertainty in comparing observation to expectation is
dominated by theoretical uncertainty.  At present, there is no
unique and reliable estimate of the systematic uncertainty of the flux
and cross section models.  The MACRO collaboration has chosen
conservatively to assign large uncertainties in the energy range
relevant for this analysis.

\subsection{Neutrino Flux}

The calculations of neutrino flux are least accurate at low energies,
especially at 1~GeV and below, where solar modulation and geomagnetic
effects complicate the calculations, and the calculations cannot be
checked against high-statistics measurements of muons at ground
level.  After conferring with the authors of the Bartol flux, MACRO
has set a systematic uncertainty of 20\% due to possible inaccuracies
in the flux calculation at the energies relevant to this analysis.

\subsection{Neutrino Cross Section}

The experimental measurement of neutrino cross section for energies
near and below 1~GeV is imprecise (recall
Figure~\ref{fig:lipariData}).  Unless one has faith in a particular
model, the uncertainty cannot be less than 10-20\%.  As described in
Section~\ref{sec:nuSigma}, although the underlying weak interaction is
well-understood at the partonic level, complications due to nuclear
and QCD effects muddle the theoretical situation considerably.  While
the simulation used in this analysis is based on the Lipari cross
section model (Section~\ref{sec:lipari}), another dataset has been
generated using the NEUGEN model (Section \ref{sec:neugen}); 3.6\%
fewer events are detected by this analysis from the NEUGEN dataset.
MACRO has assigned a systematic uncertainty of 15\% due to possible
inaccuracies in the cross section model.
\subsection{The Solar Cycle}

As detailed in Section~\ref{sec:solcyc}, the Monte Carlo dataset on
which this analysis is based was generated using the flux at solar
minimum for 77\% of the simulated livetime and the flux at solar
midcycle for the rest.  It is not clear exactly what the authors of
the flux meant by ``solar maximum,'' ``solar minimum'' and ``solar
midcycle,'' nor how the amplitude of the solar modulation during the
data-taking compared with earlier cycles on which the flux calculation
was based.  Therefore, a different proportion of solar minimum,
midcycle and maximum events might have been a better model of the
actual conditions during data-taking.

The simulation finds fewer detected events under the solar midcycle
conditions than at solar minimum: 8\% $\pm$ 1.5\% fewer when
oscillations are not simulated, and 10\% $\pm$ 2\% fewer with
oscillations.  If the proportion of simulated livetime generated with
the solar midcycle flux were halved (to 12\%) or increased by 50\% (to
35\%), the prediction would have varied by about 1\%.  Therefore, a
systematic uncertainty of 1\% is applied due to possible inaccuracies
in the modeling of the solar cycle.

\subsection{Summary}

Table~\ref{tab:errors} summarizes the corrections and uncertainties,
and computes the total experimental and theoretical systematic
uncertainties by adding all terms in quadrature.

\begin{table}[htb]
  \leavevmode
  \begin{center}
    \begin{tabular}{|l|l|l|}
      \hline
        Source & Correction & Uncertainty \\
      \hline \hline
        External Upgoing Muons       & -1 event & $\pm 1\% $     \\
        Mistimed Background          &          & -1 ev = -1.3\% \\
        $\beta\,=\,-1$ Background    &          & -3 ev = -3.9\% \\
        Nominal Detector Simulation  &          & $\pm 6\%$      \\
        Simulation of Non-ideal Detector Performance &      & $\pm 2\%$      \\
      \hline
        Total Experimental Systematic& -1 event & $\poverm{+6.4\%}{-7.6\%}$ \\
      \hline
        Counting Statistics          &          & $\pm 11.5\%$   \\
      \hline
      \hline
        Flux                         &          & $\pm 20\%$     \\
        Cross Section                &          & $\pm 15\%$     \\
        Solar Cycle                  &          & $\pm 1\%$      \\
        Simulation Statistics        &          & $\pm 1.3\%$    \\
      \hline
        Total Theoretical Systematic &          & $\pm 25.1\%$   \\
      \hline
      \hline
    \end{tabular}
  \end{center}
  \caption{\protect\label{tab:errors}Summary of corrections and
        uncertainties.}
\end{table}

The final measurement is

\[ M = 76 \pm 11.5\%_{stat}\ {\poverm{+6.4\%}{-7.6\%}}_{syst} \]

The prediction without oscillations is

\[ P_{no} = 171 \pm 25.1\%_{theor} \]

\noindent
and with oscillations at the test point $\sstt = 1, \dmm =
3.2~\times~10^{-3}~\mathrm{eV}^2$, the prediction is

\[ P_o = 96 \pm 25.1\%_{theor} \]

\section{Oscillation Analysis}

\subsection{The Number of Events}

Figure~\ref{fig:pNorm} shows the measurement and the two predictions
with errors.  The statistical and systematic uncertainties of the
measurement have been added in quadrature, yielding
$\poverm{+13.2\%}{-13.8\%}$.

\picfig{fig:pNorm}{The measurement and two predictions, with errors
  shown.  The area under each curve is unity.}{\pawfig{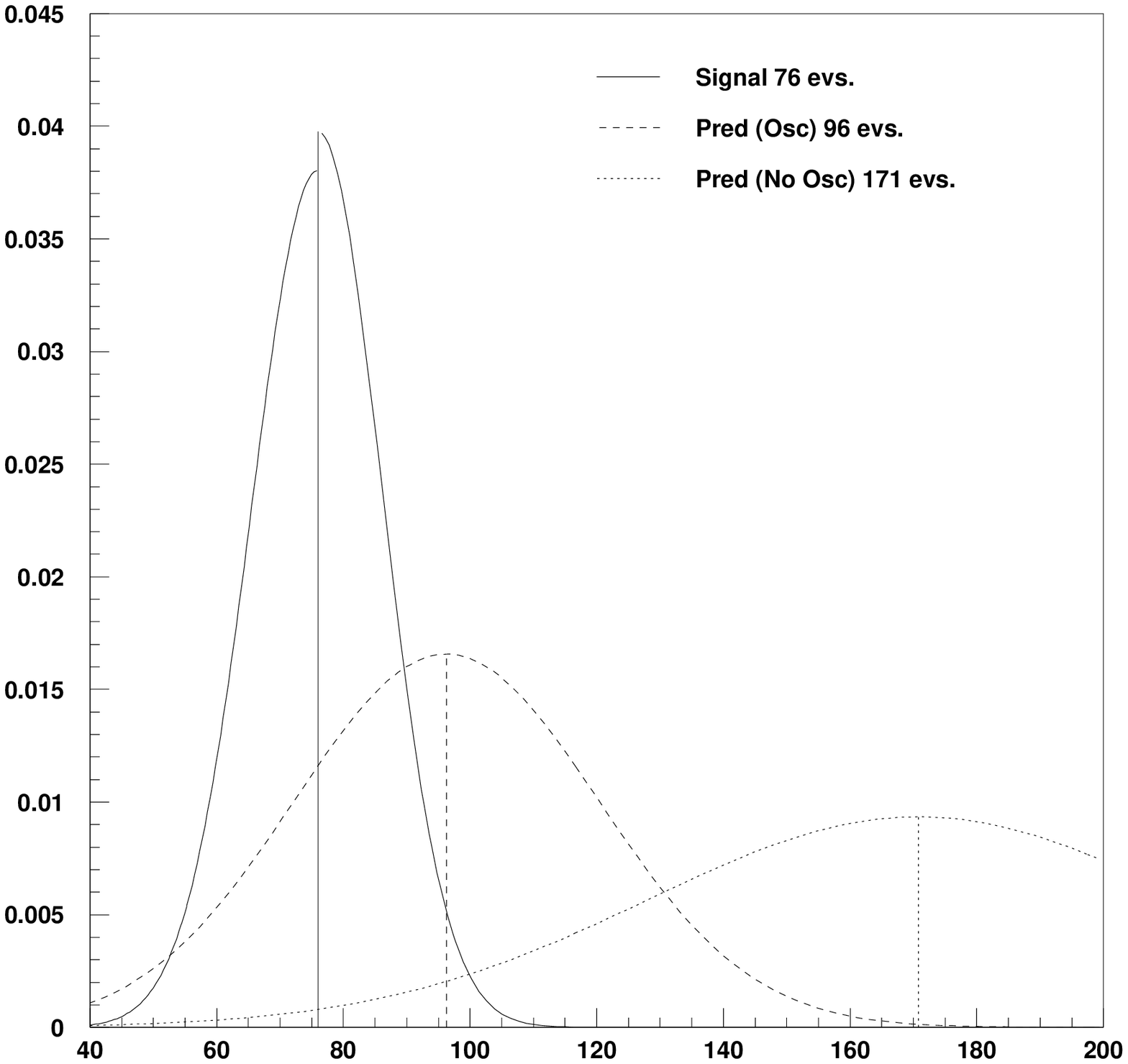}}

For an ensemble of experiments in which the true value being measured
$T$ is distributed according to the probability density function (pdf)
$P_T(T)$, and the measurement $M$ is distributed about the true value
with a pdf $P_M(M|T)$, the likelihood of a measurement M is

\[\mathcal{L} = \int dT P_T(T) P_M(M|T) \]

\noindent
which in our case is

\[ \mathcal{L} = \int dS_t P_G(S_t;1,\sigma_t)
                \int ds_m P_G^2(s_m;1,\sigma_m^+,\sigma_m^-)
                 P_P(N;S_t \times s_m \times N^0) \]

\noindent
where $S_t$ is a scaling variable representing the theoretical
uncertainty, $s_m$ is a scaling variable for the experimental
uncertainty, $P_G(x;\mu,\sigma)$ is the gaussian distribution,
$P_G^2(x;\mu,\sigma^+,\sigma^-)$ is the gaussian with different
variances in the positive and negative directions (like the solid line
in Figure~\ref{fig:pNorm} above), $P_P(N;N_0)$ is the Poisson
distribution, $N$ is the measured number of events and $N^0$ is the
prediction.  $\sigma_t$ and $\sigma_m$ are represented as percentages,
i.e.,\ $\sigma_t = 0.251$, $\sigma_m^+ = .064$ and $\sigma_m^- =
.076$.

Computing likelihood this way shows the likelihood of the real
measurement is 9.2 times greater for the oscillations prediction than
for the no-oscillations prediction.  The two-sided probability that
fluctuations would cause the measurement to be as far as it is or
farther from the oscillations prediction (including all errors) is
47\%; that is, almost half the time you did the experiment you would
get a worse agreement.  For the no-oscillations prediction, the
probability is 3.3\%.  Thus, this aspect of the experiment excludes
the no-oscillations theory at greater than 95\% confidence level.

\subsection{The Angular Distribution}

The above oscillations analysis merely counts events and ignores much
of what is known about them.  Neutrino oscillations depend on two
properties of the neutrino: its energy and the distance it travels
from the point of neutrino production to the detector.  The energy of
the muon is correlated to the energy of the parent neutrino, but MACRO
is not optimized to say anything about the muon energy.  On the other
hand, the direction the muon travels is correlated with the direction
the neutrino was traveling, and the neutrino direction determines
where on the surface of the earth it was created and thus how far it
has traveled.  Therefore, one may gain additional insight into
oscillations by considering the angular distribution of the detected
muons.

The muon does not perfectly follow the path of the neutrino,
especially at the relatively low energies of neutrinos in this
analysis.  The neutrino flux is peaked at the horizontal, but the
opening angle between the neutrino and muon directions
(Figure~\ref{fig:zenith}a) flattens the flux out
(Figure~\ref{fig:zenith}b).  Nonetheless, a correlation remains
between the measured muon direction and the length traveled by the
muon (Figure~\ref{fig:zenith}c) such that, for certain oscillation
parameters, more events disappear from vertical directions than from
horizontal (Figure~\ref{fig:zenith}d).

\picfig{fig:zenith}{Zenith angle effects.}
  {
    \makebox[6in]{
      \resizebox{3in}{!}{\includefigure{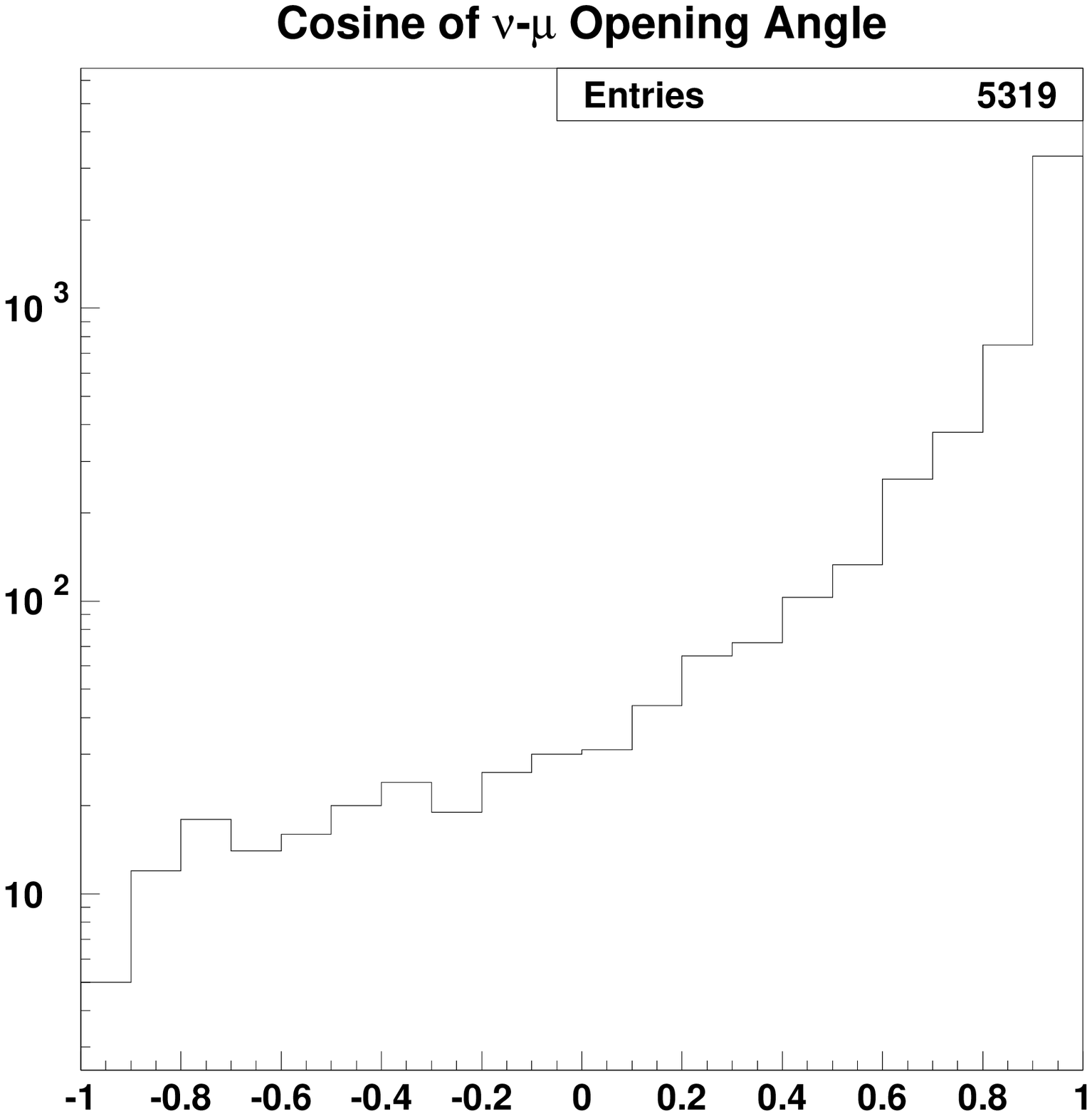}}
      \resizebox{3in}{!}{\includefigure{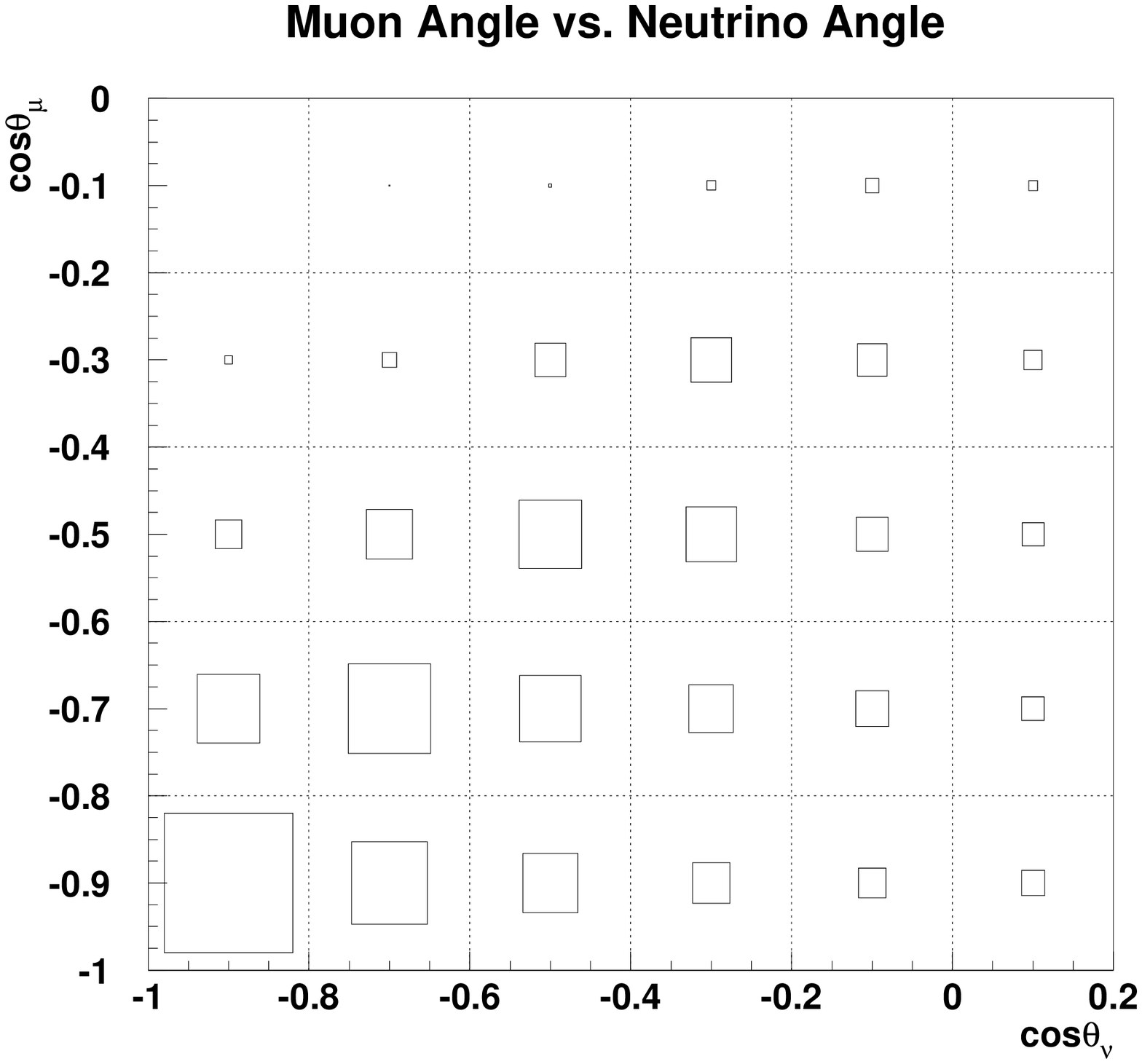}}}
    \makebox[6in]{\makebox[3in]{(a)} \makebox[3in]{(b)}}
    \makebox[6in]{\ } 
    \makebox[6in]{
        \resizebox{3in}{!}{\includefigure{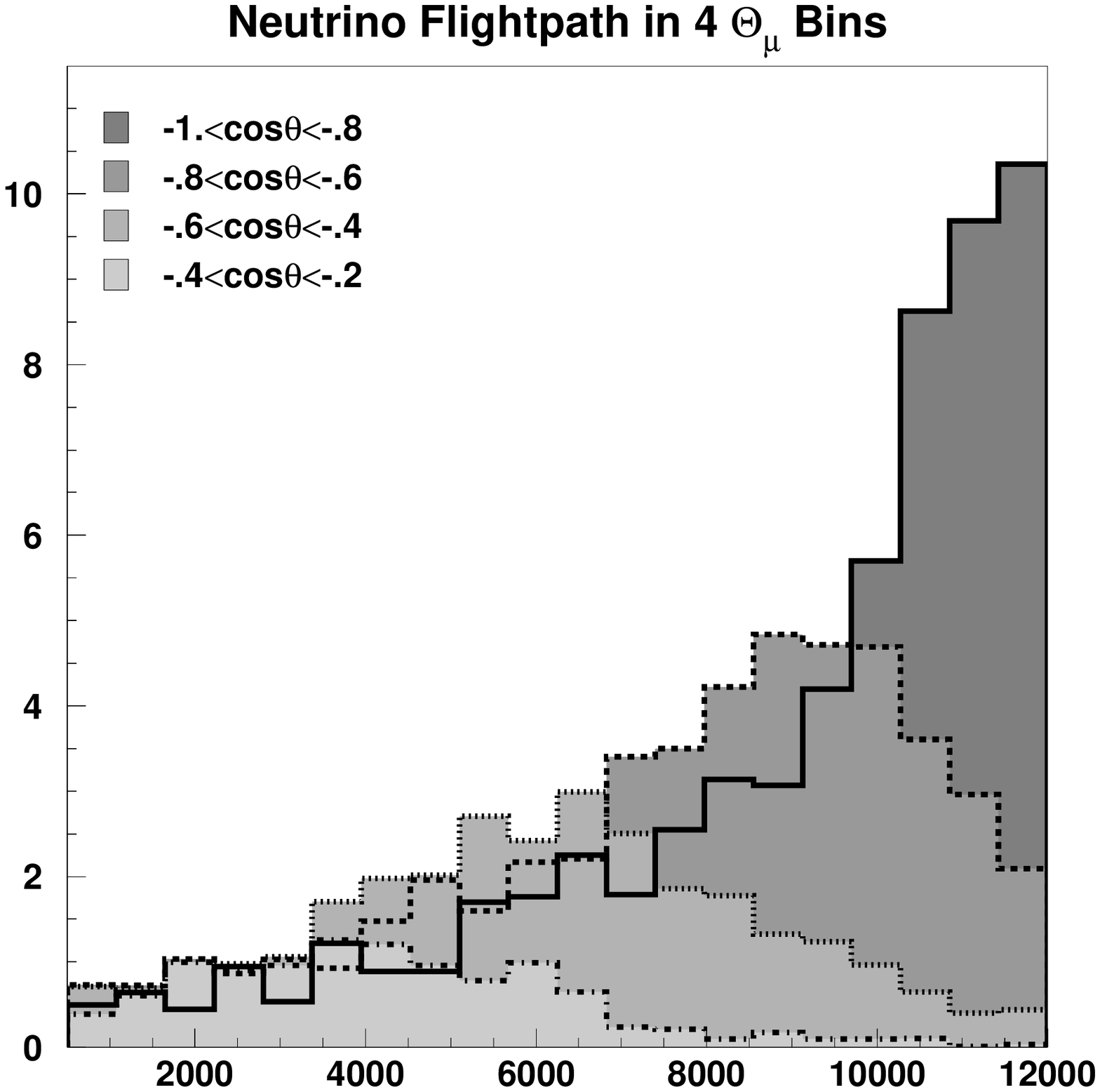}}
        \resizebox{3in}{!}{\includefigure{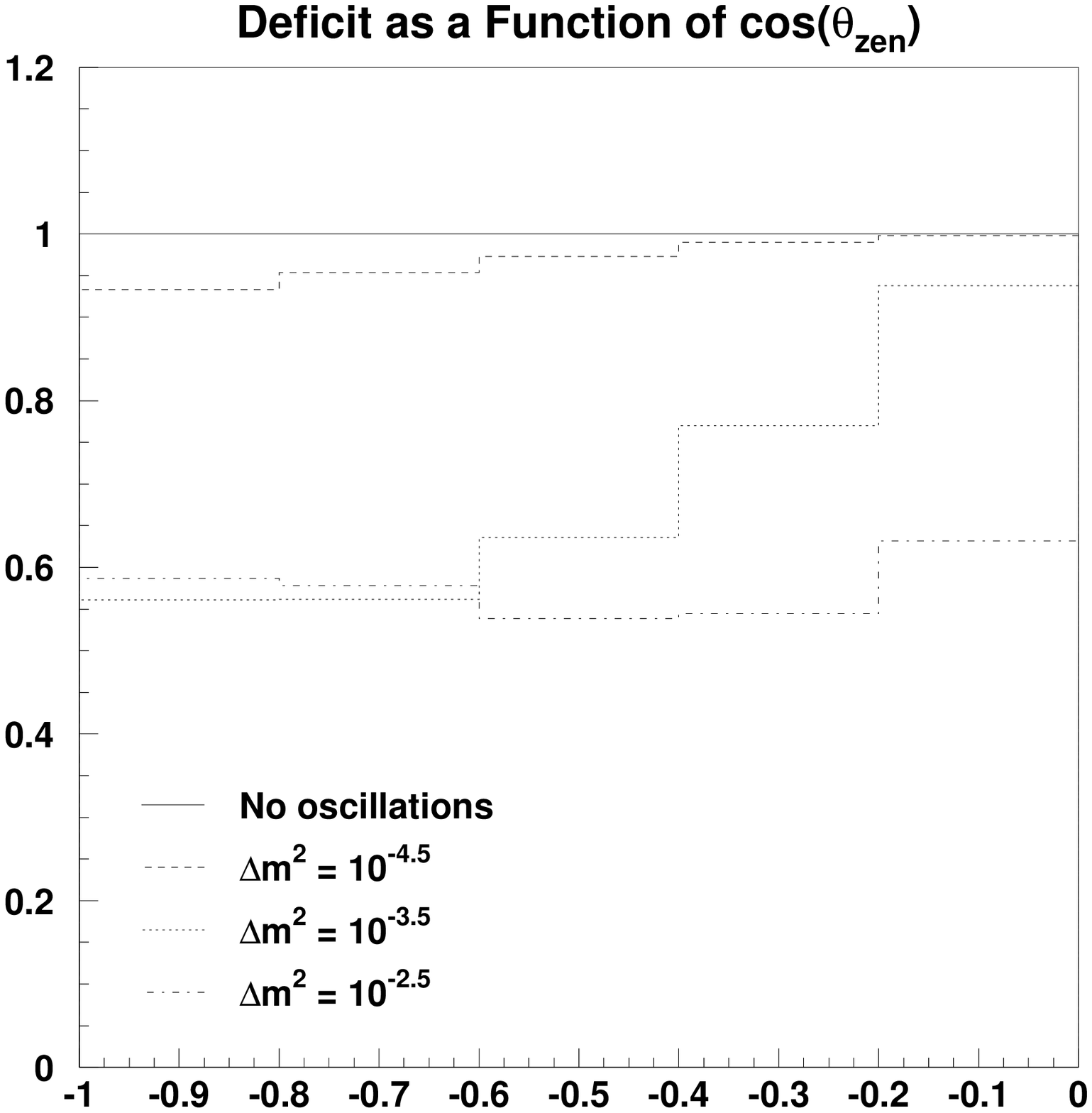}}}
    \makebox[6in]{\makebox[3in]{(c)} \makebox[3in]{(d)}}
  }

Figure~\ref{fig:mcDataZenErr} shows the measurement and two predictions for
the angular distribution of observed muons, with errors.

\picfig{fig:mcDataZenErr}{The angular distribution of observed muons.
The crosshairs give the data, with statistical and systematic errors
shown.  The darker band is the prediction for no oscillations,
including the 25.1\% theoretical error.  The lighter band is the
prediction for oscillations.}{\pawfig{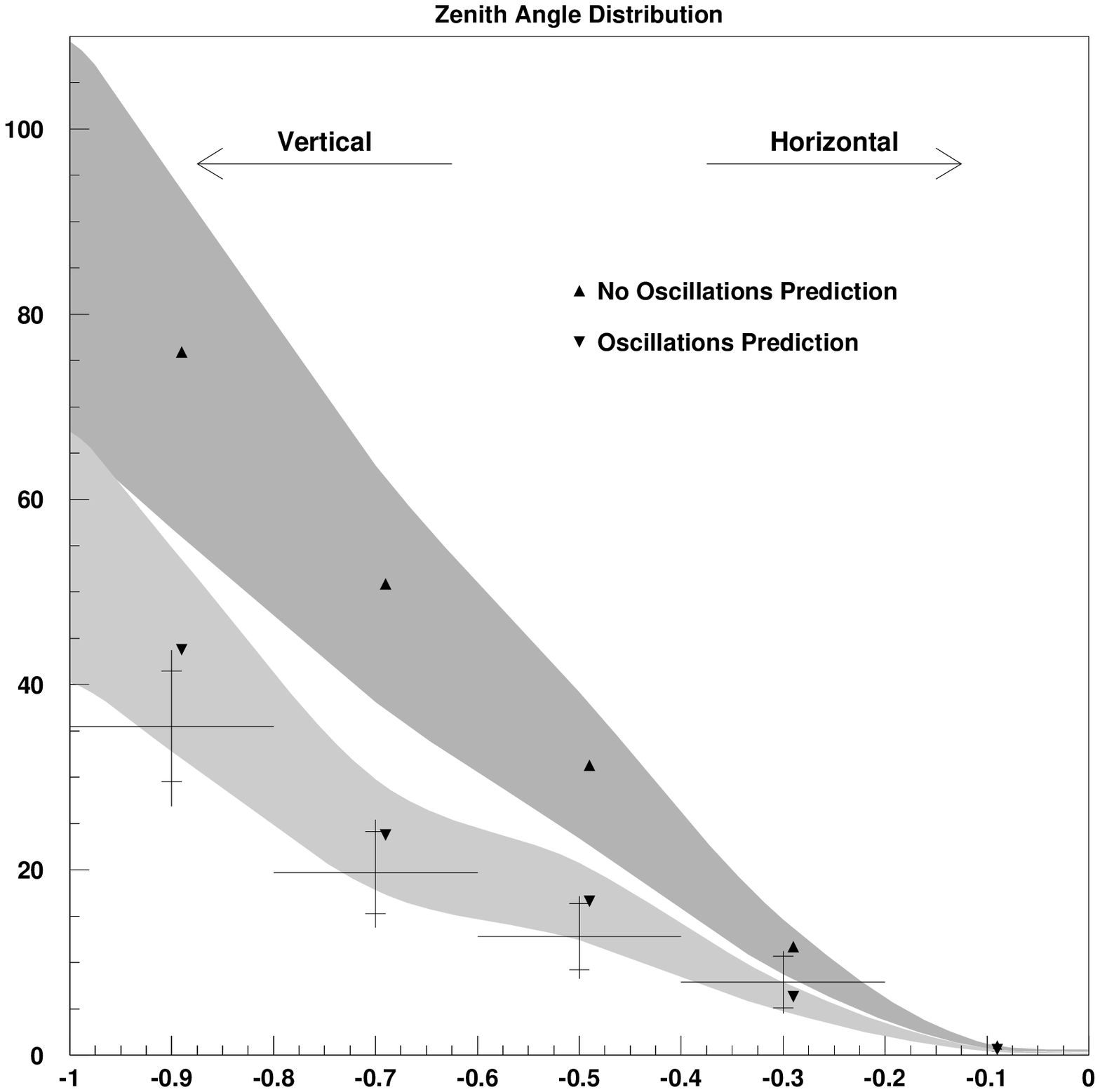}}

We may again compute the likelihood of the measurement assuming an
ensemble of experiments in which the true value is distributed about
the prediction according to the stated errors; however, we must take
care to apply certain errors in common across all bins.  For example,
if the model underestimates the neutrino cross section, that would
cause a decrease of the true value by the same factor in all bins.  We
may compute the likelihood of observing the values in the $Nbin$ bins
being $N_1,N_2,...N_{Nbin}$ as

\begin{eqnarray*}
 \mathcal{L} = &
  \int dS_G P_G(S_G;1,\sigma_t^G) \int ds_G P_G(s_G;1,\sigma_m^G)
              \times \hspace{2in} \  \\
               &
 {\displaystyle \prod_i^{Nbin}}   \int dS_i P_G(S_i;1,\sigma_t^b) \int ds_i P_G(s_i;1,\sigma_m^b)
 P_P(N_i;S_G s_G S_i s_i N_i^0)
\end{eqnarray*}

\noindent
where the theoretical uncertainty has been split into $\sigma_t^G$, a
global part that is in common across all bins, and $\sigma_t^b$ which
is an independent variation in each bin; similarly for the
experimental uncertainty $\sigma_m^G$ and $\sigma_m^b$.  I take the
uncertainties to be the same in all five bins.  The portions of the
uncertainties assigned to the global uncertainty are shown in
Table~\ref{tab:errGlobal}.

\begin{table}[htb]
  \leavevmode
  \begin{center}
    \begin{tabular}{|l|l|l|l|}
      \hline
        Source & Total Uncertainty & Global Uncertainty \\
      \hline
        External Upgoing Muons       & $\pm 1\% $  & $\pm 1\%$   \\
        Mistimed Background          & -1.3\%      & -1.3\%      \\
        $\beta\,=\,-1$ Background    & -3.9\%      & -3.9\%      \\
        Nominal Detector Simulation  & $\pm 6\%$   & $\pm 4\%$   \\
        Non-ideal Detector Performance & $\pm 2\%$ & $\pm 2\%$   \\
      \hline
        Total Experimental           &             & $\pm 5.5\%$ \\
      \hline \hline
        Flux                         & $\pm 20\%$  & $\pm 14\%$     \\
        Cross Section                & $\pm 15\%$  & $\pm 15\%$     \\
        Solar Cycle                  & $\pm 1\%$   & $\pm 1\%$      \\
        Simulation Statistics        & $\pm 1.3\%$ & $\pm 1.3\%$    \\
      \hline
        Total Theoretical            &              & $\pm 20.6\%$   \\
      \hline
    \end{tabular}
  \end{center}
  \caption{\protect\label{tab:errGlobal}Summary of corrections and
        uncertainties.}
\end{table}

From this point, we are dropping the distinction between positive and
negative experimental uncertainty.  The theoretical and experimental
bin-by-bin errors are chosen to make the variance of the total number
of events the same as in the unbinned analysis.

Following this procedure, we again find the data is more likely under
the oscillations hypothesis than under the no-oscillations hypothesis,
this time by a factor of 20.  If an ensemble of experiments were
performed as in the previous section, for the oscillations hypothesis
a result with lower likelihood than that of our real data would be
obtained 89\% of the time; for the no-oscillations hypothesis, the
number is 68\%.

The latter result may seem in contradiction to the very low
probability found in the previous section for measuring the total
number of events we measured if the no-oscillations hypothesis is
true.  However, this shows the importance of taking the angular
information into account.  There are many ways to fill the bins of the
angle histogram to add up to 76~events; and most of them are
exceptionally unlikely.  However, the way the bins are filled in our
real measurement is compatible with a suppression of the rate uniform
across all bins, which is somewhat accommodated by the theoretical and
experimental uncertainty.  If, to rank the unlikeliness of a given
measurement, instead of using the raw likelihood we use the unified
approach of Feldman and Cousins~\cite{FC}, the real measurement is
heavily penalized when considering the no-oscillations hypothesis
because there is another hypothesis that the data fits much better.
With this ranking, the probabilities of a measurement as unlikely as
our real measurement is 64\% for the oscillations hypothesis and 1.5\%
for no-oscillations.

How can the no-oscillations hypothesis be excluded at greater than
95\% confidence level by the Feldman-Cousins prescription, when the
raw likelihood of the real measured data for that hypothesis shows a
probability of greater than 50\%?  When using the raw likelihood,
which is the traditional method of evaluating a null hypothesis, the
decision whether to exclude is independent of any other hypotheses
that may be considered as alternatives.  On the other hand,
Feldman-Cousins is based on the relative probabilities of different
hypotheses.  If the no-oscillations hypothesis were true, there would
be many results (counts in the bins of the histogram that are
fluctuated from the true values) that were likely at the 68\% level;
however, the vast majority of these would not fit any other hypothesis
much better than they fit the no-oscillations hypothesis.  However,
the real measured data fits alternative hypotheses more than 20 times
better than it fits the no-oscillations hypothesis.  That is why the
Feldman-Cousins prescription rules out the no-oscillations hypothesis.

Figure~\ref{fig:FCsc} shows, for a grid of oscillation hypotheses, the
confidence level at which they are ruled out by this measurement,
using the Feldman-Cousins prescription.  The best fit of the grid
values considered is at maximal mixing and $\dmm =
3.2~\times~10^{-2}~\mathrm{eV^2}$.  Small values of $\dmm$ or of
$\sstt$ are excluded because they correspond to too small a magnitude
of $\numu$ disappearance.  Arbitrarily large values of $\dmm$ are
allowed because for hypotheses above a certain point, all upgoing
muons are fully oscillated and raising $\dmm$ does not change the
predicted number.

\picfig{fig:FCsc}{Considering the angular distribution of this
  measurement of semi-contained upgoing muons, for a grid of
  oscillation parameters, the confidence level at which those
  parameters are ruled out.  The vertical axis is $\log_{10}\dmm$
  while the horizontal axis is $\sstt$.  An approximate 90\%
  confidence level contour is shown.  Parameters below or left of the
  contour are ruled out.}{\pawfig{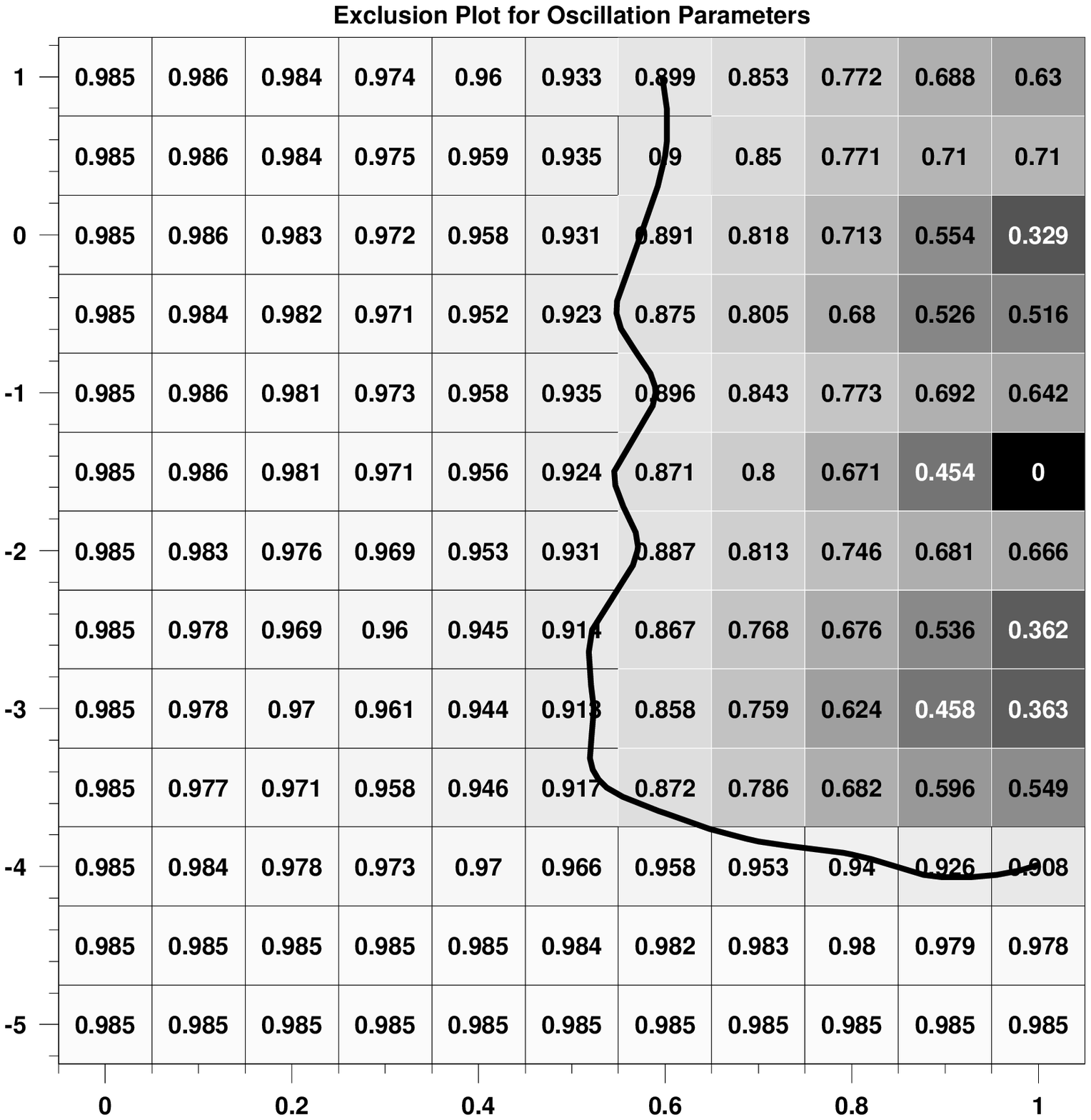}}

\section{This Analysis in the Context of Other MACRO Analyses}

This analysis is generally incompatible with the no-oscillations
hypothesis, and generally compatible with oscillations with maximal
mixing and $\dmm = 3.2~\times~10^{-3}~\mathrm{eV^2}$.  This choice of
parameters is not 
arbitrary; it is the value preferred by previously-published MACRO
neutrino analyses~\cite{upmu98,lowenu}, and is generally the same as
Super-Kamiokande's combined result~\cite{sk00}.  It is striking that a
single hypothesis is compatible with all MACRO analyses, which are
based on different events and different analysis procedures, and span
two orders of magnitude in energy distribution.  The energy
distribution is important because at the preferred oscillation
parameters, neutrinos below a few GeV have reached their maximum
deficit due to oscillations within a few tens of kilometers.  Thus, in
this analysis, at the preferred parameters we expect all neutrinos
coming from below to be fully oscillated.  However, at 100~GeV and
above, only neutrinos from farthest away, from the most vertical
downward direction, are fully oscillated.  Thus, the angular signature
of oscillations is different among the different analyses.

All oscillation analyses based on absolute measurements of atmospheric
neutrino flux are hampered by the large theoretical uncertainties on
the flux and on the cross section, a situation that is worst for
lower-energy neutrinos (i.e.,\ those considered in this analysis).
The most precise measurements of oscillation effects have always been
based on the comparison of two quantities (for example, muon neutrinos
and electron neutrinos, or upward and downward neutrinos), in which
much of the theoretical error cancels.  It is difficult to draw
conclusions comparing our measurements of lower-energy and of
higher-energy neutrinos because the uncertainties on the flux and the
cross section do not necessarily cancel when one is considering
different energy regions.

However, a very fruitful combined analysis can be made with one MACRO
analysis: the sample of both up- and down-going semi-contained
neutrinos passing the Bottom face (topologies (e) and (f) in
Figure~\ref{fig:sctop} in the previous chapter).  While MACRO cannot
distinguish the direction of the detected muon, it is known that the
sample would contain nearly equal numbers of upward and downward
particles in the absence of oscillations (recall from
Section~\ref{sec:flux} that the flux is nearly up-down symmetric in
the absence of oscillations), and an energy distribution almost
identical to the present analysis (see Figure~\ref{fig:entop}).
However, for the preferred oscillation parameters, it is expected that
most neutrinos from below (which have traveled hundreds or thousands
of kilometers) would be fully oscillated so the flux would be reduced
to about half of its nominal value, while neutrinos from above (which
have traveled only tens of kilometers) would be hardly attenuated at
all by oscillations.  Thus, the measurement reported in this work
(which I will refer to as Analysis~$A$), consisting only of
lower-energy neutrinos from below, would see about half the nominal
flux, while the other analysis (Analysis~$B$) would see about three
quarters of the nominal flux (all of the downward neutrinos and half
of the upward).  Even if there were large errors in the absolute
magnitude of the predicted interaction rate, the two results would
differ from the oscillations prediction by the same fraction.  Thus, a
combined analysis accords us some immunity from the large theoretical
uncertainties.  This general picture is borne out in Analysis~$A$
(Figure~\ref{fig:mcDataZenErr} above), which sees 76~events with 96
predicted at the oscillations test point, and Analysis~$B$
(Figure~\ref{fig:mcDataZenBO}), which sees 262~events with 288
predicted.

\picfig{fig:mcDataZenBO}{The angular distribution of events from
  Analysis~$B$ and two predictions, before the assignment of any errors
  (other than counting statistics).}{\pawfig{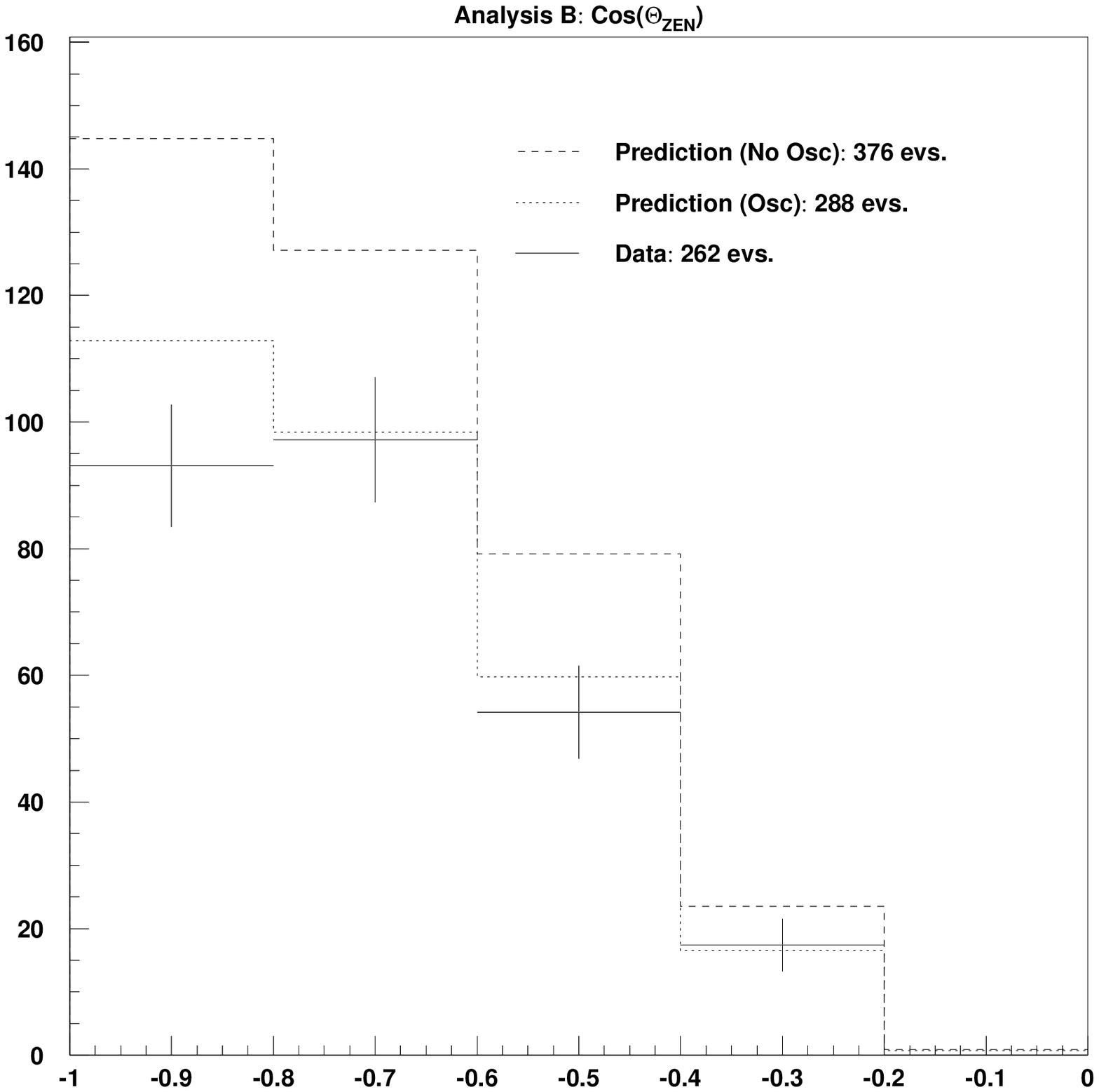}}

In the combined analysis, the various uncertainties are accounted
for by a number of scale factors, as named in Table~\ref{tab:SFnames}.
The values used for the scale factors are derived in
Table~\ref{tab:SFvalues}.  Analysis~$B$ has been published
elsewhere~\cite{lowenu}, although the measurement used here has been
updated since the publication of~\cite{lowenu}.  Systematic
uncertainties are taken from that publication, and subjective judgment
is used to partition the errors among the various scale factors.
Three of the scale factors ($f_i$, $s^a_i$ and $s^b_i$) have nothing
assigned to them, and they are discarded.

\begin{table}[htb]
  \leavevmode
  \begin{center}
    \begin{tabular}{|c|l|}
      \hline
        $S$     & Theoretical error, common to both analyses and all
                  angle bins \\
        $F$     & Experimental error, common to both analyses and all
                  angle bins \\
        $S^A$   & Theoretical error, Analysis~$A$, common to all bins \\
        $F^A$   & Experimental error, Analysis~$A$, common to all bins \\
        $S^B$   & Theoretical error, Analysis~$B$, common to all bins \\
        $F^B$   & Experimental error, Analysis~$B$, common to all bins \\
        $s_i$   & Theoretical error, common to both analyses, for
                  angle bin $i$ \\
        $f_i$   & Experimental error, common to both analyses, for
                  angle bin $i$.  NOT USED\@. \\
        $s^a_i$ & Theoretical error, Analysis~$A$, angle bin $i$.  NOT USED\@. \\
        $f^a_i$ & Experimental error, Analysis~$A$, angle bin $i$. \\
        $s^b_i$ & Theoretical error, Analysis~$B$, angle bin $i$.  NOT USED\@. \\
        $f^b_i$ & Experimental error, Analysis~$B$, angle bin $i$. \\
      \hline
    \end{tabular}
  \end{center}
  \caption{\protect\label{tab:SFnames}Scale factors in the combined
        analysis to account for systematic uncertainties.}
\end{table}

\begin{table}[htb]
  \leavevmode
  \begin{center}
    \begin{tabular}{|r|c|c||c|c|c|c|c|}
      \hline  
        EXPERIMENTAL        & Tot A & Tot B &  $F$  & $F^A$ & $F^B$ & $f^a_i$ & $f^b_i$ \\
      \hline  
        $\mu$-induced BG    & -4\%  &  4\%  &       & -4\%  &  4\%  &         &       \\
        External $\mu$ BG   &  1\%  &  0    &       &  1\%  &       &         &       \\
        Mistimed BG         & -1\%  &  0    &       & -1\%  &       &         &       \\
        Nominal Detector MC &  6\%  &  6\%  & 4.5\% & 3.8\% & 3.8\% &  2.2\%  & 2.2\% \\
        Detector Problems   &  2\%  &  3\%  &       &  2\%  &  3\%  &         &       \\
      \hline  
        Total               &       &       & 4.5\% & 5.2\% & 6.3\% &  2.2\%  & 2.2\% \\
      \hline  
      \hline  
        THEORETICAL         & Tot A & Tot B &  $S$  & $S^A$ & $S^B$ & $s_i$   &     \\
      \hline  
        Flux                & 20\%  & 20\%  & 18.3\%&       &       &  14\%   &     \\
        Cross Section       & 15\%  & 15\%  & 14\%  & 5.5\% &       &         &     \\
        Solar Cycle         &  1\%  &  1\%  &       &  1\%  &  1\%  &         &     \\
        MC Stats            & 1.3\% & 2.2\% &       & 1.3\% & 2.2\% &         &     \\
      \hline  
        Total               &       &       & 23.2\%& 5.7\% & 2.4\% &  14\%   &     \\
      \hline  
    \end{tabular}
  \end{center}
  \caption{\protect\label{tab:SFvalues}Scale factors in the combined
        analysis to account for systematic uncertainties.}
\end{table}

The no-oscillations hypothesis cannot find a decent likelihood for
both the Analysis~$A$ bins and the Analysis~$B$ bins simultaneously by
assuming a single error in the absolute normalization.  The real
measured data is 78 times more likely under the oscillations
hypothesis than under the no-oscillations hypothesis.  For an ensemble
of experiments in which the true value being measured in each bin
varies according to the errors tabulated in Table~\ref{tab:SFvalues},
the 10 bin values (five from each analysis) would be as unlikely as
the 10 actual measured bin values 83\% of the time under the
oscillations hypothesis, and less than 0.1\% under the no-oscillations
hypothesis (using the Feldman-Cousins prescription).

Figure~\ref{fig:FCmeBO} shows the exclusion plot for the combined
analysis.  The allowed region has been greatly reduced compared to
that from Analysis~$A$ alone (or from Analysis~$B$ alone).  In
particular, the Analysis~$A$ data, considered alone, are compatible
with arbitrarily large values of $\dmm$.  That is because the upgoing
neutrinos in all of the bins with significant statistics are
``fully-oscillated'' for any value of $\dmm$ above
$10^{-3}~\mathrm{eV^2}$ or so, and increasing $\dmm$ would not change
the prediction for Analysis~$A$ at all.  However, as $\dmm$ increases
above $3.2~\times~10^{-1}~\mathrm{eV^2}$, more and more of the
downgoing neutrinos in Analysis~$B$ begin to disappear.  Although
Analysis~$B$ alone could accommodate the reduction by an error in the
overall normalization, considering both analyses together shows that
upgoing neutrinos are disappearing and downgoing are not.  That
restricts $\dmm$ to a relatively small region.

\picfig{fig:FCmeBO}{The exclusion plot, analogous to
  Figure~\ref{fig:FCsc}, for the combined analysis.  Blank squares
  were below the detection limit of the software (C.L. $>$ 0.997).  The
  small region within the contour is allowed, and the large region
  outside is excluded at 90\% confidence level.}{\pawfig{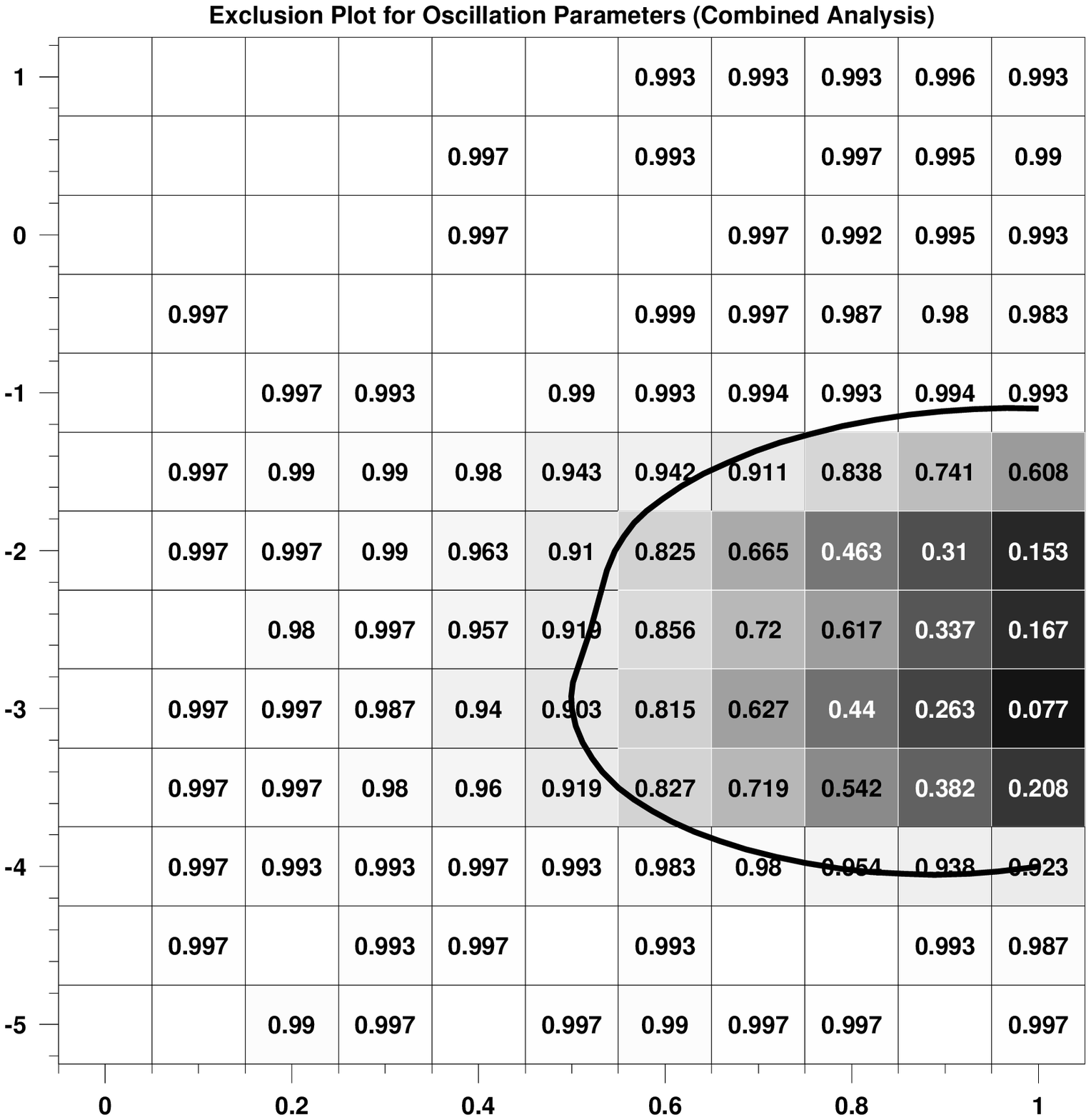}}

\chapter{Conclusion}

This thesis reports a study of atmospheric neutrinos interacting in
the MACRO detector.  A Monte Carlo calculation, based on the Bartol
flux calculation and the Lipari cross section model, predicts on the
order of 100~events could be observed in the livetime of this analysis
(4.5~years).  An analysis has been developed, taking extra care to
eliminate backgrounds due to the copious downgoing atmospheric muons.
77~events are found, a result generally consistent with neutrino
oscillations with the parameters suggested by higher-energy MACRO
analyses as well as other experiments, and inconsistent with no
oscillations.  However, due mostly to large theoretical uncertainties,
the allowed parameter space from this analysis alone is not very
precise.  Combining this analysis with another MACRO analysis, of
neutrino interactions with similar energy but different topology,
causes a partial cancellation of much of the uncertainty and allows a
more precise determination of oscillation parameters.


\bibliography{references}
\bibliographystyle{utphys_nolty}

\end{document}